# Composite Quasi-Likelihood Estimation of Dynamic Panels with Group-Specific Heterogeneity and Spatially Dependent Errors*

Ba Chu†
Carleton University

September 8, 2017

**Abstract**

This paper proposes a novel method to estimate large panel data error-correction models with stationary/non-stationary covariates and spatially dependent errors, which allows for known/unknown group-specific patterns of slope heterogeneity. Analysis is based on composite quasi-likelihood (CQL) maximization which performs estimation and classification simultaneously. The proposed CQL estimator remains unbiased in the presence of misspecification of the unobserved individual/group-specific fixed effects; therefore, neither instrumental variables nor bias corrections/reductions are required. This estimator also achieves the 'oracle' property as the estimation errors of group memberships have no effect on the asymptotic distributions of the group-specific slope parameters estimates. Classification and estimation involve a large-scale non-convex mixed-integer programming problem, which can then be solved via a new algorithm based on DC (Difference-of-Convex functions) programming - the DCA (DC Algorithm). Simulations confirm good finite-sample properties of the proposed estimator. An empirical application and a software package to implement this method are also provided.

*I gratefully acknowledge funding from the Social Science and Humanities Research Council of Canada (MBF Grant 430-2016-00682).

†*Correspondence address:* Ba Chu, Department of Economics, Carleton University, B-857 Loeb Building, 1125 Colonel By Drive, Ottawa, ON K1S 5B6, Canada. Phone: +1 (613) 520 2600 (ext. 1546). Fax: +1 (613) 520 3906. E-mail: ba.chu@carleton.ca.

# 1 Introduction

This paper proposes a novel method for estimation and inference of dynamic linear panel data models with unobserved group-specific patterns of slope heterogeneity and spatially dependent errors. Unobserved heterogeneity and spatial dependence across individuals/units have been the main focus of many econometric papers in panel data, and been well motivated from empirical economic problems, for example, in recent studies of empirical growth [see, e.g., Durlauf, Johnson, and Temple (2005), Corrado, Martin, and Weeks (2005), Meliciani and Peracchi (2006), Alexiadis (2013), Durlauf and Quah (1999), Phillips and Sul (2007, 2009)]

A panel model with grouped heterogeneity in the slopes represents a viable approach to summarize grouped data as it is a compromise between a parsimonious model and another one with too many parameters. With data clustered in units, one can estimate three different models. In the first model, one can ignore the grouped structure in the units and estimate a regression with the data pooled. The estimates from this 'pooled' model will be biased if the units differ much, but with the pooled data, the model will become the most parsimonious in terms of the number of parameters estimated. At the other extreme, one could estimate one regression model for each unit, then take the average of all the estimated slope parameters if these parameters vary randomly around a constant - this approach is called the mean-group estimator [see Pesaran and Smith (1995); Pesaran, Smith, and Im (1996) and Fotheringham, Charlton, and Brunsdon. (1997)]. Pesaran, Shin, and Smith (1999) (**PSS**, henceforth) also propose the pooled mean group estimator for autoregressive distributed lag models (ARDL) that allow for both common parameters and heterogeneous parameters. However, this option produces way more parameters, and the estimates of the slope parameters will be highly variable if there are not many observations for each unit. The grouped slope heterogeneity approach represents a middle ground between these two extremes, thus it can be viewed as a compromise between completely ignoring the structure of the data and fully taking this structure into account by estimating many different models.

To be specific, a simple linear spatial-error specification with dynamic grouped patterns of heterogeneity takes the following form as a special case of a general ARDL model defined by (3.2) in Section 3:

$$\Delta y_{i,t} = \mu_i + \phi_{g_i}\left(y_{i,t-1} - \boldsymbol{\theta}_{g_i}^{\top}\boldsymbol{x}_{i,t}\right) + \lambda_{g_i}^{*}\Delta y_{i,t-1} + \boldsymbol{\delta}_{g_i}^{*}{}^{\top}\Delta\boldsymbol{x}_{i,t-1} + \epsilon_{i,t},\ i = 1,\dots,N,\ t = 1,\dots,T, \quad (1.1)$$

where $g_i$ represents a group assignment that assigns each individual, $i$, to some specific group, say $g_i \in \{1,\dots,G\}$; here $G$ is the number of groups to be specified a priori; $\mu_i$, $i = 1,\dots,N$, are individual-specific fixed effects; $\phi_{g_i}$, $\boldsymbol{\theta}_{g_i}$, $\lambda_{g_i}^{*}$, and $\boldsymbol{\delta}_{g_i}^{*}$, $i = 1,\dots,N$, are common-group slope parameters; the explanatory variables $\boldsymbol{x}_{i,t}$ are independent of a centered random innovation, $\epsilon_{i,t}$, for each individual, $i$; moreover, $\epsilon_{i,t}$, $i \in \{1,\dots,N\}$ and $t \in \{1,\dots,T\}$, are identically distributed over space and time; for every given $i \in \{1, \ldots, N\}$, $\epsilon_{i,t}$ are serially independent, and for every given $t \in \{1, \ldots, T\}$, $\epsilon_{i,t}$ are spatially dependent [across locations], which effectively implies that $\epsilon_{i,t}$ and $\epsilon_{j,s}$, $t \neq s$, are independent if $i$ and $j$ are associated with different locations - because if $\epsilon_{i,t}$ and $\epsilon_{j,s}$ are dependent, then $\epsilon_{j,s}$ and $\epsilon_{j,t}$ are also dependent as $\epsilon_{i,t}$ are spatially dependent, this indeed leads to a contradiction.

As shown in Section 4 the proposed estimation procedure does not require any particular pattern of spatial dependence to be specified for the error terms; it merely assumes that the innovation terms $\epsilon_{i,t}$, $i = 1, \ldots, N$ and $t = 1, \ldots, T$, are realizations of a centered, stationary mixing random field so that $\sqrt{N}\epsilon_{*,t} := \sqrt{N}\frac{1}{N}\sum_{i=1}^{N}\epsilon_{i,t} = \frac{1}{\sqrt{N}}\sum_{i=1}^{N}\epsilon_{i,t}$ for $t = 1, \ldots, T$ can be approximated by serially independent Gaussian random variables under some mixing assumption on the random field $\epsilon_{i,t}$ as $N$ becomes sufficiently large, whence the composite quasi-likelihood (CQL) function can then be constructed. It is worth noting at this point that, in this estimation procedure, the normalized variance $\sigma^2_{\epsilon,N} := \frac{1}{N}Var\left(\sum_{i=1}^{N}\epsilon_{i,t}\right)$ of the spatial summation of errors and the cross-sectionally average fixed effect $\mu_* := \frac{1}{N}\sum_{i=1}^{N}\mu_i$ are treated as nuisance parameters. While $\mu_*$ is estimated by the maximum composite likelihood, $\sigma^2_{\epsilon,N}$ can also be estimated directly by using robust ('clustered') standard errors formulas (see, e.g., Arellano (1987); Conley (1999); Driscoll and Kraay (1998); Kelejian and Prucha (2007)).

**Heuristics**

Since the parameters are common within each group, say $g$, the $y_{i,t}$'s and $\boldsymbol{x}_{i,t}$'s of units within each group all have a common regression relationship, implying their common-group stochastic trends (namely time-varying cross-sectional means) also do. Suppose that these time-varying cross-sectional means look quite distinct across groups (but similar within each group). To estimate the common slope parameters within each group, $g$, one could just regress the common-group stochastic trend of all $\Delta y_{i,t}$'s in this group on its lags and the common-group stochastic trends of all the $\boldsymbol{x}_{i,t}$'s in the same group and their lags. Since these common-group stochastic trends are unobservable, they can thus be proxied by common-group cross-sectional averages - this idea is inspired from the work by Pesaran (2006) which employs cross-sectional averages to provide valid inference in stationary panel regressions with a multifactor error structure. Importantly the regressions involving common-group cross-sectional averages do not induce an endogeneity problem which is often the consequence of doing the within-group or time-differencing transformations in dynamic panel data models. Thus the estimates will still be asymptotically unbiased even for $T$ less than $N$. This intuition will be elucidated in Section 4, and formalized in Section 5.

For underlying latent group structures, estimates of the group memberships and the associated common-group slope parameters can be obtained in principle by running many regressions involving common-group cross-sectional averages for each partitioning of the set $[1, \ldots, N]$ into $G$ groups, then choose the parameter values associated with the regression that achieves the minimum sum of squared residuals amongst all the partitionings. However the number of regressions to run will be very large if $N$ is large (in fact, it is equal to a Sterling number of the second kind); this renders the so-called 'many-regressions' method infeasible. However, this method of running many regressions can be casted into a non-convex mixed integer programming problem as described in Section C.1.

**Relation to the Existing Works**

Hahn and Moon (2010) and Bester and Hansen (2016) show that the bias of grouped fixed effects (GFE) estimators asymptotically vanishes in nonlinear panel data models with finitely supported fixed effects (i.e., individual-specific fixed effects are common with each group, and differ across groups). The GFEs can be severely biased when individual-specific heterogeneity is incorrectly assumed to be constant within each group. Bonhomme, Lamadon, and Manresa's (2016) method to discretize unobserved fixed effects can only reduce the bias of GFE estimators when the number of groups is allowed to grow with the number of individuals. Therefore, for the GFEs to be asymptotically unbiased and normal when the number of groups is fixed, Bester and Hansen (2016) rely on the assumption that the maximum discrepancy between two individuals within groups goes to zero as the number of cross-sections becomes large.

In a typical dynamic linear panel, our proposed CQL estimator does not suffer from this type of bias arising due to misspecification of individual-specific heterogeneity because - unlike GFE estimators which require the individual/group-specific fixed effects to be concentrated out prior to estimation of the parameters of interest - the current estimation paradigm involves only the average fixed effect $\mu_*$ instead. Moreover, it is worth noting at this point that, since the within-group transformation in a linear dynamic panel gives rise to endogeneity, thus a least-squares estimator can be severely biased for small $T$, the proposed estimator does not rely on within-group transformation, thus it also does not suffer from an endogeneity bias. Therefore, instrumental variables (IV's) or bias correction are not required to implement our method. In a dynamic panel with long time horizon the IV estimation strategy may not be feasible as the number of lagged variables that can be used as IV's is large, thus another issue related to choice of optimal IV's needs to be dealt with. Bias correction/reduction methods (see, e.g., Hahn and Kuersteiner (2002) and Dhaene and Jochmans (2015)) require preliminary estimators of the fixed effects to estimate the bias, thus a misspecification in the fixed effects can deteriorate the quality of bias estimates.

Works on panel data models with unknown patterns of group heterogeneity are pretty recent.[1] Lin and Ng (2012) propose a conditional K-means procedure, which extends Forgy's (1965) K-means algorithm, to estimate linear panel data models, but asymptotic theory is not derived. Bonhomme and Manresa (2015) [**BM**, henceforth] propose two minimum sum-of-squares clustering

[1]In a related thematic approach, finite mixture models can be employed to model the probability that an individual belongs to a group. Thus, estimation of and inference on these membership probabilities can be performed via the mixture parameters (see, e.g., Kalai, Moitra, and Valiant (2010); Kasahara and Shimotsu (2009); Sun (2005))

algorithms based on the K-means algorithm and Hansen and Mladenović's (1997) Variable Neighborhood Search (VNS) algorithm to perform group classification and estimation in panels with time-variant grouped patterns of heterogeneity. In their asymptotic theory the GFE estimators are not influenced by the effect of group membership estimation because the probability of misclassifying at least one individual unit decays very fast as long as both $N$ and $T$ go to infinity such that $N/T^{\delta} \downarrow 0$ for some $\delta > 0$. When a lagged dependent variable is included as a covariate in a model with additive time-invariant individual fixed-effects in addition to the time-varying grouped effects the infeasible fixed-effects estimator suffers from the incidental parameters problem (Nickell (1981)); IVs are then needed to produce consistent estimates for the parameters of interest. They also demonstrate that their proposed algorithms can achieve approximately the same optimal solutions to the least-squares clustering problem as other global optimizing algorithms (for example, the *branch and bound* algorithm) in panel datasets with a small number of groups.

Su, Shi, and Phillips (2016) [**SSP**, henceforth] propose a new variant of Tibshirani's (1996) LASSO, namely classifier-LASSO, to perform group classification and estimation of regression slope coefficients simultaneously in a single step. However, this estimator often induces non-negligible asymptotic bias when it is applied to dynamic panels or panel regressions where some regressors are endogenous and the time horizon $T$ is smaller, thus bias corrections of Hahn and Kuersteiner's (2002) type are needed. Wang, Phillips, and Su (2016) propose a penalized least-squares criterion function using a new hybrid Panel-CARDS penalty function for simultaneous classification and estimation, effectively extending Ke, Fan, and Wu's (2015) CARDS procedure for cross-sectional data to panel data.

Confidence bands for group memberships are derived in Dzemski and Okui (2017). Okui and Wang (2017) employ **BM**'s clustering algorithm and group LASSO [see, e.g., Chan, Yau, and Zhang (2014); Yuan and Lin (2006)] to estimate breaks and latent group structures in panels. Ando and Bai (2016) deal with linear panel data models with grouped factor structure and a large number of explanatory variables. The group membership of each individual can be estimated along with other parameters of the model. A LASSO approach is applied to select significant explanatory variables, and optimal group memberships can be found by using the K-means algorithm.

Nonlinear panel data models with discretized fixed effects are considered in Bonhomme, Lamadon, and Manresa (2016). Druedahl, Jørgensen, and Kristensen (2016) consider a nonparametric GFE estimator for nonlinear panel data models with finitely supported fixed effects. Vogt and Linton (2017) develop methods to classify nonparametric regression functions into clusters based on the premise that there are groups of individuals who share the same regression function.

It is worth mentioning at this point that, in most of earlier works on this topic, units are often cross-sectionally independent - this is somewhat a strong assumption. Therefore, group classification is done in a purely data-driven manner by minimizing some unpenalized/penalized sum-of-squared-errors objective function. The proposed procedure is based on the premise that the response variables and covariates for all units have non-zero common-group stochastic trends, thus it is natural to let the innovation terms of dynamic panel data models have some weak cross-sectional dependence that is summarized through a cardinality-based mixing coefficient [for mixing random fields] with polynomial decay rate. Due to the presence of these cross-sectionally dependent errors the growth rate of $N$ relative to $T$ that is required for the asymptotic normality and the 'oracle' property of the slope coefficient estimates will also depend on the degree of weak cross-sectional dependence.

**Computational Consideration**

In the above-mentioned papers, various clustering techniques (notably, VNS with K-means in **BM** or classifier-LASSO in **SSP**) have been employed to partition panel data with latent group patterns while optimizing the associated objective function for estimates of group-specific coefficients. A common feature of these methods is that the problem is nonconvex and often nonsmooth (such as the K-means), thus falling into one of the most difficult areas of the optimization field. The proposed criterion function is also globally non-convex, and minimization of non-convex criterion functions of this type is a NP-hard (Non-deterministic and Polynomial-time hard) problem with possibly many local minima (Garey and Johnson, 1979). Existing methods including the VNS and the K-means algorithm can feasibly search for 'good' local solution while exact solutions are often not known for large datasets with many individuals clustered into many groups; and the K-means often performs poorly when there are outliers in the data. So far the VNS using the K-means as a local search engine [as employed in **BM**] is still a common and effective strategy for data clustering especially when the number of groups is small.

The proposed method is novel in the sense that the clustering and estimation procedure based on the CQL function can be formulated as a non-convex mixed-integer programming problem, which can then be efficiently solved by the DC (Difference-of-Convex functions) programming and the DCA (DC Algorithms) as described in Section C in the Supplemental Material. The DC programming and DCA were developed by Le Thi Hoai An and Pham Dinh Tao (2003); Pham Dinh and Souad (1988); Pham Dinh Tao and Le Thi Hoai An (1998). The DCA is one of a few algorithms to solve large non-convex optimization problems; it has a proper theoretical justification, and has been implemented to successfully solve many large-scale (smooth or non-smooth) non-convex programming problems in various fields, especially in Machine Learning where they often provide global optima and are demonstrated to be more robust and efficient than the standard methods including the K-means algorithm (see, e.g., Le Thi Hoai An, Belghiti, and Pham Dinh Tao (2007); Le Thi Hoai An, Le Hoai Minh, and Pham Dinh Tao (2014); Liu and Shen (2006) and references therein). Interested readers are referred to Le Thi Hoai An (2014); Pham Dinh Tao and Le Thi Hoai An (1997) for some background and rationale behind the DCA.

### Contributions

In light of the above discussions, we shall now summarize the main contributions. First, the proposed approach is motivated from an intuitive observation that common-group stochastic trends pertaining to the response and covariates can effectively identify common-group slope parameters. To estimate these common-group slope parameters and to select groups, one just needs to maximize a CQL function that pools up errors from regressions with unobserved common-group stochastic trends. This maximization problem has a novel form of a large-scale non-convex mixed-integer program, which can be written as a DC optimization problem. Given a starting point generated by the VNS, a DC algorithm with well-established convergence properties is then employed to maximize the CQL function. This optimization algorithm could be the most viable alternative to sequential procedures that iterate numerous large-scale convex optimization sub-problems [such as the algorithm in **SSP**], whose properties in computational complexity still remain open questions.

The second contribution sounds rather trivial, but it has an important implication for linear dynamic panels: Unlike conventional fixed effects or grouped fixed effects methods [that are often used to estimate dynamic panels], the proposed estimator is not subject to biases due to misspecification of the individual fixed effects as the form of fixed effects are completely irrelevant in the CQL function, and it is also asymptotically unbiased so that bias corrections are not required. These properties make the estimator appealing in this context whereas GFE estimators can be severely biased when individual fixed effects are in fact represented by a small number of unobserved types, and biases estimates can, as its consequence, be inconsistent.

Third, previous works on this topic do not cover the important issues of cross-sectional dependence and highly persistent covariates in panel regression models as the presence of either of these problems can have immediate consequences on inferential theory which is usually derived for the standard canonical case of cross-sectional independence and stationarity. This paper provides a unified approach to deal with all these issues of weak cross-sectional dependence and stationary/nonstationary covariates. In all these scenarios, we show that, under some regularity conditions [which we will discuss in the main text], the proposed estimator maintains its 'oracle' property in the sense that the uniform convergence rates of the group memberships estimates are so fast that the estimates of the group-specific slope coefficients have the same asymptotic distributions as in the case with known group memberships.

Fourth, this paper makes several technical contributions dealing with asymptotic analysis of spatio-temporal processes. We assume that cross-sectional dependence amongst individuals is of unknown form, and it can be modeled by using mixing random fields. While proving our main theorems, we derive a Marcinkiewicz and Zygmund (1937)-type inequality for specific types of mixing random fields considered in this study, a new functional central limit theorem for spatio-temporal processes that have a general type of weak dependence defined through a cardinality-based mixing coefficient, and other tail probability inequalities, all of which can be useful for other applications as well. It is worth noting at this point that the proof of super-consistency for the group memberships estimates based on the VNS-DCA clustering makes use of an elegant duality between the penalized DC program on smooth polyhedral sets and the primal DC program with combinatorial constraints [as pointed out by (C.7) in the Supplemental Material] together with the Karush-Kuhn-Tucker (KKT) conditions for local optima.

**Outline**

The remaining of this paper is organized as follows. Section 3 introduces the model and main assumptions leading to the formulation of the maximum CQL estimation. Section 4 explains the main CQL estimation method for dynamic linear panel data models with group-specific heterogeneity where the group structure can be left unspecified. The asymptotic properties of the proposed estimator together with a BIC-type information criterion used to select the optimal number of groups are all presented in Section 5. When the covariates are stationary the estimates of the 'true' coefficients $\phi_{0,g_i}$ and $\boldsymbol{\theta}_{g_i}$, $i = 1, \dots, N$, converge in distribution to normal random variables at rate $\sqrt{NT}$; this rate of convergence is the same as the rate that one could obtain when the parameters are homogenous, which is not a surprise as the number of groups remains fixed for any sufficiently large $N$.

When the covariates are highly persistent the rate of the distributional convergence pertaining to the long-run coefficient $\boldsymbol{\theta}_{g_i}$ is $T$ (instead of $T\sqrt{N}$ as one may think of), which is essentially similar to the rate achieved with fixed $N$ in **PSS**. This slow rate of convergence can be explained by the fact that, since the distributions of the response and covariates are not stable over time, one essentially needs more time-series observations for each individual when the number of cross-sections gets bigger in order to achieve a negligible classification error, which is then used to establish the 'oracle' property of the common-group slope parameters estimates. Derivation of the asymptotic theory is based on the premise that the spatial domain $V_N$ and the time horizon $T$ grow to infinity jointly and the sample ratio $|V_N|/T$ depends on the polynomial decay rate of a cardinality-based mixing coefficient.

An empirical study of Feldstein and Horioka's (1980) saving-investment puzzle is provided in Section 6, reconciling previous empirical findings about the long-run correlation of the saving and investment rates. Interested practitioners can find the software package to implement the proposed algorithms for clustering and estimation in Section 8. Also, to make the paper short and concise, results of technical flavour but essential for the paper are collected in three major supplemental materials: Section A in the Supplemental Material contains a summary of an extensive simulation study examining the performance of the proposed estimator in finite samples. Overall, it was found that, as long as the stability condition [cf. Assumption 4 in Section 3 below] holds, the proposed estimator enjoys relatively small finite-sample biases and mean squared errors for a variety of sample sizes and spatio-temporal error processes. Mathematical proofs of main theoretical results are collected in Section B in the Supplemental Material.

Finally, Section C in the Supplemental Material provides a detailed description of the main VNS-DCA algorithm and the derivations of the DC program used for clustering and estimation as an one-step procedure. It is important to note at this point that, since a standard DC program with mixed-integer sets can be transformed into an equivalent DC program with smooth polyhedral sets by employing a concave penalty function as in (C.7), the proposed CQL method can then be viewed as a penalty approach using some sort of combinatorial penalty function.

# 2 Notations

Some following conventional notations are commonly used: vectors/matrices and 'sites' on a sub-lattice (spatial domain) are written in boldface; $\|\cdot\|$ denotes the Euclidean norm; $\boldsymbol{A}^\top$ indicates the transpose of $\boldsymbol{A}$; $\lambda_{\min}(\boldsymbol{A})$ denotes the minimum eigenvalue of $\boldsymbol{A}$; $\iota_n$ is a $n \times 1$ column vector of ones and $\mathbb{I}_n$ is a $n$-dimensional [square] identity matrix; $\mathrm{diag}(\boldsymbol{A})$ denotes a diagonal matrix with $\boldsymbol{A}$ as the on-diagonal terms; for $\boldsymbol{x}, \boldsymbol{y} \in \mathbb{R}^d$, let $< \boldsymbol{x}, \boldsymbol{y} >$ represent the scalar product of $\boldsymbol{x}$ and $\boldsymbol{y}$; $|V|$ is the cardinality of a set, $V$; the Euclidean distance between two subsets, $A$ and $B$, is defined as $d(A, B) \coloneqq \min\{\|\boldsymbol{a} - \boldsymbol{b}\| : \ \boldsymbol{a} \in A, \ \boldsymbol{b} \in B\}$; the diameter of a set, $A$, is denoted by $\mathrm{diam}(A) \coloneqq \max\{\|\boldsymbol{a} - \boldsymbol{b}\| : \ \boldsymbol{a}, \boldsymbol{b} \in A\}$; $A^c$ denotes the complement of a set, $A$; $A \backslash B \coloneqq \{\boldsymbol{s} : \ \boldsymbol{s} \in A \text{ and } \boldsymbol{s} \notin B\}$; $C_0$ represents a generic constant that may vary from one equation to another; $\lfloor x \rfloor$ stands for the integer part of a (rational) number; $\mathbf{1}(A)$ denotes a set characteristic function that takes value 1 if $A$ is true and 0 otherwise; $\xrightarrow{d}$, $\xrightarrow{w}$, and $\xrightarrow{p}$ in order signify the distributional convergence, the weak convergence, and the convergence in probability; $o_p(\cdot)$ and $O_p(\cdot)$ are standard symbols for stochastic orders of magnitude; 'w.p.1' stands for 'with probability approaching 1'; $\|X\|_\gamma \coloneqq (E[|X|^\gamma])^{1/\gamma}$ denotes the Hölder norm; $(a, b)^+ \coloneqq \max(a, b)$ and $(a, b)^- \coloneqq \min(a, b)$; $\mathrm{vec}(A)$ denotes the vectorization of a matrix, $A$; and $\sigma^{(per)} : \ [1, G] \to P \in \mathcal{P}$ indicates a permutation operator that maps the set of original group labels, $[1, G]$, to some set, $P$, in the collection of all sets of permuted group labels $\mathcal{P}$. Also, to facilitate the reading of this paper, mathematical symbols (that are often referred to in the main text and the Supplemental Material) together with the places where they first appear are tabulated below.

| | |
|---|---|
| $\epsilon_{*,t}(\boldsymbol{\Theta})$ | composite errors with known group memberships, first defined in (4.1) |
| $\epsilon_{*,t}(\boldsymbol{\psi})$ | concentrated composite errors with known group memberships, first defined in (4.3) |
| $\underbrace{\boldsymbol{U}}_{G\times N} := (u_{i,c}) \in \mathbb{R}^{G\times N}$ | a matrix of group membership indicators, first defined right below (4.9) |
| $\Delta_S$ | a $(G-1)$-dimensional simplex in $\mathbb{R}^G$, first defined right above (4.10) |
| $g_{*,c} := \frac{\lvert V_{N,c}\rvert}{N} \in (0,1)$ for $c = 1,\ldots,G$ | group sizes, first defined right above (4.1) |
| $\epsilon_{*,t}(\boldsymbol{\Theta}, \boldsymbol{U})$ | composite errors with unknown group memberships, $\boldsymbol{U}$, first defined in (4.10) |
| $\epsilon_{*,t}(\boldsymbol{\psi}, \boldsymbol{U})$ | concentrated composite errors with unknown group memberships, $\boldsymbol{U}$, first defined in (4.12) |
| $\xi_{*,t}(\boldsymbol{\theta}_c)$ | common-group cross-sectional mean of $\xi_{i,t}$, $i = 1,\ldots,N$, in a known group, $c$, first defined right above (4.1) |
| $\xi_{*,t}(\boldsymbol{\theta}_c, \boldsymbol{u}_c)$ | common-group cross-sectional mean of $\xi_{i,t}$, $i = 1,\ldots,N$, in an unknown group, $c$, first defined at the beginning of Section 5.2 |
| $\boldsymbol{x}_{*,t}^{(c)}$ | common-group cross-sectional means of $\boldsymbol{x}_{i,t}$, $i = 1,\ldots,N$, in a known group, $c$, first defined in (4.4) |
| $\boldsymbol{x}_{*,t}(\boldsymbol{u}_c)$ | common-group cross-sectional means of $\boldsymbol{x}_{i,t}$, $i = 1,\ldots,N$, in an unknown group, $c$, first defined at the beginning of Section 5.2 |
| $\alpha(\cdot)$ | the cardinality-based mixing coefficient for random fields, first mentioned in Definition B.1 in the Supplemental Material |

# 3 Model and Assumptions

Consider the following autoregressive distributed lag model for panel data observed on $T$ time periods, $t = 1,\ldots,T$, and $N$ individuals (units), $i = 1,\ldots,N$ :

$$y_{i,t} = \sum_{j=1}^{p} \lambda_{g_i,j} y_{i,t-j} + \sum_{j=0}^{q} \boldsymbol{\delta}_{g_i,j}^{\top} \boldsymbol{x}_{i,t-j} + \mu_i + \epsilon_{i,t}, \tag{3.1}$$

where the $d_x$ covariates ($\boldsymbol{x}_{i,t}$) and the $p$ lags of $y_{i,t}$ (viz. $y_{i,t-1},\ldots,y_{i,t-p}$) are contemporaneously uncorrelated with the errors $\epsilon_{i,t}$; $\lambda_{g_i,j}$ for $i = 1,\ldots,N$ and $j = 1,\ldots,p$ and $\boldsymbol{\delta}_{g_i,j}$ for $i = 1,\ldots,N$ and $j = 0,\ldots,q$ are group-specific autoregression and regression coefficients respectively. Units are divided into $G$ mutually exclusive, exhaustive groups; and the group membership variables $g_i \in \{1, \ldots, G\}$ are defined via an onto mapping $g: \{1, \ldots, N\} \to \{1, \ldots, G\}$. To study the potential long-run relationship between $y_{i,t}$ and $\boldsymbol{x}_{i,t}$ within each group, we shall rewrite (3.1) in the following error-correction form:

$$\Delta y_{i,t} = \mu_i + \phi_{g_i}(y_{i,t-1} - \boldsymbol{\theta}_{g_i}^\top \boldsymbol{x}_{i,t}) + \sum_{j=1}^{p-1} \lambda_{g_i,j}^* \Delta y_{i,t-j} + \sum_{j=0}^{q-1} \boldsymbol{\delta}_{g_i,j}^{*\top} \Delta \boldsymbol{x}_{i,t-j} + \epsilon_{i,t}, \tag{3.2}$$

where $\phi_{g_i} := -\left(1 - \sum_{j=1}^{p} \lambda_{g_i,j}\right)$, $\boldsymbol{\theta}_{g_i} := -\frac{\sum_{j=0}^{q} \boldsymbol{\delta}_{g_i,j}}{\phi_{g_i}}$, $\lambda_{g_i,j}^* := -\sum_{m=j+1}^{p} \lambda_{g_i,m}$ for $j = 1, \ldots, p-1$, and $\boldsymbol{\delta}_{g_i,j}^* := -\sum_{m=j+1}^{q} \boldsymbol{\delta}_{g_i,m}$ for $j = 1, \ldots, q-1$.

Suppose that each unit ,$i$, is associated with a location, say $s_i$, on a $d_v$-dimensional Euclidean space, $V_N$, equipped with the Euclidean metric $\|\cdot\|$ measuring the distance between any two locations in $V_N$. Here, for clarity of exposition, $V_N$ is assumed to be a sublattice [of the standard $d_v$-dimensional integer lattice $\mathbb{Z}^{d_v}$] indexed by $N$; the other cases where $V_N$ is some sublattice of $\mathbb{R}^{d_v}$ merely require a slight modification of the proofs with more notations as long as the distance between any two points in $V_N$ is greater than or equal to one (see, e.g., Jenish and Prucha (2012)).

Random variables are spatially dependent at some point in time, $t$, if their measurements at two different locations depend on each other, and this dependence is assumed to be weaker as the distance between the locations becomes further. For the model to remain parsimonious and tractable, we can allow for spatial dependence in the relationship between $y$ and $\boldsymbol{x}$ by assuming that, at a specific time period, the errors $\epsilon_{i,t}$, $i = 1, \ldots, N$ and $t = 1, \ldots, T$, at two different locations are dependent whilst they are independent at different points in time. First, we make the following assumptions:

**Assumption 1** *The errors, $\epsilon_{i,t}$ with $i = 1, \ldots, N$ and $t = 1, \ldots, T$, defined in (3.2), are independent across time and, at some given point in time, they are dependent across locations such that $\epsilon_{i,t} \sim N(0, \sigma_i)$.*

It is important to note that the normality of cross-sectional error terms, $\epsilon_{s_i,t}$, $i = 1, \ldots, N$, in Assumption 1 could be relaxed when $N$ is sufficiently large since the CLT for strongly mixing random fields (see, e.g., Bulinski and Shashkin (2007)) warrants that $\frac{1}{\sqrt{N}} \sum_{i=1}^{N} \epsilon_{i,t}$ converges in distribution to a normal random variable (i.e., $\frac{1}{\sqrt{N}} \sum_{i=1}^{N} \epsilon_{i,t}$ for $t = 1, \ldots, T$ can be approximated by independent normal random variables.)

**Assumption 2** *The model (3.1) is stable in that the roots of*

$$\sum_{j=1}^{p} \lambda_{g_i,j} z^j = 1, \quad i = 1, \ldots, N$$

*lie outside the unit circle.*

Assumption 2 is originally employed in **PSS** to ensure that the order of integration of $y_{i,t}$ is at most equal to the maximum of the orders of integration of the elements of the vector $\boldsymbol{x}_{i,t}$. This condition also warrants the existence of a long-run relationship between $y_{i,t}$ and $\boldsymbol{x}_{i,t}$ within each group. Let $\boldsymbol{w}_{i,t} \coloneqq (\Delta y_{i,t-1}, \ldots, \Delta y_{i,t-p+1}, \Delta \boldsymbol{x}_{i,t}^\top, \ldots, \Delta \boldsymbol{x}_{i,t-q+1}^\top)^\top$ denote a vector of $d_w = p + d_x q - 1$ auxiliary covariates, and let $\boldsymbol{\lambda}_{g_i} \coloneqq (\lambda_{g_i,1}, \ldots, \lambda_{g_i,p-1}, \boldsymbol{\delta}_{g_i}^\top, \ldots, \boldsymbol{\delta}_{g_i,q-1}^\top)^\top$ be their coefficients. We can also rewrite (3.2) as

$$\Delta y_{i,t} = \mu_i + \phi_{g_i} \xi_{i,t}(\boldsymbol{\theta}_{g_i}) + \boldsymbol{\lambda}_{g_i}^\top \boldsymbol{w}_{i,t} + \epsilon_{i,t}, \tag{3.3}$$

where $\xi_{i,t}(\boldsymbol{\theta}_{g_i}) \coloneqq y_{i,t-1} - \boldsymbol{\theta}_{g_i}^\top \boldsymbol{x}_{i,t}$. Our objects for inference are the long-run coefficients $\boldsymbol{\theta}_{g_i}$ and the long-run adjustment speed parameters $\phi_{g_i}$ with $i = 1, \ldots, N$.

It is important to note that, since the joint likelihood of the model is not the same as the product of likelihoods for each unit (or group), standard ML procedures will therefore involve a large unknown spatial variance-covariance matrix of $\epsilon_{i,t}$, $i = 1, \ldots, N$, thus they inevitably become infeasible. Moreover the expectations of the score functions of the concentrated joint log-likelihood function are not zero due to the absence of the complete orthogonality between $\epsilon_{i,t}$ and $\xi_{i,t}(\boldsymbol{\theta}_{g_i})$, thus resulting in significant biases. Therefore, here we shall instead construct the composite quasi-likelihood function. To simplify notations, we assume that the nuisance parameters are group-invariant (i.e., $\boldsymbol{\lambda}_{g_1} = \cdots \boldsymbol{\lambda}_{g_N} = \boldsymbol{\lambda}$.) Note that this simplification does not much change our mathematical arguments, thus our asymptotic results will still remain valid even when these nuisance parameters vary over groups. To see this, notice that in the representation of the composite errors (4.3), the projections of $\boldsymbol{x}_{*,t}$'s and $\xi_{*,t}$'s on the span of $\{\boldsymbol{w}_{*,t} :\ t = 1, \ldots, T\}$ can be of lower asymptotic orders when $T$ goes to infinity and the cluster sizes grow sufficiently large. In fact, simulation results reported in Section A of the Supplemental Material confirm that the algorithm for clustering and estimation based on the objective function imposing group-invariant nuisance parameters performs well even when data are generated from a data generating process with group-variant nuisance parameters.

# 4 Estimation with Known/Unknown Group Memberships

As discussed in the Introduction the dynamic panel data model defined via (3.3) is estimated by using a composite quasi-likelihood method. The general principle of composite likelihood methods is to simplify complex dependence relationships by computing marginal or conditional distributions associated with some subsets of data, and multiplying these together to form an inference function. Employing composite likelihood methods can reduce the computational complexity so that it is possible to deal with large datasets and even very complex models where the use of standard likelihood or Bayesian methods is not feasible. Composite likelihood estimators also have good theoretical properties, and behave well in many complex applications (see, e.g., Reid (2013); Varin, Reid, and Firth (2011) for recent reviews of this subject matter.) Following Lindsay (1988), let $\{f(\boldsymbol{y};\boldsymbol{\theta}),\ \boldsymbol{y} \in \mathcal{Y}, \boldsymbol{\theta} \in \Theta\}$, where $\mathcal{Y} \subset \mathbb{R}^n$ and $\Theta \subset \mathbb{R}^d$ with $n \geq 1$ and $d \geq 1$, be a parametric model. Consider a set, $\{\mathcal{A}_1, \dots, \mathcal{A}_k, \dots, \mathcal{A}_K\}$, of marginal or conditional events associated with likelihoods, $\mathcal{L}_k(\boldsymbol{\theta};\boldsymbol{y}) \propto f(\boldsymbol{y} \in \mathcal{A}_k;\boldsymbol{\theta})$. A composite likelihood is formally defined as a weighted product $\prod_{k=1}^{K} \mathcal{L}_k(\boldsymbol{\theta};\boldsymbol{y})^{w_k}$, where $w_k$, $k = 1, \dots, K$, represent some non-negative composite weights to be chosen.

We first present the main procedure based on composite quasi-likelihood to estimate Model (3.1) when group memberships of individuals/units are given (i.e., each individual belongs to a specified group.) By Assumption 1, $\sqrt{N}\epsilon_{*,t} = \sqrt{N}\frac{1}{N}\sum_{i=1}^{N} \epsilon_{s_i,t} \overset{i.i.d.}{\approx} N(0, \sigma_\epsilon^2)$, where $\sigma_\epsilon^2 := \lim_{N\uparrow\infty} \frac{1}{N} Var\left(\sum_{i=1}^{N} \epsilon_{s_i,t}\right) < \infty$ if the spatial dependence among $\epsilon_{i,t}$, $i = 1, \dots, N$, is weak for units that are far apart. Therefore, all the likelihoods associated with conditional events, $\mathcal{A}_t(x) := \{(\epsilon_{1,t}, \dots, \epsilon_{N,t}) \in \mathbb{R}^T :\ \sqrt{N}\epsilon_{*,t} = x\}$ with $t = 1, \dots, T$, are approximately Gaussian, thus they are referred to as quasi-likelihoods.

Let $V_{N,c}$ represent a set of locations for units in group $c \in \{1, \dots, G\}$ so that $V_N := \bigcup_{c=1}^{G} V_{N,c}$, $L_{N,c} := |V_{N,c}|$, and $g_{*,c} := \frac{L_{N,c}}{N}$ for $c = 1, \dots, G$. Define $\Delta y_{*,t} := \frac{1}{N}\sum_{i=1}^{N} \Delta y_{i,t}$, $\boldsymbol{w}_{*,t} := \frac{1}{N}\sum_{i=1}^{N} \boldsymbol{w}_{i,t}$, $\mu_* := \begin{cases} \sum_{c=1}^{G} g_{*,c}\mu_c & \text{if } \mu_i's \text{ are group-specific,} \\ \frac{1}{N}\sum_{i=1}^{N} \mu_i & \text{if } \mu_i's \text{ are individual-specific,} \end{cases}$ and $\xi_{*,t}(\boldsymbol{\theta}_c) := \frac{1}{L_{N,c}} \sum_{\boldsymbol{j} \in V_{N,c}} \xi_{\boldsymbol{j},t}(\boldsymbol{\theta}_c)$, where $\mu_c := \mu_{g(V_{N,c})}$ and $\boldsymbol{\theta}_c := \boldsymbol{\theta}_{g(V_{N,c})}$, $c = 1, \dots, G$. Collecting all the unknown parameters into a vector, say $\boldsymbol{\Theta} := \left(\boldsymbol{\theta}^\top, \boldsymbol{\phi}^\top, \boldsymbol{\lambda}^\top, \mu_*\right)$, where $\boldsymbol{\theta} := \left(\boldsymbol{\theta}_1^\top, \dots, \boldsymbol{\theta}_g^\top, \dots, \boldsymbol{\theta}_G^\top\right)^\top$ and $\boldsymbol{\phi} := (\phi_1, \dots, \phi_c, \dots, \phi_G)^\top$ with $\phi_c := \phi_{g(V_{N,c})}$. Setting all the composite weights $\{w_t,\ t = 1, \dots, T\}$ to ones the composite quasi-log-likelihood function (*or* composite quasi-likelihood function) can then be written as

$$Q_{N,T}(\boldsymbol{\Theta}, \sigma_\epsilon^2) := -\frac{T}{2}\log 2\pi - \frac{T}{2}\log \sigma_\epsilon^2 - \frac{N}{2\sigma_\epsilon^2}\sum_{t=1}^{T} \epsilon_{*,t}^2(\boldsymbol{\Theta}), \tag{4.1}$$

where $\epsilon_{*,t}(\boldsymbol{\Theta}) := \Delta y_{*,t} - \mu_* - \sum_{i=1}^{G} g_{*,i}\phi_i\xi_{*,t}(\boldsymbol{\theta}_i) - \boldsymbol{\lambda}^\top \boldsymbol{w}_{*,t}$.

**Remark 4.1** *Intuitively, while clustering with the least squares criterion function (as in* ***BM****) is based on the premise that - for given values of the parameters - an individual is assigned to a group if its temporal summation of squared errors associated with that group is less than its temporal summations of squared errors associated with all other groups, the CQL criterion assigns a subset of individuals, say $\mathfrak{C}$, to a group if the temporal summation of squared $\mathfrak{C}$-mean errors (or* centroids *in the language of machine learning) associated with this group is less than the temporal summations of squared $\mathfrak{C}$-mean errors associated with all other groups whilst maintaining that the mean errors* *of any pair of groups are as little correlated as possible. To see this point, notice that*

$$\begin{aligned} \frac{1}{T}\sum_{t=1}^{T}\epsilon_{*,t}^2(\boldsymbol{\kappa}) &= \frac{1}{T}\sum_{t=1}^{T}\left(\sum_{c=1}^{G}\frac{1}{N}\sum_{i=1}^{N}u_{i,c}\epsilon_{i,t}(\boldsymbol{\kappa}_c)\right)^2 \\ &= \sum_{c=1}^{G}\sum_{t=1}^{T}\underbrace{\left(\frac{1}{N}\sum_{i=1}^{N}u_{i,c}\epsilon_{i,t}(\boldsymbol{\kappa}_c)\right)^2}_{\text{squared mean error of group } c} \\ &\quad + \sum_{c=1}^{G}\sum_{g\neq c}^{G}\underbrace{\frac{1}{T}\sum_{t=1}^{T}\left(\frac{1}{N}\sum_{i=1}^{N}u_{i,c}\epsilon_{i,t}(\boldsymbol{\kappa}_c)\right)\left(\frac{1}{N}\sum_{i=1}^{N}u_{i,g}\epsilon_{i,t}(\boldsymbol{\kappa}_g)\right)}_{\text{correlation between the mean errors of two groups}}, \end{aligned}$$

*where* $\boldsymbol{\kappa} = (\boldsymbol{\kappa}_1^\top, \ldots, \boldsymbol{\kappa}_G^\top)^\top$. *When groups are mutually independent, the CQL criterion function is the same as the summation of all the sums of squared errors obtained from G regressions of common-group stochastic trends.*

To proceed, we concentrate out the nuisance parameters $\boldsymbol{\lambda}$, and let $\boldsymbol{\Omega} := (\boldsymbol{\psi}^\top, \sigma_\epsilon^2)^\top$ with $\boldsymbol{\psi} := (\boldsymbol{\theta}^\top, \boldsymbol{\phi}^\top, \mu_*)^\top$. We can then obtain the concentrated composite quasi-log-likelihood function:

$$Q_{N,T}(\boldsymbol{\Omega}) := -\frac{T}{2}\log 2\pi - \frac{T}{2}\log\sigma_\epsilon^2 - \frac{N}{2\sigma_\epsilon^2}\sum_{t=1}^{T}\epsilon_{*,t}^2(\boldsymbol{\psi}), \tag{4.2}$$

where

$$\begin{aligned} \epsilon_{*,t}(\boldsymbol{\psi}) := \Delta y_{*,t} &- \left(\sum_{s=1}^{T}\Delta y_{*,s}\boldsymbol{w}_{*,s}^\top\right)\left(\sum_{s=1}^{T}\boldsymbol{w}_{*,s}\boldsymbol{w}_{*,s}^\top\right)^{-1}\boldsymbol{w}_{*,t} \\ &- \sum_{c=1}^{G}g_{*,c}\phi_c\left(\xi_{*,t}(\boldsymbol{\theta}_c) - \left(\sum_{s=1}^{T}\xi_{*,s}^{(c)}(\boldsymbol{\theta}_c)\boldsymbol{w}_{*,s}^\top\right)\left(\sum_{s=1}^{T}\boldsymbol{w}_{*,s}\boldsymbol{w}_{*,s}^\top\right)^{-1}\boldsymbol{w}_{*,t}\right) \\ &- \mu_*\left(1 - \sum_{s=1}^{T}\boldsymbol{w}_{*,s}^\top\left(\sum_{s=1}^{T}\boldsymbol{w}_{*,s}\boldsymbol{w}_{*,s}^\top\right)^{-1}\boldsymbol{w}_{*,t}\right). \end{aligned} \tag{4.3}$$

To derive the first-order conditions for the CQL maximization, we first need to define the following quantities:

$$\underbrace{\boldsymbol{A}_t}_{Gd_x\times 1} := \left(\boldsymbol{x}_{*,t}^{(1)\top} - \boldsymbol{w}_{*,t}^\top\left(\sum_{s=1}^{T}\boldsymbol{w}_{*,s}\boldsymbol{w}_{*,s}^\top\right)^{-1}\sum_{s=1}^{T}\boldsymbol{w}_{*,s}\boldsymbol{x}_{*,s}^{(1)\top}, \ldots, \boldsymbol{x}_{*,t}^{(G)\top}\right.$$

$$- \boldsymbol{w}_{*,t}^{\top} \left( \sum_{s=1}^{T} \boldsymbol{w}_{*,s} \boldsymbol{w}_{*,s}^{\top} \right)^{-1} \sum_{s=1}^{T} \boldsymbol{w}_{*,s} \boldsymbol{x}_{*,s}^{(G)\top} \Bigg)^{\top}, \tag{4.4}$$

$$\text{where } \boldsymbol{x}_{*,t}^{(c)} := \frac{1}{L_{N,c}} \sum_{\boldsymbol{j} \in V_{N,c}} \boldsymbol{x}_{\boldsymbol{j},t} \text{ for } i = 1, \dots, G,$$

$$\underbrace{\boldsymbol{B}_t(\boldsymbol{\theta})}_{G \times 1} := \Bigg( \left( \sum_{s=1}^{T} \xi_{*,s}^{(1)}(\boldsymbol{\theta}_1) \boldsymbol{w}_{*,s}^{\top} \right) \left( \sum_{s=1}^{T} \boldsymbol{w}_{*,s} \boldsymbol{w}_{*,s}^{\top} \right)^{-1} \boldsymbol{w}_{*,t} - \xi_{*,t}^{(1)}(\boldsymbol{\theta}_1), \dots,$$

$$\left( \sum_{s=1}^{T} \xi_{*,s}^{(G)}(\boldsymbol{\theta}_G) \boldsymbol{w}_{*,s}^{\top} \right) \left( \sum_{s=1}^{T} \boldsymbol{w}_{*,s} \boldsymbol{w}_{*,s}^{\top} \right)^{-1} \boldsymbol{w}_{*,t} - \xi_{*,t}^{(G)}(\boldsymbol{\theta}_G) \Bigg)^{\top}, \tag{4.5}$$

and

$$C_t := \sum_{s=1}^{T} \boldsymbol{w}_{*,s}^{\top} \left( \sum_{s=1}^{T} \boldsymbol{w}_{*,s} \boldsymbol{w}_{*,s}^{\top} \right)^{-1} \boldsymbol{w}_{*,t} - 1. \tag{4.6}$$

Some algebraic manipulations yield

$$\frac{\partial Q_{N,T}(\boldsymbol{\Omega})}{\partial \boldsymbol{\theta}} = -\mathrm{diag}(g_{*,c} \phi_c \boldsymbol{I}_{d_x},\ c = 1, \dots, G) \frac{N}{\sigma_\epsilon^2} \sum_{t=1}^{T} \epsilon_{*,t}(\boldsymbol{\psi}) \boldsymbol{A}_t, \tag{4.7}$$

$$\frac{\partial Q_{N,T}(\boldsymbol{\Omega})}{\partial \boldsymbol{\phi}} = -\mathrm{diag}(g_{*,c},\ c = 1, \dots, G) \frac{N}{\sigma_\epsilon^2} \sum_{t=1}^{T} \epsilon_{*,t}(\boldsymbol{\psi}) \boldsymbol{B}_t(\boldsymbol{\theta}), \tag{4.8}$$

and

$$\frac{\partial Q_{N,T}(\boldsymbol{\Omega})}{\partial \mu_*} = -\frac{N}{\sigma_\epsilon^2} \sum_{t=1}^{T} \epsilon_{*,t}(\boldsymbol{\psi}) C_t. \tag{4.9}$$

One can readily obtain the 'oracle' estimates $\widetilde{\boldsymbol{\psi}} := \left( \widetilde{\boldsymbol{\theta}}, \widetilde{\boldsymbol{\phi}}, \widetilde{\mu_*} \right)$ of the 'true' parameters $\boldsymbol{\psi}_0 := (\boldsymbol{\theta}_0, \boldsymbol{\phi}_0, \mu_{*0})$ by finding the roots of (4.7)-(4.9).

Now we shall adapt the CQL procedure described above to the case where group memberships of individuals are not specified a priori. Suppose that the number of groups (*or* clusters) $G$ is given. Let $\boldsymbol{U} := (u_{i,c}) \in \mathbb{R}^{G \times N}$, $i = 1, \dots, N$ and $c = 1, \dots, G$, be a $G \times N$ matrix whose elements are defined by $u_{i,c} = 1$ if individual $i \in [1, N]$ belongs to group $c \in [1, G]$, and $u_{i,c} = 0$ otherwise. Because each individual can only be assigned to one group, we need to impose the constraint that $\sum_{c=1}^{G} u_{i,c} = 1$ for every $i = 1, \dots, N$. Moreover, let $\Delta_S := \{\boldsymbol{u} \in [0,1]^G : \sum_{c=1}^{G} u_c = 1\}$ represent a $(G-1)$-simplex in $\mathbb{R}^G$, and $\Delta_S^N$ indicates the Cartesian product of $N$ simplices, $\Delta_S$; thus, $\boldsymbol{U} \in \Delta_S^N \bigcap \{0,1\}^{G \times N}$.

With this matrix of group membership indicators, define the composite error as

$$\epsilon_{*,t}(\boldsymbol{\Theta}, \boldsymbol{U}) := \Delta y_{*,t} - \mu_{*} - \boldsymbol{\lambda}^{\top}\boldsymbol{w}_{*,t} - \sum_{c=1}^{G}\frac{1}{N}\sum_{i=1}^{N}u_{i,c}\phi_c\xi_{i,t}(\boldsymbol{\theta}_c). \tag{4.10}$$

The CQL function is then given by

$$Q_{N,T}(\boldsymbol{\theta}, \boldsymbol{\phi}, \boldsymbol{\lambda}, \mu_{*}, \sigma_{\epsilon}^{2}, \boldsymbol{U}) := -\frac{T}{2}\log 2\pi - \frac{T}{2}\log\sigma_{\epsilon}^{2} - \frac{N}{2\sigma_{\epsilon}^{2}}\sum_{t=1}^{T}\epsilon_{*,t}^{2}(\boldsymbol{\Theta}, \boldsymbol{U}). \tag{4.11}$$

By concentrating the nuisance parameters $\boldsymbol{\lambda}$ out, one readily obtains that

$$\epsilon_{*,t}(\boldsymbol{\psi}, \boldsymbol{U}) := \Delta y_{*,t}^{(w)} - \sum_{c=1}^{G}\frac{1}{N}\sum_{i=1}^{N}u_{i,c}\phi_c\left(y_{i,t-1}^{(w)} - \boldsymbol{\theta}_c^{\top}\boldsymbol{x}_{i,t}^{(w)}\right) - \mu_{*}1_t^{(w)}, \tag{4.12}$$

where

$$Z_t^{(w)} := Z_t - \left(\sum_{t=1}^{T}Z_t\boldsymbol{w}_{*,t}^{\top}\right)\left(\sum_{t=1}^{T}\boldsymbol{w}_{*,t}\boldsymbol{w}_{*,t}^{\top}\right)^{-1}\boldsymbol{w}_{*,t}$$

with $Z_t$ being either $y_{i,t-1}$ or $\boldsymbol{x}_{i,t}$, and

$$1_t^{(w)} := 1 - \left(\sum_{t=1}^{T}\boldsymbol{w}_{*,t}^{\top}\right)\left(\sum_{t=1}^{T}\boldsymbol{w}_{*,t}\boldsymbol{w}_{*,t}^{\top}\right)^{-1}\boldsymbol{w}_{*,t}.$$

Moreover, notice that

$$\Delta y_{*,t}^{(w)} = \sum_{c=1}^{G}\frac{1}{N}\sum_{i=1}^{N}u_{0,i,c}\phi_{0,c}\xi_{i,t}^{(w)}(\boldsymbol{\theta}_{0,c}) + \mu_{*0}1_t^{(w)},$$

where all the subscripts '0' signify the true [unknown] quantities for the rest of the paper. The objective CQL function $Q_{N,T}$ defined by (4.11) is invariant with respect to all permutations of the group labels; let $\sigma^{(per)} : [1, G] \to P \in \mathcal{P}$ denote a permutation operator, which is a bijective mapping from the set, $[1, G]$, of the original group labels to some set, $P$, of permuted group labels, and $\mathcal{P}$ is the collection of all the sets of permuted group labels. It then follows that the concentrated composite error $\epsilon_{*,t}(\boldsymbol{\psi}, \boldsymbol{U})$ can also be written as

$$\begin{aligned}\epsilon_{*,t}(\boldsymbol{\psi}, \boldsymbol{U}) = &\sum_{c=1}^{G}\phi_{\sigma^{(per)}(c)}(\boldsymbol{\theta}_{\sigma^{(per)}(c)} - \boldsymbol{\theta}_{0,c})\frac{1}{N}\sum_{i=1}^{N}u_{i,\sigma^{(per)}(c)}\boldsymbol{x}_{i,t}^{(w)} \\ &+ \sum_{c=1}^{G}(\phi_{0,c} - \phi_{\sigma^{(per)}(c)})\frac{1}{N}\sum_{i=1}^{N}u_{i,\sigma^{(per)}(c)}\xi_{i,t}^{(w)}(\boldsymbol{\theta}_{0,c}) + (\mu_{*0} - \mu_{*})1_t^{(w)}\end{aligned}$$

$$+\sum_{c=1}^{G}\phi_{0,c}\frac{1}{N}\sum_{i=1}^{N}(u_{0,i,c}-u_{i,\sigma^{(per)}(c)})\xi_{i,t}^{(w)}(\boldsymbol{\theta}_{0,c})+\epsilon_{0,*,t}^{(w)}, \tag{4.13}$$

where $\epsilon_{0,*,t} := \epsilon_{*,t}(\boldsymbol{\Theta}_0, \boldsymbol{U}_0)$. The concentrated CQL function is then given by

$$Q_{N,T}(\boldsymbol{\psi},\sigma_\epsilon^2,\boldsymbol{U}) := -\frac{T}{2}\log 2\pi - \frac{T}{2}\log\sigma_\epsilon^2 - \frac{N}{2\sigma_\epsilon^2}\sum_{t=1}^{T}\epsilon_{*,t}^2(\boldsymbol{\psi},\boldsymbol{U}).$$

Consequentially, the maximum CQL estimates $\widehat{\boldsymbol{\psi}}$, $\widehat{\sigma}_\epsilon^2$, and $\widehat{\boldsymbol{U}}$ of $\boldsymbol{\psi}_0$, $\sigma_{\epsilon,0}^2$, and $\boldsymbol{U}_0$ respectively are defined as the solutions to the following large-scale non-convex mixed-integer programming problem:

$$\min\left\{Q_{N,T}(\boldsymbol{\psi},\sigma_\epsilon^2,\boldsymbol{U}):\ \boldsymbol{\psi}\in\Theta_\psi\subset\mathbb{R}^{G(d_x+1)+1},\ \sigma_\epsilon^2\in\Theta_\sigma\subset\mathbb{R},\ \text{and } \boldsymbol{U}\in\Delta_S^N\bigcap\{0,1\}^{G\times N}\right\}. \tag{4.14}$$

**Intuition behind the proposed CQL estimator.** Suppose that $y_{i,t}$ and $\boldsymbol{x}_{i,t}$ share a common relationship in a group, $c \in [1,\dots,G]$. The common-group stochastic trends that can be reasonably proxied by the observable vector of groupwise cross-sectional averages $(y_{*,t}^{(c)}, \boldsymbol{x}_{*,t}^{(c)})$, $c \in [1,\dots,G]$, also obey the same relationship, i.e., $\Delta y_{*,t}^{(c)} = \mu_c + \phi_c\xi_{*,t}^{(c)}(\boldsymbol{\theta}_c) + \boldsymbol{\lambda}_c^\top \boldsymbol{w}_{*,t}^{(c)} + \epsilon_{*,t}^{(c)}$. This point is illustrated in Figure 1 below. Since $\epsilon_{*,t}^{(c)}$ will be close to zero as the group size is sufficiently large, one needs to blow it up by $\sqrt{N}$ so that $\sqrt{N}\sum_{c=1}^{G}\epsilon_{*,t}^{(c)} \sim N(0,\sigma_\epsilon^2)$. Therefore the CQL estimator can be viewed as the minimizer of the temporal average of the squares of the errors from regressions involving the common stochastic trends of $y_{i,t}$ and $\boldsymbol{x}_{i,t}$ in $G$ groups. For given $N$ units, there are many ways to partition these $N$ units into $G$ groups. The estimated group memberships are associated with the group partitioning that minimizes the sum of squared residuals obtained from $G$ regressions involving common-group stochastic trends.

# 5 Asymptotic Theory

## 5.1 Known Group Membership

First of all, it is important to note that the parameter spaces ($\Theta_\theta$, $\Theta_\phi$, $\Theta_\mu$, and $\Theta_\sigma$) of $(\boldsymbol{\theta}_0^\top, \boldsymbol{\phi}_0^\top, \mu_{*0}, \sigma_{\epsilon,0}^2)^\top$ are assumed to be compact throughout the paper. We study the asymptotic behaviour of the 'oracle' estimate $\widetilde{\boldsymbol{\psi}}$ in two different cases. In the first case, it is assumed that, for each $\boldsymbol{i} \in V_N$, $\boldsymbol{x}_{\boldsymbol{i},t}$ is a stationary time series; and in addition the spatio-temporal processes $\{\boldsymbol{x}_{\boldsymbol{j},t} :\ \boldsymbol{j} \in V_{N,c} \text{ and } t \in [1,T]\}$, $c = 1,\dots,G$, are mixing and satisfy the following assumption:

**Assumption 3** *Within each group, $c$, the random variables $\{\boldsymbol{x}_{\boldsymbol{j},t} :\ \boldsymbol{j} \in V_{N,c} \text{ and } t \in [1,T]\}$ are identically distributed across time and space. Moreover,*

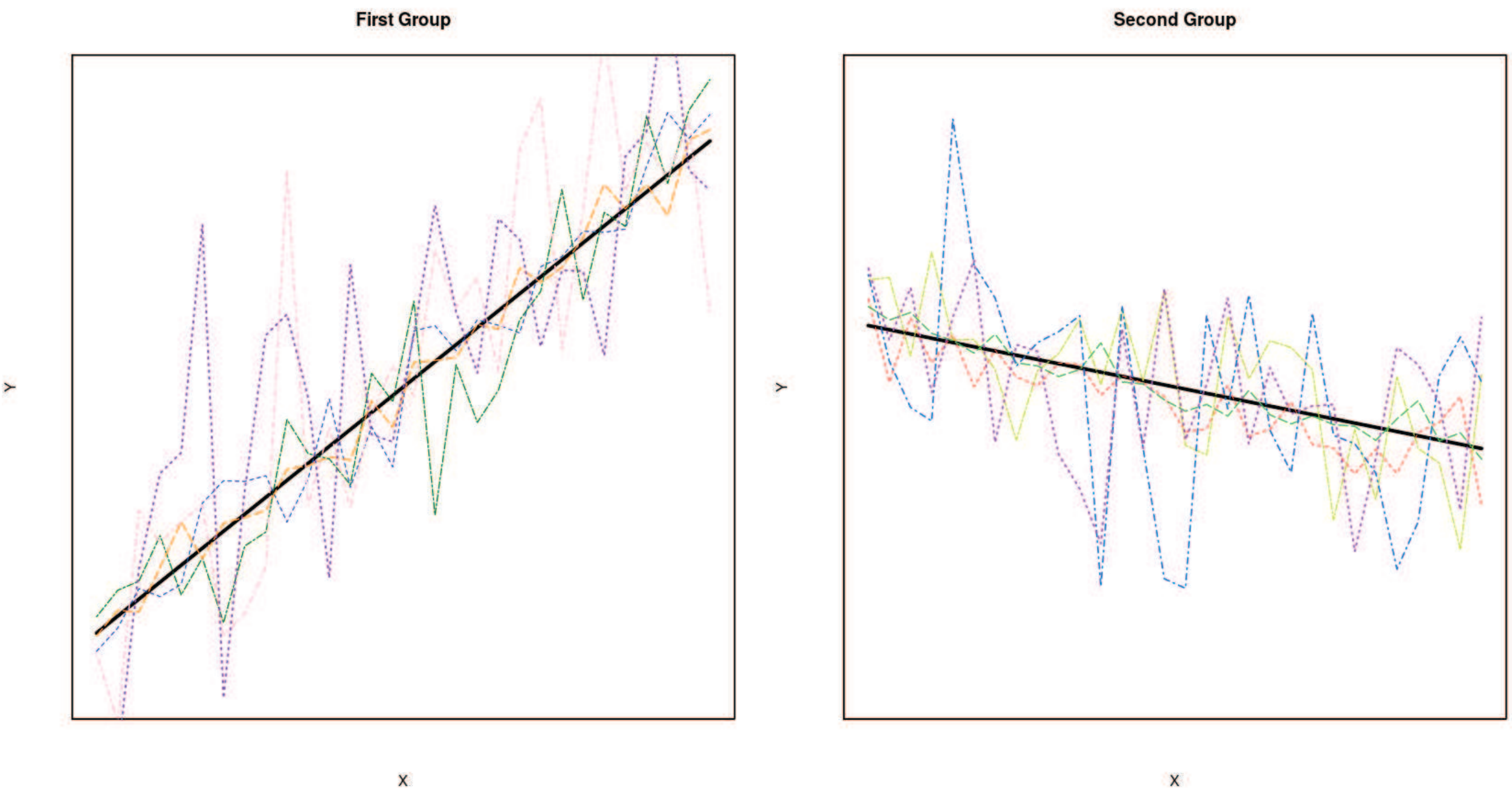

Figure 1: Two groups of regression relationships with two different patterns of slopes. In all subplots: the bold dark line represents the exact relationship between common-group stochastic trends, $y_t^{(c)}$ and $x_t^{(c)}$ with $c = 1, 2$.

**(a)** *the mixing coefficient - as represented by $\alpha(\cdot)$ in Definition B.1 [in the Supplemental Material] - for $\{(\boldsymbol{x}_t^{(c)}, \epsilon_{\boldsymbol{j},t}) : \boldsymbol{j} \in V_{N,c}$ and $t \in [1, T]\}$, $c = 1, \ldots, G$, where $\boldsymbol{x}_t^{(c)}$ are the common-group stochastic trends of all the $\boldsymbol{x}_{\boldsymbol{j},t}$'s in group $c$, decays to zero at some rate such that $\alpha(\tau) \leq C_\theta \tau^{-\theta_\alpha}$ for some*

$$\theta_\alpha \geq \max\left(\frac{p(d_v+1)\gamma_\eta}{(p-q)(\gamma_\eta-2)} + (d_v+1)\gamma_M, \frac{d_v+1}{1-\frac{2}{\gamma_\eta}}, \frac{2p}{p-4} - \gamma_M\right),$$

*where the generic constants $C_\theta > 0$, $\gamma_\eta > 2$, $p > 4$, $q = \frac{2p}{p-2}$, $d_v$ is the dimension of $V_N$ with $\gamma_M \geq 1$ provided in Definition B.1;*

**(b)** $\max_{\boldsymbol{j}} \left(E\|\boldsymbol{x}_{\boldsymbol{j},1}\|^p, E\|\boldsymbol{x}_{\boldsymbol{j},1}\|^{\gamma_\eta}, E\|\boldsymbol{x}_{\boldsymbol{j},1}\|^4\right) < \infty$;

**(c)** $\max\left(|V_N|^{\gamma_M} T^{\gamma_M+1-\theta_\alpha}, |V_N|^{\gamma_M(1-2/\gamma_\eta)} T^{\epsilon-1/2}, T^{(\gamma_M+\theta_\alpha-1)\epsilon-\frac{1}{2}(\theta_\alpha-\gamma_M-1)}|V_N|^{\gamma_M}\right) \downarrow 0$ *for some* $\epsilon \in \left(0, \min\left(\frac{1}{2}, \frac{\theta_\alpha-\gamma_M-1}{2(\gamma_M+\theta_\alpha-1)}\right)\right)$.

**Assumption 4** *Let $\boldsymbol{X}_{N,T,t}(\boldsymbol{\theta}) := \left(\boldsymbol{x}_{*,t}^{(1)\top}, \ldots, \boldsymbol{x}_{*,t}^{(G)\top}, -\xi_{*,t}^{(1)}(\boldsymbol{\theta}_1), \ldots, -\xi_{*,t}^{(G)}(\boldsymbol{\theta}_G), -1\right)^\top$. It is assumed that the minimum eigenvalue of the non-stochastic limiting matrix $\boldsymbol{D}_g \boldsymbol{Q}_{zz} \boldsymbol{D}_g^\top$ of the Gram ma-* *trix* $\boldsymbol{D}_g\left(\frac{1}{T}\sum_{t=1}^{T}\boldsymbol{X}_t(\boldsymbol{\theta}_0)\boldsymbol{X}_t(\boldsymbol{\theta}_0)^\top\right)\boldsymbol{D}_g^\top$, *where* $\underbrace{\boldsymbol{D}_g}_{(G(d_x+1)+1)\times(G(d_x+1)+1)} := diag(\boldsymbol{g}\otimes\mathbb{I}_{d_x},\boldsymbol{g},1)$ *with* $\boldsymbol{g} := (g_{*,1},\dots,g_{*,G})^\top \in (0,1)^G$, *is strictly positive.*

A few remarks are now in order. Condition 3(a) imposes a specific degree of weak spatio-temporal dependence on the covariates and the error term. This type of cardinality-based mixing conditions is often used to quantify the notion of weak dependence in random fields, which includes time-series dependence as a special case (see, e.g., Doukhan (1994, Sec. 1.3.1) and also Bradley (2010) for an analytic treatment of various mixing conditions for random fields, including the mixing condition used in this paper).

It is important to note that, throughout this paper, we assume polynomial decay rates for the mixing coefficient for the reason that the mixing rate in stationary random fields is most of the time quite slow. Nevertheless, assuming the exponential rate can significantly simplify analytical arguments. As suggested in Bradley (1993) and Bradley (2007, Chap. 29) the mixing rate $\alpha_s(\tau)$ [in Definition B.1] can be of order $O(1/\tau)$ or some arbitrarily slower rate. Many well-known spatial processes - including but not limited to linear fields (see, e.g., Guyon (1987)), the Cliff and Ord (1973) type spatial autoregressive processes using sparse spatial weight matrices, Volterra fields (see, e.g., Casti (1985)), and Markovian fields (see, e.g., Arbia (2006, Sec. 2.4.2)) - can verify this type of condition.

Condition 3(b) is rather standard - it requires some moments of the Euclidean norm of the vector of covariates to be bounded. Condition 3(c) allows both $T$ and $N$ go to infinity and the divergence speed of $N$ relative to $T$ depends on the structure of the spatio-temporal processes $\{(\boldsymbol{x}_t^{(c)},\epsilon_{\boldsymbol{j},t}):\ \ \boldsymbol{j}\in V_{N,c}$ and $t\in[1,T]$, $c=1,\dots,G$, and the decay rate of the mixing coefficient. This condition is weaker than the condition [proposed in Hahn and Moon (2010)] that allows $N$ to be some exponential function of $T$ (as such, $N$ needs to be much greater than $T$) under some common types of weak serial dependence.

Assumption 4 is reminiscent of the standard assumption [employed in the OLS regression] about the positive-definiteness of the square matrix involving regressors. It basically requires that the covariates $\boldsymbol{x}_{i,t}$ (or at least one of them) have non-zero cross-sectional means that may vary within each group, but these variations must be remarkably heterogeneous across groups. Therefore, if the probability limits of the common-group cross-sectional means $g_{*,c}x_{*,t}^{(c)} = \frac{1}{N}\sum_{i=1}^{N}u_{0,i,c}x_{i,t}$, $c=1,\dots,G$, are similar across groups, or one of them is zero in some group, then this assumption as well as the 'well-separated groups' assumption (Assumption 7 below) become invalid.

We now present the consistency of $\widetilde{\boldsymbol{\psi}}$ for the stationarity case:

**Theorem 1 (Consistency)** *Suppose that Assumptions 1, 2, 3, and 4 hold. Then,* $|\widetilde{\sigma}_\epsilon^2-\sigma_\epsilon^2| = o_p(1)$ *and* $\|\widetilde{\boldsymbol{\psi}}-\boldsymbol{\psi}_0\| = o_p\left(N^{-1/2}\right)$.

**Theorem 2 (Asymptotic Normality)** *Let the conditions for Theorem 1 hold. Then,*

$$\sqrt{NT}(\widetilde{\boldsymbol{\psi}} - \boldsymbol{\psi}_0) \xrightarrow{d} \sigma_\epsilon N\left(\mathbf{0}, [\boldsymbol{D}_{\phi_0}\boldsymbol{D}_g\boldsymbol{Q}_{zz}\boldsymbol{D}_g\boldsymbol{D}_{\phi_0}]^{-1}\right),$$

*where* $\underbrace{\boldsymbol{D}_\phi}_{(G(d_x+1)+1)\times(G(d_x+1)+1)} := diag(\boldsymbol{\phi}\otimes \mathbb{I}_{d_x}, \mathbb{I}_{G+1}).$

**Remark 5.1** *Since the CQL criterion function is nonlinear in the coefficients $\boldsymbol{\theta}$ and $\boldsymbol{\phi}$ of the error-correction representation defined via Eq. (3.2), it is not obvious to see the $\sqrt{NT}$-consistency of the CL estimators. To get some intuition about Theorem 2, we consider a linear panel data model with fixed effects and a group-specific slope coefficient: $y_{i,t} = \mu_i + \theta_g x_{i,t} + \epsilon_{i,t}$ for all $i$ in group $g \in [1, 2, \dots, G]$. Define $\mu_* := \frac{1}{N}\sum_{i=1}^N \mu_i$, $\overline{x}_g := \frac{1}{T}\sum_{t=1}^T x_{*,t}^{(g)}$, $\overline{y} := \frac{1}{T}\sum_{t=1}^T y_{*,t}$, $\overline{z}_g := \frac{1}{T}\sum_{t=1}^T x_{*,t}^{(g)} y_{*,t}$, and $\overline{x}_{g,c} := \frac{1}{T}\sum_{t=1}^T x_{*,t}^{(g)} x_{*,t}^{(c)}$ for $g, c \in [1, G]$, where $x_{*,t}^{(c)} := \frac{1}{N}\sum_{i=1}^N u_{i,c} x_{i,t}$ and $y_{*,t} := \frac{1}{N}\sum_{i=1}^N y_{i,t}$. The 'oracle' CQL estimate of $\boldsymbol{\psi}_0 := (\mu_{*0}, \theta_{0,1}, \dots, \theta_{0,G})^\top$ is given by*

$$\widetilde{\boldsymbol{\psi}} := \begin{bmatrix} 1 & \overline{x}_1 & \cdots & \overline{x}_G \\ 1 & \overline{x}_{1,1} & \cdots & \overline{x}_{1,G} \\ \vdots & \vdots & \ddots & \vdots \\ 1 & \overline{x}_{G,1} & \cdots & \overline{x}_{G,G} \end{bmatrix}^{-1} \begin{bmatrix} \overline{y} \\ \overline{z}_1 \\ \vdots \\ \overline{z}_G \end{bmatrix}.$$

*One can then obtain that:*

$$\sqrt{NT}\left(\widetilde{\boldsymbol{\psi}} - \boldsymbol{\psi}_0\right) = \underbrace{\begin{bmatrix} 1 & \overline{x}_1 & \cdots & \overline{x}_G \\ 1 & \overline{x}_{1,1} & \cdots & \overline{x}_{1,G} \\ \vdots & \vdots & \ddots & \vdots \\ 1 & \overline{x}_{G,1} & \cdots & \overline{x}_{G,G} \end{bmatrix}^{-1}}_{\mathcal{A}} \begin{bmatrix} \sqrt{\frac{N}{T}}\sum_{t=1}^T \epsilon_{*,t} \\ \sqrt{\frac{N}{T}}\sum_{t=1}^T x_{*,t}^{(1)}\epsilon_{*,t} \\ \vdots \\ \sqrt{\frac{N}{T}}\sum_{t=1}^T x_{*,t}^{(G)}\epsilon_{*,t} \end{bmatrix}, \tag{5.1}$$

*where $\epsilon_{*,t} := \frac{1}{N}\sum_{i=1}^N \epsilon_{i,t}$. Cross-sectional means can well approximate stochastic trends. Therefore, by naively assuming $x_{i,t}$ to have an additive structure: $x_{i,t} = x_{g,t} + x_i$ with $E[x_i] = 0$ for each unit $i$ in group $g$, one can obtain from law of large numbers that $x_{*,t}^{(g)} \approx x_{g,t}$. Moreover, since $\epsilon_{i,t}$ is a centered random error, we previously argued that $\sqrt{N}\epsilon_{*,t}$ can be approximated by a normal random variable, say $\mathcal{N}_t$. By applying a central limit theorem, it then follows that $\sqrt{\frac{N}{T}}\sum_{t=1}^T \epsilon_{*,t}$ and $\sqrt{\frac{N}{T}}\sum_{t=1}^T x_{*,t}^{(g)}\epsilon_{*,t}$, $g = 1, \dots, G$, can be approximated by mean-zero normal random variables as long as $x_{*,t}^{(g)}$ and $\epsilon_{*,t}$ are uncorrelated. The Gram matrix $\mathcal{A}$ can converge to a finite positive definite matrix provided that $x_{g,t}$, $g = 1, \dots, G$, are heterogeneous across groups. We can therefore obtain the $\sqrt{NT}$-consistency. The same intuition carries over to general error-correction models.*

In the second case when one assumes that, at each location, $\boldsymbol{i} \in V_N$, $\boldsymbol{x}_{\boldsymbol{i},t}$ is an integrated process of order 1; besides, the spatio-temporal processes $\{\boldsymbol{x}_{\boldsymbol{j},t} : \boldsymbol{j} \in V_{N,c} \text{ and } t \in [1, T]\}$, $c = 1, \dots, G$, are heterogeneous across groups. To be precise, we state the following assumption:

**Assumption 5** *Let $\boldsymbol{x}_{\boldsymbol{i},t} = \sum_{s=1}^{t} \boldsymbol{\eta}_{\boldsymbol{i},s}$, where $\boldsymbol{\eta}_{\boldsymbol{i},s}$ is a mixing centered spatio-temporal process; and within each group, $c \in [1, G]$, the random variables $\{\boldsymbol{\eta}_{\boldsymbol{j},t},\ \boldsymbol{j} \in V_{N,c}$ and $t \in [1, T]\}$ are identically distributed across space and time. Moreover,*

**(a)** *the mixing coefficient $\alpha(\cdot)$ for $\{(\boldsymbol{\eta}_{\boldsymbol{j},t}, \epsilon_{\boldsymbol{j},t}) :\ \boldsymbol{j} \in V_{N,c}$ and $t \in [1, T]\}$, $c = 1, \dots, G$, decays to zero at a certain rate such that $\alpha(\tau) \le C_\theta \tau^{-\theta_\alpha}$ with some*

$$\theta_\alpha \ge \max\left(\frac{p(d_v+1)\gamma_\eta}{(p-q)(\gamma_\eta - 2)} + (d_v+1)\gamma_M, \frac{d_v+1}{1-\frac{2}{\gamma_\eta}}, \frac{2p}{p-4} - \gamma_M\right),$$

*where the generic constants $C_\theta > 0$, $\gamma_\eta > 2$, $p > 4$, $q = \frac{2p}{p-2}$, $d_v$ is the dimension of $V_N$, and $\gamma_M \ge 1$ is provided in Definition B.1;*

**(b)** $\max_{\boldsymbol{j}} \left(E\|\boldsymbol{\eta}_{\boldsymbol{j},1}\|^p, E\|\boldsymbol{\eta}_{\boldsymbol{j},1}\|^{\gamma_\eta}, E\|\boldsymbol{\eta}_{\boldsymbol{j},1}\|^4\right) < \infty$;

**(c)** $\max\left(|V_N|^{\gamma_M} T^{\gamma_M+1-\theta_\alpha}, |V_N|^{\gamma_M(1-2/\gamma_\eta)} T^{\epsilon-1/2}, T^{(\gamma_M+\theta_\alpha-1)\epsilon-\frac{1}{2}(\theta_\alpha-\gamma_M-1)}|V_N|^{\gamma_M}\right) \downarrow 0$ *for some* $\epsilon \in \left(0, \min\left(\frac{1}{2}, \frac{\theta_\alpha-\gamma_M-1}{2(\gamma_M+\theta_\alpha-1)}\right)\right)$.

**Assumption 6** *Let $\boldsymbol{X}_{N,T,t} \equiv \boldsymbol{X}_{N,T,t}(\boldsymbol{\theta}_0)$ be the same as in Assumption 4. It is assumed that the minimum eigenvalue of the stochastic limiting matrix of the normalized Gram matrix:*

$$\begin{aligned}\boldsymbol{Q}_{zz} \coloneqq plim_{N,T\uparrow\infty}\Bigg\{ & diag\left(T^{-1/2}\mathbb{I}_{G\times d_x}, N^{-1/2}\mathbb{I}_{G+1}\right)\boldsymbol{D}_g\left(\frac{N}{T}\sum_{t=1}^{T}\boldsymbol{X}_{N,T,t}\boldsymbol{X}_{N,T,t}^\top\right) \\ & \times \boldsymbol{D}_g^\top diag\left(T^{-1/2}\mathbb{I}_{G\times d_x}, N^{-1/2}\mathbb{I}_{G+1}\right)^\top\Bigg\}\end{aligned}$$

*is positive.*

A few remarks are now in order. It is worth noting that, when the covariates have common-group stochastic trends, say $\boldsymbol{x}_{\boldsymbol{i},t} = \boldsymbol{x}_t^{(c)} + \sum_{s=1}^{t} \boldsymbol{\eta}_{\boldsymbol{i},s}$ for every $\boldsymbol{i} \in V_{N,c}$ with $c = 1, \dots, G$, the asymptotic results developed here still hold for the most part [while the proofs may require slight modification involving cumbersome notations] as long as $\boldsymbol{x}_t^{(c)}$ is a stationary centered vector-valued process, thus it is dominated by the partial sum of noises $\sum_{s=1}^{t} \boldsymbol{\eta}_{\boldsymbol{i},s}$. Assumption 5(a) requires that the mixing coefficient should vanish at a rate depending on the orders of the moments specified in Assumption 5(b). Assumption 5(c) refers to the growth rates of $V_N$ and $T$, which also depend on the dimension and structure of $V_N$ as well as the decay rate of the mixing coefficient. Assumption 6 again bears some congruence with the standard assumption [employed in the OLS regression] about the positive-definiteness of the square matrix involving regressors

**Theorem 3 (Consistency)** *Suppose that Assumptions 1, 2, 5, and 6 hold. Then,* $|\widetilde{\sigma}_\epsilon^2 - \sigma_{\epsilon,0}^2| = o_p(1)$, $\|\widetilde{\boldsymbol{\theta}} - \boldsymbol{\theta}_0\| = o_p\left(T^{-1/2}\right)$, $\|\widetilde{\boldsymbol{\phi}} - \boldsymbol{\phi}_0\| = o_p\left(N^{-1/2}\right)$, *and* $|\widetilde{\mu}_* - \mu_{*0}| = o_p\left(N^{-1/2}\right)$.

To derive the limiting distribution of $\widetilde{\boldsymbol{\psi}}$, we define some further notations.

$$\underbrace{\mathcal{H}_{N,T}^{(ab)}(\boldsymbol{\phi})}_{G\times d_x\times G} := \text{diag}(\boldsymbol{g}\mathbb{I}_{d_x})\text{diag}(\boldsymbol{\phi}\mathbb{I}_{d_x})\left\{\frac{N^{1/2}}{T^{3/2}}\sum_{t=1}^{T}\boldsymbol{A}_t\boldsymbol{B}_t(\boldsymbol{\theta}_0)^\top\right\}\text{diag}(\boldsymbol{g}),$$

$$\underbrace{\mathcal{H}_{N,T}^{(ac)}(\boldsymbol{\phi})}_{G\times d_x\times 1} := \text{diag}(\boldsymbol{g}\mathbb{I}_{d_x})\text{diag}(\boldsymbol{\phi}\mathbb{I}_{d_x})\left\{\frac{N^{1/2}}{T^{3/2}}\sum_{t=1}^{T}\boldsymbol{A}_tC_t\right\},$$

$$\underbrace{\mathcal{H}_{N,T}^{(bc)}}_{G\times 1} := \text{diag}(\boldsymbol{g})\left\{\frac{1}{T}\sum_{t=1}^{T}\boldsymbol{B}_t(\boldsymbol{\theta}_0)C_t\right\},$$

$$\underbrace{\mathcal{H}_{N,T}^{(aa)}(\boldsymbol{\phi})}_{G\times d_x\times G\times d_x} := \text{diag}(\boldsymbol{g}\mathbb{I}_{d_x})\text{diag}(\boldsymbol{\phi}\mathbb{I}_{d_x})\left\{\frac{N}{T^2}\sum_{t=1}^{T}\boldsymbol{A}_t\boldsymbol{A}_t^\top\right\}\text{diag}(\boldsymbol{g}\mathbb{I}_{d_x})\text{diag}(\boldsymbol{\phi}\mathbb{I}_{d_x}),$$

and

$$\underbrace{\mathcal{H}_{N,T}^{(bb)}}_{G\times G} := \text{diag}(\boldsymbol{g})\left\{\frac{1}{T}\sum_{t=1}^{T}\boldsymbol{B}_t(\boldsymbol{\theta}_0)\boldsymbol{B}_t(\boldsymbol{\theta}_0)^\top\right\}\text{diag}(\boldsymbol{g}).$$

Let $\mathcal{H}_{N,T}(\boldsymbol{\phi}_0) := \begin{pmatrix} \mathcal{H}_{N,T}^{(aa)}(\boldsymbol{\phi}_0) & \mathcal{H}_{N,T}^{(ab)}(\boldsymbol{\phi}_0) & \mathcal{H}_{N,T}^{(ac)}(\boldsymbol{\phi}_0) \\ \mathcal{H}_{N,T}^{(ab)}(\boldsymbol{\phi}_0)^\top & \mathcal{H}_{N,T}^{(bb)} & \mathcal{H}_{N,T}^{(bc)} \\ \mathcal{H}_{N,T}^{(ac)}(\boldsymbol{\phi}_0)^\top & \mathcal{H}_{N,T}^{(bc)\top} & 1 \end{pmatrix}$. Lemma 5 in the Supplemental Material effectively implies that

$$\lim_{N,T\uparrow\infty}\mathcal{H}_{N,T}(\boldsymbol{\phi}_0) = \mathcal{H}(\boldsymbol{\phi}_0),$$

where $\mathcal{H}(\boldsymbol{\phi}_0)$ is a positive-definite stochastic matrix.

**Theorem 4 (Asymptotic Normality)** *Let Assumptions 1, 2, 5, and 6 hold. Then,*

$$\begin{pmatrix} T(\widetilde{\boldsymbol{\theta}} - \boldsymbol{\theta}_0) \\ \sqrt{NT}(\widetilde{\boldsymbol{\phi}} - \boldsymbol{\phi}_0) \\ \sqrt{NT}(\widetilde{\mu}_* - \mu_{*0}) \end{pmatrix} \xrightarrow{w} MN\left(\mathbf{0}, \mathcal{H}(\boldsymbol{\phi}_0)^{-1}\right),$$

*where* $MN(\cdot,\cdot)$ *stands for a mixed-normal random variable.*

### 5.2 Unknown Group Membership

We start by defining some common notations that will be used specifically for the rest of this section. Let $\boldsymbol{u}_c = (u_{1,c}, \dots, u_{N,c})^\top$ be a $N \times 1$ vector of group membership indicators associated with group $c \in [1, G]$. In addition, with a slight abuse of notation, some of the symbols to be defined below may look the same as in Section 5.1.

$$
\begin{aligned}
\xi^{(w)}_{*,t}(\boldsymbol{u}_c) \equiv \xi^{(w)}_{*,t}(\boldsymbol{\theta}_{0,c}, \boldsymbol{u}_c) &\coloneqq \frac{1}{N}\sum_{i=1}^{N} u_{i,c}\xi^{(w)}_{i,t}(\boldsymbol{\theta}_{0,c}), \text{ where } \xi^{(w)}_{i,t}(\boldsymbol{\theta}_c) \coloneqq y^{(w)}_{i,t-1} - \boldsymbol{\theta}_c^\top \boldsymbol{x}^{(w)}_{i,t}, \\
\boldsymbol{x}^{(w)}_{*,t}(\boldsymbol{u}_c) &\coloneqq \frac{1}{N}\sum_{i=1}^{N} u_{i,c}\boldsymbol{x}^{(w)}_{i,t}, \\
\boldsymbol{\xi}^{(w)}_{*,t}(\boldsymbol{U}, \sigma^{(per)}) &\coloneqq \left(\xi^{(w)}_{*,t}(\boldsymbol{u}_{\sigma^{(per)}(1)}), \dots, \xi^{(w)}_{*,t}(\boldsymbol{u}_{\sigma^{(per)}(G)})\right)^\top, \\
\boldsymbol{x}^{(w)}_{*,t}(\boldsymbol{U}, \sigma^{(per)}) &\coloneqq \left(\boldsymbol{x}^{(w)\top}_{*,t}(\boldsymbol{u}_{\sigma^{(per)}(1)}), \dots, \boldsymbol{x}^{(w)\top}_{*,t}(\boldsymbol{u}_{\sigma^{(per)}(G)})\right)^\top, \\
\boldsymbol{F}_t(\boldsymbol{U}, \boldsymbol{U}_0) &\coloneqq \left(\boldsymbol{x}^{(w)\top}_{*,t}(\boldsymbol{U}, \widetilde{\sigma}^{(per)}), \boldsymbol{\xi}^{(w)\top}_{*,t}(\boldsymbol{U}, \widetilde{\sigma}^{(per)}), 1^{(w)}_t, \boldsymbol{\xi}^{(w)\top}_{*,t}(\boldsymbol{U}_0, \sigma^{(per)}) - \boldsymbol{\xi}^{(w)\top}_{*,t}(\boldsymbol{U}, \widetilde{\sigma}^{(per)})\right)^\top, \\
\boldsymbol{D}_\phi(\widetilde{\sigma}^{(per)}) &\coloneqq \operatorname{diag}\left(\phi_{\widetilde{\sigma}^{(per)}(1)}, \dots, \phi_{\widetilde{\sigma}^{(per)}(G)}\right), \\
\boldsymbol{\theta}^{(\sigma^{(per)})} &\coloneqq \left(\boldsymbol{\theta}^\top_{\sigma^{(per)}(1)}, \dots, \boldsymbol{\theta}^\top_{\sigma^{(per)}(G)}\right)^\top, \\
\boldsymbol{\phi}^{(\sigma^{(per)})} &\coloneqq \left(\phi_{\sigma^{(per)}(1)}, \dots, \phi_{\sigma^{(per)}(G)}\right)^\top.
\end{aligned}
$$

Therefore, in view of (4.13), one obtains that

$$
\begin{aligned}
\epsilon_{*,t}(\boldsymbol{\psi}, \boldsymbol{U}) - \epsilon^{(w)}_{0,*,t} = \Big((\boldsymbol{\theta}^{(\widetilde{\sigma}^{(per)})} - \boldsymbol{\theta}_0^{(\sigma^{(per)})})^\top, (\boldsymbol{\phi}_0^{(\sigma^{(per)})} - \boldsymbol{\phi}^{(\widetilde{\sigma}^{(per)})})^\top, \mu_{*0} - \mu_*, \boldsymbol{\phi}_0^{(\sigma^{(per)})\top}\Big) \\
\times \operatorname{diag}\left(\boldsymbol{D}_\phi(\widetilde{\sigma}^{(per)}), \mathbb{I}_{2G+1}\right) \boldsymbol{F}_t(\boldsymbol{U}, \boldsymbol{U}_0).
\end{aligned}
$$

Moreover, let

$$
\begin{aligned}
H(\widehat{\boldsymbol{U}}, \boldsymbol{U}_0) \coloneqq &\Bigg(\max_{\sigma^{(per)} \in \sigma(\mathcal{P})} \min_{\widetilde{\sigma}^{(per)} \in \sigma(\mathcal{P})} \frac{1}{N}\sum_{c=1}^{G}\sum_{i=1}^{N} |\widehat{u}_{i,\widetilde{\sigma}^{(per)}(c)} - u_{0,i,\sigma^{(per)}(c)}|, \\
&\max_{\widetilde{\sigma}^{(per)} \in \sigma(\mathcal{P})} \min_{\sigma^{(per)} \in \sigma(\mathcal{P})} \frac{1}{N}\sum_{c=1}^{G}\sum_{i=1}^{N} |\widehat{u}_{i,\widetilde{\sigma}^{(per)}(c)} - u_{0,i,\sigma^{(per)}(c)}|\Bigg)^+ \\
= &\left(\min_{\widetilde{\sigma}^{(per)} \in \sigma(\mathcal{P})} \frac{1}{N}\sum_{c=1}^{G}\sum_{i=1}^{N} |\widehat{u}_{i,\widetilde{\sigma}^{(per)}(c)} - u_{0,i,c}|, \min_{\sigma^{(per)} \in \sigma(\mathcal{P})} \frac{1}{N}\sum_{c=1}^{G}\sum_{i=1}^{N} |\widehat{u}_{i,c} - u_{0,i,\sigma^{(per)}(c)}|\right)^+,
\end{aligned}
$$

where $\sigma(\mathcal{P})$ is the set of all permutation operators operating on $\mathcal{P}$, denote the *optimal matching* distance between $\widehat{\boldsymbol{U}}$ and $\boldsymbol{U}_0$; and

$$
\begin{aligned}
H(\widehat{\boldsymbol{\psi}}, \boldsymbol{\psi}_0) &:= \Bigg( \max_{\sigma^{(per)} \in \sigma(\mathcal{P})} \min_{\widetilde{\sigma}^{(per)} \in \sigma(\mathcal{P})} \left( \sum_{c=1}^{G} \left\| \widehat{\boldsymbol{\psi}}_{\widetilde{\sigma}^{(per)}(c)} - \boldsymbol{\psi}_{0,\sigma^{(per)}(c)} \right\|^2 \right)^{\frac{1}{2}}, \\
&\qquad \max_{\widetilde{\sigma}^{(per)} \in \sigma(\mathcal{P})} \min_{\sigma^{(per)} \in \sigma(\mathcal{P})} \left( \sum_{c=1}^{G} \left\| \widehat{\boldsymbol{\psi}}_{\widetilde{\sigma}^{(per)}(c)} - \boldsymbol{\psi}_{0,\sigma^{(per)}(c)} \right\|^2 \right)^{\frac{1}{2}} \Bigg)^{+} \\
&= \left( \min_{\widetilde{\sigma}^{(per)} \in \sigma(\mathcal{P})} \left( \sum_{c=1}^{G} \left\| \widehat{\boldsymbol{\psi}}_{\widetilde{\sigma}^{(per)}(c)} - \boldsymbol{\psi}_{0,c} \right\|^2 \right)^{\frac{1}{2}}, \min_{\sigma^{(per)} \in \sigma(\mathcal{P})} \left( \sum_{c=1}^{G} \left\| \widehat{\boldsymbol{\psi}}_{c} - \boldsymbol{\psi}_{0,\sigma^{(per)}(c)} \right\|^2 \right)^{\frac{1}{2}} \right)^{+},
\end{aligned}
$$

where $\widehat{\boldsymbol{\psi}}_c := (\widehat{\boldsymbol{\theta}}_c^\top, \widehat{\phi}_c, \widehat{\mu}_*)^\top$ and $\boldsymbol{\psi}_{0,c} := (\boldsymbol{\theta}_{0,c}^\top, \phi_{0,c}, \mu_{*0})^\top$, be the *optimal matching* distance between $\widehat{\boldsymbol{\psi}}$ and $\boldsymbol{\psi}_0$.

For the *stationary* case, we first need to state the following assumption:

**Assumption 7** *Suppose that* $\lim_{N\uparrow\infty, T\uparrow\infty} \inf_{H(\boldsymbol{U},\boldsymbol{U}_0)>\eta_u} \lambda_{\min}\left(\frac{1}{T}\sum_{t=1}^{T} \boldsymbol{F}_t(\boldsymbol{U},\boldsymbol{U}_0)\boldsymbol{F}_t(\boldsymbol{U},\boldsymbol{U}_0)^\top\right) > 0$. *Moreover, let* $\boldsymbol{F}_t^{(1)}(\boldsymbol{U}) := \left(\boldsymbol{x}_{*,t}^{(w)\top}(\boldsymbol{U}, \widetilde{\sigma}^{(per)}), \boldsymbol{\xi}_{*,t}^{(w)\top}(\boldsymbol{U}, \widetilde{\sigma}^{(per)}), 1_t^{(w)}\right)^\top \subset \boldsymbol{F}_t(\boldsymbol{U},\boldsymbol{U}_0)$, *assume that*

$$
\lim_{N\uparrow\infty, T\uparrow\infty} \inf_{H(\boldsymbol{U},\boldsymbol{U}_0)<\eta_u} \lambda_{\min}\left(\frac{1}{T}\sum_{t=1}^{T} \boldsymbol{F}_t^{(1)}(\boldsymbol{U})\boldsymbol{F}_t^{(1)\top}(\boldsymbol{U})\right) > 0.
$$

Assumption 7 states that groups must be well-separated in the sense that, if two matrices of group membership indicators, $\boldsymbol{U}$ and $\boldsymbol{U}_0$, are mismatched, then the Gram matrix involving differences, $\boldsymbol{\xi}_{*,t}^{(w)\top}(\boldsymbol{U}_0, \sigma^{(per)}) - \boldsymbol{\xi}_{*,t}^{(w)\top}(\boldsymbol{U}, \widetilde{\sigma}^{(per)})$, will be a positive-definite matrix. The second part of Assumption 7 is similar to the standard assumption employed in the OLS regression.

**Theorem 5** *Under Assumptions 3, 4 and 7, it holds that* $\sqrt{N} H(\widehat{\boldsymbol{\psi}}, \boldsymbol{\psi}_0) \xrightarrow{p} 0$, $H(\widehat{\boldsymbol{U}}, \boldsymbol{U}_0) \xrightarrow{p} 0$, *and* $|\widehat{\sigma}_\epsilon^2 - \sigma_{\epsilon,0}^2| \xrightarrow{p} 0$.

Theorem 6 below gives the expected bias [in terms of the *optimal matching* distance] of the estimates $\widehat{\boldsymbol{U}}(\boldsymbol{\psi})$ uniformly over all $\boldsymbol{\psi}$ in a local neighborhood of $\boldsymbol{\psi}_0$. The rate at which this expected bias goes to zero also depends on the decay rate of the mixing coefficient.

**Theorem 6** *Let* $\{(\boldsymbol{x}_t^{(c)}, \epsilon_{\boldsymbol{i},t}) : \ \boldsymbol{i} \in V_{N,c},\ t \in [1,T]\}$ *with* $c = 1, \dots, G$ *represent a mixing vector-valued spatio-temporal process and* $\widehat{\boldsymbol{U}}(\boldsymbol{\psi}) := \text{argmin}_{\boldsymbol{U} \in \Delta_S^N \bigcap \{0,1\}^{G\times N}} \frac{N}{T}\sum_{t=1}^{T} \epsilon_{*,t}^2(\boldsymbol{\psi}, \boldsymbol{U})$. *Suppose that (a) within each group,* $c \in [1,G]$, $\{\boldsymbol{x}_{\boldsymbol{i},t}, \epsilon_{\boldsymbol{i},t}\}$, $\boldsymbol{i} \in V_N$ *and* $t \in [1,T]$ *are identically distributed over time and space; (b) the mixing coefficient* $\alpha(\tau) < C_0 \tau^{-\theta_\alpha}$, $\theta_\alpha > \left(\frac{4\gamma_M}{3}, \frac{2d_v+1}{1-2/\delta_\alpha}\right)^+$ *for some* $\delta_\alpha > 2$; *(c)* $\|\boldsymbol{x}_{\boldsymbol{i},t}\epsilon_{\boldsymbol{i},t}\|_{\delta_\alpha} < \infty$; *(d)* $\max_{\boldsymbol{j}} E[\exp(\ell|\epsilon_{\boldsymbol{j},t}|)] \le C_\ell$ *and* $\max_{\boldsymbol{j}} E[\exp(\ell\,\|\boldsymbol{x}_{\boldsymbol{j},t}\|)] \le C_\ell$ *for a positive constant,* $C_\ell > 0$, *and* $\ell > 0$ *small enough. Then, it holds that*

$$E\left[\sup_{\psi\in\mathcal{B}(\psi_0,\eta_\psi)} H\left(\widehat{\boldsymbol{U}}(\boldsymbol{\psi}),\boldsymbol{U}_0\right)\right] \;\leq\; C_0\left\{T^{-C_\alpha}+N^{2\gamma_M}\log^2(T)T^{\gamma_M-\frac{3}{4}\theta_\alpha}+\exp\left(-C_M\frac{T^{1/4}}{\log^2(T)}\right)\right\},$$

*where $\mathcal{B}(\boldsymbol{\psi}_0,\eta_\psi)$ is an open ball centered at $\boldsymbol{\psi}_0$ with an arbitrarily small radius, $\eta_\psi$, in terms of the optimal matching distance; and $C_\alpha$ and $C_M$ denote some sufficiently large constants that do not depend on $N$ or $T$.*

**Remark 5.2** *It is important to note that most conditions in Theorem 6 above involve standard bounded moments and decay rates of the cardinality-based mixing coefficient, except Condition (d). The sub-exponential tails of $\epsilon_{\boldsymbol{j},t}$ and $\boldsymbol{x}_{\boldsymbol{j},t}$ assumed there are needed to apply the truncation technique that yields the first term $T^{-C_\alpha}$ in the decay rate. The same condition is employed in Mammen, Rothe, and Schienle (2012). This condition can be satisfied if the stationary density functions of $\epsilon_{\boldsymbol{j},t}$ and $\boldsymbol{x}_{\boldsymbol{j},t}$ have compact supports.*

In light of Theorem 6, we readily derive the decay rate of the *optimal matching* distance between the CQL estimates $\widehat{\boldsymbol{\psi}}$ and the 'oracle' CQL estimates (i.e. the estimates constructed by maximizing the CQL function using the 'true' unknown groups $\boldsymbol{U}_0$) $\widetilde{\boldsymbol{\psi}}$. The main result is stated in Theorem 7. Again, this decay rate depends on the decay rate of the mixing coefficient.

**Theorem 7** *Let all the conditions in Theorem 6 hold. Under Assumptions 3, 4 and 7, it holds that*

$$H\left(\widehat{\boldsymbol{\psi}},\widetilde{\boldsymbol{\psi}}\right)=O_p\left(NT^{-C_\alpha}+N^{\gamma_M+1}\log(T)T^{\frac{\gamma_M}{2}-\frac{3}{8}\theta_\alpha}+N\exp\left(-C_M\frac{T^{1/4}}{\log^2(T)}\right)\right),$$

*where $C_\alpha$ and $C_M$ are some sufficiently large constants that do not depend on $N$ or $T$.*

**Remark 5.3** *Theorem 7 above shows that, if the mixing exponent $\theta_\alpha$ is sufficiently large, the* optimal matching *distance between the CQL estimate and the 'oracle' CQL estimate of $\boldsymbol{\psi}_0$ can decay to zero faster than $O_p\left(1/\sqrt{NT}\right)$ so that they are distributionally equivalent in the limit. From a practical point of view the availability of panel data permits the possibility of re-sampling each individual $T$ times so that the probability of making a misclassification error is quite small even when $N$ is quite large.*

For the *nonstationary* case, let

$$\boldsymbol{\Lambda}_{N,T}(\boldsymbol{U},\boldsymbol{U}_0):=\operatorname{diag}\left(\sqrt{\frac{N}{T}}\mathbb{I}_{G\times d_x},\mathbb{I}_{2G+1}\right)\frac{1}{T}\sum_{t=1}^{T}\boldsymbol{F}_t(\boldsymbol{U},\boldsymbol{U}_0)\boldsymbol{F}_t(\boldsymbol{U},\boldsymbol{U}_0)^\top\operatorname{diag}\left(\sqrt{\frac{N}{T}}\mathbb{I}_{G\times d_x},\mathbb{I}_{2G+1}\right)$$

represent a normalized Gram matrix. It then follows from Lemma 5 in the Supplemental Material that

$$\boldsymbol{\Lambda}_{N,T}(\boldsymbol{U},\boldsymbol{U}_0)\xrightarrow{w}\boldsymbol{\Lambda}(\boldsymbol{U},\boldsymbol{U}_0),$$

where the limit $\mathbf{\Lambda}(\boldsymbol{U}, \boldsymbol{U}_0)$ is a stochastic matrix. We first state a variant of Assumption 7 about group well-separability in Assumption 8 below.

**Assumption 8** $\liminf_{H(\boldsymbol{U},\boldsymbol{U}_0)>\eta_u} \mathbf{\Lambda}(\boldsymbol{U}, \boldsymbol{U}_0) > 0$ *for every* $\eta_u > 0$.

**Theorem 8** *Let* $\boldsymbol{x}_{\boldsymbol{j},t} = \sum_{s=1}^{t} \boldsymbol{\eta}_{\boldsymbol{j},s}$, *where* $\{(\boldsymbol{\eta}_{\boldsymbol{j},t}, \epsilon_{\boldsymbol{j},t}) : \ \boldsymbol{j} \in V_N, \ t \in [1, T]\}$ *is a mixing vector-valued spatio-temporal process. Suppose that (a) within each group, say* $c \in [1, G]$, $\{\boldsymbol{\eta}_{\boldsymbol{i},t}, \epsilon_{\boldsymbol{i},t}\}$, $\boldsymbol{i} \in V_{N,c}$ *and* $t \in [1, T]$ *are identically distributed over space and time; (b) the mixing coefficient* $\alpha(\tau) < C_0\tau^{-\theta_\alpha}$, $\theta_\alpha > \left(\frac{4\gamma_M}{3}, \frac{2d_v+1}{1-2/\delta_\alpha}\right)^+$ *for some* $\delta_\alpha > 2$; *(c)* $\max_{\boldsymbol{j}} \|\boldsymbol{\eta}_{\boldsymbol{j},t}\epsilon_{\boldsymbol{i},t}\|_{\delta_\alpha} < \infty$; *(d)* (sub-exponential tails) $\max_{\boldsymbol{j}} E[\exp(\ell|\epsilon_{\boldsymbol{j},t}|)] \leq C_\ell$ *and* $\max_{\boldsymbol{j}} E[\exp(\ell \|\boldsymbol{\eta}_{\boldsymbol{j},t}\|)] \leq C_\ell$ *for a positive constant,* $C_\ell > 0$, *and* $\ell > 0$ *small enough; (e)* $N/T \longrightarrow const.$ *Then, under Assumptions 1, 2, 5, and 8, it holds that* $\sqrt{T}H\left(\widehat{\boldsymbol{\theta}}, \boldsymbol{\theta}_0\right) = o_p(1)$, $\sqrt{N}H\left(\widehat{\boldsymbol{\phi}}, \boldsymbol{\phi}_0\right) = o_p(1)$, $\sqrt{N}|\widehat{\mu}_* - \mu_{*0}| = o_p(1)$, *and* $|\widehat{\sigma}^2_\epsilon - \sigma^2_{\epsilon,0}| = o_p(1)$.

**Remark 5.4** *Conditions (a) - (d) in Theorem 8 are similar to those in Theorem 6, except Condition (e). This condition basically requires that* $N$ *and* $T$ *should grow at the same speed, or* $N$ *must grow much more slowly than* $T$ *while this growth rate of* $N$ *relative to* $T$ *is not needed in the stationary case. A simple explanation for this requirement is that, when covariates follow unit-root processes, a large number of time periods is needed to uncover a long-run relationship. Also, as the distributions of* $y_{i,t}$ *and* $\boldsymbol{x}_{i,t}$ *are not stable, one may need in principle even more time-series observations when the number of individuals gets bigger in order to achieve a negligible classification error.*

Theorems 9 and 10 below provide the decay rates for the expected uniform bias of the estimates of the 'true' group membership indicators, and for the *optimal matching* distance between the CQL estimates $\widehat{\boldsymbol{\psi}}$ and the 'oracle' CQL estimates $\widetilde{\boldsymbol{\psi}}$ of $\boldsymbol{\psi}_0$. These theorems are the 'nonstationary' versions of Theorems 6 and 7 (given above) respectively.

**Theorem 9** *Let* $\widehat{\boldsymbol{U}}(\boldsymbol{\psi}) := argmin_{\boldsymbol{U}\in\Delta_S^N \bigcap\{0,1\}^{G\times N}} \frac{N}{T} \sum_{t=1}^{T} \epsilon^2_{*,t}(\boldsymbol{\psi}, \boldsymbol{U})$. *Suppose that the conditions of Theorem 8 hold. Then,*

$$E\left[\sup_{\boldsymbol{\psi}\in\mathcal{N}_{N,T}(\boldsymbol{\psi}_0,\eta_\psi)} H\left(\widehat{\boldsymbol{U}}(\boldsymbol{\psi}), \boldsymbol{U}_0\right)\right] \leq C_0\left\{N^{-C_\alpha} + T^{-C_\alpha} + N^{2\gamma_M}\log^2(T)T^{\gamma_M - \frac{3}{4}\theta_\alpha} + \exp\left(-C_M\frac{T^{1/4}}{\log^2(T)}\right)\right\}$$

*for some sufficiently large constants,* $C_\alpha$ *and* $C_M$, *that do not depend on* $N$ *or* $T$, *where* $\mathcal{N}_{N,T}(\boldsymbol{\psi}_0, \eta_\psi) := \bigcap_{\substack{\eta_\theta>0,\eta_\phi>0,\eta_\mu>0 \\ \left(\eta_\theta^2+\eta_\phi^2+\eta_\mu^2\right)^{\frac{1}{2}}=\eta_\psi}} \mathcal{B}_T(\boldsymbol{\theta}_0, \eta_\theta)\times\mathcal{B}_N(\boldsymbol{\phi}_0, \eta_\phi)\times\mathcal{B}_N(\mu_{*0}, \eta_\mu)$ *with* $\mathcal{B}_T(\boldsymbol{\theta}_0, \eta_\theta) := \{\boldsymbol{\theta} \in \Theta_\theta : \ \sqrt{T}H(\boldsymbol{\theta}, \boldsymbol{\theta}_0) < \eta_\theta\}$, $\mathcal{B}_N(\boldsymbol{\phi}_0, \eta_\phi) := \{\boldsymbol{\phi} \in \Theta_\phi : \ \sqrt{N}H(\boldsymbol{\phi}, \boldsymbol{\phi}_0) < \eta_\phi\}$, *and* $\mathcal{B}_N(\mu_{*0}, \eta_\mu) := \{\mu_* \in \Theta_\mu : \ \sqrt{N}|\mu_* - \mu_{*0}| < \eta_\mu\}$.

**Theorem 10** *Assume all the conditions presented in Theorem 9. Then, it holds under Assumption 6 that*

$$H\left(diag\left(\sqrt{T}\mathbb{I}_{G\times d_x}, \sqrt{N}\mathbb{I}_{G+1}\right)\widehat{\boldsymbol{\psi}}, diag\left(\sqrt{T}\mathbb{I}_{G\times d_x}, \sqrt{N}\mathbb{I}_{G+1}\right)\widetilde{\boldsymbol{\psi}}\right)$$
$$= O_p\left(N^{\frac{1-C_\alpha}{2}} + N^{1/2}T^{-\frac{C_\alpha}{2}} + N^{\gamma_M+\frac{1}{2}}\log(T)T^{\frac{\gamma_M}{2}-\frac{3}{8}\theta_\alpha} + N^{\frac{1}{2}}\exp\left(-C_M\frac{T^{1/4}}{2\log^2(T)}\right)\right),$$

*where $C_\alpha$ and $C_M$ are some sufficiently large constants that do not depend on $N$ or $T$.*

**Empirical choice of the optimal number of groups.** In the present maximum CQL paradigm, the optimal selection of the number of groups can be implemented by employing the following BIC-type information criterion. An information criterion is typically a sum of a goodness-of-fit measure and a penalty term used to account for the model complexity.

$$IC(G) := \frac{N}{T}\sum_{t=1}^{T}\epsilon_{*,t}^2\left(\widehat{\boldsymbol{\psi}}, \widehat{\boldsymbol{U}}\right) + G\times\omega_N, \tag{5.2}$$

where $\underbrace{\widehat{\boldsymbol{U}}}_{G\times N}$ consists of CQL estimates for the 'true' group membership indicators $\boldsymbol{U}_0$ given a number of groups, $G$; $\widehat{\boldsymbol{\psi}}$ is the vector containing CQL estimates for the model parameters associated with the group classification provided by $\widehat{\boldsymbol{U}}$; both $\widehat{\boldsymbol{\psi}}$ and $\widehat{\boldsymbol{U}}$ are found by maximizing the CQL function in (4.11); and $\omega_N$ is a penalty term that diverges with $N$ at a certain rate (to be specified in Assumption 9 below).

Theorem 11 below suggests that $G^* = \mathrm{argmin}_G IC(G)$ can consistently estimate the true number of groups $G_0$. To state this theorem, define some further notations.

$$\boldsymbol{X}_{G,P,K,t}(\boldsymbol{\theta}) := \left(1_t^{(w)}, \xi_{0,*,t}^{(w)}(\boldsymbol{\theta}_{0,1}), \dots, \xi_{0,*,t}^{(w)}(\boldsymbol{\theta}_{0,G}), \boldsymbol{x}_{*,t}^{(w)}(\boldsymbol{u}_{0,1})^\top, \dots, \boldsymbol{x}_{*,t}^{(w)}(\boldsymbol{u}_{0,P})^\top, \underbrace{\boldsymbol{\xi}_t^{(w)}(\boldsymbol{\theta}_1)^\top}_{1\times N}, \dots, \boldsymbol{\xi}_t^{(w)}(\boldsymbol{\theta}_K)^\top\right)^\top,$$
$$\text{where}\quad \boldsymbol{\xi}_t^{(w)}(\boldsymbol{\theta}_c) := \left(\xi_{1,t}^{(w)}(\boldsymbol{\theta}_c), \dots, \xi_{N,t}^{(w)}(\boldsymbol{\theta}_c)\right),$$
$$\boldsymbol{Z}_{G,P,t}(\boldsymbol{\theta}, \boldsymbol{U}) := \Big(1_t^{(w)}, \xi_{0,*,t}^{(w)}(\boldsymbol{\theta}_{0,1}), \dots, \xi_{0,*,t}^{(w)}(\boldsymbol{\theta}_{0,G_0}), \boldsymbol{x}_{*,t}^{(w)}(\boldsymbol{u}_1)^\top, \dots, \boldsymbol{x}_{*,t}^{(w)}(\boldsymbol{u}_G)^\top, \boldsymbol{\xi}_t^{(w)}(\boldsymbol{\theta}_{0,1})^\top, \dots, \boldsymbol{\xi}_t^{(w)}(\boldsymbol{\theta}_{0,p})^\top,$$
$$\boldsymbol{\eta}_{P,t}(\boldsymbol{\theta})^\top\Big)^\top,$$
$$\text{where}\quad \boldsymbol{\eta}_{P,t}(\boldsymbol{\theta}) := \begin{cases} \left(\boldsymbol{\xi}_t^{(w)}(\boldsymbol{\theta}_{0,p+1})^\top, \dots, \boldsymbol{\xi}_t^{(w)}(\boldsymbol{\theta}_{0,P})^\top\right)^\top & \text{if} \quad P \le G_0, \\ \left(\boldsymbol{\xi}_t^{(w)}(\boldsymbol{\theta}_{0,p+1})^\top, \dots, \boldsymbol{\xi}_t^{(w)}(\boldsymbol{\theta}_{0,G_0})^\top, \boldsymbol{\xi}_t^{(w)}(\boldsymbol{\theta}_{G_0+1})^\top, \boldsymbol{\xi}_t^{(w)}(\boldsymbol{\theta}_P)^\top\right)^\top & \text{if} \quad P > G_0; \end{cases}$$
$$\boldsymbol{\ell}_{G,P,N,T} := \begin{cases} \mathrm{diag}\left(1, \mathbb{I}_{G_0}, \sqrt{\frac{N}{T}}\mathbb{I}_{G\times d_x}, \mathbb{I}_{P\times N}\right) & \text{if} \quad P \le G_0, \\ \mathrm{diag}\left(1, \mathbb{I}_{G_0}, \sqrt{\frac{N}{T}}\mathbb{I}_{G_0\times d_x}, \mathbb{I}_{G\times N}, \frac{1}{T}\mathbb{I}_{(P-G_0)\times N}\right) & \text{if} \quad P > G_0; \end{cases}$$
$$\boldsymbol{\mathcal{X}}_{G,N,T}^*(\boldsymbol{\theta}) := \frac{1}{T}\sum_{t=1}^{T}\boldsymbol{X}_{G_0,G,G,t}(\boldsymbol{\theta})\boldsymbol{X}_{G_0,G,G,t}(\boldsymbol{\theta})^\top,$$

where all of its elements converge in probability by the weak law of large numbers,

and

$$\boldsymbol{\mathcal{X}}^{**}_{G,P,N,T}(\boldsymbol{\theta},\boldsymbol{U}) \coloneqq \frac{1}{T}\sum_{t=1}^{T}\boldsymbol{Z}_{G,P,t}(\boldsymbol{U},\boldsymbol{\theta})\boldsymbol{Z}_{G,P,t}(\boldsymbol{U},\boldsymbol{\theta})^{\top}.$$

We also need the following assumption:

**Assumption 9** *Assume either of the following conditions:*

(a) *The vector of covariates $\boldsymbol{x}_{i,t}$ is stationary for each $i \in [1,N]$, and satisfies Assumption 3. Also the minimum eigenvalue of $\boldsymbol{\mathcal{X}}^{*}_{N,T}(\boldsymbol{\theta})$ is strictly positive for every vector of pairwise different sub-vectors, $\boldsymbol{\theta}^* \coloneqq \{\boldsymbol{\theta}_1,\ldots,\boldsymbol{\theta}_G\}$, as $T$ and $N$ become large. That is,*

$$\lim_{T,N\uparrow\infty}\inf_{\substack{G\leq\overline{G}\\ \boldsymbol{\theta}^*}}\lambda_{\min}\left(\boldsymbol{\mathcal{X}}^{*}_{G,N,T}(\boldsymbol{\theta}^*)\right) > 0,$$

*where $\overline{G}$ is the maximum number of groups to cluster individuals.*

(b) *The covariates $\boldsymbol{x}_{i,t}$ follow unit-root processes for each $i \in [1,N]$, and satisfies Assumption 5. Besides, for every vector of pairwise different sub-vectors, $\boldsymbol{\theta}^{\diamond} \coloneqq \{\boldsymbol{\theta}_{G_0+1},\ldots,\boldsymbol{\theta}_P\}$, and groups with sizes that increase with $N$ (i.e., $\lim_{N\uparrow\infty}\frac{1}{N}\sum_{i=1}^{N}u_{i,c} > 0$ for every $c \in [1,G_0]$), it holds that $\lim_{T,N\uparrow\infty}\inf_{\substack{G\leq G_0\\ P<\overline{G}}}\inf_{\boldsymbol{U},\boldsymbol{\theta}^{\diamond}}\lambda_{\min}\left(\boldsymbol{\ell}_{G,P,N,T}\boldsymbol{\mathcal{X}}^{**}_{G,P,N,T}(\boldsymbol{\theta},\boldsymbol{U})\boldsymbol{\ell}_{G,P,N,T}\right) > 0$.*

Assumption 9 requires that the Gram matrix $\boldsymbol{\mathcal{X}}^{*}_{G,N,T}(\boldsymbol{\theta})$ has a positive minimum eigenvalue uniformly in $G$ and $\boldsymbol{\theta}^*$ for the stationary case. In the nonstationary case the elements of the normalized Gram matrix $\boldsymbol{\ell}_{G,P,N,T}\boldsymbol{\mathcal{X}}^{**}_{G,P,N,T}(\boldsymbol{\theta},\boldsymbol{U})\boldsymbol{\ell}_{G,P,N,T}$ are shown to converge in probability to non-zero stochastic limits for every $\boldsymbol{U}$ and $\boldsymbol{\theta}^{\diamond}$, and the matrix of these limiting elements also has a positive minimum eigenvalue. Since it is well known that the Gram matrices can become degenerate when $N$ is much greater than $T$, the regularity condition involving the positivity of the minimum eigenvalues can certainly be met if $N/T$ goes to a constant.

**Theorem 11** *Let $\omega_N = o(N)$ (for example, $\omega_N = \log N$). Then, under Assumption 9,*

$$Prob\left(\min_{G\neq G_0} IC(G) > IC(G_0)\right) \longrightarrow 1.$$

Some other information criteria using penalty functions of this form are Nishii's (1984) generalized information criterion (GIC) and Andrews's (1999) GMM-BIC. This BIC-type information criterion imposes a higher penalty for overfitting compared with the standard AIC, thus it is more suitable for selection of groups in moderate or large samples.

# 6 Empirical Application

An open economy can effectively finance its investment by borrowing abroad since domestic saving [as the main source of funds for investment] flow to wherever there are profitable investment projects. Therefore, high correlation between domestic saving and investment - both measured as percentages of gross domestic product (GDP) - empirically established in a regression model for open economies is well known as the Feldstein-Horioka puzzle (henceforth FHP). This puzzle started when Feldstein and Horioka (1980) (FH) showed, by using the cross-section data of 16 Organization for Economic Cooperation and Development (OECD) economies for the period 1960-1974, that temporally averaged national saving and domestic investment were highly correlated. They interpreted this high long-run correlation as an evidence of low international capital mobility. The FHP - which Obstfeld and Rogoff (2001) view as one of the six major puzzles in international macroeconomics - still persists as estimates of the saving-investment (SI) association for small open economies have remained quite high despite ongoing financial market integration and globalization over recent decades (see, e.g., Chang and Smith (2014)). The question as to whether the apparently high capital mobility is a chimera or an elusive reality is still attracting much attention because capital mobility is critical both for the efficient allocation of capital to the most productive locations and for consumption smoothing. It is also relevant for policy issues such as large current account deficits or the role of net overseas balances.

Another plausible interpretation of the close long-run relationship between the investment and saving ratios [first established by Feldstein and Horioka (1980)] is provided by Coakley, Kulasi, and Smith (1996); Jansen (1996). They argued that, since saving and investment behave like unit-root processes, the long-run SI correlation should reflect the intertemporal budget constraint or solvency constraint, which essentially requires that the current account (saving minus investment) must be a stationary process as debt cannot explode. This solvency constraint in turn implies that saving and investment are cointegrated with a unit cointegrating vector. As a result the long-run SI correlation in a cross-section regression should be equal to one. Thus, it may well be that the FH coefficient is not a puzzle, but merely a consequence of the solvency constraint. Jansen (1998) deems that the long-run correlation can provide a test of the relevance of the intertemporal budget constraint, which is one of the cornerstones of modern open-economy macroeconomics. Non-binding of the intertemporal budget constraint implies that the saving and investment rates are not cointegrated (i.e., saving and investment are not correlated in the long-run). This constitutes evidence in favour of international capital mobility by the Feldstein-Horioka criterion.

A variety of econometric specifications has been employed to estimate the SI-correlation . Jansen (1996) applies a vector error-correction model (ECM) - which is consistent with intertemporal general equilibrium models - to the OECD countries, and find that saving and investment are cointegrated across countries. However the degrees of long-run SI-correlation display some variety across countries when including more recent observations into the sample. This heterogeneity [in the SI relationship] between countries can be explained by differences in their economic structures, sizes, cyclical positions, government policies, and macroeconomic openness. To control for the potentially important effects of heterogeneity in saving and investment ratios, panel estimation techniques (such as the dynamic fixed-effects estimator, Pesaran and Smith's (1995) mean-group estimator, and **PSS**'s pooled mean-group estimator) are commonly employed (see, e.g., Coakley, Fuertes, and Spagnolo (2004); Pelgrin and Schich (2008)).

This empirical study revisits the long-run SI-relationship by applying the proposed CQL estimation approach to a quarterly dataset consisting of 27 OECD countries (Australia , Belgium, Canada, Czech Republic, Denmark, Estonia, Finland, France, Germany, Greece, Hungary, Israel, Italy, Japan, South Korea, Mexico, Netherlands, New Zealand, Norway, Portugal, Slovak Republic, Slovenia, Spain, Sweden, Switzerland, UK, USA) and four non-OECD countries and organizations (South Africa, the European Union (EU-28), Latvia, Costa Rica) from 1995 Q1 to 2015 Q2.[2] The period covered in this dataset is associated with the era when international capital movements and deregulation of domestic financial markets become more and more popular. Thus, one can expect that, for the countries under our study, the long-run relationship between saving and investment rates has been rather deteriorated.

In view of Jansen (1996) and Pelgrin and Schich (2008), we shall consider the following group-wise ECM with a maximum lag of one:

$$\Delta I_{i,t} = \alpha_i + \sum_{c=1}^{G} \phi_c u_{i,c} \left( I_{i,t-1} - \theta_c S_{i,t} \right) + \sum_{c=1}^{G} \gamma_c u_{i,c} \Delta S_{i,t} + \epsilon_{i,t}, \tag{6.1}$$

where $I_{i,t}$ and $S_{i,t}$ represent the investment rate and the saving rate of country $i$ in period $t$; and $\epsilon_{i,t} \sim N(0, \sigma_i^2)$; $\alpha_i$ is the country-specific fixed effect; $\phi_c$ is the error-correction coefficient associated with group $c$; $\theta_c$ is the long-run SI-correlation coefficient associated with group $c$; $\gamma_c$ is the short-run SI-correlation coefficient associated with group $c$; and $u_{i,c}$, $i = 1, \ldots, 31$ and $c = 1, \ldots, G$, are indicators of group memberships. The model (6.1) takes into consideration possible heterogeneity between groups of countries with common characteristics, economic policies, and structures by allowing for group variations in the SI-correlation coefficients, whereas, in many other studies, these coefficients are assumed to be either equal across countries when temporally pooling observations together (Feldstein and Horioka, 1980; Jansen, 1998), or completely different across countries (Coakley, Fuertes, and Spagnolo, 2004; Pelgrin and Schich, 2008). Other types of country-specific heterogeneity can also be accounted for by including fixed effects and error variances that differ across countries.

[2] All the data used for this empirical study are downloaded from the OECD data bank at http://www.oecd-ilibrary.org.

We start by examining the persistence property of $I_{i,t}$ and $S_{i,t}$. Table 2 presents the augmented Dickey-Fuller (ADF) test results. The $p$-values reported are greater than the 5 percent level for virtually all series. These findings are consistent with the existing evidence that saving and investment ratios have their dynamics indistinguishable from unit-root processes.

Next, we conduct estimation and inference of the error-correction model (6.1). The VNS-DCA procedure searches for globally optimal points of the CL criterion function over the domains $[-2, 2]$ of $\theta$'s, $[-2, -0.001]$ of $\phi$'s, and $[-1, 1]$ of $\alpha$'s and $\gamma$'s. The estimation results are reported in Tables 3 and 4. Since the normalized sum of squared composite errors becomes quite small when the number of groups $G$ increases to 4, we shall then consider the case where there are 4 groups. The estimate of the EC coefficient $\widehat{\phi}_4 \approx 0$ means that, in the fourth group of countries the saving and investment rates are not cointegrated in the long-run, this implies high international capital mobility by the initial interpretation of Feldstein and Horioka (1980). Since $\widehat{\phi}_4$ is very close to zero the estimate of the true $\theta_4$ then becomes irrelevant; thus, this results in a wide confidence interval, $[-435.43, 444.79]$ (cf. Table 3). The estimates of the other EC coefficients $\phi_1$, $\phi_2$, and $\phi_3$ are significantly different from zero, there exists a long-run relationship between the saving rate and investment rate. In the second group the estimate of the cointegrating vector is not much different from $(1, -1)$ the current account is stationary in the long-run. Therefore, according to Coakley, Kulasi, and Smith (1996) and Jansen (1996) the close long-run relationship between the saving and investment rates should be viewed as a solvency condition that must be satisfied rather than as evidence against capital immobility, thus no conclusion about capital mobility can be drawn for the countries in this group. In the first and third groups the estimates of the cointegrating vectors are significantly different from $(1, -1)$ the current account is nonstationary in the long-run. This result is evidence in favour of international capital mobility. Moreover, we can conclude - by inspecting the estimates for the short-run correlation coefficients $\gamma_1$ and $\gamma_3$ - that low short-run correlations imply that capital is sufficiently mobile in the countries belonging to the first and third groups. However the degree of capital mobility in these groups is less than in the fourth group.

In addition the geographical sketch of countries on the world map (cf. Figure 2) shows that there is little neighborhood effect in the long-run SI-relationship. Nowadays, many countries can have common economic structure or fiscal policy due to trade linkages or globalization, not necessarily due to geographical closeness. As noticed in Figure 3, countries in the fourth group have the lowest capital control - this is consistent with our finding that there is no long-run relationship between the investment and saving ratios, thus one could expect high capital mobility in this group. In the first and second groups the capital control indices became rather high after the year of 2004, which provides moderate evidence in favour of capital mobility. Therefore, high average capital control indices for the first and second groups suggest that there are long-run SI-relationships, which is consistent with our finding reported earlier that capital is mobile to a certain degree in the first group while it still remains inconclusive for the second group.

# 7 Conclusion

This paper proposes a novel one-step procedure to estimate dynamic panel regression models where the slope coefficients can display some latent group structure and the error terms can be spatially dependent. The idea of this estimator is to run regressions involving cross-sectional averages as proxies for the common-group stochastic trends of the response and covariates in every possible group - we call it the composite quasi-likelihood (CQL) estimator. This 'many-regression' problem can also be rewritten in the form of a large-scale non-convex mixed-integer programming problem, which is then solved by employing the efficient algorithm DCA. The proposed estimator can avoid potential problems while doing inference due to misspecification in the unobserved fixed effects. Besides, in the stationary case, it is asymptotically unbiased even when $N \gg T$, thus it does not require using instrumental variables or bias corrections. Asymptotic theory is developed for both stationary and nonstationary covariates while allowing both $N$ and $T$ diverge at a certain rate.

It is demonstrated [through asymptotic theory and Monte Carlo simulations] that the proposed estimator is asymptotically valid and has a good finite-sample performance. We also provide a BIC-type information criterion (IC) to select the optimal number of groups and the optimal way of grouping, and verify that the proposed IC is indeed asymptotically consistent. An empirical application shows the usefulness of the proposed approach by shedding new light on the famous Feldstein-Horioka puzzle. Possible extensions include a mixture of both common slopes and group-specific slopes in a dynamic panel model or a multifactor structure in the error terms. We believe that it is rather straightforward to extend the proposed CQL estimator to accommodate the first possibility while the second possibility requires a different treatment in another asymptotic framework that has not been considered in this study.

# 8 Software

A GUI software package to implement the method proposed in this paper can be downloaded from http://http-server.carleton.ca/~bchu/ecmg.htm (source code available upon request).

Table 2: Unit Root Tests

| Country | Unit root ADF test[*] | |
|---|---|---|
| | $I_{i,t}$ | $S_{i,t}$ |
| Australia | 0.1529 | 0.3528 |
| Belgium | 0.0482 | 0.8366 |
| Canada | 0.4372 | 0.2867 |
| Costa Rica | 0.4589 | 0.0477 |
| Czech Republic | 0.3651 | 0.2809 |
| Denmark | 0.2294 | 0.4473 |
| Estonia | 0.2478 | 0.5324 |
| European Union (EU-28) | 0.5762 | 0.1357 |
| Finland | 0.1356 | 0.9741 |
| France | 0.2363 | 0.3993 |
| Germany | 0.3279 | 0.0857 |
| Greece | 0.9345 | 0.6244 |
| Hungary | 0.4449 | 0.0895 |
| Israel | 0.0810 | 0.0455 |
| Italy | 0.6776 | 0.5049 |
| Japan | 0.3629 | 0.5700 |
| Korea | 0.0197 | 0.1168 |
| Latvia | 0.1844 | 0.5216 |
| Mexico | 0.0527 | 0.2478 |
| Netherlands | 0.4776 | 0.2230 |
| New Zealand | 0.0128 | 0.0845 |
| Norway | 0.3402 | 0.1501 |
| Portugal | 0.9577 | 0.5390 |
| Slovak Republic | 0.1069 | 0.0060 |
| Slovenia | 0.7136 | 0.0508 |
| South Africa | 0.4728 | 0.3390 |
| Spain | 0.8018 | 0.4308 |
| Sweden | 0.0452 | 0.1721 |
| Switzerland | 0.0000 | 0.7048 |
| UK | 0.2104 | 0.3151 |
| USA | 0.5978 | 0.6530 |

[*] ADF is the Augmented Dickey-Fuller test with optimal lag orders ($\leq$ 11) selected by the Schwartz information criterion. Numbers reported are MacKinnon (1996) one-sided $p$-values.

Table 3: Investment-Saving Error-Correction Model (ECM) Estimates

| $G$ | $\widehat{\phi}_1$ | $\widehat{\phi}_2$ | $\widehat{\phi}_3$ | $\widehat{\phi}_4$ | $\widehat{\phi}_5$ | $\widehat{\phi}_6$ | $\widehat{\theta}_1$ | $\widehat{\theta}_2$ | $\widehat{\theta}_3$ | $\widehat{\theta}_4$ | $\widehat{\theta}_5$ | $\widehat{\theta}_6$ | $\widehat{\gamma}_1$ | $\widehat{\gamma}_2$ | $\widehat{\gamma}_3$ | $\widehat{\gamma}_4$ | $\widehat{\gamma}_5$ | $\widehat{\gamma}_6$ | Normalized SSCE$^{\diamond}$ |
|---|---|---|---|---|---|---|---|---|---|---|---|---|---|---|---|---|---|---|---|
| 2 | -0.001<br>[-0.4148,0.4168]* | -1.0089<br>[-1.6152,-0.4025] | – | – | – | – | -0.7310<br>[-551.74,550.28] | 0.4660<br>[0.0735,0.8586] | – | – | – | – | 0.1812 | -0.0112 | – | – | – | – | 0.0026 |
| 3 | -0.3898<br>[-1.7557,0.9760] | -0.001<br>[-0.4039,0.4059] | -0.4608<br>[-0.8393,-0.0824] | – | – | – | -0.4127<br>[-5.6886, 4.8631] | -0.4679<br>[-565.57,564.63] | -0.0444<br>[-1.1959,1.1070] | – | – | – | -0.1833 | 0.3406 | 0.3617 | – | – | – | 0.0019 |
| 4 | -0.8441<br>[-1.4443,-0.2439] | -0.9785<br>[-1.3922,-0.5647] | -0.1365<br>[-0.8329,0.5597] | -0.001<br>[-0.2497, 0.2517] | – | – | 0.1701<br>[-0.4510,0.7913] | 1.0573<br>[0.6879,1.4267] | 0.4336<br>[-8.8299,9.6971] | -0.8177<br>[-435.43,444.79] | – | – | 0.0784 | 0.4491 | -0.0750 | -0.0931 | – | – | 0.0005 |
| 5 | -0.4407<br>[-1.0298,0.1484] | -0.0465<br>[-1.3475,1.2544] | -0.1560<br>[-0.8147,0.5025] | -0.001<br>[-0.6292,0.6312] | -1.2015<br>[-1.6539,-0.7490] | – | -0.2933<br>[-1.7819,1.1951] | -0.0851<br>[-32.36,32.19] | -0.3379<br>[-6.2712,5.5953] | -0.0647<br>[-697.95,697.83] | 0.8283<br>[0.4359,1.2207] | – | 0.2212 | 0.0855 | 0.2959 | -0.2268 | 0.3686 | – | 0.0008 |
| 6 | 0.5167 | -0.3255 | 0.0277 | -0.3094 | -1.1471 | -0.0873 | -0.7834 | -0.3776 | -0.4112 | 0.0010 | 0.8522 | 0.0665 | 0.1107 | -0.0149 | 0.5724 | -0.0758 | 0.1516 | -0.4504 | 0.0006 |

**Note:** (*) Script-size numbers in the square brackets are the bounds of asymptotic 95% confidence intervals.
($\diamond$) The normalized sum of squared composite errors (SSCE) is defined as $\frac{N}{T}\sum_{t=1}^{T}\widehat{\epsilon}^{2}_{*,t}$.

Table 4: Investment-Saving Error-Correction Model (ECM) Group Classifications

| $G$ | Group estimates |
|---|---|
| 2 | *Group 1:* Australia, France, Germany, Italy, Korea, Mexico, Netherlands, New Zealand, Portugal, Slovak Republic, Spain, UK, USA, Estonia, Israel, South Africa, Euro28, Costa Rica<br>*Group 2:* Belgium, Canada, Czech Republic, Denmark, Finland, Greece, Hungary, Japan, Norway, Sweden, Switzerland, Slovenia, Latvia |
| 3 | *Group 1:* Spain, Israel,South Africa<br>*Group 2:* Australia, Belgium, Canada, Czech Republic, Greece, Italy, Korea, Netherlands, New Zealand, Portugal, Sweden, UK, USA, Estonia, Costa Rica<br>*Group 3:* Denmark, Finland, France, Germany, Hungary, Japan, Mexico, Norway, Slovak Republic, Switzerland, Slovenia, Euro28, Latvia |
| 4 | *Group 1:* Denmark, Hungary, Italy, New Zealand, Spain, Sweden, South Africa<br>*Group 2:* Australia, Belgium, Canada, Czech Republic, Finland, France, Germany, Japan, Mexico, Norway, Switzerland, UK, Euro28<br>*Group 3:* Latvia<br>*Group 4:* Greece, Korea, Netherlands, Portugal, Slovak Republic, USA, Estonia, Israel, Slovenia, Costa Rica |
| 5 | *Group 1:* Denmark, Hungary, Korea, New Zealand, Portugal, South Africa, Latvia<br>*Group 2:* Netherlands, Slovenia<br>*Group 3:* Greece, Italy, Slovak Republic<br>*Group 4:* Germany, Spain, Estonia, Israel, Costa Rica<br>*Group 5:* Australia, Belgium, Canada, Czech Republic, Finland, France, Japan, Mexico, Norway, Switzerland, UK, USA, Euro28 |

Figure 2: Geographical Locations by Group

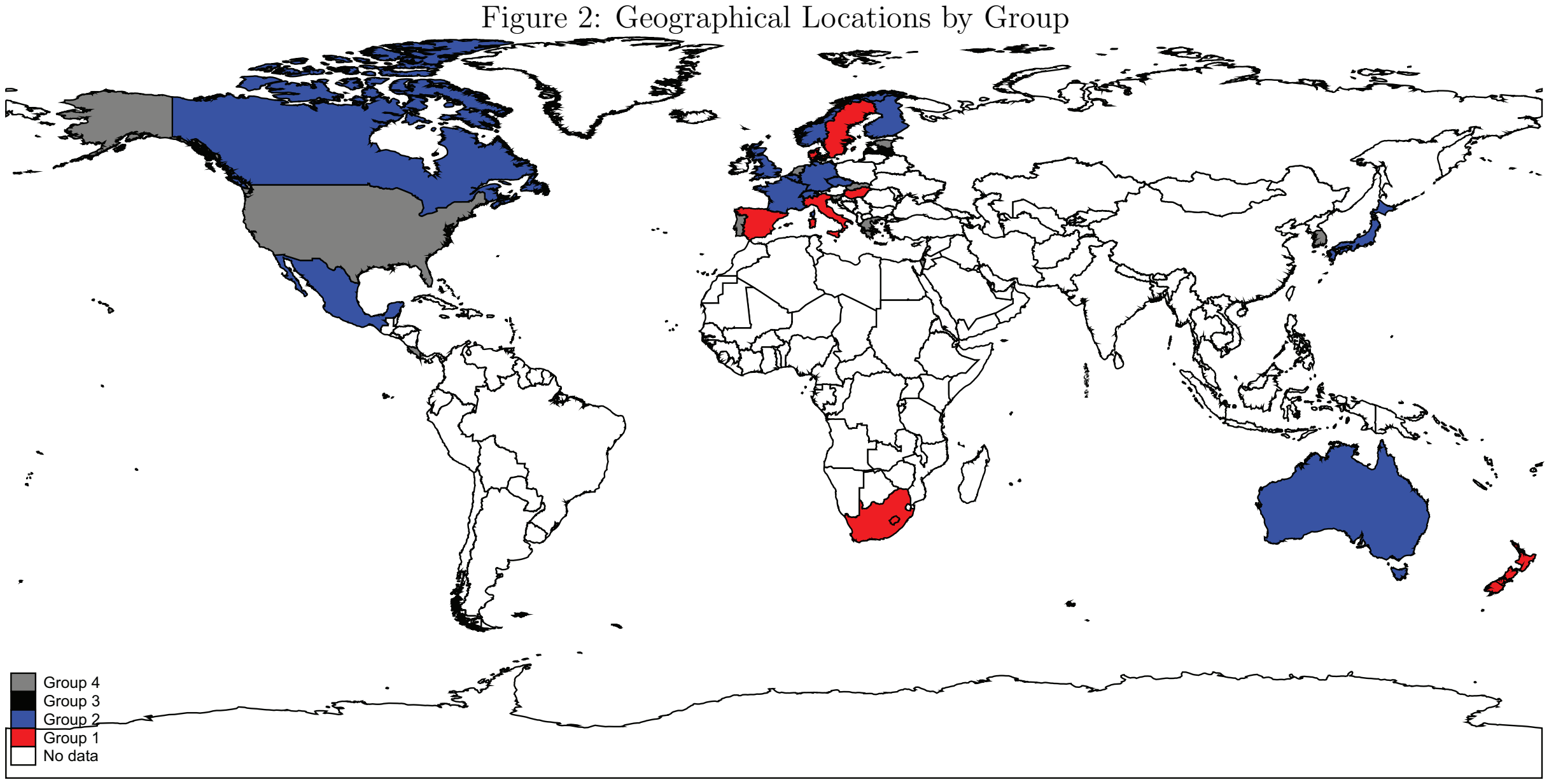

**Note:** This group map is sketched using Bonhomme and Manresa's (2015) Stata code.

Figure 3: Average Capital Control Indices by Group

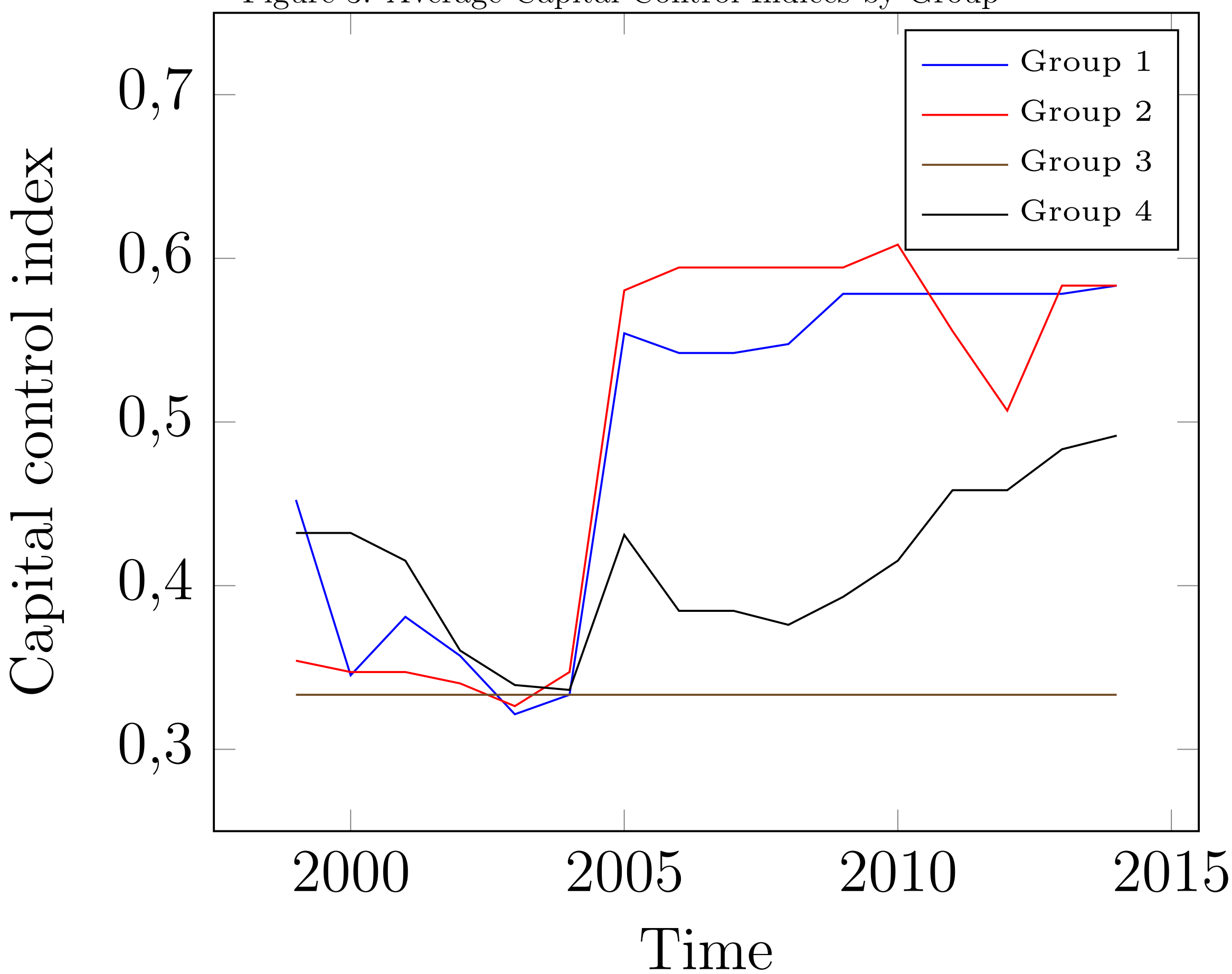

**Note:** The indices reported here are constructed based on Miniane's (2004) methodology.

# Composite Quasi-Likelihood Estimation of Dynamic Panels with Group-Specific Heterogeneity and Spatially Dependent Errors*

# – Supplemental Material –

Ba Chu†

Carleton University

September 8, 2017

**Abstract**

This supplemental material contains a Monte-Carlo simulation study (Section A), proofs of the theoretical results provided in the main text (Section B), and a detailed description of the numerical algorithms used to implement the proposed method (Section C). Numerical outputs from Monte-Carlo experiments are tabulated in Section D.

*I gratefully acknowledge funding from the Social Science and Humanities Research Council of Canada (MBF Grant 430-2016-00682).

†*Correspondence address:* Ba Chu, Department of Economics, Carleton University, B-857 Loeb Building, 1125 Colonel By Drive, Ottawa, ON K1S 5B6, Canada. Phone: +1 (613) 520 2600 (ext. 1546). Fax: +1 (613) 520 3906. E-mail: ba.chu@carleton.ca.

# Contents

# A Monte Carlo Simulations

## A.1 Monte Carlo Design

This simulation study provides some evidence for the small-sample performance of the proposed CQL estimator. The design is based on an autoregressive distributed lag (ARDL) model of order one, as denoted by ARDL(1,1), with a covariate that follows $I(0)$ or $I(1)$ process and errors being generated using linear/nonlinear SAR processes. Suppose that the covariate is $I(0)$, for a given error-generating process, two different sets of parameters are imposed on this ARDL model in order to examine the impact of the stability condition on the finite-sample performance of the estimator; the same experiment is also replicated for the case where the covariate is $I(1)$. To be specific, we consider the following data generating process (d.g.p.) with four heterogeneous groups:

$$\Delta y_{\boldsymbol{s}_g,t} = \phi_g(y_{\boldsymbol{s}_g,t-1} - \theta_g x_{\boldsymbol{s}_g,t}) + \lambda_g \Delta y_{\boldsymbol{s}_g,t-1} + \gamma_g \Delta x_{\boldsymbol{s}_g,t} + \mu_g + \epsilon_{\boldsymbol{s}_g,t},\ g = 1,\dots,4, \tag{A.1}$$

where $\boldsymbol{s}_g = (s_{g,1}, s_{g,2})$ indicates the location of each unit on a rectangular region, say $V_g$, for Group $g$, and the covariate $x_{\boldsymbol{s}_g,t}$ can take either

$$x_{\boldsymbol{s}_g,t} = \begin{cases} 0.6x_{\boldsymbol{s}_g,t-1}+\eta_{\boldsymbol{s}_g,t} & \text{if } |x_{\boldsymbol{s}_g,t-1}|<1, \\ -0.6x_{\boldsymbol{s}_g,t-1}+\eta_{\boldsymbol{s}_g,t} & \text{if } |x_{\boldsymbol{s}_g,t-1}|\geq 1 \end{cases} \tag{A.2}$$

or the unit-root process

$$x_{\boldsymbol{s}_g,t} = x_{\boldsymbol{s}_g,t-1} + \eta_{\boldsymbol{s}_g,t}. \tag{A.3}$$

for every $\boldsymbol{s}_g \in V_g$ and $t \in [1, T]$.

In the first scenario, it is assumed that the errors are generated by linear SAR processes. To specify the error-generating processes, note that the lattice $V_g$ has a lexicographical order, thus there exists a bijection between the elements of $V_g$ and the counting set $\{1, 2, \dots, |V_g|\}$. The errors $\epsilon_{\boldsymbol{s}_g,t}$ for $\boldsymbol{s}_g \in V_g$ and $t = 1,\dots,T$ are generated by a linear SAR process, which can then be represented as

$$\epsilon_{\ell_g,t} = \rho_g \sum_{\substack{h_g=1 \\ h_g\neq \ell_g}}^{|V_g|} w_{\ell_g,h_g}\epsilon_{h_g,t} + e_{\ell_g,t},\ \ell_g, h_g \in \{1, 2, \dots, |V_g|\},\ g = 1,\dots,4, \tag{A.4}$$

where $e_{\ell_g,t} \overset{i.i.d.}{\sim} N(0, \sigma_g^2)$ with $\sigma_g^2 \overset{i.i.d.}{\sim}$ Uniform$(0.5, 1.5)$ and $\boldsymbol{W}_g = \{w_{\ell_g,h_g}\}$ for $\ell_g \in \{1, 2, \dots, |V_g|\}$ and $h_g \in \{1, 2, \dots, |V_g|\}$ is a spatial weight matrix for each group, $g$; similarly, the d.g.p. for

$\eta_{\boldsymbol{s}_g,t},\ \boldsymbol{s}_g \in V_g,\ t = 1, \dots, T$, is defined by

$$\eta_{\ell_g,t} = -\rho_g \sum_{\substack{h_g=1\\ h_g \neq \ell_g}}^{|V_g|} w_{\ell_g,h_g} \eta_{h_g,t} + \xi_{\ell_g,t},\ \ell_g \in \{1, 2, \dots, |V_g|\},\ g = 1, \dots, 4, \tag{A.5}$$

where $\xi_{\ell_g,t} \overset{i.i.d.}{\sim} N(0, \sigma_g^2)$ with $\sigma_g^2 \overset{i.i.d.}{\sim}$ Uniform$(0.5, 1)$.

In the second scenario, it is assumed that the errors are generated by nonlinear spatial autoregressions; the following d.g.p.'s are similar to the one used by Hallin, Lu, and Tran (2004):

$$\epsilon_{s_{g,1},s_{g,2},t} = \sin\left(\epsilon_{s_{g,1}-1,s_{g,2},t} + \epsilon_{s_{g,1},s_{g,2}-1,t} + \epsilon_{s_{g,1}+1,s_{g,2},t} + \epsilon_{s_{g,1},s_{g,2}+1,t}\right) + e_{s_{g,1},s_{g,2},t} \tag{A.6}$$

and

$$\eta_{s_{g,1},s_{g,2},t} = \sin\left(\eta_{s_{g,1}-1,s_{g,2},t} + \eta_{s_{g,1},s_{g,2}-1,t} + \eta_{s_{g,1}+1,s_{g,2},t} + \eta_{s_{g,1},s_{g,2}+1,t}\right) + \xi_{s_{g,1},s_{g,2},t}, \tag{A.7}$$

where, for every $\boldsymbol{s}_g = (s_{g,1}, s_{g,2}) \in V_g$, $e_{s_{g,1},s_{g,2},t} \overset{i.i.d.}{\sim} N(0, \sigma_{\boldsymbol{s}_g}^2)$ with $\sigma_{\boldsymbol{s}_g} \overset{i.i.d.}{\sim}$ Uniform$(0.5, 1)$ and $\xi_{s_{g,1},s_{g,2},t} \overset{i.i.d.}{\sim} N(0, \sigma_{\boldsymbol{s}_g}^2)$ with $\sigma_{\boldsymbol{s}_g} \overset{i.i.d.}{\sim}$ Uniform$(0.5, 1.5)$ are independently generated.

With the d.g.p.'s defined above in mind, we conduct the following Monte Carlo experiments:

**Experiment 1:** Data are generated according to (A.1), (A.2), (A.4) and (A.5). The spatial weight matrices $\boldsymbol{W}_g,\ g = 1, \dots, 4$, are of rook-contiguity (or queen-contiguity) type, constructed from actual maps of counties in the four U.S. states: Georga ($g = 1$), Kansas ($g = 2$), Missouri ($g = 3$), and Texas ($g = 4$). In this experiment the numbers of 'neighbouring' counties in these states are set to $N_1 = |V_1| = 45$, $N_2 = |V_2| = 30$, $N_3 = |V_3| = 30$, and $N_4 = |V_4| = 70$ respectively, and the following sets of parameters will be used:

$$\begin{aligned}
\{\phi_1, \phi_2, \phi_3, \phi_4\} &= \{-0.9, -0.5, -0.2, -0.7\},\\
\{\theta_1, \theta_2, \theta_3, \theta_4\} &= \{-2., -1., 1., 8.\},\\
\{\lambda_1, \lambda_2, \lambda_3, \lambda_4\} &= \{-1., -0.05, 0.05, 1.\},\\
\{\gamma_1, \gamma_2, \gamma_3, \gamma_4\} &= \{-1., -0.04, 0.04, 1.\},\\
\{\mu_1, \mu_2, \mu_3, \mu_4\} &= \{-0.05, 0.05, -1., 1.\},\\
\{\rho_1, \rho_2, \rho_3, \rho_4\} &= \{0.4, 0.05, 0.6, 0.1\}.
\end{aligned}$$

This experiment illustrates the situation whereby the stability condition nearly breaks down.

**Experiment 2:** This experiment is similar to Experiment 1 except that the numbers of 'neighbouring' counties in the above-mentioned states are now set equal to $|V_1| = 100$, $|V_2| = 60$, $|V_3| = 65$, and $|V_4| = 150$. The experiment demonstrates how the proposed estimator performs as the cross-sectional dimension grows relative to the number of time periods.

**Experiment 3:** This experiment is similar to Experiment 1 except for the set of parameters $\{\lambda_1, \lambda_2, \lambda_3, \lambda_4\} = \{-0.5, -0.05, 0.05, 0.5\}$. This experiment illustrates the situation whereby the stability condition certainly holds true.

**Experiment 4:** This experiment is the same as Experiment 3 except that the numbers of 'neighbouring' counties specified in Experiment 2 are being used.

**Experiment 5:** Data are generated according to (A.1), (A.2), (A.6) and (A.7). The same sets of parameters specified in Experiments 1 and 2 are being used. This experiment illustrates the robustness of the proposed estimator when the error-generating d.g.p.'s change. It is important to note at this point that simulating sample paths from a nonlinear SAR, such as (A.6) or (A.7), is not a straight-forward task. Since the principle of contraction mapping warrants that the trigonometric sine function has a fixed point, one could simulate the processes (A.6) and (A.7) using the fixed-point iteration method. We shall briefly describe the algorithm to simulate (A.6) as (A.7) can be simulated in the same way. For each $g \in \{1, 2, 3, 4\}$ and $t \in \{1, \ldots, T\}$, to generate $m_g \times n_g$ observations of $\epsilon_{\boldsymbol{s}_g,t}$ on a rectangular region, one can perform the following steps.

**Step 1:** Set all the initial values of $\epsilon_{\boldsymbol{s}_g,t}$ to zero and generate an array, $\{e_{s_{g,1},s_{g,2},t}\}_{\substack{s_{g,1}=1,\ldots,100+m_g \\ s_{g,2}=1,\ldots,100+n_g}}$, of mixed-normal random variables.

**Step 2:** Start from the values generated in Step 1 the process is iterated, say 30 times, for example, $\epsilon^{(k+1)}_{s_{g,1},s_{g,2},t} = \sin\left(\epsilon^{(k)}_{s_{g,1}-1,s_{g,2},t} + \epsilon^{(k)}_{s_{g,1},s_{g,2}-1,t} + \epsilon^{(k)}_{s_{g,1}+1,s_{g,2},t} + \epsilon^{(k)}_{s_{g,1},s_{g,2}+1,t}\right) + e_{s_{g,1},s_{g,2},t}$, $k = 1, \ldots, 29$.

**Step 3:** Take $\{\epsilon^{(30)}_{s_{g,1},s_{g,2},t}\}_{\substack{s_{g,1}=75,\ldots,74+m_g \\ s_{g,2}=75,\ldots,74+n_g}}$ as the simulated sample, and discard $\{\epsilon^{(30)}_{s_{g,1},s_{g,2},t}\}_{\substack{s_{g,1}=1,\ldots,74 \\ s_{g,2}=1,\ldots,74}}$ to allow for a warming-up zone.

**Experiment 6:** We repeat all the above experiments with (A.2) being replaced by (A.3) to examine the finite-sample performance of the estimator when the covariate is nonstationary.

## A.2 Monte Carlo Results

### A.2.1 Known group memberships

In this case, group memberships of individuals are known, thus no group classification is needed. *Stationary Covariate:* The vector of the 'true' parameters defined in Experiment 1 indicates that the stability condition does not hold in Group $g = 4$. Both the simulated biases and MSEs of the estimates shrink to zero slowly even for a large number of time periods in both small and large spatial groups; and the estimates in Groups $g = 2$ and 3 seem to be much less biased than in Groups $g = 1$ and 4 where the stability condition does not strictly hold. This pattern seems to persist for a variety of spatial error processes (cf. the first two panels in Tables 1-6).

For the 'true' parameters defined in Experiment 3, the simulated biases and MSEs are small for relatively large numbers of time periods in both small and large groups. However, comparing the last two panels of Tables 4 and 5 the biases are clearly less severe for the case with nonlinear SAR errors than the case with linear SAR errors, especially when the group sizes are large.

*Nonstationary Covariate:* The simulated biases and MSEs of the estimates of the long-run slope coefficients in Groups $g = 1, 2$ and 3 seem not much affected by the failure of the stability condition in Group $g = 4$. This is particularly true for the case with nonlinear SAR errors (cf. Tables 7-12). When the d.g.p.'s for all the groups are stable, the simulated biases and MSEs become smaller for the case where the spatial errors follow nonlinear SAR processes.

### A.2.2 Unknown group memberships

We implement the VNS-DCA procedure to minimize the criterion function. To measure the performance of the VNS-DCA as a clustering algorithm, we report the Rand index in Table 13. The Rank index (named after William M. Rand) measures the number of pairwise agreements. For each unit $i \in \{1, \dots, N\}$, let $G_I(i)$ represent its initial group label, and $G_C(i)$ represent its group label obtained from a clustering algorithm. According to Rand (1971) the Rand index is defined, in mathematical terms, as

$$\text{RandI} := \frac{a+d}{a+b+c+d},$$

where

$$\begin{aligned}
a &:= |\{i, j \in [1, N] : \ G_I(i) = G_I(j) \text{ and } G_C(i) = G_C(j)\}|, \\
b &:= |\{i, j \in [1, N] : \ G_I(i) = G_I(j) \text{ and } G_C(i) \neq G_C(j)\}|, \\
c &:= |\{i, j \in [1, N] : \ G_I(i) \neq G_I(j) \text{ and } G_C(i) = G_C(j)\}|, \\
d &:= |\{i, j \in [1, N] : \ G_I(i) \neq G_I(j) \text{ and } G_C(i) \neq G_C(j)\}|.
\end{aligned}$$

Note that $\text{RandI} \in [0, 1]$, where '0' indicates that two clusters of data do not agree on any pair of points whilst '1' indicates that two clusters bearing possibly different labels are exactly the same.

Suppose that data are generated by the d.g.p. defined via (A.1) and (A.2) with SAR errors using queen-contiguity weights. Table 13 reports improved values for the simulated RandI's as the number of sampled locations increases. Therefore the VNS-DCA performs clustering computations efficiently in Experiments 3 and 4. The number of repetitions in each simulation is 500; and most of the computational time is spent on finding 'good' starting points through implementing the VNS algorithm while the DCA performs quite efficiently (it usually converges to an optimum after about 800 to 1500 iterations). The computational time increases polynomially with the number of time periods. According to Tables 13 and 14, the proposed procedure performs well in terms of both biases and mean squared errors.

In addition, Tables 17 and 18 report the finite-sample performance of the proposed procedure when data are generated by (A.1) with the covariate following a unit-root process (A.3) and SAR errors using queen-continuity weights. The Rand index clearly improves as the number of time periods increases, compared to the case when the covariate follows a stationary process (cf. Tables 15 and 16). Besides the empirical biases and MSE's have much faster decay rates in this case, especially for big clusters. Therefore, this method could perform really well when covariates are nonstationary. The same simulations using SAR errors with rook-continuity weights are repeated for data generated by the d.g.p. (A.1) and (A.3); Results in Tables 19 and 20 show even smaller biases and MSE's, confirming that the rates of convergence can significantly depend on the degree of spatial dependence as asserted in the main theorems.

# B Mathematical Proofs

## B.1 Proofs of Results in Section 5.1

To start with, we define some notations: $\overline{Q}_{N,T}(\boldsymbol{\Omega}) \coloneqq \frac{1}{T}Q_{N,T}(\boldsymbol{\Omega})$; $\underbrace{\boldsymbol{F}_t(\boldsymbol{\theta})}_{(G(d_x+1)+1)\times 1} \coloneqq (\boldsymbol{A}_t^\top, \boldsymbol{B}_t(\boldsymbol{\theta})^\top, C_t)^\top$, where $\boldsymbol{A}_t$, $\boldsymbol{B}_t(\boldsymbol{\theta})$, and $C_t$ are defined by (4.4), (4.5), and (4.6) respectively in the main text; $\epsilon_{0,*,t} \coloneqq \frac{1}{N}\sum_{i=1}^N \epsilon_{s_i,t}(\boldsymbol{\Theta}_0)$; and $\epsilon_{*,t}(\boldsymbol{\psi}_0) \coloneqq \epsilon_{0,*,t} - \sum_{s=1}^T \epsilon_{0,*,t}\boldsymbol{w}_{*,s}^\top \left(\sum_{s=1}^T \boldsymbol{w}_{*,s}\boldsymbol{w}_{*,s}^\top\right)^{-1} \boldsymbol{w}_{*,t}$. Some algebraic manipulations then lead to

$$
\begin{aligned}
\overline{Q}_{N,T}(\boldsymbol{\Omega}) - \overline{Q}_{N,T}(\boldsymbol{\Omega}_0) &= \frac{1}{2}\left(\frac{\sigma_{\epsilon,0}^2}{\sigma_\epsilon^2} - \log\frac{\sigma_{\epsilon,0}^2}{\sigma_\epsilon^2} - 1\right) + \frac{1}{2}\left(\frac{1}{\sigma_\epsilon^2} - \frac{1}{\sigma_{\epsilon,0}^2}\right)\left(\frac{N}{T}\sum_{t=1}^T \epsilon_{0,*,t}^2 - \sigma_{\epsilon,0}^2\right) \\
&\quad + \frac{1}{\sigma_\epsilon^2}(\boldsymbol{\psi} - \boldsymbol{\psi}_0)^\top \boldsymbol{D}_\phi \boldsymbol{D}_g \frac{N}{T}\sum_{t=1}^T \boldsymbol{F}_t(\boldsymbol{\theta}_0)\epsilon_{*,t}(\boldsymbol{\psi}_0) \\
&\quad + \frac{1}{2\sigma_\epsilon^2}(\boldsymbol{\psi} - \boldsymbol{\psi}_0)^\top \boldsymbol{D}_\phi \boldsymbol{D}_g \left(\frac{N}{T}\sum_{t=1}^T \boldsymbol{F}_t(\boldsymbol{\theta}_0)\boldsymbol{F}_t(\boldsymbol{\theta}_0)^\top\right) \boldsymbol{D}_\phi \boldsymbol{D}_g (\boldsymbol{\psi} - \boldsymbol{\psi}_0) \\
&=: \mathcal{T}_1 + \mathcal{T}_2 + \mathcal{T}_3 + \mathcal{T}_4. \qquad \text{(B.1)}
\end{aligned}
$$

The proofs of the main theorems in Section 5.1 require the following lemmas:

**Lemma 1** *Suppose that, for each site $\boldsymbol{i} \in V_N$, $\boldsymbol{x}_{\boldsymbol{i},t}$ is a stationary process. Let Assumptions 1, 2,*

*and 3 hold. Then,*

$$\sqrt{\frac{N}{T}}\sum_{t=1}^{T}\boldsymbol{F}_t(\boldsymbol{\theta}_0)\epsilon_{*,t}(\boldsymbol{\psi}_0) = O_p(1).$$

**Proof of Lemma 1.** First of all, note that $\sqrt{\frac{N}{T}}\sum_{t=1}^{T}\boldsymbol{F}_t(\boldsymbol{\theta}_0)\epsilon_{*,t}(\boldsymbol{\psi}_0) = \sqrt{\frac{N}{T}}\sum_{t=1}^{T}\boldsymbol{F}_t(\boldsymbol{\theta}_0)\epsilon_{0,*,t}$. Define a $(G(d_x+1)+1)\times 1$ vector, $\boldsymbol{Z}_{N,T,t} \coloneqq \frac{1}{\sqrt{T}}\left(\boldsymbol{x}_{*,t}^{(1)\top},\ldots,\boldsymbol{x}_{*,t}^{(G)\top},-\xi_{*,t}(\boldsymbol{\theta}_{0,1}),\ldots,-\xi_{*,t}(\boldsymbol{\theta}_{0,G}),-1\right)^{\top}$, and $u_{0,*,t} \coloneqq \sqrt{N}\epsilon_{0,*,t}$. Next, one needs to prove that

$$\left\|\sqrt{\frac{N}{T}}\sum_{t=1}^{T}\boldsymbol{F}_t(\boldsymbol{\theta}_0)\epsilon_{0,*,t} - \sum_{t=1}^{T}\boldsymbol{Z}_{N,T,t}u_{0,*,t}\right\| = O_p\left(N^{-1/2}\right). \tag{B.2}$$

Notice that

$$\left|\sqrt{\frac{N}{T}}\sum_{t=1}^{T}\boldsymbol{F}_t(\boldsymbol{\theta}_0)\epsilon_{0,*,t} - \sum_{t=1}^{T}\boldsymbol{Z}_{N,T,t}u_{0,*,t}\right| = \left|\sum_{s=1}^{T}\boldsymbol{Z}_{N,T,s}\boldsymbol{w}_{*,s}^{\top}\left(\sum_{s=1}^{T}\boldsymbol{w}_{*,s}\boldsymbol{w}_{*,s}^{\top}\right)^{-1}\sum_{t=1}^{T}\boldsymbol{w}_{*,t}u_{0,*,t}\right|.$$

Since $\frac{1}{T}\sum_{s=1}^{T}E\left|\boldsymbol{w}_{*,s}\boldsymbol{w}_{*,s}^{\top}\right| = \frac{1}{T|V_N|^2}\sum_{s=1}^{T}\sum_{\boldsymbol{i},\boldsymbol{j}\in V_N}\left|\boldsymbol{w}_{\boldsymbol{i},s}\boldsymbol{w}_{\boldsymbol{j},s}^{\top}\right| < E[\|\boldsymbol{w}_{\boldsymbol{i},s}\|^2] < \infty$ in view of Assumption 3(c), an application of Lemma 22 yields $\left|\sum_{s=1}^{T}\boldsymbol{w}_{*,s}\boldsymbol{w}_{*,s}^{\top}\right| = O_{a.s.}(T)$. In addition, by the same argument, one also obtains that $\left|\sum_{s=1}^{T}\boldsymbol{Z}_{N,T,s}\boldsymbol{w}_{*,s}^{\top}\right| \le \left|\sum_{s=1}^{T}\boldsymbol{Z}_{N,T,s}\right|\max_{1\le t\le T}\left|\boldsymbol{w}_{*,t}^{\top}\right| = O_p(\sqrt{T})O_p(1) = O_p(\sqrt{T})$. Invoking Lemma 10 together with Assumption 3 yields $\left|\sum_{t=1}^{T}\boldsymbol{w}_{*,t}u_{0,*,t}\right| = O_p\left(\sqrt{\frac{T}{N}}\right)$. Therefore, we can obtain (B.2).

Assumption 2 ensures that, for each $i\in[1,G]$ and $\boldsymbol{j}\in V_{N,i}$, the time series $\xi_{\boldsymbol{j},t}(\boldsymbol{\theta}_{0,i})$ is stationary. Hence, by applying Lemma 22, it is not hard to show that

$$\sum_{t=1}^{T}\boldsymbol{Z}_{N,T,t}\boldsymbol{Z}_{N,T,t}^{\top} \xrightarrow{a.s.} \boldsymbol{Q}_{zz}, \tag{B.3}$$

where the limiting matrix $\boldsymbol{Q}_{zz}$ is non-stochastic. Furthermore, note that every element of the vector $\boldsymbol{Z}_{N,T,t}$ has the $(2+\delta)$-th moment being bounded by $T^{-1-\delta/2}$; for example the $k$-the element, $x_{k,*,t}^{(1)}$, of $\boldsymbol{x}_{*,t}^{(1)}$ has the $(2+\delta)$-th moment satisfying

$$\begin{aligned}\frac{1}{|V_{N,1}|^{2+\delta}}E\left|\sum_{\boldsymbol{j}\in V_{N,1}}x_{k,\boldsymbol{j},t}\right|^{2+\delta} &\le \frac{1}{|V_{N,1}|^{2+\delta}}\left\{2^{\delta+1}E\left|\sum_{\boldsymbol{j}\in V_{N,1}}(x_{k,\boldsymbol{j},t}-E[x_{k,\boldsymbol{j},t}])\right|^{2+\delta}\right.\\ &\quad \left.+2^{\delta+1}|V_{N,1}|^{2+\delta}(E[x_{k,\boldsymbol{j},t}])^{2+\delta}\right\}\\ &< \infty,\end{aligned}$$

where the last inequality follows because $E\left|\sum_{\boldsymbol{j}\in V_{N,1}}(x_{k,\boldsymbol{j},t}-E[x_{k,\boldsymbol{j},t}])\right|^{2+\delta}\leq C_*|V_{N,1}|^{1+\delta/2}$ by Lemma 8, implying that $E\left|\frac{1}{\sqrt{T}}x^{(1)}_{k,*,t}\right|^{2+\delta}<C_0\frac{1}{T^{1+\delta/2}}$. Therefore, one has

$$E[\|\boldsymbol{Z}_{N,T,t}\|^{2+\delta}]\leq E[\|\boldsymbol{Z}_{N,T,t}\|_1^{2+\delta}]<C_0\frac{1}{T^{1+\delta/2}}. \tag{B.4}$$

The conditions in Lemma 21 hold because of (B.3) and (B.4); it then follows that

$$\sum_{t=1}^{T}\boldsymbol{Z}_{N,T,t}u_{0,*,t}\xrightarrow{d}\sigma_\epsilon N(\boldsymbol{0},\boldsymbol{Q}_{zz}). \tag{B.5}$$

The lemma then follows from (B.4) and (B.5). ■

**Lemma 2** *Let the assumptions of Lemma 1 hold. Then,*

$$\frac{1}{T}\sum_{t=1}^{T}\boldsymbol{F}_t(\boldsymbol{\theta}_0)\boldsymbol{F}_t(\boldsymbol{\theta}_0)^\top=O_p(1).$$

**Proof of Lemma 2.** We need to show that the block matrices on the diagonal are stochastically bounded as the same argument can also be applied to the other block matrices off the diagonal. Define $\boldsymbol{X}_{*,t}:=\left(\boldsymbol{x}^{(1)\top}_{*,t},\ldots,\boldsymbol{x}^{(G)\top}_{*,t}\right)^\top$. By the same argument as in the proof of Lemma 1, one immediately shows that

$$\left|\frac{1}{T}\sum_{t=1}^{T}\boldsymbol{A}_t\boldsymbol{A}_t^\top-\frac{1}{T}\sum_{t=1}^{T}\boldsymbol{X}_{*,t}\boldsymbol{X}_{*,t}^\top\right|=\left|\left(\frac{1}{T}\sum_{t=1}^{T}\boldsymbol{X}_{*,t}\boldsymbol{w}_{*,t}^\top\right)\left(\frac{1}{T}\sum_{t=1}^{T}\boldsymbol{w}_{*,t}\boldsymbol{w}_{*,t}^\top\right)^{-1}\left(\frac{1}{T}\sum_{t=1}^{T}\boldsymbol{w}_{*,t}\boldsymbol{X}_{*,t}^\top\right)\right|$$
$$=O_p(1).$$

An application of Lemma 22 yields

$$\frac{1}{T}\sum_{t=1}^{T}\boldsymbol{X}_{*,t}\boldsymbol{X}_{*,t}^\top=O_p(1).$$

It then follows that

$$\frac{1}{T}\sum_{t=1}^{T}\boldsymbol{A}_t\boldsymbol{A}_t^\top=O_p(1).$$

■

**Lemma 3** *Under Assumptions 1 and 5, we have*

$$\frac{N}{T}\sum_{t=1}^{T}\boldsymbol{A}_t\epsilon_{0,*,t}\xrightarrow{w}\sigma_\epsilon\boldsymbol{K}_G\boldsymbol{\Sigma}_\eta^{1/2}\int_0^1\boldsymbol{W}_\eta(\tau)dW_\epsilon(\tau),$$

*where*

$$\boldsymbol{K}_G\coloneqq diag\left(\frac{1}{\sqrt{g_{*,c}}}\mathbb{I}_{d_x},\ c=1,\dots,G\right)$$

*and*

$$\boldsymbol{\Sigma}_\eta\coloneqq\left\{\boldsymbol{\Sigma}_\eta^{(g,c)}\right\}_{g=1,\dots,G;c=1,\dots,G}$$

*with*

$$\boldsymbol{\Sigma}_\eta^{(g,c)}\coloneqq plim_{N,T\uparrow\infty}\frac{1}{T\sqrt{|V_{N,g}||V_{N,c}|}}E\left[\boldsymbol{S}_\eta(V_{N,g},T)\boldsymbol{S}_\eta(V_{N,c},T)^\top\right],\ \textit{where}\ \boldsymbol{S}_\eta(V_{N,g},t)\coloneqq\sum_{s=1}^{t}\sum_{\boldsymbol{j}\in V_{N,g}}\boldsymbol{\eta}_{\boldsymbol{j},s};$$

*and* $\boldsymbol{W}_\eta(\tau)$ *is a* $G.d_x\times 1$ *vector of Brownian motions with the covariance kernel* $E[\boldsymbol{W}_\eta(\tau)\boldsymbol{W}_\eta(\kappa)^\top]=\min(\tau,\kappa)\mathbb{I}_{G.d_x}$, *which are also independent of* $W_\epsilon(\tau)$.

**Proof of Lemma 3.** We merely need to study the limiting distribution of each term in the random vector $\frac{N}{T}\sum_{t=1}^{T}\boldsymbol{A}_t\epsilon_{0,*,t}$ as the limiting joint distribution of the random vector per se can be derived by applying the Cramér-Wold device. First, noticing that the $c$-th term of $\frac{N}{T}\sum_{t=1}^{T}\boldsymbol{A}_t\epsilon_{0,*,t}$ can be written as

$$\begin{aligned}\mathfrak{A}_{c,N,T}&\coloneqq\frac{N}{T}\sum_{t=1}^{T}\left(\boldsymbol{x}_{*,t}^{(c)}-\sum_{s=1}^{T}\boldsymbol{x}_{*,s}^{(c)}\boldsymbol{w}_{*,s}^\top\left(\sum_{s=1}^{T}\boldsymbol{w}_{*,s}\boldsymbol{w}_{*,s}^\top\right)^{-1}\boldsymbol{w}_{*,t}\right)\epsilon_{0,*,t}\\&=\frac{N}{T}\sum_{t=1}^{T}\boldsymbol{x}_{*,t}^{(c)}\epsilon_{0,*,t}-\frac{N}{T}\sum_{s=1}^{T}\boldsymbol{x}_{*,s}^{(c)}\boldsymbol{w}_{*,s}^\top\left(\sum_{s=1}^{T}\boldsymbol{w}_{*,s}\boldsymbol{w}_{*,s}^\top\right)^{-1}\sum_{t=1}^{T}\boldsymbol{w}_{*,t}\epsilon_{0,*,t}\\&\eqqcolon\mathcal{A}_{c,N,T}+\mathcal{B}_{c,N,T}.\end{aligned}\tag{B.6}$$

Define $S_\epsilon(V_N,t)\coloneqq\sum_{s=1}^{t}\sum_{\boldsymbol{j}\in V_N}\epsilon_{\boldsymbol{j},s}$. Invoking Lemma 9 together with the Cramér-Wold device, one obtains that

$$\begin{aligned}\mathcal{A}_{c,N,T}&=\frac{1}{L_{N,c}T}\sum_{t=1}^{T}\boldsymbol{S}_\eta(V_{N,c},t)[S_\epsilon(V_N,t)-S_\epsilon(V_N,t-1)]\\&\approx\frac{1}{\sqrt{g_{*,c}L_{N,c}NT}}\sum_{t=1}^{T}\int_{t/T}^{\frac{t+1}{T}}\boldsymbol{S}_\eta(V_{N,c},\lfloor T\tau\rfloor)\Delta S_\epsilon(V_N,\lfloor T\tau\rfloor)\end{aligned}$$

$$\xrightarrow{w} \frac{1}{\sqrt{g_{*,c}}}\sigma_\epsilon \boldsymbol{\Sigma}_\eta^{(c)\,1/2}\int_0^1 \boldsymbol{W}_\eta^{(i)}(\tau)dW_\epsilon(\tau),$$

where $\boldsymbol{\Sigma}_\eta^{(c)} \coloneqq \operatorname{plim}_{N,T\uparrow\infty}\frac{1}{T|V_{N,c}|}E\left[\boldsymbol{S}_\eta(V_{N,c},T)\boldsymbol{S}_\eta(V_{N,c},T)^\top\right] < \infty$, and $\boldsymbol{W}_\eta^{(c)}(\tau)$ is a $d_x \times 1$ vector of Brownian motions with the covariance kernel $E\left[\boldsymbol{W}_\eta^{(c)}(\tau)\boldsymbol{W}_\eta^{(c)}(\kappa)^\top\right] = \min(\tau,\kappa)\mathbb{I}_{d_x}$, which are also independent of $W_\epsilon(\tau)$. To bound $\mathcal{B}_{c,N,T}$, note that

$$\begin{aligned}\sum_{s=1}^T \boldsymbol{x}_{*,s}^{(c)}\boldsymbol{w}_{*,s}^\top &= \left(\sum_{s=1}^T |\boldsymbol{x}_{*,s}^{(c)}|\right)\max_{1\le t\le T}|\boldsymbol{w}_{*,t}|^\top \\ &\approx \frac{T^{\frac{3}{2}}}{|V_{N,c}|^{\frac{1}{2}}}\left\{\sum_{s=1}^T\int_{\frac{s}{T}}^{\frac{s+1}{T}}\left(\frac{1}{\sqrt{T|V_{N,c}|}}\boldsymbol{S}_\eta(V_{N,c},\lfloor T\tau\rfloor)\right)\frac{1}{T}\right\}\max_{1\le t\le T}|\boldsymbol{w}_{*,t}|^\top,\end{aligned}$$

where $\sum_{s=1}^T\int_{\frac{s}{T}}^{\frac{s+1}{T}}\left(\frac{1}{\sqrt{T|V_{N,c}|}}\boldsymbol{S}_\eta(V_{N,c},\lfloor T\tau\rfloor)\right)\frac{1}{T} \xrightarrow{w} \boldsymbol{\Sigma}_\eta^{(c)\,\frac{1}{2}}\int_0^1 \boldsymbol{W}_\eta^{(c)}(\tau)d\tau$ by Lemma 9 and the continuous mapping theorem; and, for every $t\in[1,T]$, $\boldsymbol{w}_{*,t} = o_p(1)$, which can immediately be shown by employing Lemma 8. One then has

$$\sum_{s=1}^T \boldsymbol{x}_{*,s}^{(c)}\boldsymbol{w}_{*,s}^\top = o_p\left(T^{3/2}N^{-1/2}\right).$$

Furthermore, by the same argument as in the proof of Lemma 1, one obtains that

$$\sum_{t=1}^T \boldsymbol{w}_{*,t}\epsilon_{0,*,t} = O_p\left(T^{1/2}N^{-1}\right)$$

and

$$\sum_{t=1}^T \boldsymbol{w}_{*,t}\boldsymbol{w}_{*,t}^\top = O_p(T).$$

It then follows that $\mathcal{B}_{c,N,T} = o_p(N^{-1/2})$. Therefore, it has been shown that

$$\mathfrak{A}_{c,N,T} \xrightarrow{w} \frac{1}{\sqrt{g_{*,c}}}\sigma_\epsilon\boldsymbol{\Sigma}_\eta^{(c)\,1/2}\int_0^1 \boldsymbol{W}_\eta^{(c)}(\tau)dW_\epsilon(\tau).$$

■

**Lemma 4** *Let* $\boldsymbol{\xi}_{N,T,t}(\boldsymbol{\theta}_0) \coloneqq (-\xi_{*,t}(\boldsymbol{\theta}_{0,1}),\ldots,-\xi_{*,t}(\boldsymbol{\theta}_{0,G}),-1)^\top$. *Under Assumptions 1 and 5, we have*

$$\sqrt{\frac{N}{T}}\sum_{t=1}^T(\boldsymbol{B}_t(\boldsymbol{\theta}_0)^\top, C_t)^\top\epsilon_{0,*,t} \xrightarrow{w} \sigma_\epsilon N\left(\boldsymbol{0},\boldsymbol{Q}_{\xi\xi}\right),$$

*where* $\boldsymbol{Q}_{\xi\xi} := plim_{N,T\uparrow\infty}\frac{1}{T}\sum_{t=1}^{T}\boldsymbol{\xi}_{N,T,t}(\boldsymbol{\theta}_0)\boldsymbol{\xi}_{N,T,t}(\boldsymbol{\theta}_0)^{\top} < \infty$ *w.p.1.*

**Proof of Lemma 4.** By the same argument used to verify (B.6), one can show that

$$\sqrt{\frac{N}{T}}\left\|\sum_{t=1}^{T}(\boldsymbol{B}_t(\boldsymbol{\theta}_0)^{\top}, C_t)^{\top}\epsilon_{0,*,t} - \sum_{t=1}^{T}\boldsymbol{\xi}_{N,T,t}(\boldsymbol{\theta}_0)\epsilon_{0,*,t}\right\| = O_p\left(N^{-1/2}\right). \tag{B.7}$$

In the same spirit as Lemma 2, an application of the central limit theorem for martingale difference sequences yields

$$\sqrt{\frac{N}{T}}\sum_{t=1}^{T}\boldsymbol{\xi}_{N,T,t}(\boldsymbol{\theta}_0)\epsilon_{0,*,t} \xrightarrow{d} \sigma_{\epsilon}N\left(\boldsymbol{0}, \boldsymbol{Q}_{\xi\xi}\right). \tag{B.8}$$

The lemma then follows from (B.7) and (B.8). ■

**Lemma 5** *Under Assumptions 1 and 5, we have*

$$\frac{N}{T^2}\sum_{t=1}^{T}\boldsymbol{A}_t\boldsymbol{A}_t^{\top} \xrightarrow{w} (\boldsymbol{g}\otimes\boldsymbol{\iota}_{d_x})(\boldsymbol{g}\otimes\boldsymbol{\iota}_{d_x})^{\top}\boldsymbol{\Sigma}_{\eta}\int_0^1\boldsymbol{W}_{\eta}(\tau)\boldsymbol{W}_{\eta}(\tau)^{\top}d\tau, \tag{B.9}$$

*where* $\boldsymbol{\iota}_{d_x}$ *is the* $d_x\times 1$ *vector of ones,* $\boldsymbol{\Sigma}_{\eta}$ *and* $\boldsymbol{W}_{\eta}(\tau)$ *are as defined in Lemma 3;*

$$\frac{N^{1/2}}{T^{3/2}}\sum_{t=1}^{T}\boldsymbol{A}_t\boldsymbol{B}_t(\boldsymbol{\theta}_0)^{\top} = O_p(1); \tag{B.10}$$

$$\frac{N^{1/2}}{T^{3/2}}\sum_{t=1}^{T}\boldsymbol{A}_tC_t = O_p(1); \tag{B.11}$$

*and*

$$\frac{1}{T}\sum_{t=1}^{T}(\boldsymbol{B}_t(\boldsymbol{\theta}_0)^{\top}, C_t)^{\top}(\boldsymbol{B}_t(\boldsymbol{\theta}_0)^{\top}, C_t) \xrightarrow{p} E\left[\boldsymbol{\xi}_{N,T,t}(\boldsymbol{\theta}_0)\boldsymbol{\xi}_{N,T,t}(\boldsymbol{\theta}_0)^{\top}\right], \tag{B.12}$$

*where* $\boldsymbol{\xi}_{N,T,t}(\boldsymbol{\theta}_0)$ *is defined in Lemma 4.*

**Proof of Lemma 5.** We shall show (B.9), (B.10) and (B.11) as the proof for (B.12) is pretty similar to Lemma 4. As in Lemma 2, let $\boldsymbol{X}_{*,t} := \left(\boldsymbol{x}_{*,t}^{(1)\top}, \ldots, \boldsymbol{x}_{*,t}^{(G)\top}\right)^{\top}$. Write

$$\frac{N}{T^2}\sum_{t=1}^{T}\boldsymbol{A}_t\boldsymbol{A}_t^{\top} = \frac{N}{T^2}\sum_{t=1}^{T}\boldsymbol{X}_{*,t}\boldsymbol{X}_{*,t}^{\top} - \frac{N}{T^2}\sum_{t=1}^{T}\boldsymbol{X}_{*,t}\boldsymbol{w}_{*,t}^{\top}\left(\sum_{t=1}^{T}\boldsymbol{w}_{*,t}\boldsymbol{w}_{*,t}^{\top}\right)^{-1}\sum_{t=1}^{T}\boldsymbol{w}_{*,t}\boldsymbol{X}_{*,t}^{\top} =: \mathcal{T}_{N,T,1} + \mathcal{T}_{N,T,2}.$$

First, notice that

$$\frac{N}{T^2}\sum_{t=1}^{T}\boldsymbol{x}_{*,t}^{(g)}\boldsymbol{x}_{*,t}^{(c)\top} = \frac{N}{T\sqrt{L_{N,g}L_{N,c}}}\sum_{t=1}^{T}\left\{\frac{1}{\sqrt{TL_{N,g}}}\boldsymbol{S}_{\eta}(V_{N,g},t)\right\}\left\{\frac{1}{\sqrt{TL_{N,c}}}\boldsymbol{S}_{\eta}(V_{N,c},t)^{\top}\right\}$$

$$\approx \frac{1}{\sqrt{g_{*,g}g_{*,c}}}\sum_{t=1}^{T}\int_{\frac{t}{T}}^{\frac{t+1}{T}}\left\{\frac{1}{\sqrt{TL_{N,g}}}\boldsymbol{S}_\eta(V_{N,g},\lfloor T\tau\rfloor)\right\}\left\{\frac{1}{\sqrt{TL_{N,c}}}\boldsymbol{S}_\eta(V_{N,c},\lfloor T\tau\rfloor)^\top\right\}\frac{1}{T}$$
$$\xrightarrow{w} \frac{1}{\sqrt{g_{*,g}g_{*,c}}}\boldsymbol{\Sigma}_\eta^{(g)^{1/2}}\boldsymbol{\Sigma}_\eta^{(c)^{1/2}}\int_0^1 \boldsymbol{W}_\eta^{(g)}(\tau)\boldsymbol{W}_\eta^{(c)}(\tau)d\tau,$$

where $\boldsymbol{W}_\eta^{(c)}(\tau)$ is a $d_x \times 1$ vector of Brownian motions (as defined during the proof of Lemma 3) and both $\boldsymbol{W}_\eta^{(g)}(\tau)$ and $\boldsymbol{W}_\eta^{(c)}(\tau)$ for $g \neq c$ are possibly correlated. Invoking the Cramér-Wold device, one immediately obtains that

$$\mathcal{T}_{N,T,1} \xrightarrow{w} (\boldsymbol{g}\otimes\boldsymbol{\iota}_{d_x})(\boldsymbol{g}\otimes\boldsymbol{\iota}_{d_x})^\top\boldsymbol{\Sigma}_\eta\int_0^1 \boldsymbol{W}_\eta(\tau)\boldsymbol{W}_\eta(\tau)^\top d\tau. \tag{B.13}$$

Moreover, by the same argument as in the proof of Lemma 3, it follows that $\sum_{t=1}^T \boldsymbol{X}_{*,t}\boldsymbol{w}_{*,t}^\top = o_p\left(T^{3/2}N^{-1/2}\right)$ and $\sum_{t=1}^T \boldsymbol{w}_{*,t}\boldsymbol{w}_{*,t}^\top = O_p(T)$. Therefore, one has $\mathcal{T}_{N,T,2} = o_p(1)$ and (B.9) has been verified.

To show (B.10), write

$$\begin{aligned}\frac{N^{1/2}}{T^{3/2}}\sum_{t=1}^T \boldsymbol{A}_t\boldsymbol{B}_t(\boldsymbol{\theta}_0)^\top &= \frac{N^{1/2}}{T^{3/2}}\sum_{t=1}^T \boldsymbol{X}_{*,t}\boldsymbol{\xi}_{N,T,t}(\boldsymbol{\theta}_0)^\top \\ &\quad - \frac{N^{1/2}}{T^{3/2}}\sum_{t=1}^T \boldsymbol{X}_{*,t}\boldsymbol{w}_{*,t}^\top\left(\sum_{t=1}^T \boldsymbol{w}_{*,t}\boldsymbol{w}_{*,t}^\top\right)^{-1}\sum_{t=1}^T \boldsymbol{w}_{*,t}\boldsymbol{\xi}_{N,T,t}(\boldsymbol{\theta}_0)^\top \\ &=: \mathfrak{T}_{N,T,1} + \mathfrak{T}_{N,T,2}.\end{aligned}$$

To bound each element of $\mathfrak{T}_{N,T,1}$, an application of Lemmas 8 and 9 immediately yields

$$\begin{aligned}\left|\sum_{s=1}^T \boldsymbol{x}_{*,s}^{(c)}\boldsymbol{\xi}_{N,T,s}(\boldsymbol{\theta}_0)^\top\right| &\leq \left(\sum_{s=1}^T |\boldsymbol{x}_{*,s}^{(c)}|\right)\max_{1\leq t\leq T}|\boldsymbol{\xi}_{N,T,t}(\boldsymbol{\theta}_0)^\top| \\ &\approx \frac{T^{\frac{3}{2}}}{|V_{N,c}|^{\frac{1}{2}}}\left\{\sum_{s=1}^T\int_{\frac{s}{T}}^{\frac{s+1}{T}}\left(\frac{1}{\sqrt{T|V_{N,c}|}}\boldsymbol{S}_\eta(V_{N,c},\lfloor T\tau\rfloor)\right)\frac{1}{T}\right\}\max_{1\leq t\leq T}|\boldsymbol{\xi}_{N,T,t}(\boldsymbol{\theta}_0)|^\top \\ &= O_p\left(\frac{T^{3/2}}{N^{1/2}}\right)O_p(1) \text{ by invoking the continuous mapping theorem.}\end{aligned}$$

Therefore, one has that $\mathfrak{T}_{N,T,1} = O_p(1)$, and $\mathfrak{T}_{N,T,2} = o_p(1)$ by the same argument. Similarly, one could prove (B.11). ■

The proofs of Theorems 1 - 4 in the main text now follow:

**Proof of Theorem 1.** Since $\overline{Q}_{N,T}(\widehat{\boldsymbol{\Omega}}) - \overline{Q}_{N,T}(\boldsymbol{\Omega}_0) \leq 0$ and $\overline{Q}_{N,T}(\boldsymbol{\Omega}_0) = O_p(1)$, it must be the case that $\overline{Q}_{N,T}(\widehat{\boldsymbol{\Omega}}) = O_p(1)$ as well. Intuitively, this theorem then follows if we can argue that $\overline{Q}_{N,T}(\widehat{\boldsymbol{\Omega}}) - \overline{Q}_{N,T}(\boldsymbol{\Omega}_0) \geq O_p\left(\left\|\sqrt{N}(\widehat{\boldsymbol{\psi}} - \boldsymbol{\psi}_0)\right\|^2\right) + o_p\left(\left\|\sqrt{N}(\widehat{\boldsymbol{\psi}} - \boldsymbol{\psi}_0)\right\|^2\right)$. This is what we are now going to do.

Introduce the following open balls centered at the true parameters: $B(\sigma^2_{\epsilon,0}, \delta_\sigma) := \{\sigma^2_\epsilon \in \Theta_\sigma : |\sigma^2_\epsilon - \sigma^2_{\epsilon,0}| < \delta_\sigma\}$, $B_N(\boldsymbol{\theta}_0, \delta_\theta) := \{\boldsymbol{\theta} \in \Theta_\theta : \sqrt{N}\|\boldsymbol{\theta} - \boldsymbol{\theta}_0\| < \delta_\theta\}$, $B_N(\boldsymbol{\phi}_0, \delta_\phi) := \{\boldsymbol{\phi} \in \Theta_\phi : \sqrt{N}\|\boldsymbol{\phi} - \boldsymbol{\phi}_0\| < \delta_\phi\}$, and $B_N(\mu_{*0}, \delta_\mu) := \{\mu_* \in \Theta_\mu : \sqrt{N}|\mu_* - \mu_{*0}| < \delta_\mu\}$, where $\delta_\sigma$, $\delta_\theta$, $\delta_\phi$, and $\delta_\mu$ are the radiuses. Let $B^c$ denote the complement of any ball, $B$, in a parameter space, define

$$A(\boldsymbol{\psi}_0, \delta_\psi) := \bigcup_{\substack{0<\delta_\theta,\delta_\phi,\delta_\mu<\infty \\ (\delta^2_\theta+\delta^2_\phi+\delta^2_\mu)^{1/2}=\delta_\psi}} B^c_N(\boldsymbol{\theta}_0, \delta_\theta) \times B^c_N(\boldsymbol{\phi}_0, \delta_\phi) \times B^c_N(\mu_{*0}, \delta_\mu).$$

It then follows that the coverage probability $P\Big(\widetilde{\sigma}^2_\epsilon \in B^c(\sigma^2_{\epsilon,0}, \delta_\sigma),\ \widetilde{\boldsymbol{\theta}} \in B^c_N(\boldsymbol{\theta}_0, \delta_\theta),\ \widetilde{\boldsymbol{\phi}} \in B^c_N(\boldsymbol{\phi}_0, \delta_\phi),\ \widetilde{\mu}_* \in B^c_N(\mu_{*0}, \delta_\mu)$ for every $0 < \delta_\sigma, \delta_\theta, \delta_\phi, \delta_\mu < \infty\Big)$ is bounded by $P\left(\sup_{\substack{\sigma^2_\epsilon \in B^c(\sigma^2_{\epsilon,0},\delta_\sigma) \\ \boldsymbol{\psi} \in A(\boldsymbol{\psi}_0,\delta_\psi)}} \overline{Q}_{N,T}(\boldsymbol{\Omega}) \geq \overline{Q}_{N,T}(\boldsymbol{\Omega}_0)\right)$. Therefore, one needs to verify that either

$$\lim_{N,T\uparrow\infty} P\left(\sup_{\substack{\sigma^2_\epsilon \in B^c(\sigma^2_{\epsilon,0},\delta_\sigma) \\ \boldsymbol{\psi} \in A(\boldsymbol{\psi}_0,\delta_\psi)}} \overline{Q}_{N,T}(\boldsymbol{\Omega}) \geq \overline{Q}_{N,T}(\boldsymbol{\Omega}_0)\right) = 0$$

or

$$\lim_{N,T\uparrow\infty} P\left(\inf_{\substack{\sigma^2_\epsilon \in B^c(\sigma^2_{\epsilon,0},\delta_\sigma) \\ \boldsymbol{\psi} \in A(\boldsymbol{\psi}_0,\delta_\psi)}} \left[\overline{Q}_{N,T}(\boldsymbol{\Omega}_0) - \overline{Q}_{N,T}(\boldsymbol{\Omega})\right] \geq 0\right) = 1 \tag{B.14}$$

holds.

We examine the terms defined in (B.1). First, note that $\inf_{\sigma^2_\epsilon \in B^c(\sigma^2_{\epsilon,0},\delta_\sigma)} \mathcal{T}_1 > 0$ and $\inf_{\sigma^2_\epsilon \in B^c(\sigma^2_{\epsilon,0},\delta_\sigma)} \mathcal{T}_2 = o_p(1)$ for every $\delta_\sigma > 0$ by Assumption 1 and the weak law of large numbers. Moreover, since $B^c_N(\boldsymbol{\theta}_0, \delta_\theta)$, $B^c_N(\boldsymbol{\phi}_0, \delta_\phi)$, and $B^c_N(\mu_{*0}, \delta_\mu)$ are compact sets, then, for each triplet, $0 < \delta_\theta, \delta_\phi, \delta\mu < \infty$, there exist a vector, $\boldsymbol{\psi}^* := \left(\boldsymbol{\theta}^{*\top}, \boldsymbol{\phi}^{*\top}, \mu^*_*\right)^\top \in B^c_N(\boldsymbol{\theta}_0, \delta_\theta) \times B^c_N(\boldsymbol{\phi}_0, \delta_\phi) \times B^c_N(\mu_{*0}, \delta_\mu)$, such that $\boldsymbol{\theta}^* = \boldsymbol{\theta}_0 + N^{-1/2}\boldsymbol{d}_\theta$, $\boldsymbol{\phi}^* = \boldsymbol{\phi}_0 + N^{-1/2}\boldsymbol{d}_\phi$, and $\mu^*_* = \mu_{*0} + N^{-1/2}d_\mu$ respectively; thus, in view of Lemma 1, one has that

$$\begin{aligned}\inf_{\substack{\sigma^2_\epsilon \in B^c(\sigma^2_{\epsilon,0},\delta_\sigma) \\ \boldsymbol{\psi} \in A(\boldsymbol{\psi}_0,\delta_\psi)}} \mathcal{T}_3 &\geq \frac{1}{\sup_{\sigma^2_\epsilon \in B(\sigma^2_{\epsilon,0},\delta_\sigma)} \sigma^2_\epsilon} (\boldsymbol{d}^\top_\theta, \boldsymbol{d}^\top_\phi, d_\mu) \inf_{\boldsymbol{\phi} \in B_N(\boldsymbol{\phi}_0,\delta_\phi)} \boldsymbol{D}_\phi \boldsymbol{D}_g \frac{\sqrt{N}}{T} \sum_{t=1}^T \boldsymbol{F}_t(\boldsymbol{\theta}_0)\epsilon_{*,t}(\boldsymbol{\psi}_0) \\ &= O_p(T^{-1/2}),\end{aligned}$$

where $\boldsymbol{d}_\theta$, $\boldsymbol{d}^\top_\phi$, and $d_\mu$ do not vary with $N$ because if they do, then, for some arbitrarily small constant, $\nu$, there exist sufficiently large integers, $T_0 = T_0(\nu)$ and $N_0 = N_0(\nu)$, such that

$$\left| \frac{\sqrt{N_0}}{T_0} \sum_{t=1}^{T_0} \boldsymbol{F}_t(\boldsymbol{\theta}_0)\epsilon_{*,t}(\boldsymbol{\psi}_0) \right| < \nu$$

with probability 1. Since $\delta_\psi = (\delta_\theta^2 + \delta_\phi^2 + \delta_\mu^2)^{1/2}$ is an arbitrarily positive constant (neither depending on $N$ nor $T$), it may happen that $\left\|(\boldsymbol{d}_{\theta,N_0}^\top, \boldsymbol{d}_{\phi,N_0}^\top, d_{\mu,N_0})\right\| < \delta_\psi$, where the subscript $N_0$ emphasizes the dependence on $N_0$, so that $\boldsymbol{\psi}^* \notin B_N^c(\boldsymbol{\theta}_0, \delta_\theta) \times B_N^c(\boldsymbol{\phi}_0, \delta_\phi) \times B_N^c(\mu_{*0}, \delta_\mu)$; we then have a contradiction.

Finally, an application of the minimum eigenvalue inequality yields

$$\begin{aligned}
\inf_{\substack{\sigma_\epsilon^2 \in B^c(\sigma_{\epsilon,0}^2, \delta_\sigma) \\ \boldsymbol{\psi} \in A(\boldsymbol{\psi}_0, \delta_\psi)}} \mathcal{T}_4 &\geq \frac{1}{2 \sup_{\sigma_\epsilon^2 \in B(\sigma_{\epsilon,0}^2, \delta_\sigma)} \sigma_\epsilon^2} \inf_{\boldsymbol{\psi} \in A(\boldsymbol{\psi}_0, \delta_\psi)} \|\sqrt{N}(\boldsymbol{\psi} - \boldsymbol{\psi}_0)\boldsymbol{D}_\phi \boldsymbol{D}_g\|^2 \lambda_{\min} \left( \frac{1}{T} \sum_{t=1}^T \boldsymbol{F}_t(\boldsymbol{\theta}_0)\boldsymbol{F}_t(\boldsymbol{\theta}_0)^\top \right) \\
&> \frac{1}{2 \sup_{\sigma_\epsilon^2 \in B(\sigma_{\epsilon,0}^2, \delta_\sigma)} \sigma_\epsilon^2} \delta_\psi^2 \inf_{\boldsymbol{\phi} \in B_N(\boldsymbol{\phi}_0, \delta_\phi)} \lambda_{\min} \left( \boldsymbol{D}_\phi \boldsymbol{D}_g \boldsymbol{D}_g^\top \boldsymbol{D}_\phi^\top \right) \lambda_{\min} \left( \frac{1}{T} \sum_{t=1}^T \boldsymbol{F}_t(\boldsymbol{\theta}_0)\boldsymbol{F}_t(\boldsymbol{\theta}_0)^\top \right) \\
&> 0,
\end{aligned}$$

where the last inequality follows because of Lemma 2 and Assumption 4. Collecting all the above arguments, we have proved (B.14). ■

**Proof of Theorem 2.** Let $\mathcal{Q}_{N,T}(\boldsymbol{\psi}) := \sigma_\epsilon^2 \frac{\partial \overline{Q}_{N,T}(\boldsymbol{\Omega})}{\partial \boldsymbol{\psi}}$, where $\frac{\partial \overline{Q}_{N,T}(\boldsymbol{\Omega})}{\partial \boldsymbol{\psi}}$ is defined by (4.7)-(4.9). In view of the consistency of $\widetilde{\boldsymbol{\Omega}} = (\widetilde{\boldsymbol{\psi}}^\top, \widetilde{\sigma}_\epsilon^2)^\top$ (as established in Theorem 1), by applying the mean-value expansion of $\mathcal{Q}_{N,T}(\boldsymbol{\psi})$ around $\boldsymbol{\psi}_0$, we have

$$\mathbf{0} = \frac{\partial \mathcal{Q}_{N,T}(\widetilde{\boldsymbol{\psi}})}{\partial \boldsymbol{\psi}} = \frac{\partial \mathcal{Q}_{N,T}(\boldsymbol{\psi}_0)}{\partial \boldsymbol{\psi}} + \frac{\partial^2 \mathcal{Q}_{N,T}(\boldsymbol{\psi}^*)}{\partial \boldsymbol{\psi} \partial \boldsymbol{\psi}^\top}(\widetilde{\boldsymbol{\psi}} - \boldsymbol{\psi}_0),$$

where $\boldsymbol{\psi}^* := \left( {\boldsymbol{\theta}_{N,T}^*}^\top, {\boldsymbol{\phi}_{N,T}^*}^\top, \mu_*^* \right)^\top$ is a vector of the mean values such that

$$P\left(\boldsymbol{\theta}^* \in B_N(\boldsymbol{\theta}_0, \delta_\theta),\ \boldsymbol{\phi}^* \in B_N(\boldsymbol{\phi}_0, \delta_\phi),\ \text{and } \mu_*^* \in B_N(\mu_{*0}, \delta_\mu)\right) \approx 1$$

for sufficiently large integers, $N$ and $T$, where $B_N(\boldsymbol{\theta}_0, \delta_\theta)$, $B_N(\boldsymbol{\phi}_0, \delta_\phi)$, and $B_N(\mu_{*0}, \delta_\mu)$ are the open balls defined in the proof of Theorem 1. It then follows that

$$\sqrt{NT}(\widetilde{\boldsymbol{\psi}} - \boldsymbol{\psi}_0) = \left( -\frac{1}{N} \frac{\partial^2 \mathcal{Q}_{N,T}(\boldsymbol{\psi}^*)}{\partial \boldsymbol{\psi} \partial \boldsymbol{\psi}^\top} \right)^{-1} \left( \sqrt{\frac{T}{N}} \frac{\partial \mathcal{Q}_{N,T}(\boldsymbol{\psi}_0)}{\partial \boldsymbol{\psi}} \right). \tag{B.15}$$

By the same argument as in in the proof of Lemma 1, one can immediately show that

$$\sqrt{\frac{T}{N}}\frac{\partial \mathcal{Q}_{N,T}(\boldsymbol{\psi}_0)}{\partial \boldsymbol{\psi}} = -\boldsymbol{D}_{\phi_0}\boldsymbol{D}_g\sqrt{\frac{N}{T}}\sum_{t=1}^{T}\boldsymbol{F}_t(\boldsymbol{\theta}_0)\epsilon_{*,t}(\boldsymbol{\psi}_0)$$
$$\xrightarrow{d} \sigma_\epsilon N(\boldsymbol{0}, \boldsymbol{D}_{\phi_0}\boldsymbol{D}_g\boldsymbol{Q}_{zz}\boldsymbol{D}_{\phi_0}\boldsymbol{D}_g). \tag{B.16}$$

In addition, we have that

$$\frac{1}{T}\sum_{t=1}^{T}\boldsymbol{F}_t(\boldsymbol{\theta}^*)\boldsymbol{F}_t(\boldsymbol{\theta}^*)^\top = \frac{1}{T}\sum_{t=1}^{T}\boldsymbol{F}_t(\boldsymbol{\theta}_0)\boldsymbol{F}_t(\boldsymbol{\theta}_0)^\top + \frac{1}{T}\sum_{t=1}^{T}\boldsymbol{F}_t(\boldsymbol{\theta}_0)(\boldsymbol{F}_t(\boldsymbol{\theta}^*) - \boldsymbol{F}_t(\boldsymbol{\theta}_0))^\top$$
$$+ \frac{1}{T}\sum_{t=1}^{T}(\boldsymbol{F}_t(\boldsymbol{\theta}^*) - \boldsymbol{F}_t(\boldsymbol{\theta}_0))\boldsymbol{F}_t(\boldsymbol{\theta}_0)^\top + \frac{1}{T}\sum_{t=1}^{T}\big(\boldsymbol{F}_t(\boldsymbol{\theta}^*) - \boldsymbol{F}_t(\boldsymbol{\theta}_0)\big)\big(\boldsymbol{F}_t(\boldsymbol{\theta}^*) - \boldsymbol{F}_t(\boldsymbol{\theta}_0)\big)^\top,$$

where $\boldsymbol{F}_t(\boldsymbol{\theta}^*) - \boldsymbol{F}_t(\boldsymbol{\theta}_0) = \left(\underbrace{\boldsymbol{0}}_{Gd_x\times 1}{}^\top, \underbrace{\boldsymbol{A}_t^\top \mathrm{diag}((\boldsymbol{\theta}_c^* - \boldsymbol{\theta}_{0,c}),\ c = 1, \dots, G)}_{Gd_x\times G}, 0\right)^\top$. Therefore, by Theorem 1 and the same argument used in the proof of Lemma 1, one can show that $\max_{1\le t\le T}|\boldsymbol{F}_t(\boldsymbol{\theta}^*) - \boldsymbol{F}_t(\boldsymbol{\theta}_0)| = O_p(N^{-1/2})$ and $\max_{1\le t\le T}|\boldsymbol{F}_t(\boldsymbol{\theta}_0)| = O_p(1)$. This then implies that

$$\frac{1}{T}\sum_{t=1}^{T}\boldsymbol{F}_t(\boldsymbol{\theta}^*)\boldsymbol{F}_t(\boldsymbol{\theta}^*)^\top = \frac{1}{T}\sum_{t=1}^{T}\boldsymbol{F}_t(\boldsymbol{\theta}_0)\boldsymbol{F}_t(\boldsymbol{\theta}_0)^\top + O_p(N^{-1/2}).$$

Since $\frac{1}{N}\frac{\partial^2\mathcal{Q}_{N,T}(\boldsymbol{\psi}^*)}{\partial\boldsymbol{\psi}\partial\boldsymbol{\psi}^\top} = \boldsymbol{D}_g\boldsymbol{D}_{\phi^*}\left(\frac{1}{T}\sum_{t=1}^{T}\boldsymbol{F}_t(\boldsymbol{\theta}^*)\boldsymbol{F}_t(\boldsymbol{\theta}^*)^\top\right)\boldsymbol{D}_{\phi^*}\boldsymbol{D}_g$, where $\|\boldsymbol{D}_{\phi^*} - \boldsymbol{D}_{\phi_0}\| = o_p(N^{-1/2})$, an application of Lemma 2 yields

$$\frac{1}{N}\frac{\partial^2\mathcal{Q}_{N,T}(\boldsymbol{\psi}^*)}{\partial\boldsymbol{\psi}\partial\boldsymbol{\psi}^\top} \xrightarrow{p} \boldsymbol{D}_{\phi_0}\boldsymbol{D}_g\boldsymbol{Q}_{zz}\boldsymbol{D}_{\phi_0}\boldsymbol{D}_g.$$

The theorem then follows by the continuous mapping theorem. ■

**Proof of Theorem 3.** Define open balls centered at the true parameters, $B(\sigma^2_{\epsilon,0}, \delta_\sigma) := \{\sigma^2_\epsilon \in \Theta_\sigma : |\sigma^2_\epsilon - \sigma_{\epsilon,0}| < \delta_\sigma\}$, $B_T(\boldsymbol{\theta}_0, \delta_\theta) := \{\boldsymbol{\theta} \in \Theta_\theta :\ \sqrt{T}\|\boldsymbol{\theta} - \boldsymbol{\theta}_0\| < \delta_\theta\}$, $B_N(\boldsymbol{\phi}_0, \delta_\phi) := \{\boldsymbol{\phi} \in \Theta_\phi :\ \sqrt{N}\|\boldsymbol{\phi} - \boldsymbol{\phi}_0\| < \delta_\phi\}$, and $B_N(\mu_{*0}, \delta_\mu) := \{\mu_* \in \Theta_\mu :\ \sqrt{N}|\mu_* - \mu_{*0}| < \delta_\mu\}$, where $\delta_\sigma$, $\delta_\theta$, $\delta_\phi$, and $\delta_\mu$ are the radiuses of the respective balls. Let $B^c$ denote the complement of any ball, $B$, in a parameter space, define

$$A(\boldsymbol{\psi}_0, \delta_\psi) = \bigcup_{\substack{0<\delta_\theta,\delta_\phi,\delta_\mu<\infty \\ (\delta_\theta^2+\delta_\phi^2+\delta_\mu^2)^{1/2}=\delta_\psi}} B_T^c(\boldsymbol{\theta}_0, \delta_\theta) \times B_N^c(\boldsymbol{\phi}_0, \delta_\phi) \times B_N^c(\mu_{*0}, \delta_\mu).$$

We need to prove along the lines of the proof of Theorem 1 that

$$\lim_{N,T\uparrow\infty} P\left(\inf_{\substack{\sigma_\epsilon^2\in B^c(\sigma_{\epsilon,0}^2,\delta_\sigma)\\ \boldsymbol{\psi}\in A(\boldsymbol{\psi}_0,\delta_\psi)}} \left[\overline{Q}_{N,T}(\boldsymbol{\Omega}_0)-\overline{Q}_{N,T}(\boldsymbol{\Omega})\right]\geq 0\right)=1. \tag{B.17}$$

To examine the terms defined in (B.1), $\inf_{\sigma_\epsilon^2\in B^c(\sigma_{\epsilon,0}^2,\delta_\sigma)}\mathcal{T}_1>0$ and $\inf_{\sigma_\epsilon^2\in B^c(\sigma_{\epsilon,0}^2,\delta_\sigma)}\mathcal{T}_2=o_p(1)$ as in the proof of Theorem 1. For the third term, notice that

$$\begin{aligned}
\inf_{\substack{\sigma_\epsilon^2\in B^c(\sigma_{\epsilon,0}^2,\delta_\sigma)\\ \boldsymbol{\psi}\in A(\boldsymbol{\psi}_0,\delta_\psi)}}\mathcal{T}_3 \geq{}& \frac{1}{2\inf_{\sigma_\epsilon^2\in B^c(\sigma_{\epsilon,0}^2,\delta_\sigma)}\sigma_\epsilon^2}\left\{\inf_{\substack{\boldsymbol{\theta}\in B_T^c(\boldsymbol{\theta}_0,\delta_\theta)\\ \boldsymbol{\phi}\in B_N^c(\boldsymbol{\phi}_0,\delta_\phi)}}(\boldsymbol{\theta}-\boldsymbol{\theta}_0)^\top\mathrm{diag}(\boldsymbol{g}\otimes\mathbb{I}_{d_x})\mathrm{diag}(\boldsymbol{\phi}\otimes\mathbb{I}_{d_x})\frac{N}{T}\sum_{t=1}^T\boldsymbol{A}_t\epsilon_{*,t}(\boldsymbol{\phi}_0)\right.\\
&+\inf_{\substack{\boldsymbol{\theta}\in B_T^c(\boldsymbol{\theta}_0,\delta_\theta)\\ \boldsymbol{\phi}\in B_N^c(\boldsymbol{\phi}_0,\delta_\phi)}}(\boldsymbol{\phi}-\boldsymbol{\phi}_0)^\top\mathrm{diag}(\boldsymbol{g})\frac{N}{T}\sum_{t=1}^T\boldsymbol{B}_t(\boldsymbol{\theta}_0)\epsilon_{*,t}(\boldsymbol{\phi}_0)\\
&\left.+\inf_{\mu_*\in B_N^c(\mu_{*0},\delta_\mu)}(\mu_*-\mu_{*0})\frac{N}{T}\sum_{t=1}^T C_t\epsilon_{*,t}(\boldsymbol{\phi}_0)\right\}\\
=:{}&\frac{1}{2\inf_{\sigma_\epsilon^2\in B^c(\sigma_{\epsilon,0}^2,\delta_\sigma)}\sigma_\epsilon^2}(\mathcal{T}_{3,a}+\mathcal{T}_{3,b}+\mathcal{T}_{3,c}).
\end{aligned}$$

Because $B_N^c(\boldsymbol{\theta}_0,\delta_\theta)$, $B_N^c(\boldsymbol{\phi}_0,\delta_\phi)$, and $B_N^c(\mu_{*0},\delta_\mu)$ are compact sets, then, for each triplet, $0<\delta_\theta,\delta_\phi,\delta\mu<\infty$, there exist a vector, $\boldsymbol{\psi}^*:=\left(\boldsymbol{\theta}^{*\top},\boldsymbol{\phi}^{*\top},\mu_*^*\right)^\top\in B_N^c(\boldsymbol{\theta}_0,\delta_\theta)\times B_N^c(\boldsymbol{\phi}_0,\delta_\phi)\times B_N^c(\mu_{*0},\delta_\mu)$, such that $\boldsymbol{\theta}^*=\boldsymbol{\theta}_0+T^{-1/2}\boldsymbol{d}_\theta$, $\boldsymbol{\phi}^*=\boldsymbol{\phi}_0+N^{-1/2}\boldsymbol{d}_\phi$, and $\mu_*^*=\mu_{*0}+N^{-1/2}d_\mu$ respectively to satisfy

$$\begin{aligned}
\mathcal{T}_{3,a} &= \boldsymbol{d}_\theta^\top\mathrm{diag}(\boldsymbol{g}\otimes\mathbb{I}_{d_x})\boldsymbol{\phi}\in B_N^c(\boldsymbol{\phi}_0,\delta_\phi)\mathrm{diag}(\boldsymbol{\phi}\otimes\mathbb{I}_{d_x})\frac{N}{T^{3/2}}\sum_{t=1}^T\boldsymbol{A}_t\epsilon_{*,t}(\boldsymbol{\phi}_0),\\
\mathcal{T}_{3,b} &= \boldsymbol{d}_\phi^\top\mathrm{diag}(\boldsymbol{g})\frac{\sqrt{N}}{T}\sum_{t=1}^T\boldsymbol{B}_t(\boldsymbol{\theta}_0)\epsilon_{*,t}(\boldsymbol{\phi}_0),\\
\mathcal{T}_{3,c} &= d_\mu\frac{\sqrt{N}}{T}\sum_{t=1}^T C_t\epsilon_{*,t}(\boldsymbol{\phi}_0).
\end{aligned}$$

As in the proof of Theorem 1, one can argue that $\boldsymbol{d}_\theta$, $\boldsymbol{d}_\phi$, and $d_mu$ do not vary with $T$ and $N$. Invoking Lemmas 3 and 4, one readily has $\mathcal{T}_{3,a}=O_p(T^{-1/2})$, $\mathcal{T}_{3,b}=O_p(T^{-1/2})$, and $\mathcal{T}_{3,c}=O_p(T^{-1/2})$. Therefore, $\inf_{\substack{\sigma_\epsilon^2\in B^c(\sigma_{\epsilon,0}^2,\delta_\sigma)\\ \boldsymbol{\psi}\in A(\boldsymbol{\psi}_0,\delta_\psi)}}\mathcal{T}_3\geq 0$ in probability. Finally, to bound $\mathcal{T}_4$, define the normalization matrix $\boldsymbol{K}_{N,T}=\mathrm{diag}\left(T^{1/2}\mathbb{I}_{G.d_x},N^{1/2}\mathbb{I}_{G+1}\right)$. By the minimum eigenvalue inequality together with

Assumption 6, one can obtain that, as $N$ and $T$ become large,

$$\begin{aligned}
\inf_{\substack{\sigma_\epsilon^2 \in B^c(\sigma_{\epsilon,0}^2,\delta_\sigma) \\ \boldsymbol{\psi}\in A(\boldsymbol{\psi}_0,\delta_\psi)}} \mathcal{T}_4 &\geq \frac{1}{2\inf_{\sigma_\epsilon^2 \in B^c(\sigma_{\epsilon,0}^2,\delta_\sigma)} \sigma_\epsilon^2} \left(\boldsymbol{K}_{N,T}(\boldsymbol{\psi}-\boldsymbol{\psi}_0)\right)^\top \boldsymbol{D}_\phi \boldsymbol{D}_g \boldsymbol{K}_{N,T}^{-1} \left(\frac{N}{T}\sum_{t=1}^T \boldsymbol{F}_t(\boldsymbol{\theta}_0)\boldsymbol{F}_t(\boldsymbol{\theta}_0)^\top\right) \\
&\quad \times \boldsymbol{K}_{N,T}^{-1}\boldsymbol{D}_\phi \boldsymbol{D}_g \boldsymbol{K}_{N,T}(\boldsymbol{\psi}-\boldsymbol{\psi}_0) \\
&\geq \frac{1}{2\inf_{\sigma_\epsilon^2 \in B^c(\sigma_{\epsilon,0}^2,\delta_\sigma)} \sigma_\epsilon^2} \inf_{\boldsymbol{\psi}\in A(\boldsymbol{\psi}_0,\delta_\psi)} \left\| \left(\boldsymbol{K}_{N,T}(\boldsymbol{\psi}-\boldsymbol{\psi}_0)\right)^\top \boldsymbol{D}_\phi \boldsymbol{D}_g \right\|^2 \\
&\quad \times \lambda_{\min}\left(\boldsymbol{K}_{N,T}^{-1}\left(\frac{N}{T}\sum_{t=1}^T \boldsymbol{F}_t(\boldsymbol{\theta}_0)\boldsymbol{F}_t(\boldsymbol{\theta}_0)^\top\right)\boldsymbol{K}_{N,T}^{-1}\right) \\
&\geq \frac{1}{2\inf_{\sigma_\epsilon^2 \in B^c(\sigma_{\epsilon,0}^2,\delta_\sigma)} \sigma_\epsilon^2} \delta_\psi^2 \inf_{\boldsymbol{\phi}\in B_N^c(\boldsymbol{\phi}_0,\delta_\phi)} \lambda_{\min}\left(\boldsymbol{D}_\phi \boldsymbol{D}_g \boldsymbol{D}_g \boldsymbol{D}_\phi\right) \lambda_{\min}(\boldsymbol{Q}_{zz}) \\
&> 0,
\end{aligned}$$

where the stochastic limiting matrix $\boldsymbol{Q}_{zz}$ exists because of Lemma 5. We have verified (B.17). ∎

**Proof of Theorem 4.** By using the same notations as in the proof of Theorem 2, in view of the consistency of $\widetilde{\boldsymbol{\Omega}} = (\widetilde{\boldsymbol{\psi}}^\top, \widetilde{\sigma}_\epsilon^2)^\top$ established in Theorem 3, an application of the first-order Taylor expansion of $\mathcal{Q}_{N,T}(\boldsymbol{\psi})$ around $\boldsymbol{\theta}_0$ yields

$$\mathbf{0} = \frac{\partial \mathcal{Q}_{N,T}(\widetilde{\boldsymbol{\psi}})}{\partial \boldsymbol{\theta}} = \frac{\partial \mathcal{Q}_{N,T}(\boldsymbol{\theta}_0, \widetilde{\boldsymbol{\phi}}, \widetilde{\mu}_*)}{\partial \boldsymbol{\theta}} + \frac{\partial^2 \mathcal{Q}_{N,T}(\boldsymbol{\theta}_T^*, \widetilde{\boldsymbol{\phi}}, \widetilde{\mu}_*)}{\partial \boldsymbol{\theta}\partial \boldsymbol{\theta}^\top}(\widetilde{\boldsymbol{\theta}} - \boldsymbol{\theta}_0),$$

where $\boldsymbol{\theta}_T^*$ is some point lying in the ball $B_T(\boldsymbol{\theta}_0, \delta_\theta)$. Thus,

$$\widetilde{\boldsymbol{\theta}} - \boldsymbol{\theta}_0 = \left(-\frac{\partial^2 \mathcal{Q}_{N,T}(\boldsymbol{\theta}_T^*, \widetilde{\boldsymbol{\phi}}, \widetilde{\mu}_*)}{\partial \boldsymbol{\theta}\partial \boldsymbol{\theta}^\top}\right)^{-1} \frac{\partial \mathcal{Q}_{N,T}(\boldsymbol{\theta}_0, \widetilde{\boldsymbol{\phi}}, \widetilde{\mu}_*)}{\partial \boldsymbol{\theta}}. \tag{B.18}$$

Note that

$$\begin{aligned}
\epsilon_{*,t}(\boldsymbol{\psi}) &= \epsilon_{*,t}(\boldsymbol{\psi}_0) + (\boldsymbol{\theta}-\boldsymbol{\theta}_0)^\top \mathrm{diag}(\boldsymbol{g}\mathbb{I}_{d_x})\mathrm{diag}(\boldsymbol{\phi}\mathbb{I}_{d_x})\boldsymbol{A}_t + (\boldsymbol{\phi}-\boldsymbol{\phi}_0)^\top \mathrm{diag}(\boldsymbol{g})\boldsymbol{B}_t(\boldsymbol{\theta}_0) \\
&\quad + (\mu_* - \mu_{*0})C_t.
\end{aligned} \tag{B.19}$$

One then obtains, in view of (4.7), that

$$\begin{aligned}
\frac{\partial \mathcal{Q}_{N,T}(\boldsymbol{\theta}_0, \widetilde{\boldsymbol{\phi}}, \widetilde{\mu}_*)}{\partial \boldsymbol{\theta}} &= -\mathrm{diag}(\boldsymbol{g}\mathbb{I}_{d_x})\mathrm{diag}(\widetilde{\boldsymbol{\phi}}\mathbb{I}_{d_x})\frac{N}{T}\sum_{t=1}^T \boldsymbol{A}_t \epsilon_{*,t}(\boldsymbol{\psi}_0) \\
&\quad - \mathrm{diag}(\boldsymbol{g}\mathbb{I}_{d_x})\mathrm{diag}(\widetilde{\boldsymbol{\phi}}\mathbb{I}_{d_x})\left\{\frac{N^{1/2}}{T^{3/2}}\sum_{t=1}^T \boldsymbol{A}_t \boldsymbol{B}_t(\boldsymbol{\theta}_0)^\top\right\}\mathrm{diag}(\boldsymbol{g})\sqrt{NT}(\widetilde{\boldsymbol{\phi}} - \boldsymbol{\phi}_0)
\end{aligned}$$

$$- \operatorname{diag}(\boldsymbol{g}\mathbb{I}_{d_x})\operatorname{diag}(\widetilde{\boldsymbol{\phi}}\mathbb{I}_{d_x}) \left\{ \frac{N^{1/2}}{T^{3/2}} \sum_{t=1}^{T} \boldsymbol{A}_t C_t \right\} \sqrt{NT}(\widetilde{\mu}_* - \mu_{*0}).$$

In addition, from Lemma 5 and Theorem 3, we also have that

$$\begin{aligned}
-\frac{1}{T}\frac{\partial^2 \mathcal{Q}_{N,T}(\boldsymbol{\theta}_T^*, \widetilde{\boldsymbol{\phi}}, \widetilde{\mu}_*)}{\partial \boldsymbol{\theta} \partial \boldsymbol{\theta}^\top} &= \operatorname{diag}(\boldsymbol{g}\mathbb{I}_{d_x})\operatorname{diag}(\widetilde{\boldsymbol{\phi}}\mathbb{I}_{d_x}) \left\{ \frac{N}{T^2} \sum_{t=1}^{T} \boldsymbol{A}_t \boldsymbol{A}_t^\top \right\} \operatorname{diag}(\boldsymbol{g}\mathbb{I}_{d_x})\operatorname{diag}(\widetilde{\boldsymbol{\phi}}\mathbb{I}_{d_x}) \\
&= \mathcal{H}_{N,T}^{(aa)}(\boldsymbol{\phi}_0) + o_p(1).
\end{aligned}$$

It then follows from (B.18) and (B.19) that, as $\widetilde{\boldsymbol{\phi}}$ is consistent by Theorem 3,

$$\begin{aligned}
(\mathcal{H}_{N,T}^{(aa)}(\boldsymbol{\phi}_0)+o_p(1))T(\widetilde{\boldsymbol{\theta}}-\boldsymbol{\theta}_0)+(\mathcal{H}_{N,T}^{(ab)}(\boldsymbol{\phi}_0)+o_p(1))\sqrt{NT}(\widetilde{\boldsymbol{\phi}}-\boldsymbol{\phi}_0)+(\mathcal{H}_{N,T}^{(ac)}(\boldsymbol{\phi}_0)+o_p(1))\sqrt{NT}(\widetilde{\mu}_*-\mu_{*0}) \\
= -\operatorname{diag}(\boldsymbol{g}\mathbb{I}_{d_x})\operatorname{diag}(\boldsymbol{\phi}\mathbb{I}_{d_x})\frac{N}{T}\sum_{t=1}^{T} \boldsymbol{A}_t \epsilon_{*,t}(\boldsymbol{\psi}_0) + o_p(1), \quad \text{(B.20)}
\end{aligned}$$

where $\mathcal{H}_{N,T}^{(aa)}(\boldsymbol{\phi}_0) = O_p(1)$, $\mathcal{H}_{N,T}^{(ab)}(\boldsymbol{\phi}_0) = O_p(1)$, and $\mathcal{H}_{N,T}^{(ac)}(\boldsymbol{\phi}_0) = O_p(1)$ in view of Lemma 5. By the same argument leading to (B.18), one can derive that

$$\widetilde{\boldsymbol{\phi}} - \boldsymbol{\phi}_0 = \left( -\frac{\partial^2 \mathcal{Q}_{N,T}(\widetilde{\boldsymbol{\theta}}, \boldsymbol{\phi}_N^*, \widetilde{\mu}_*)}{\partial \boldsymbol{\phi} \partial \boldsymbol{\phi}^\top} \right)^{-1} \frac{\partial \mathcal{Q}_{N,T}(\widetilde{\boldsymbol{\theta}}, \boldsymbol{\phi}_0, \widetilde{\mu}_*)}{\partial \boldsymbol{\phi}}, \quad \text{(B.21)}$$

where $\boldsymbol{\phi}_N^*$ is lying in an open ball, $B_N(\boldsymbol{\phi}_0, \delta_\phi)$, centered at $\boldsymbol{\phi}_0$; and

$$\begin{aligned}
-\frac{1}{N}\frac{\partial^2 \mathcal{Q}_{N,T}(\widetilde{\boldsymbol{\theta}}, \boldsymbol{\phi}_N^*, \widetilde{\mu}_*)}{\partial \boldsymbol{\phi} \partial \boldsymbol{\phi}^\top} &= \operatorname{diag}(\boldsymbol{g}) \left\{ \frac{1}{T} \sum_{t=1}^{T} \boldsymbol{B}_t(\widetilde{\boldsymbol{\theta}}) \boldsymbol{B}_t(\widetilde{\boldsymbol{\theta}})^\top \right\} \operatorname{diag}(\boldsymbol{g}) \\
&= \operatorname{diag}(\boldsymbol{g}) \left\{ \frac{1}{T} \sum_{t=1}^{T} \boldsymbol{B}_t(\boldsymbol{\theta}_0) \boldsymbol{B}_t(\boldsymbol{\theta}_0)^\top \right\} \operatorname{diag}(\boldsymbol{g}) \\
&\quad + \operatorname{diag}(\boldsymbol{g}) \left\{ \frac{1}{T} \sum_{t=1}^{T} \boldsymbol{B}_t(\boldsymbol{\theta}_0) \left( \boldsymbol{B}_t(\widetilde{\boldsymbol{\theta}}) - \boldsymbol{B}_t(\boldsymbol{\theta}_0) \right)^\top \right\} \operatorname{diag}(\boldsymbol{g}) \\
&\quad + \operatorname{diag}(\boldsymbol{g}) \left\{ \frac{1}{T} \sum_{t=1}^{T} \left( \boldsymbol{B}_t(\widetilde{\boldsymbol{\theta}}) - \boldsymbol{B}_t(\boldsymbol{\theta}_0) \right) \boldsymbol{B}_t(\boldsymbol{\theta}_0)^\top \right\} \operatorname{diag}(\boldsymbol{g}) \\
&\quad + \operatorname{diag}(\boldsymbol{g}) \left\{ \frac{1}{T} \sum_{t=1}^{T} \left( \boldsymbol{B}_t(\widetilde{\boldsymbol{\theta}}) - \boldsymbol{B}_t(\boldsymbol{\theta}_0) \right) \left( \boldsymbol{B}_t(\widetilde{\boldsymbol{\theta}}) - \boldsymbol{B}_t(\boldsymbol{\theta}_0) \right)^\top \right\} \operatorname{diag}(\boldsymbol{g}) \\
&=: \mathcal{H}_{N,T}^{(bb)} + \mathcal{J}_1 + \mathcal{J}_2 + \mathcal{J}_3.
\end{aligned}$$

Since $\boldsymbol{B}_t(\widetilde{\boldsymbol{\theta}}) - \boldsymbol{B}_t(\boldsymbol{\theta}_0) = \operatorname{diag}\left( (\widetilde{\boldsymbol{\theta}}_c - \boldsymbol{\theta}_{c,0})^\top,\ c = 1, \ldots, G \right) \boldsymbol{A}_t$, using the same argument as in Lemma

3 together with Theorem 3 yields that

$$\begin{aligned}
\mathcal{J}_1 &= \operatorname{diag}(\boldsymbol{g}) \left\{ \frac{1}{T} \sum_{t=1}^{T} \boldsymbol{B}_t(\boldsymbol{\theta}_0) \boldsymbol{A}_t^\top \right\} \operatorname{diag}\left( (\widetilde{\boldsymbol{\theta}}_c - \boldsymbol{\theta}_{c,0}),\ c = 1, \ldots, G \right) \operatorname{diag}(\boldsymbol{g}) \\
&= O_p\left(T^{1/2} N^{-1/2}\right) o_p\left(T^{-1/2}\right) \\
&= o_p(N^{-1/2}).
\end{aligned}$$

Analogously, one also obtains that $\mathcal{J}_2 = o_p(N^{-1/2})$ and $\mathcal{J}_3 = \operatorname{diag}(\boldsymbol{g})\operatorname{diag}\left( (\widetilde{\boldsymbol{\theta}}_c - \boldsymbol{\theta}_{c,0})^\top,\ c = 1, \ldots, G \right)$ $\left\{ \frac{1}{T} \sum_{t=1}^{T} \boldsymbol{A}_t \boldsymbol{A}_t^\top \right\} \operatorname{diag}\left( (\widetilde{\boldsymbol{\theta}}_c - \boldsymbol{\theta}_{c,0}),\ c = 1, \ldots, G \right) \operatorname{diag}(\boldsymbol{g}) = o_p(T^{-1}) O_p(TN^{-1}) = o_p(N^{-1})$. It then follows that

$$-\frac{1}{N} \frac{\partial^2 \mathcal{Q}_{N,T}(\widetilde{\boldsymbol{\theta}}, \boldsymbol{\phi}_N^*, \widetilde{\mu}_*)}{\partial \boldsymbol{\phi} \partial \boldsymbol{\phi}^\top} = \mathcal{H}_{N,T}^{(bb)} + o_p(1). \tag{B.22}$$

In view of (4.8), we have

$$\frac{\partial \mathcal{Q}_{N,T}(\widetilde{\boldsymbol{\theta}}, \boldsymbol{\phi}_0, \widetilde{\mu}_*)}{\partial \boldsymbol{\phi}} = -\operatorname{diag}(\boldsymbol{g}) \frac{N}{T} \sum_{t=1}^{T} \boldsymbol{B}_t(\widetilde{\boldsymbol{\theta}}) \epsilon_{*,t}(\widetilde{\boldsymbol{\theta}}, \boldsymbol{\phi}_0, \widetilde{\mu}_*),$$

where

$$\begin{aligned}
\boldsymbol{B}_t(\widetilde{\boldsymbol{\theta}}) \epsilon_{*,t}(\widetilde{\boldsymbol{\theta}}, \boldsymbol{\phi}_0, \widetilde{\mu}_*) &= \boldsymbol{B}_t(\boldsymbol{\theta}_0) \epsilon_{*,t}(\boldsymbol{\psi}_0) + \left( \boldsymbol{B}_t(\widetilde{\boldsymbol{\theta}}) - \boldsymbol{B}_t(\boldsymbol{\theta}_0) \right) \epsilon_{*,t}(\boldsymbol{\psi}_0) + \boldsymbol{B}_t(\widetilde{\boldsymbol{\theta}}) \left( \epsilon_{*,t}(\widetilde{\boldsymbol{\theta}}, \boldsymbol{\phi}_0, \widetilde{\mu}_*) - \epsilon_{*,t}(\boldsymbol{\psi}_0) \right) \\
&= \boldsymbol{B}_t(\boldsymbol{\theta}_0) \epsilon_{*,t}(\boldsymbol{\psi}_0) + \operatorname{diag}\left( (\widetilde{\boldsymbol{\theta}}_c - \boldsymbol{\theta}_{c,0})^\top,\ c = 1, \ldots, G \right) \boldsymbol{A}_t \epsilon_{*,t}(\boldsymbol{\psi}_0) \\
&\quad + \boldsymbol{B}_t(\widetilde{\boldsymbol{\theta}}) \boldsymbol{A}_t^\top \operatorname{diag}(\boldsymbol{\phi}_0 \mathbb{I}_{d_x}) \operatorname{diag}(\boldsymbol{g} \mathbb{I}_{d_x}) (\widetilde{\boldsymbol{\theta}} - \boldsymbol{\theta}_0) + (\widetilde{\mu}_* - \mu_{*0}) \boldsymbol{B}_t(\widetilde{\boldsymbol{\theta}}) C_t \\
&= \boldsymbol{B}_t(\boldsymbol{\theta}_0) \epsilon_{*,t}(\boldsymbol{\psi}_0) + \boldsymbol{B}_t(\boldsymbol{\theta}_0) \boldsymbol{A}_t^\top \operatorname{diag}(\boldsymbol{\phi}_0 \mathbb{I}_{d_x}) \operatorname{diag}(\boldsymbol{g} \mathbb{I}_{d_x}) (\widetilde{\boldsymbol{\theta}} - \boldsymbol{\theta}_0) \\
&\quad + \boldsymbol{B}_t(\boldsymbol{\theta}_0) C_t (\widetilde{\mu}_* - \mu_{*0}) + \operatorname{diag}\left( (\widetilde{\boldsymbol{\theta}}_c - \boldsymbol{\theta}_{c,0})^\top,\ c = 1, \ldots, G \right) \boldsymbol{A}_t \epsilon_{*,t}(\boldsymbol{\psi}_0) \\
&\quad + \operatorname{diag}\left( (\widetilde{\boldsymbol{\theta}}_c - \boldsymbol{\theta}_{c,0})^\top,\ c = 1, \ldots, G \right) \boldsymbol{A}_t C_t (\widetilde{\mu}_* - \mu_{*0}) \\
&\quad + \operatorname{diag}\left( (\widetilde{\boldsymbol{\theta}}_c - \boldsymbol{\theta}_{c,0})^\top,\ c = 1, \ldots, G \right) \boldsymbol{A}_t \boldsymbol{A}_t^\top \operatorname{diag}(\boldsymbol{\phi}_0 \mathbb{I}_{d_x}) \operatorname{diag}(\boldsymbol{g} \mathbb{I}_{d_x}) (\widetilde{\boldsymbol{\theta}} - \boldsymbol{\theta}_0).
\end{aligned}$$

By Lemma 3 and Theorem 3, one obtains in view of (B.20) that

$$\begin{aligned}
\operatorname{diag}\left( (\widetilde{\boldsymbol{\theta}}_c - \boldsymbol{\theta}_{c,0})^\top,\ c = 1, \ldots, G \right) \frac{N^{1/2}}{T^{1/2}} \sum_{t=1}^{T} \boldsymbol{A}_t \epsilon_{*,t}(\boldsymbol{\psi}_0) &= o_p\left(N^{-1/2}\right), \\
\operatorname{diag}\left( (\widetilde{\boldsymbol{\theta}}_c - \boldsymbol{\theta}_{c,0})^\top,\ c = 1, \ldots, G \right) \left\{ \frac{N^{1/2}}{T^{1/2}} \sum_{t=1}^{T} \boldsymbol{A}_t C_t \right\} (\widetilde{\mu}_* - \mu_{*0}) &= O_p\left(N^{-1/2}\right),
\end{aligned}$$

$$\mathrm{diag}\left((\widetilde{\boldsymbol{\theta}}_c - \boldsymbol{\theta}_{c,0})^\top,\ c = 1, \ldots, G\right) \left\{ \frac{N^{1/2}}{T^{1/2}} \sum_{t=1}^{T} \boldsymbol{A}_t \boldsymbol{A}_t^\top \right\} \mathrm{diag}(\boldsymbol{\phi}_0 \mathbb{I}_{d_x}) \mathrm{diag}(\boldsymbol{g} \mathbb{I}_{d_x})(\widetilde{\boldsymbol{\theta}} - \boldsymbol{\theta}_0) = O_p\left(N^{-1/2} T^{-1/2}\right).$$

It then follows that

$$\begin{aligned}
\sqrt{\frac{T}{N}} \frac{\partial \mathcal{Q}_{N,T}(\widetilde{\boldsymbol{\theta}}, \boldsymbol{\phi}_0, \widetilde{\mu}_*)}{\partial \boldsymbol{\phi}} = &-\mathrm{diag}(\boldsymbol{g}) \sqrt{\frac{N}{T}} \sum_{t=1}^{T} \boldsymbol{B}_t(\boldsymbol{\theta}_0) \epsilon_{*,t}(\boldsymbol{\psi}_0) \\
&- \mathrm{diag}(\boldsymbol{g}) \left( \frac{N^{1/2}}{T^{3/2}} \sum_{t=1}^{T} \boldsymbol{B}_t(\boldsymbol{\theta}_0) \boldsymbol{A}_t^\top \right) \mathrm{diag}(\boldsymbol{\phi}_0 \mathbb{I}_{d_x}) \mathrm{diag}(\boldsymbol{g} \mathbb{I}_{d_x}) T(\widetilde{\boldsymbol{\theta}} - \boldsymbol{\theta}_0) \\
&- \mathrm{diag}(\boldsymbol{g}) \left\{ \frac{1}{T} \sum_{t=1}^{T} \boldsymbol{B}_t(\boldsymbol{\theta}_0) C_t \right\} \sqrt{NT}(\widetilde{\mu}_* - \mu_{*0}) + o_p(1). \qquad \text{(B.23)}
\end{aligned}$$

Therefore, in view of (B.21), we have

$$\begin{aligned}
&(\mathcal{H}_{N,T}^{(bb)} + o_p(1)) \sqrt{NT}(\widetilde{\boldsymbol{\phi}} - \boldsymbol{\phi}_0) + \mathcal{H}_{N,T}^{(ab)}(\boldsymbol{\phi}_0)^\top T(\widetilde{\boldsymbol{\theta}} - \boldsymbol{\theta}_0) + \mathcal{H}_{N,T}^{(bc)} \sqrt{NT}(\widetilde{\mu}_* - \mu_{*0}) \\
&\qquad = -\mathrm{diag}(\boldsymbol{g}) \sqrt{\frac{N}{T}} \sum_{t=1}^{T} \boldsymbol{B}_t(\boldsymbol{\theta}_0) \epsilon_{*,t}(\boldsymbol{\psi}_0) + o_p(1). \qquad \text{(B.24)}
\end{aligned}$$

By the same argument leading to (B.21), it can be shown that

$$\widetilde{\mu}_* - \mu_{*0} = \left( -\frac{\partial^2 \mathcal{Q}_{N,T}(\widetilde{\boldsymbol{\theta}}, \widetilde{\boldsymbol{\phi}}, \mu_{*,N}^*)}{\partial \mu_*^2} \right)^{-1} \frac{\partial \mathcal{Q}_{N,T}(\widetilde{\boldsymbol{\theta}}, \widetilde{\boldsymbol{\phi}}, \mu_{*0})}{\partial \mu_*}, \qquad \text{(B.25)}$$

where $\mu_{*,N}^*$ is some point in an open ball, $B_N(\mu_{*0}, \delta_\mu)$, centered at $\mu_{*0}$, and

$$-\frac{1}{N} \frac{\partial^2 \mathcal{Q}_{N,T}(\widetilde{\boldsymbol{\theta}}, \widetilde{\boldsymbol{\phi}}, \mu_{*,N}^*)}{\partial \mu_*^2} = 1 + o_p(1).$$

By the same argument leading to (B.24), one readily obtains that

$$\begin{aligned}
\sqrt{\frac{T}{N}} \frac{\partial \mathcal{Q}_{N,T}(\widetilde{\boldsymbol{\theta}}, \widetilde{\boldsymbol{\phi}}, \mu_{*0})}{\partial \mu_*} = &-\sqrt{\frac{N}{T}} \sum_{t=1}^{T} C_t \epsilon_{*,t}(\boldsymbol{\psi}_0) - \left( \frac{N^{1/2}}{T^{3/2}} \sum_{t=1}^{T} C_t \boldsymbol{A}_t^\top \right) \mathrm{diag}(\widetilde{\boldsymbol{\phi}} \mathbb{I}_{d_x}) \mathrm{diag}(\boldsymbol{g} \mathbb{I}_{d_x}) T(\widetilde{\boldsymbol{\theta}} - \boldsymbol{\theta}_0) \\
&- \left( \frac{1}{T} \sum_{t=1}^{T} C_t \boldsymbol{B}_t(\boldsymbol{\theta}_0)^\top \right) \mathrm{diag}(\boldsymbol{g}) \sqrt{NT}(\widetilde{\boldsymbol{\phi}} - \boldsymbol{\phi}_0).
\end{aligned}$$

Therefore, from (B.25), one obtains that

$$(\mathcal{H}_{N,T}^{(ac)}(\boldsymbol{\phi}_0)^\top + o_p(1)) T(\widetilde{\boldsymbol{\theta}} - \boldsymbol{\theta}_0) + \mathcal{H}_{N,T}^{(bc)\top} \sqrt{NT}(\widetilde{\boldsymbol{\phi}} - \boldsymbol{\phi}_0) + \sqrt{NT}(\widetilde{\mu}_* - \mu_{*0})$$

$$= -\sqrt{\frac{N}{T}} \sum_{t=1}^{T} C_t \epsilon_{*,t}(\boldsymbol{\psi}_0) + o_p(1). \quad \text{(B.26)}$$

Collecting up the terms defined by (B.21), (B.24), and (B.26), we have

$$\begin{aligned}
\mathcal{H}_{N,T}(\boldsymbol{\phi}_0) \begin{pmatrix} T(\widetilde{\boldsymbol{\theta}} - \boldsymbol{\theta}_0) \\ \sqrt{NT}(\widetilde{\boldsymbol{\phi}} - \boldsymbol{\phi}_0) \\ \sqrt{NT}(\widetilde{\mu}_* - \mu_{*0}) \end{pmatrix} &= -\mathrm{diag}\left(\mathrm{diag}(\boldsymbol{g}\mathbb{I}_{d_x})\mathrm{diag}(\boldsymbol{\phi}_0\mathbb{I}_{d_x}), \mathrm{diag}(\boldsymbol{g}), 1\right) \sum_{t=1}^{T} \begin{pmatrix} \frac{\sqrt{N}}{T}\boldsymbol{A}_t \\ \frac{1}{\sqrt{T}}\boldsymbol{B}_t \\ \frac{1}{\sqrt{T}}C_t \end{pmatrix} \sqrt{N}\epsilon_{*,t}(\boldsymbol{\psi}_0) \\
&= \mathcal{M}_{N,T}(\boldsymbol{\phi}_0).
\end{aligned}$$

It thus follows that

$$\begin{pmatrix} T(\widetilde{\boldsymbol{\theta}} - \boldsymbol{\theta}_0) \\ \sqrt{NT}(\widetilde{\boldsymbol{\phi}} - \boldsymbol{\phi}_0) \\ \sqrt{NT}(\widetilde{\mu}_* - \mu_{*0}) \end{pmatrix} = -\mathcal{H}_{N,T}(\boldsymbol{\phi}_0)^{-1}\mathcal{M}_{N,T}(\boldsymbol{\phi}_0).$$

Invoking Lemmas 3, 4, and 5, one can prove that $\mathcal{M}_{N,T}(\boldsymbol{\phi}_0) \xrightarrow{w} MN\left(\mathbf{0}, \mathcal{H}(\boldsymbol{\phi}_0)\right)$. The main theorem then follows by applying the continuous mapping theorem. ■

## B.2 Proof of Results in Section 5.2

We start by defining some common notations that will be used for the rest of this section. Let $\boldsymbol{u}_c := (u_{1,c}, \ldots, u_{N,c})^\top$ be a $N \times 1$ vector of group membership indicators associated with group labelled $c \in [1, G]$, $\xi^{(w)}_{0,*,t}(\boldsymbol{u}_c) \equiv \xi^{(w)}_{*,t}(\boldsymbol{\theta}_{0,c}, \boldsymbol{u}_c) := \frac{1}{N}\sum_{i=1}^{N} u_{i,c}\xi^{(w)}_{i,t}(\boldsymbol{\theta}_{0,c})$, $\boldsymbol{x}^{(w)}_{*,t}(\boldsymbol{u}_c) := \frac{1}{N}\sum_{i=1}^{N} u_{i,c}\boldsymbol{x}^{(w)}_{i,t}$.
**Proof of Theorem 5.** An application of Lemma 1 yields

$$\sqrt{\frac{N}{T}} \sum_{t=1}^{T} \boldsymbol{F}_t(\boldsymbol{U}, \boldsymbol{U}_0)\epsilon^{(w)}_{0,*,t} = O_p(1).$$

Therefore, it follows that

$$\begin{aligned}
\frac{1}{T}\sum_{t=1}^{T}\{\epsilon^2_{*,t}(\boldsymbol{\psi}, \boldsymbol{U}) - \epsilon^{(w)2}_{0,*,t}\} &= \left((\boldsymbol{\theta}^{(\widetilde{\sigma}^{(per)})} - \boldsymbol{\theta}_0^{(\sigma^{(per)})})^\top, (\boldsymbol{\phi}_0^{(\sigma^{(per)})} - \boldsymbol{\phi}^{(\widetilde{\sigma}^{(per)})})^\top, \mu_{*0} - \mu_*, \boldsymbol{\phi}_0^{(\sigma^{(per)})\top}\right) \\
&\quad \times \mathrm{diag}\left(\boldsymbol{D}_\phi(\widetilde{\sigma}^{(per)}), \mathbb{I}_{2G+1}\right) \frac{1}{T}\sum_{t=1}^{T} \boldsymbol{F}_t(\boldsymbol{U}, \boldsymbol{U}_0)\boldsymbol{F}_t(\boldsymbol{U}, \boldsymbol{U}_0)^\top \\
&\quad \times \mathrm{diag}\left(\boldsymbol{D}_\phi(\widetilde{\sigma}^{(per)}), \mathbb{I}_{2G+1}\right) \\
&\quad \times \left((\boldsymbol{\theta}^{(\widetilde{\sigma}^{(per)})} - \boldsymbol{\theta}_0^{(\sigma^{(per)})})^\top, (\boldsymbol{\phi}_0^{(\sigma^{(per)})} - \boldsymbol{\phi}^{(\widetilde{\sigma}^{(per)})})^\top, \mu_{*0} - \mu_*, \boldsymbol{\phi}_0^{(\sigma^{(per)})\top}\right)^\top \\
&\quad + O_p\left((NT)^{-1/2}\right). \qquad \text{(B.27)}
\end{aligned}$$

Let $\mathcal{B}(\boldsymbol{\psi}_0,\eta_\psi) \coloneqq \{\boldsymbol{\psi} \in \Theta_\psi : \ H(\boldsymbol{\psi},\boldsymbol{\psi}_0) < \eta_\psi\}$ represent an open ball centered at $\boldsymbol{\psi}_0$ with radius $\eta_\psi$, and $\mathcal{B}(\boldsymbol{U}_0,\eta_u) \coloneqq \{\boldsymbol{U} \in \Delta_S^N \bigcap \{0,1\}^{G\times N} : \ H(\boldsymbol{U},\boldsymbol{U}_0) < \eta_u\}$ be an open ball centered at $\boldsymbol{U}_0$ with radius $\eta_u$. We denote by $\mathcal{B}^c(\boldsymbol{\psi}_0,\eta_\psi)$ and $\mathcal{B}^c(\boldsymbol{U}_0,\eta_u)$ the complements of $\mathcal{B}(\boldsymbol{\psi}_0,\eta_\psi)$ and $\mathcal{B}(\boldsymbol{U}_0,\eta_u)$ respectively. Since $(\widehat{\boldsymbol{\psi}},\widehat{\boldsymbol{U}})$ are the minimum values of $\frac{1}{T}\sum_{t=1}^T \epsilon_{*,t}^2(\boldsymbol{\psi},\boldsymbol{U})$, it then follows that

$$P\left(\widehat{\boldsymbol{\psi}} \in \mathcal{B}^c(\boldsymbol{\psi}_0,\eta_\psi), \widehat{\boldsymbol{U}} \in \mathcal{B}(\boldsymbol{U}_0,\eta_u)\right) \leq P\left(\inf_{\substack{\widehat{\boldsymbol{\psi}}\in\mathcal{B}^c(\boldsymbol{\psi}_0,\eta_\psi)\\ \widehat{\boldsymbol{U}}\in\mathcal{B}(\boldsymbol{U}_0,\eta_u)}} \frac{1}{T}\sum_{t=1}^T \{\epsilon_{*,t}^2(\boldsymbol{\psi},\boldsymbol{U}) - \epsilon_{0,*,t}^{(w)2}\} \leq 0\right). \tag{B.28}$$

In view of (B.27), an application of the eigenvalue inequality yields

$$\begin{aligned}\frac{1}{T}\sum_{t=1}^T \{\epsilon_{*,t}^2(\boldsymbol{\psi},\boldsymbol{U}) - \epsilon_{0,*,t}^{(w)2}\} &\geq C_0 \inf_{H(\boldsymbol{U},\boldsymbol{U}_0)>\eta_u} \lambda_{\min}\left(\frac{1}{T}\sum_{t=1}^T \boldsymbol{F}_t(\boldsymbol{U},\boldsymbol{U}_0)\boldsymbol{F}_t(\boldsymbol{U},\boldsymbol{U}_0)^\top\right) H(\widehat{\boldsymbol{\psi}},\boldsymbol{\psi}_0)^2 \\ &\quad + O_p\left((NT)^{-1/2}\right) > 0\end{aligned}$$

by Assumption 7. In view of (B.28), one obtains that $H(\widehat{\boldsymbol{\psi}},\boldsymbol{\psi}_0) < \eta_\psi$ and $H(\widehat{\boldsymbol{U}},\boldsymbol{U}_0) < \eta_u$ w.p.1 for some arbitrarily small constants, $\eta_\psi$ and $\eta_u$.

One can now refine the rates that $H(\widehat{\boldsymbol{\psi}},\boldsymbol{\psi}_0) \xrightarrow{p} 0$. First let's define new open ball nested in $\mathcal{B}(\boldsymbol{\psi}_0,\eta_\psi)$, i.e., $\mathfrak{B}(\boldsymbol{\psi}_0,\eta_\psi'/\sqrt{N}) \subset \mathcal{B}(\boldsymbol{\psi}_0,\eta_\psi)$. Some algebra yields

$$\begin{aligned}\overline{Q}_{N,T}(\boldsymbol{\psi}_0,\sigma_{\epsilon,0}^2,\boldsymbol{U}_0) - \overline{Q}_{N,T}(\boldsymbol{\psi},\sigma_\epsilon^2,\boldsymbol{U}) &= \frac{1}{2}\left(\frac{\sigma_{\epsilon,0}^2}{\sigma_\epsilon^2} - \log\frac{\sigma_{\epsilon,0}^2}{\sigma_\epsilon^2} - 1\right) \\ &\quad + \frac{1}{2}\left(\frac{1}{\sigma_\epsilon^2} - \frac{1}{\sigma_{\epsilon,0}^2}\right)\left(\frac{N}{T}\sum_{t=1}^T \epsilon_{0,*,t}^2 - \sigma_{\epsilon,0}^2\right) \\ &\quad + \frac{1}{2\sigma_\epsilon^2}\frac{N}{T}\sum_{t=1}^T \{\epsilon_{*,t}^2(\boldsymbol{\psi},\boldsymbol{U}) - \epsilon_{0,*,t}^{(w)2}\} \\ &=: \mathcal{T}_1 + \mathcal{T}_2(N,T) + \mathcal{T}_3(N,T),\end{aligned} \tag{B.29}$$

where $\mathcal{T}_1 > 0$ for every $|\sigma_\epsilon^2 - \sigma_{\epsilon,0}^2| > \eta_\sigma$ with some arbitrarily small $\eta_\sigma > 0$, and $\mathcal{T}_2(N,T) = o_p(1)$ by the same argument in Theorem 1. Moreover, by Lemma 1, we have

$$\sqrt{\frac{N}{T}}\sum_{t=1}^T \underbrace{\left(\boldsymbol{x}_{*,t}^{(w)\top}(\boldsymbol{U},\widetilde{\sigma}^{(per)}), \boldsymbol{\xi}_{*,t}^{(w)\top}(\boldsymbol{U},\widetilde{\sigma}^{(per)}), 1_t^{(w)}\right)}_{\boldsymbol{F}_t^{(1)}(\boldsymbol{U})^\top}\epsilon_{0,*,t}^{(w)} = O_p(1)$$

for every $\widetilde{\sigma}^{(per)} \in \sigma(\mathcal{P})$; and

$$\left(\max_{\widetilde{\sigma}^{(per)}\in\sigma(\mathcal{P})}\inf_{\sigma^{(per)}\in\sigma(\mathcal{P})}\left|\frac{N}{T}\sum_{t=1}^{T}\left(\boldsymbol{\xi}_{*,t}^{(w)\top}(\boldsymbol{U}_0,\sigma^{(per)})-\boldsymbol{\xi}_{*,t}^{(w)\top}(\boldsymbol{U},\widetilde{\sigma}^{(per)})\right)\epsilon_{0,*,t}^{(w)}\right|,\right.$$
$$\left.\max_{\sigma^{(per)}\in\sigma(\mathcal{P})}\inf_{\widetilde{\sigma}^{(per)}\in\sigma(\mathcal{P})}\left|\frac{N}{T}\sum_{t=1}^{T}\left(\boldsymbol{\xi}_{*,t}^{(w)\top}(\boldsymbol{U}_0,\sigma^{(per)})-\boldsymbol{\xi}_{*,t}^{(w)\top}(\boldsymbol{U},\widetilde{\sigma}^{(per)})\right)\epsilon_{0,*,t}^{(w)}\right|\right)^{+}=o_p(1)$$

for every $\boldsymbol{U}\in\mathcal{B}(\boldsymbol{U}_0,\eta_u)$. Therefore, we have that, for every $\boldsymbol{U}\in\mathcal{B}(\boldsymbol{U}_0,\eta_u)$ and $\boldsymbol{\psi}\in\mathcal{B}(\boldsymbol{\psi}_0,\eta_\psi)$,

$$\left(\max_{\widetilde{\sigma}^{(per)}\in\sigma(\mathcal{P})}\inf_{\sigma^{(per)}\in\sigma(\mathcal{P})}\left|\left((\boldsymbol{\theta}^{(\widetilde{\sigma}^{(per)})}-\boldsymbol{\theta}_0^{(\sigma^{(per)})})^\top,(\boldsymbol{\phi}_0^{(\sigma^{(per)})}-\boldsymbol{\phi}^{(\widetilde{\sigma}^{(per)})})^\top,\mu_{*0}-\mu_*,\boldsymbol{\phi}_0^{(\sigma^{(per)})\top}\right)\right.\right.$$
$$\left.\times\text{diag}\left(\boldsymbol{D}_\phi(\widetilde{\sigma}^{(per)}),\mathbb{I}_{2G+1}\right)\sqrt{\frac{N}{T}}\sum_{t=1}^{T}\boldsymbol{F}_t(\boldsymbol{U},\boldsymbol{U}_0)\epsilon_{0,*,t}^{(w)}\right|,$$
$$\max_{\sigma^{(per)}\in\sigma(\mathcal{P})}\inf_{\widetilde{\sigma}^{(per)}\in\sigma(\mathcal{P})}\left|\left((\boldsymbol{\theta}^{(\widetilde{\sigma}^{(per)})}-\boldsymbol{\theta}_0^{(\sigma^{(per)})})^\top,(\boldsymbol{\phi}_0^{(\sigma^{(per)})}-\boldsymbol{\phi}^{(\widetilde{\sigma}^{(per)})})^\top,\mu_{*0}-\mu_*,\boldsymbol{\phi}_0^{(\sigma^{(per)})\top}\right)\right.$$
$$\left.\left.\times\text{diag}\left(\boldsymbol{D}_\phi(\widetilde{\sigma}^{(per)}),\mathbb{I}_{2G+1}\right)\sqrt{\frac{N}{T}}\sum_{t=1}^{T}\boldsymbol{F}_t(\boldsymbol{U},\boldsymbol{U}_0)\epsilon_{0,*,t}^{(w)}\right|\right)^{+}=o_p(1).$$

It then follows from (B.29) that

$$\overline{Q}_{N,T}(\boldsymbol{\psi}_0,\sigma_{\epsilon,0}^2,\boldsymbol{U}_0)-\overline{Q}_{N,T}(\boldsymbol{\psi},\sigma_\epsilon^2,\boldsymbol{U})\geq C_0\inf_{H(\boldsymbol{U},\boldsymbol{U}_0)<\eta_u}\lambda_{\min}\left(\frac{1}{T}\sum_{t=1}^{T}\boldsymbol{F}_t^{(1)}(\boldsymbol{U})\boldsymbol{F}_t^{(1)}(\boldsymbol{U})^\top\right)NH(\boldsymbol{\psi},\boldsymbol{\psi}_0)^2$$
$$+o_p(1)$$

for every $\boldsymbol{\psi}\in\mathcal{B}(\boldsymbol{\psi}_0,\eta_\psi)$ and $\boldsymbol{U}\in\mathcal{B}(\boldsymbol{U}_0,\eta_u)$. It then follows that

$$P\left(|\widehat{\sigma}_\epsilon^2-\sigma_{\epsilon,0}^2|>\eta_\sigma,\widehat{\boldsymbol{\psi}}\in\mathfrak{B}\left(\boldsymbol{\psi}_0,\eta'_\psi/\sqrt{N}\right)\right)$$
$$\leq P\left(\inf_{\boldsymbol{\psi}\in\mathfrak{B}^c\left(\boldsymbol{\psi}_0,\eta'_\psi/\sqrt{N}\right),\boldsymbol{U}\in\mathcal{B}(\boldsymbol{U}_0,\eta_u)}\left\{\overline{Q}_{N,T}(\boldsymbol{\psi}_0,\sigma_{\epsilon,0}^2,\boldsymbol{U}_0)-\overline{Q}_{N,T}(\boldsymbol{\psi},\sigma_\epsilon^2,\boldsymbol{U})\right\}\leq 0\right)$$

because $\inf_{\boldsymbol{\psi}\in\mathfrak{B}^c\left(\boldsymbol{\psi}_0,\eta'_\psi/\sqrt{N}\right),\boldsymbol{U}\in\mathcal{B}(\boldsymbol{U}_0,\eta_u)}\left\{\overline{Q}_{N,T}(\boldsymbol{\psi}_0,\sigma_{\epsilon,0}^2,\boldsymbol{U}_0)-\overline{Q}_{N,T}(\boldsymbol{\psi},\sigma_\epsilon^2,\boldsymbol{U})\right\}\geq\eta_\psi'^2>0$. This completes the proof. ■

**Proof of Theorem 6.** First, note that discrete constraints of the form $\boldsymbol{U}\in\{0,1\}^{G\times N}$ in the combinatorial optimization problem (4.14) are equivalent to a system of d.c. constraints: $\boldsymbol{U}\in[0,1]^{G\times N}$, $g(\boldsymbol{U}):=\sum_{c=1}^{G}\sum_{i=1}^{N}u_{i,c}(1-u_{i,c})\leq 0$. Clearly, $g(\boldsymbol{U})$ is finitely concave on $\mathbb{R}^{G\times N}$, nonnegative on $\Delta_S^N$. It immediately follows that $\Delta_S^N\bigcap\{0,1\}^{G\times N}=\{\boldsymbol{U}\in\Delta_S^N:\ g(U)=0\}=\{\boldsymbol{U}\in$

$\Delta_S^N: \ g(U) \leq 0\}$. By Lemma 23 the following programs are equivalent:

$$(P_\Delta) \ \inf\left\{\frac{N}{T}\sum_{t=1}^{T}\epsilon_{*,t}^2(\boldsymbol{\psi},\boldsymbol{U}): \ \boldsymbol{\psi}\in\Theta_\psi\subset\mathbb{R}^{G(d_x+1)+1}, \ \boldsymbol{U}\in\Delta_S^N\bigcap\{0,1\}^{G\times N}\right\},$$
$$(P_\gamma) \ \inf\left\{\frac{N}{T}\sum_{t=1}^{T}\epsilon_{*,t}^2(\boldsymbol{\psi},\boldsymbol{U})+\gamma g(\boldsymbol{U}): \ \boldsymbol{\psi}\in\Theta_\psi\subset\mathbb{R}^{G(d_x+1)+1}, \ \boldsymbol{U}\in\Delta_S^N\right\}$$

for all $\gamma > \gamma_0$, where $\gamma_0$ is some positive constant.

For given $\boldsymbol{\psi}\in\mathcal{B}(\boldsymbol{\psi}_0,\eta_\psi)$ with $\mathcal{B}(\boldsymbol{\psi}_0,\eta_\psi)$ being an open ball centered at $\boldsymbol{\psi}_0$ with an arbitrarily small radius, $\eta_\psi$, one can obtain that

$$\widehat{\boldsymbol{U}}(\boldsymbol{\psi}) := \operatorname{argmin}_{\boldsymbol{U}\in\Delta_S^N\bigcap\{0,1\}^{G\times N}}\frac{N}{T}\sum_{t=1}^{T}\epsilon_{*,t}^2(\boldsymbol{\psi},\boldsymbol{U}) = \operatorname{argmin}_{\boldsymbol{U}\in\Delta_S^N}\left\{\frac{N}{T}\sum_{t=1}^{T}\epsilon_{*,t}^2(\boldsymbol{\psi},\boldsymbol{U})+\gamma g(\boldsymbol{U})\right\}.$$

Then, $\widehat{\boldsymbol{U}}(\boldsymbol{\psi})$ satisfies the Karush-Kuhn-Tucker (KKT) conditions (see, e.g., Bonnans and Shapiro (2000, p. 146)):

$$-\nabla\mathcal{Q}_{N,T}(\boldsymbol{\psi},\widehat{\boldsymbol{U}}(\boldsymbol{\psi}))\in N_{\Delta_S^N}(\widehat{\boldsymbol{U}}(\boldsymbol{\psi})), \tag{B.30}$$

where $\mathcal{Q}_{N,T}(\boldsymbol{\psi},\boldsymbol{U}) := \frac{N}{T}\sum_{t=1}^{T}\epsilon_{*,t}^2(\boldsymbol{\psi},\boldsymbol{U})+\gamma g(\boldsymbol{U})$ and $N_{\Delta_S^N}(\boldsymbol{U})$ is the normal cone of $\Delta_S^N$ at $\operatorname{vec}(\boldsymbol{U})$;

$$\boldsymbol{u}^\top\nabla^2\mathcal{Q}_{N,T}(\boldsymbol{\psi},\widehat{\boldsymbol{U}}(\boldsymbol{\psi}))\boldsymbol{u}\geq 0 \text{ for every } \boldsymbol{u}\in T_{\Delta_S^N}(\widehat{\boldsymbol{U}}(\boldsymbol{\psi})), \tag{B.31}$$

where $T_{\Delta_S^N}(\boldsymbol{U})$ is the tangent cone of $\Delta_S^N$ at $\operatorname{vec}(\boldsymbol{U})$. These KKT conditions basically imply that, if the vector of optimal group memberships $\widehat{\boldsymbol{U}}(\boldsymbol{\psi})$ is not the same as $\boldsymbol{U}_0$, then $\mathcal{Q}_{N,T}(\boldsymbol{\psi},\boldsymbol{U})$ must have some descent direction at $\boldsymbol{U}_0$. The rest of this proof proceeds along the line of this intuition.

Because $\boldsymbol{U}_0$ and $\widehat{\boldsymbol{U}}(\boldsymbol{\psi})$ are matrices of binary variables, it follows from (B.30) and (B.31) that

$$\begin{aligned}
&E\left[\sup_{\boldsymbol{\psi}\in\mathcal{B}(\boldsymbol{\psi}_0,\eta_\psi)}H\left(\widehat{\boldsymbol{U}}(\boldsymbol{\psi}),\boldsymbol{U}_0\right)\right] = \int_0^2 P\left(\sup_{\boldsymbol{\psi}\in\mathcal{B}(\boldsymbol{\psi}_0,\eta_\psi)}H\left(\widehat{\boldsymbol{U}}(\boldsymbol{\psi}),\boldsymbol{U}_0\right)>\tau\right)d\tau \\
&\leq C_0 P\Bigg(\Bigg(\min_{\sigma^{(per)}\in\sigma(\mathcal{P})}\frac{1}{N}\sum_{i=1}^{N}|\widehat{u}_{i,\sigma^{(per)}(c)}(\boldsymbol{\psi})-u_{0,i,c}|, \min_{\sigma^{(per)}\in\sigma(\mathcal{P})}\frac{1}{N}\sum_{i=1}^{N}|\widehat{u}_{i,c}(\boldsymbol{\psi})-u_{0,i,\sigma^{(per)}(c)}|\Bigg)^+\neq 0 \\
&\qquad \text{for every } \boldsymbol{\psi}\in\mathcal{B}(\boldsymbol{\psi}_0,\eta_\psi) \text{ and at least one } c\in[1,G]\Bigg) \qquad\qquad (\text{B.32}) \\
&\leq P\Bigg(\Bigg|\underbrace{\sum_{c=1}^{G}\Bigg\{\min_{\sigma^{(per)}\in\sigma(\mathcal{P})}\frac{1}{T}\sum_{t=1}^{T}\epsilon_{*,t}(\boldsymbol{\psi},\boldsymbol{U}_0)\phi_c\frac{1}{N}\sum_{i=1}^{N}(u_{0,i,\sigma^{(per)}(c)}-u_{i,c}) \times\left\{(\boldsymbol{\theta}_c-\boldsymbol{\theta}_{0,\sigma^{(per)}(c)})^\top\boldsymbol{x}_{i,t}^{(w)}-\xi_{i,t}^{(w)}(\boldsymbol{\theta}_{0,\sigma^{(per)}(c)})\right\}+\gamma\left(\frac{1}{N}\sum_{i=1}^{N}u_{0,i,\sigma^{(per)}(c)}u_{i,c}-1\right)\Bigg\}\Bigg|>0}_{\text{by (B.30)}}
\end{aligned}$$

$$\text{for every } \gamma < \underbrace{\min_{\sigma^{(per)}\in\sigma(\mathcal{P})}\sum_{c=1}^{G}\phi_c^2\frac{1}{NT}\sum_{t=1}^{T}\left((\boldsymbol{\theta}_c-\boldsymbol{\theta}_{0,\sigma^{(per)}(c)})^\top\boldsymbol{x}_{i,t}^{(w)}-\xi_{i,t}^{(w)}(\boldsymbol{\theta}_{0,\sigma^{(per)}(c)})\right)^2,}_{\text{by (B.31)}}$$

$$\left.\boldsymbol{\psi}\in\mathcal{B}(\boldsymbol{\psi}_0,\eta_\psi),\ \text{and } \boldsymbol{u}\in\Delta_S^N\right)$$

$$+P\Bigg(\underbrace{\left|\sum_{c=1}^{G}\left\{\min_{\sigma^{(per)}\in\sigma(\mathcal{P})}\frac{1}{T}\sum_{t=1}^{T}\epsilon_{*,t}(\boldsymbol{\psi},\boldsymbol{U}_0)\phi_{\sigma^{(per)}(c)}\frac{1}{N}\sum_{i=1}^{N}(u_{0,i,c}-u_{i,\sigma^{(per)}(c)})\times\left\{(\boldsymbol{\theta}_{\sigma^{(per)}(c)}-\boldsymbol{\theta}_{0,c})^\top\boldsymbol{x}_{i,t}^{(w)}-\xi_{i,t}^{(w)}(\boldsymbol{\theta}_{0,c})\right\}+\gamma\left(\frac{1}{N}\sum_{i=1}^{N}u_{0,i\sigma^{(per)}(c)}u_{i,c}-1\right)\right\}\right|>0}_{\text{by (B.30)}}$$

$$\text{for every } \gamma < \underbrace{\min_{\sigma^{(per)}\in\sigma(\mathcal{P})}\sum_{c=1}^{G}\phi_{\sigma^{(per)}(c)}^2\frac{1}{NT}\left((\boldsymbol{\theta}_{\sigma^{(per)}(c)}-\boldsymbol{\theta}_{0,c})^\top\boldsymbol{x}_{i,t}^{(w)}-\xi_{i,t}^{(w)}(\boldsymbol{\theta}_{0,c})\right)^2,}_{\text{by (B.31)}}$$

$$\left.\boldsymbol{\psi}\in\mathcal{B}(\boldsymbol{\psi}_0,\eta_\psi)\ \text{and } \boldsymbol{u}\in\Delta_S^N\right)$$

$$=:\mathcal{T}_{1,N,T}+\mathcal{T}_{2,N,T}. \tag{B.33}$$

To bound $E\left[\sup_{\boldsymbol{\psi}\in\mathcal{B}(\boldsymbol{\psi}_0,\eta_\psi)}H\left(\widehat{\boldsymbol{U}}(\boldsymbol{\psi}),\boldsymbol{U}_0\right)\right]$, we shall bound $\mathcal{T}_{1,N,T}$ since $\mathcal{T}_{2,N,T}$ can be bounded in the same manner. As the cardinality of $\sigma(\mathcal{P})$ is finite, we only need to work out the rate of convergence for

$$\mathcal{T}_{1,N,T}' := P\Bigg(\sum_{c=1}^{G}\min_{\sigma^{(per)}\in\sigma(\mathcal{P})}\left\{\frac{1}{T}\sum_{t=1}^{T}\epsilon_{*,t}(\boldsymbol{\psi},\boldsymbol{U}_0)\phi_c\frac{1}{N}\sum_{i=1}^{N}(u_{0,i,c}-u_{i,c})\left\{(\boldsymbol{\theta}_{\sigma^{(per)}(c)}-\boldsymbol{\theta}_{0,c})^\top\boldsymbol{x}_{i,t}^{(w)}-\xi_{i,t}^{(w)}(\boldsymbol{\theta}_{0,c})\right\}+\gamma\left(\frac{1}{N}\sum_{i=1}^{N}u_{0,i,c}u_{i,c}-1\right)\right\}>\epsilon_\eta \text{ for every } \gamma<\min_{\sigma^{(per)}\in\sigma(\mathcal{P})}\sum_{c=1}^{G}\phi_c^2\frac{1}{NT}\sum_{t=1}^{T}\left((\boldsymbol{\theta}_c-\boldsymbol{\theta}_{0,\sigma^{(per)}(c)})^\top\boldsymbol{x}_{i,t}^{(w)}-\xi_{i,t}^{(w)}(\boldsymbol{\theta}_{0,\sigma^{(per)}(c)})\right)^2,\ \boldsymbol{\psi}\in\mathcal{B}(\boldsymbol{\psi}_0,\eta_\psi),\ \text{and } \boldsymbol{u}\in\Delta_S^N\Bigg),$$

where $\epsilon_\eta$ is some arbitrarily small positive constant. Notice that $\boldsymbol{U}$ is bounded and $\gamma = O_{a.s.}(N^{-1})$ by the strong law of large numbers and the compactness of the parameter spaces. An application of Boole's inequality yields that

$$\mathcal{T}_{1,N,T}' \le P\left(\sup_{\boldsymbol{U}\in\Delta_S^N}\frac{1}{NT}\sum_{c=1}^{G}\min_{\sigma^{(per)}\in\sigma(\mathcal{P})}\left|\sum_{i=1}^{N}\sum_{t=1}^{T}\phi_c(u_{0,i,c}-u_{i,c})\left\{(\boldsymbol{\theta}_{\sigma^{(per)}(c)}-\boldsymbol{\theta}_{0,c})^\top\boldsymbol{x}_{i,t}^{(w)}-\xi_{i,t}^{(w)}(\boldsymbol{\theta}_{0,c})\right\}\times(\mu_{*0}-\mu_*)1_t^{(w)}\right|>\frac{\epsilon_\eta}{4}\right)$$

$$
\begin{aligned}
&+P\Bigg(\sup_{\boldsymbol{U}\in\Delta_S^N}\frac{1}{NT}\sum_{c=1}^{G}\min_{\sigma^{(per)}\in\sigma(\mathcal{P})}\Bigg|\sum_{i=1}^{N}\sum_{t=1}^{T}(\phi_{0,c}-\phi_{\sigma^{(per)}(c)})(u_{0,i,c}-u_{i,c})\big\{(\boldsymbol{\theta}_{\sigma^{(per)}(c)}-\boldsymbol{\theta}_{0,c})^\top\boldsymbol{x}_{i,t}^{(w)}\\
&\qquad\qquad -\xi_{i,t}^{(w)}(\boldsymbol{\theta}_{0,c})\big\}\xi_{*,t}^{(w)}(\boldsymbol{\theta}_{0,c},\boldsymbol{u}_{0,c})\Bigg|>\epsilon_\eta/4\Bigg)\\
&+P\Bigg(\sup_{\boldsymbol{U}\in\Delta_S^N}\frac{1}{NT}\sum_{c=1}^{G}\min_{\sigma^{(per)}\in\sigma(\mathcal{P})}\Bigg|\sum_{i=1}^{N}\sum_{t=1}^{T}\phi_{\sigma^{(per)}(c)}(u_{0,i,c}-u_{i,\sigma^{(per)}(c)})\left(\boldsymbol{\theta}_{\sigma^{(per)}(c)}-\boldsymbol{\theta}_{0,c}\right)^\top\boldsymbol{x}_{*,t}^{(w)}(\boldsymbol{u}_{0,c})\\
&\qquad\qquad\times\left\{(\boldsymbol{\theta}_{\sigma^{(per)}(c)}-\boldsymbol{\theta}_{0,c})^\top\boldsymbol{x}_{i,t}^{(w)}-\xi_{i,t}^{(w)}(\boldsymbol{\theta}_{0,c})\right\}\Bigg|>\epsilon_\eta/4\Bigg)\\
&+P\Bigg(\sup_{\boldsymbol{U}\in\Delta_S^N}\frac{1}{NT}\sum_{c=1}^{G}\min_{\sigma^{(per)}\in\sigma(\mathcal{P})}\Bigg|\sum_{i=1}^{N}\sum_{t=1}^{T}\phi_{\sigma^{(per)}(c)}(u_{0,i,c}-u_{i,\sigma^{(per)}(c)})\big\{(\boldsymbol{\theta}_{\sigma^{(per)}(c)}-\boldsymbol{\theta}_{0,c})^\top\boldsymbol{x}_{i,t}^{(w)}\\
&\qquad\qquad-\xi_{i,t}^{(w)}(\boldsymbol{\theta}_{0,c})\big\}\epsilon_{0,*,t}^{(w)}\Bigg|>\epsilon_\eta/4\Bigg)\\
&=:\mathcal{T}_{1,N,T}^{(a)}+\mathcal{T}_{1,N,T}^{(b)}+\mathcal{T}_{1,N,T}^{(c)}+\mathcal{T}_{1,N,T}^{(d)}.
\end{aligned}
$$

As $T$ becomes large, using the same argument in Lemma 1, it can be verified that $\boldsymbol{x}_{i,t}^{(w)}=\boldsymbol{x}_{i,t}+o_p(1)$, $\xi_{i,t}^{(w)}(\boldsymbol{\theta}_{0,c})=\xi_{i,t}(\boldsymbol{\theta}_{0,c})+o_p(1)$, and $1_t^{(w)}=1+o_p(1)$. Therefore, by Boole's inequality, one has that

$$
\begin{aligned}
\mathcal{T}_{1,N,T}^{(a)}\le\ &P\Bigg(\sup_{\boldsymbol{U}\in\Delta_S^N}\frac{1}{NT}\sum_{c=1}^{G}\min_{\sigma^{(per)}\in\sigma(\mathcal{P})}\Bigg|\sum_{i=1}^{N}\sum_{t=1}^{T}\phi_{\sigma^{(per)}(c)}(u_{0,i,c}-u_{i,c})(\mu_{*0}-\mu_*)(\boldsymbol{\theta}_{\sigma^{(per)}(c)}-\boldsymbol{\theta}_{0,c})\boldsymbol{x}_{i,t}\Bigg|\\
&\qquad>\epsilon_\eta/12\Bigg)\\
&+P\Bigg(\sup_{\boldsymbol{U}\in\Delta_S^N}\frac{1}{NT}\sum_{c=1}^{G}\min_{\sigma^{(per)}\in\sigma(\mathcal{P})}\Bigg|\sum_{i=1}^{N}\sum_{t=1}^{T}\phi_{\sigma^{(per)}(c)}(u_{0,i,c}-u_{i,c})(\mu_{*0}-\mu_*)(\xi_{i,t}(\boldsymbol{\theta}_{0,c})-E[\xi_{i,t}(\boldsymbol{\theta}_{0,c})])\Bigg|\\
&\qquad>\epsilon_\eta/12\Bigg)\\
&+P\left(\sup_{\boldsymbol{U}\in\Delta_S^N}\frac{1}{NT}\sum_{c=1}^{G}\min_{\sigma^{(per)}\in\sigma(\mathcal{P})}\left|\sum_{i=1}^{N}\sum_{t=1}^{T}\phi_{\sigma^{(per)}(c)}\left(u_{0,i,c}-u_{i,c}\right)\left(\mu_{*0}-\mu_*\right)E[\xi_{i,t}(\boldsymbol{\theta}_{0,c})]\right|>\epsilon_\eta/12\right)\\
=:\ &\mathcal{T}_{1,N,T}^{(a*)}+\mathcal{T}_{1,N,T}^{(a**)}+\mathcal{T}_{1,N,T}^{(a***)}.
\end{aligned}\tag{B.34}
$$

Since $\left|\phi_{\sigma^{(per)}(c)}(u_{0,i,c}-u_{i,c})(\mu_{*0}-\mu_*)(\boldsymbol{\theta}_{\sigma^{(per)}(c)}-\boldsymbol{\theta}_{0,c})\right|<\infty$, an application of Lemma 11 yields

$$
\max(\mathcal{T}_{1,N,T}^{(a*)},\mathcal{T}_{1,N,T}^{(a**)})<C_0\left\{T^{-C_\alpha}+N^{\gamma_M}\log(T)T^{\gamma_M-\frac{3}{4}\theta_\alpha}+\max\left(\exp\left(-C_{\epsilon_\eta}NT\right),\exp\left(-C_{\epsilon_\eta}\frac{T^{1/4}}{\log(T)}\right)\right)\right\},
$$

where $C_\alpha$ is a sufficiently large constant, and $C_{\epsilon_\eta}$ is some generic constant depending on $\epsilon_\eta$. More-

over, notice that

$$\sup_{\boldsymbol{U}\in\Delta_S^N}\frac{1}{NT}\sum_{c=1}^{G}\min_{\sigma^{(per)}\in\sigma(\mathcal{P})}\left|\sum_{i=1}^{N}\sum_{t=1}^{T}\phi_{\sigma^{(per)}(c)}\left(u_{0,i,c}-u_{i,c}\right)(\mu_{*0}-\mu_*)E[\xi_{i,t}(\boldsymbol{\theta}_{0,c})]\right|$$
$$\leq \eta_\psi \max_{c,i,t}|E[\xi_{i,t}(\boldsymbol{\theta}_{0,c})]|\sup_{\boldsymbol{U}\in\Delta_S^N}\frac{1}{NT}\sum_{c=1}^{G}\min_{\sigma^{(per)}\in\sigma(\mathcal{P})}\left|\sum_{i=1}^{N}\sum_{t=1}^{T}\phi_{\sigma^{(per)}(c)}\left(u_{0,i,c}-u_{i,c}\right)\right|.$$

Choosing $\eta_\psi \leq K_{\eta,1}$ with $K_{\eta,1} := \frac{\epsilon_\eta}{12\lim_{\substack{N\uparrow\infty\\T\uparrow\infty}}|E[\xi_{i,t}(\boldsymbol{\theta}_{0,c})]|\sup_{\boldsymbol{U}\in\Delta_S^N}\frac{1}{NT}\sum_{c=1}^{G}\min_{\sigma^{(per)}\in\sigma(\mathcal{P})}\left|\sum_{i=1}^{N}\sum_{t=1}^{T}\phi_{\sigma^{(per)}(c)}(u_{0,i,c}-u_{i,c})\right|}$,
one has $\mathcal{T}_{1,N,T}^{(a***)}=0$ for sufficiently large $N$ and $T$. From (B.34), we can conclude that

$$\mathcal{T}_{1,N,T}^{(a)} \;<\; C_0\left\{T^{-C_\alpha}+N^{\gamma_M}\log(T)T^{\gamma_M-\frac{3}{4}\theta_\alpha}+\max\left(\exp\left(-C_{\epsilon_\eta}NT\right),\exp\left(-C_{\epsilon_\eta}\frac{T^{1/4}}{\log(T)}\right)\right)\right\}. \quad \text{(B.35)}$$

Since $\xi_{*,t}^{(w)}(\boldsymbol{\theta}_{0,c},\boldsymbol{u}_{0,c}) = g_cE[\xi_{i,t}(\boldsymbol{\theta}_{0,c})]+o_p(1)$ with $g_c=\frac{1}{N}\sum_{i=1}^{N}u_{0,i,c}$, one has that

$$\mathcal{T}_{1,N,T}^{(b)} < P\Bigg(\sup_{\boldsymbol{U}\in\Delta_S^N}\frac{1}{NT}\sum_{c=1}^{G}g_cE[\xi_{i,t}(\boldsymbol{\theta}_{0,c})]\min_{\sigma^{(per)}\in\sigma(\mathcal{P})}\Bigg|\sum_{i=1}^{N}\sum_{t=1}^{T}(\phi_{0,c}-\phi_{\sigma^{(per)}(c)})(u_{0,i,c}-u_{i,\sigma^{(per)}(c)})$$
$$\times(\boldsymbol{\theta}_{\sigma^{(per)}(c)}-\boldsymbol{\theta}_{0,c})^\top\boldsymbol{x}_{i,t}^{(w)}\Bigg|>\frac{\epsilon_\eta}{12}\Bigg)$$
$$+P\Bigg(\sup_{\boldsymbol{U}\in\Delta_S^N}\frac{1}{NT}\sum_{c=1}^{G}g_cE[\xi_{i,t}(\boldsymbol{\theta}_{0,c})]\min_{\sigma^{(per)}\in\sigma(\mathcal{P})}\Bigg|\sum_{i=1}^{N}\sum_{t=1}^{T}(\phi_{0,c}-\phi_{\sigma^{(per)}(c)})(u_{0,i,c}-u_{i,\sigma^{(per)}(c)})\big(\xi_{i,t}(\boldsymbol{\theta}_{0,c})$$
$$-E[\xi_{i,t}(\boldsymbol{\theta}_{0,c})]\big)\Bigg|>\frac{\epsilon_\eta}{12}\Bigg)$$
$$+P\Bigg(\sup_{\boldsymbol{U}\in\Delta_S^N}\frac{1}{NT}\sum_{c=1}^{G}g_cE^2[\xi_{i,t}(\boldsymbol{\theta}_{0,c})]\min_{\sigma^{(per)}\in\sigma(\mathcal{P})}\Bigg|\sum_{i=1}^{N}\sum_{t=1}^{T}(\phi_{0,c}-\phi_{\sigma^{(per)}(c)})(u_{0,i,c}-u_{i,\sigma^{(per)}(c)})\Bigg|>\frac{\epsilon_\eta}{12}\Bigg),$$

where the first two terms can be bounded in the same manner as (B.35) and the last term can be made arbitrarily close to zero by choosing $\eta_\psi < (K_{\eta,1},K_{\eta,2})^-$ with

$$K_{\eta,2} := \frac{\epsilon_\eta}{12\sup_{\boldsymbol{U}\in\Delta_S^N}\frac{1}{NT}\sum_{c=1}^{G}g_cE^2[\xi_{i,t}(\boldsymbol{\theta}_{0,c})]\min_{\sigma^{(per)}\in\sigma(\mathcal{P})}\left|\sum_{i=1}^{N}\sum_{t=1}^{T}(u_{0,i,c}-u_{i,\sigma^{(per)}(c)})\right|}.$$

It then follows that

$$\mathcal{T}_{1,N,T}^{(b)} \;<\; C_0\left\{T^{-C_\alpha}+N^{\gamma_M}\log(T)T^{\gamma_M-\frac{3}{4}\theta_\alpha}+\max\left(\exp\left(-C_{\epsilon_\eta}NT\right),\exp\left(-C_{\epsilon_\eta}\frac{T^{1/4}}{\log(T)}\right)\right)\right\}. \quad \text{(B.36)}$$

Using exactly the same argument, one can also obtain that

$$\mathcal{T}_{1,N,T}^{(c)} \;<\; C_0 \left\{ T^{-C_\alpha} + N^{\gamma_M} \log(T) T^{\gamma_M - \frac{3}{4}\theta_\alpha} + \max\left( \exp\left(-C_{\epsilon_\eta} NT\right), \exp\left( -C_{\epsilon_\eta} \frac{T^{1/4}}{\log(T)} \right) \right) \right\}. \quad \text{(B.37)}$$

Now, to bound the last term of $\mathcal{T}'_{1,N,T}$. Notice that

$$\begin{aligned}
\mathcal{T}_{1,N,T}^{(d)} \le{}& P\Bigg( \sup_{\boldsymbol{U} \in \Delta_S^N} \frac{1}{NT} \sum_{c=1}^{G} \min_{\sigma^{(per)} \in \sigma(\mathcal{P})} \left| \phi_{\sigma^{(per)}(c)} \left( \boldsymbol{\theta}_{\sigma^{(per)}(c)} - \boldsymbol{\theta}_{0,c} \right)^\top \right| \left| \sum_{i=1}^{N} \sum_{t=1}^{T} (u_{0,i,c} - u_{i,\sigma^{(per)}(c)}) \boldsymbol{x}_{i,t}^{(w)}(\boldsymbol{\theta}_{0,c}) \epsilon_{0,*,t}^{(w)} \right| \\
&> \frac{\epsilon_\eta}{8} \Bigg) \\
&+ P\left( \sup_{\boldsymbol{U} \in \Delta_S^N} \frac{1}{NT} \sum_{c=1}^{G} \min_{\sigma^{(per)} \in \sigma(\mathcal{P})} \left| \sum_{i=1}^{N} \sum_{t=1}^{T} \phi_{\sigma^{(per)}(c)} (u_{0,i,c} - u_{i,\sigma^{(per)}(c)}) \xi_{i,t}^{(w)}(\boldsymbol{\theta}_{0,c}) \epsilon_{0,*,t}^{(w)} \right| > \frac{\epsilon_\eta}{8} \right) \\
=:{}& \mathcal{T}_{1,N,T}^{(d)*} + \mathcal{T}_{1,N,T}^{(d)**}. \qquad \text{(B.38)}
\end{aligned}$$

Since $\boldsymbol{x}_{i,t}^{(w)} = \boldsymbol{x}_{i,t} + o_p(1)$ and $\epsilon_{0,*,t}^{(w)} = \epsilon_{0,*,t} + o_p(1)$, it then follows that

$$\begin{aligned}
\mathcal{T}_{1,N,T}^{(d)*} \le{}& P\Bigg( \sup_{\boldsymbol{U} \in \Delta_S^N} \frac{1}{NT} \sum_{c=1}^{G} \min_{\sigma^{(per)} \in \sigma(\mathcal{P})} \left| \phi_{\sigma^{(per)}(c)} \left( \boldsymbol{\theta}_{\sigma^{(per)}(c)} - \boldsymbol{\theta}_{0,c} \right)^\top \right| \\
&\times \left| \sum_{i=1}^{N} \sum_{t=1}^{T} (u_{0,i,c} - u_{i,\sigma^{(per)}(c)}) \left\{ \boldsymbol{x}_{i,t}(\boldsymbol{\theta}_{0,c}) \epsilon_{0,*,t} - E[\boldsymbol{x}_{i,t}(\boldsymbol{\theta}_{0,c}) \epsilon_{0,*,t}] \right\} \right| > \frac{\epsilon_\eta}{16} \Bigg) \\
&+ P\Bigg( \sup_{\boldsymbol{U} \in \Delta_S^N} \frac{1}{NT} \sum_{c=1}^{G} \min_{\sigma^{(per)} \in \sigma(\mathcal{P})} \left| \phi_{\sigma^{(per)}(c)} \left( \boldsymbol{\theta}_{\sigma^{(per)}(c)} - \boldsymbol{\theta}_{0,c} \right)^\top \right| \\
&\times \left| \sum_{i=1}^{N} \sum_{t=1}^{T} (u_{0,i,c} - u_{i,\sigma^{(per)}(c)}) E\left[ \boldsymbol{x}_{i,t}(\boldsymbol{\theta}_{0,c}) \epsilon_{0,*,t} \right] \right| > \frac{\epsilon_\eta}{16} \Bigg),
\end{aligned}$$

where the last term can be made arbitrarily close to zero by choosing $\eta_\psi < (K_{\eta,1}, K_{\eta,2}, K_{\eta,3})^-$ with $\eta_\psi < \frac{\epsilon_\eta}{16 \sup_{\boldsymbol{U} \in \Delta_S^N} \frac{1}{NT} \sum_{c=1}^{G} \min_{\sigma^{(per)} \in \sigma(\mathcal{P})} \left\| \phi_{\sigma^{(per)}(c)} \sum_{i=1}^{N} \sum_{t=1}^{T} \left( u_{0,i,c} - u_{i,\sigma^{(per)}(c)} \right) E[\boldsymbol{x}_{i,t}(\boldsymbol{\theta}_{0,c}) \epsilon_{0,*,t}] \right\|}$, and the first term can be bounded by invoking Lemma 12. Therefore, we have that

$$\mathcal{T}_{1,N,T}^{(d)*} \le C_0 \left( T^{-C_\alpha} + N^{2\gamma_M} \log^2(T) T^{\gamma_M - \frac{3}{4}\theta_\alpha} + \max\left\{ \exp\left( -C_\sigma N^2 \log^2(T) T^{7/4} \right), \exp\left( -C_M \frac{T^{1/4}}{\log^2(T)} \right) \right\} \right).$$

By exactly the same argument, one can also show that

$$\mathcal{T}_{1,N,T}^{(d)**} \le C_0 \left( T^{-C_\alpha} + N^{2\gamma_M} \log^2(T) T^{\gamma_M - \frac{3}{4}\theta_\alpha} + \max\left\{ \exp\left( -C_\sigma N^2 \log^2(T) T^{7/4} \right), \exp\left( -C_M \frac{T^{1/4}}{\log^2(T)} \right) \right\} \right).$$

It then follows from (B.38) that

$$\mathcal{T}^{(d)}_{1,N,T} \leq C_0 \Big( T^{-C_\alpha} + N^{2\gamma_M} \log^2(T) T^{\gamma_M - \frac{3}{4}\theta_\alpha} + \max\left\{ \exp\left(-C_\sigma N^2 \log^2(T) T^{7/4}\right), \exp\left(-C_M \frac{T^{1/4}}{\log^2(T)}\right)\right\}\Big). \qquad \text{(B.39)}$$

In view of (B.34)-(B.39) the main theorem then follows. ■

**Proof of Theorem 7.** The proof proceeds in the following three main steps:

STEP 1: It can immediately be verified that $\epsilon_{*,t}(\boldsymbol{\psi}, \boldsymbol{U}_0) = (\boldsymbol{\psi} - \widetilde{\boldsymbol{\psi}})^\top \boldsymbol{D}_\phi \boldsymbol{D}_g \boldsymbol{X}_{N,T,t}(\boldsymbol{\theta}) + \epsilon_{*,t}(\widetilde{\boldsymbol{\psi}}, \boldsymbol{U}_0)$. Since $\widetilde{\boldsymbol{\psi}}$ is the minimum value of $\widetilde{\mathcal{Q}}_{N,T}(\boldsymbol{\psi}) \coloneqq \frac{N}{T}\sum_{t=1}^T \epsilon^2_{*,t}(\boldsymbol{\psi}, \boldsymbol{U}_0)$, it must satisfy the equations:

$$\begin{aligned} \frac{N}{T}\sum_{t=1}^T \xi^{(w)}_{*,t}(\widetilde{\boldsymbol{\theta}}_c, \boldsymbol{u}_c)\epsilon_{*,t}(\widetilde{\boldsymbol{\psi}}, \boldsymbol{U}_0) &= 0, \\ \frac{N}{T}\sum_{t=1}^T \boldsymbol{x}^{(w)}_{*,t}(\boldsymbol{u}_{0,c})\epsilon_{*,t}(\widetilde{\boldsymbol{\psi}}, \boldsymbol{U}_0) &= 0, \\ \frac{N}{T}\sum_{t=1}^T 1^{(w)}_t \epsilon_{*,t}(\widetilde{\boldsymbol{\psi}}, \boldsymbol{U}_0) &= 0. \end{aligned}$$

Therefore, an application of the eigenvalue inequality and Theorem 5 yields

$$\widetilde{\mathcal{Q}}_{N,T}(\widehat{\boldsymbol{\psi}}) - \widetilde{\mathcal{Q}}_{N,T}(\widetilde{\boldsymbol{\psi}}) \geq \lambda_{\min}\left(\frac{N}{T}\sum_{t=1}^T \boldsymbol{X}_{N,T,t}(\boldsymbol{\theta}_0)\boldsymbol{X}_{N,T,t}(\boldsymbol{\theta}_0)^\top\right) H(\widehat{\boldsymbol{\psi}}, \widetilde{\boldsymbol{\psi}}) > C_0 H(\widehat{\boldsymbol{\psi}}, \widetilde{\boldsymbol{\psi}}), \qquad \text{(B.40)}$$

where the last inequality follows from Assumption 4.

STEP 2: Let $\widehat{\mathcal{Q}}_{N,T}(\boldsymbol{\psi}) \coloneqq \frac{N}{T}\sum_{t=1}^T \epsilon^2_{*,t}(\boldsymbol{\psi}, \widehat{\boldsymbol{U}})$, where $\widehat{\boldsymbol{U}} \equiv \widehat{\boldsymbol{U}}(\boldsymbol{\psi}) \coloneqq \operatorname{argmin}_{\boldsymbol{U}\in\Delta^N_S \bigcap \{0,1\}^{G\times N}} \frac{N}{T}\sum_{t=1}^T \epsilon^2_{*,t}(\boldsymbol{\psi}, \boldsymbol{U})$. One can show that

$$\begin{aligned} \widehat{\mathcal{Q}}_{N,T}(\boldsymbol{\psi}) - \widetilde{\mathcal{Q}}_{N,T}(\boldsymbol{\psi}) = {}& \sum_{c=1}^G \phi_c \frac{1}{N}\sum_{i=1}^N (u_{0,i,c} - \widehat{u}_{i,\sigma^{(per)}(c)}) \frac{N}{T}\sum_{t=1}^T \xi^{(w)}_{i,t}(\boldsymbol{\theta}_{0,c}) \left(\epsilon_{*,t}(\boldsymbol{\psi}, \widehat{\boldsymbol{U}}) + \epsilon_{*,t}(\boldsymbol{\psi}, \boldsymbol{U}_0)\right) \\ & + \sum_{c=1}^G \phi_c (\boldsymbol{\theta}_{\sigma^{(per)}(c)} - \boldsymbol{\theta}_{0,c})^\top \frac{1}{N}\sum_{i=1}^N (\widehat{u}_{i,\sigma^{(per)}(c)} - u_{0,i,c}) \frac{N}{T}\sum_{t=1}^T \boldsymbol{x}^{(w)}_{i,t} \left(\epsilon_{*,t}(\boldsymbol{\psi}, \widehat{\boldsymbol{U}})\right. \\ & \left. + \epsilon_{*,t}(\boldsymbol{\psi}, \boldsymbol{U}_0)\right). \end{aligned}$$

By Hölder's inequality, one obtains that

$$\begin{aligned}\left|\widehat{\mathcal{Q}}_{N,T}(\boldsymbol{\psi})-\widetilde{\mathcal{Q}}_{N,T}(\boldsymbol{\psi})\right| \leq & \sum_{c=1}^{G}|\phi_c|\left\{\frac{1}{N}\sum_{i=1}^{N}(u_{0,i,c}-\widehat{u}_{i,\sigma^{(per)}(c)})^2\right\}^{\frac{1}{2}} \\ & \times\left\{\frac{N}{T^2}\sum_{i=1}^{N}\left(\sum_{t=1}^{T}\xi_{i,t}^{(w)}(\boldsymbol{\theta}_{0,c})\left(\epsilon_{*,t}(\boldsymbol{\psi},\widehat{\boldsymbol{U}})+\epsilon_{*,t}(\boldsymbol{\psi},\boldsymbol{U}_0)\right)\right)^2\right\}^{\frac{1}{2}} \\ & +\sum_{c=1}^{G}\left|\phi_c(\boldsymbol{\theta}_{\sigma^{(per)}(c)}-\boldsymbol{\theta}_{0,c})^\top\right|\left\{\frac{1}{N}\sum_{i=1}^{N}(\widehat{u}_{i,\sigma^{(per)}(c)}-u_{0,i,c})^2\right\}^{\frac{1}{2}} \\ & \times\left\{\frac{N}{T^2}\sum_{i=1}^{N}\left(\sum_{t=1}^{T}\boldsymbol{x}_{i,t}^{(w)}\left(\epsilon_{*,t}(\boldsymbol{\psi},\widehat{\boldsymbol{U}})+\epsilon_{*,t}(\boldsymbol{\psi},\boldsymbol{U}_0)\right)\right)^2\right\}^{\frac{1}{2}}. \end{aligned} \tag{B.41}$$

Because all the clusters are sufficiently large and $\frac{1}{T}\sum_{t=1}^{T}\boldsymbol{x}_{i,t}\xi_{i,t}(\boldsymbol{\theta}_{0,c}) = O_p(1)$ for every $i \in [1,N]$ and $c \in [1,G]$, one can verify that

$$\frac{1}{NT^2}\sum_{i=1}^{N}\left(\sum_{t=1}^{T}\xi_{i,t}^{(w)}(\boldsymbol{\theta}_{0,c})\left(\epsilon_{*,t}(\boldsymbol{\psi},\widehat{\boldsymbol{U}})+\epsilon_{*,t}(\boldsymbol{\psi},\boldsymbol{U}_0)\right)\right)^2 = O_p(1)$$

and

$$\frac{1}{NT^2}\sum_{i=1}^{N}\left(\sum_{t=1}^{T}\boldsymbol{x}_{i,t}^{(w)}(\boldsymbol{\theta}_{0,c})\left(\epsilon_{*,t}(\boldsymbol{\psi},\widehat{\boldsymbol{U}})+\epsilon_{*,t}(\boldsymbol{\psi},\boldsymbol{U}_0)\right)\right)^2 = O_p(1).$$

Since the objective functions are invariant with respect to relabelling of groups, by Theorem 6, we obtain that

$$\begin{aligned}&\sup_{\boldsymbol{\psi}\in\mathcal{B}(\boldsymbol{\psi}_0,\eta_\psi)}\left|\widehat{\mathcal{Q}}_{N,T}(\boldsymbol{\psi})-\widetilde{\mathcal{Q}}_{N,T}(\boldsymbol{\psi})\right| \\ &\qquad = O_p\left(NT^{-C_\alpha}+N^{\gamma_M+1}\log(T)T^{\frac{\gamma_M}{2}-\frac{3}{8}\theta_\alpha}+N\exp\left(-C_M\frac{T^{1/4}}{\log^2(T)}\right)\right). \end{aligned} \tag{B.42}$$

STEP 3: Notice that

$$\begin{aligned}\widetilde{\mathcal{Q}}_{N,T}(\widehat{\boldsymbol{\psi}})-\widetilde{\mathcal{Q}}_{N,T}(\widetilde{\boldsymbol{\psi}}) &= \widetilde{\mathcal{Q}}_{N,T}(\widehat{\boldsymbol{\psi}})-\widehat{\mathcal{Q}}_{N,T}(\widehat{\boldsymbol{\psi}})+\underbrace{\widehat{\mathcal{Q}}_{N,T}(\widehat{\boldsymbol{\psi}})-\widehat{\mathcal{Q}}_{N,T}(\widetilde{\boldsymbol{\psi}})}_{\leq 0}+\widehat{\mathcal{Q}}_{N,T}(\widetilde{\boldsymbol{\psi}})-\widetilde{\mathcal{Q}}_{N,T}(\widetilde{\boldsymbol{\psi}}) \\ &\leq \left\{\widetilde{\mathcal{Q}}_{N,T}(\widehat{\boldsymbol{\psi}})-\widehat{\mathcal{Q}}_{N,T}(\widehat{\boldsymbol{\psi}})\right\}+\left\{\widehat{\mathcal{Q}}_{N,T}(\widetilde{\boldsymbol{\psi}})-\widetilde{\mathcal{Q}}_{N,T}(\widetilde{\boldsymbol{\psi}})\right\}. \end{aligned} \tag{B.43}$$

Some probability event computations yield that

$$\left\{\left|\widetilde{\mathcal{Q}}_{N,T}(\widehat{\boldsymbol{\psi}})-\widehat{\mathcal{Q}}_{N,T}(\widehat{\boldsymbol{\psi}})\right|>\epsilon\right\}\subset\left\{\widehat{\boldsymbol{\psi}}\notin\mathcal{B}(\boldsymbol{\psi}_0,\eta_\psi)\right\}\bigcup\left\{\widehat{\boldsymbol{\psi}}\in\mathcal{B}(\boldsymbol{\psi}_0,\eta_\psi),\left|\widetilde{\mathcal{Q}}_{N,T}(\widehat{\boldsymbol{\psi}})-\widehat{\mathcal{Q}}_{N,T}(\widehat{\boldsymbol{\psi}})\right|>\epsilon\right\}$$

and

$$\left\{\left|\widetilde{\mathcal{Q}}_{N,T}(\widetilde{\boldsymbol{\psi}})-\widehat{\mathcal{Q}}_{N,T}(\widetilde{\boldsymbol{\psi}})\right|>\epsilon\right\}\subset\left\{\widetilde{\boldsymbol{\psi}}\notin\mathcal{B}(\boldsymbol{\psi}_0,\eta_\psi)\right\}\bigcup\left\{\widetilde{\boldsymbol{\psi}}\in\mathcal{B}(\boldsymbol{\psi}_0,\eta_\psi),\left|\widetilde{\mathcal{Q}}_{N,T}(\widetilde{\boldsymbol{\psi}})-\widehat{\mathcal{Q}}_{N,T}(\widetilde{\boldsymbol{\psi}})\right|>\epsilon\right\}.$$

It then follows from (B.43) that

$$\begin{aligned}P\left(\widetilde{\mathcal{Q}}_{N,T}(\widehat{\boldsymbol{\psi}})-\widetilde{\mathcal{Q}}_{N,T}(\widetilde{\boldsymbol{\psi}})>\epsilon\right)&\leq P\left(\widehat{\boldsymbol{\psi}}\notin\mathcal{B}(\boldsymbol{\psi}_0,\eta_\psi)\right)+P\left(\widetilde{\boldsymbol{\psi}}\notin\mathcal{B}(\boldsymbol{\psi}_0,\eta_\psi)\right)\\&\quad+2P\left(\sup_{\boldsymbol{\psi}\in\mathcal{B}(\boldsymbol{\psi}_0,\eta_\psi)}\left|\widetilde{\mathcal{Q}}_{N,T}(\boldsymbol{\psi})-\widehat{\mathcal{Q}}_{N,T}(\boldsymbol{\psi})\right|>\frac{\epsilon}{2}\right).\end{aligned}$$

Invoking Theorem 6 together with (B.42) and by letting

$$\epsilon:=C_0\left(NT^{-C_\alpha}+N^{\gamma_M+1}\log(T)T^{\frac{\gamma_M}{2}-\frac{3}{8}\theta_\alpha}+N\exp\left(-C_M\frac{T^{1/4}}{\log^2(T)}\right)\right),$$

one can obtain that

$$0\leq\widetilde{\mathcal{Q}}_{N,T}(\widehat{\boldsymbol{\psi}})-\widetilde{\mathcal{Q}}_{N,T}(\widetilde{\boldsymbol{\psi}})=O_p\left(NT^{-C_\alpha}+N^{\gamma_M+1}\log(T)T^{\frac{\gamma_M}{2}-\frac{3}{8}\theta_\alpha}+N\exp\left(-C_M\frac{T^{1/4}}{\log^2(T)}\right)\right).\quad\text{(B.44)}$$

Combining (B.40) and (B.44), we obtain the main theorem. ■

**Proof of Theorem 8.** We proceed in the following four main steps:

STEP 1: First, by arguing along the lines of the proof of Lemma 3, one obtains that

$$\begin{aligned}\frac{N}{T}\sum_{t=1}^{T}\boldsymbol{x}_{*,t}^{(w)}(\boldsymbol{u}_c)\epsilon_{0,*,t}^{(w)}&=\frac{N}{T}\sum_{t=1}^{T}\boldsymbol{x}_{*,t}^{(w)}(\boldsymbol{u}_c)\epsilon_{0,*,t}\\&=\frac{N}{T}\sum_{t=1}^{T}\boldsymbol{x}_{*,t}(\boldsymbol{u}_c)\epsilon_{0,*,t}-\frac{N}{T}\left(\sum_{t=1}^{T}\boldsymbol{x}_{*,t}(\boldsymbol{u}_c)\boldsymbol{w}_{*,t}^\top\right)\left(\sum_{s=1}^{T}\boldsymbol{w}_{*,s}\boldsymbol{w}_{*,s}^\top\right)^{-1}\left(\sum_{t=1}^{T}\boldsymbol{w}_{*,t}\epsilon_{0,*,t}\right)\\&=:\mathcal{T}_{1,N,T}+\mathcal{T}_{2,N,T}.\end{aligned}$$

Define $N(\boldsymbol{u}_c):=\sum_{i=1}^{N}u_{i,c}$ and $g_c\equiv g(\boldsymbol{u}_c):=\frac{N(\boldsymbol{u}_c)}{N}$. One then obtains that

$$\mathcal{T}_{1,N,T}=\frac{1}{NT}\sum_{t=1}^{T}\sum_{i=1}^{N}u_{i,c}\boldsymbol{x}_{i,t}\sum_{i=1}^{N}\epsilon_{0,i,t}$$

$$= \sqrt{g_c}\frac{1}{T}\sum_{t=1}^{T}\left(\frac{1}{\sqrt{N(\boldsymbol{u}_c)}}\sum_{i=1}^{N}u_{i,c}\boldsymbol{x}_{i,t}\right)\left(\frac{1}{\sqrt{N}}\sum_{i=1}^{N}\epsilon_{0,i,t}\right)$$
$$\xrightarrow{w} \sigma_\epsilon\sqrt{g_c}\left(\boldsymbol{\Sigma}_\eta^{(c,c)}\right)^{\frac{1}{2}}\int_0^1 \boldsymbol{W}_\eta^{(c)}(\tau)dW_\epsilon(\tau),$$

where $\boldsymbol{\Sigma}_\eta^{(c,c)}$ is defined as in Lemma 3 and $\boldsymbol{W}_\eta^{(c)}(\tau)$ is a $d_x \times 1$ vector of Brownian motions with the covariance kernel $E[\boldsymbol{W}_\eta^{(c)}(\tau)\boldsymbol{W}_\eta^{(c)}(\kappa)^\top] = \min(\kappa,\tau)\mathbb{I}_{d_x}$. Moreover, note that $\frac{N}{T}\sum_{t=1}^T \boldsymbol{x}_{*,t}(\boldsymbol{u}_c)\boldsymbol{w}_{*,t}^\top \leq \left(\frac{N}{T}\sum_{t=1}^T |\boldsymbol{x}_{*,t}(\boldsymbol{u}_c)|\right)\max_{t\in[1,T]}|\boldsymbol{w}_{*,t}^\top| = o_p(1)$ by Lemma 9. It then follows from the weak law of large numbers that $\mathcal{T}_{2,N,T} = o_p(1)$. Therefore, we obtain

$$\frac{N}{T}\sum_{t=1}^{T}\boldsymbol{x}_{*,t}^{(w)}(\boldsymbol{u}_c)\epsilon_{0,*,t}^{(w)} \xrightarrow{w} \sigma_\epsilon\sqrt{g_c}\left(\boldsymbol{\Sigma}_\eta^{(c,c)}\right)^{\frac{1}{2}}\int_0^1 \boldsymbol{W}_\eta^{(c)}(\tau)dW_\epsilon(\tau). \tag{B.45}$$

Using the same argument as in the proof of Lemma 5, we can also show that

$$\sqrt{\frac{N}{T}}\sum_{t=1}^{T}\xi_{*,t}^{(w)}(\boldsymbol{\theta}_{0,c}\boldsymbol{u}_c)\epsilon_{0,*,t}^{(w)} = O_p(1), \tag{B.46}$$
$$\frac{N}{T^2}\sum_{t=1}^{T}\boldsymbol{x}_{*,t}^{(w)}(\boldsymbol{u}_c)\boldsymbol{x}_{*,t}^{(w)}(\boldsymbol{u}_k)^\top \xrightarrow{w} g_c g_k\left(\boldsymbol{\Sigma}_\eta^{(c,c)}\right)^{\frac{1}{2}}\left(\boldsymbol{\Sigma}_\eta^{(k,k)}\right)^{\frac{1}{2}}\int_0^1 \boldsymbol{W}_\eta^{(c)}(\tau)\boldsymbol{W}_\eta^{(c)}(\tau)^\top d\tau, \tag{B.47}$$
$$\frac{N^{1/2}}{T^{3/2}}\sum_{t=1}^{T}\boldsymbol{x}_{*,t}^{(w)}(\boldsymbol{u}_c)\xi_{*,t}^{(w)}(\boldsymbol{\theta}_{0,c},\boldsymbol{u}_d) = O_p(1), \tag{B.48}$$
$$\frac{1}{T}\sum_{t=1}^{T}\xi_{*,t}^{(w)}(\boldsymbol{\theta}_{0,c},\boldsymbol{u}_c)\xi_{*,t}^{(w)}(\boldsymbol{\theta}_{0,c},\boldsymbol{u}_d) = O_p(1), \tag{B.49}$$

where all the terms in the limits are stochastic.

STEP 2: By the definitions of $\boldsymbol{F}_t(\boldsymbol{U},\boldsymbol{U}_0)$, one can write

$$\frac{1}{T}\sum_{t=1}^{T}(\epsilon_{*,t}^2(\boldsymbol{\psi},\boldsymbol{U}) - \epsilon_{0,*,t}^{(w)2}) = 2\left(\sqrt{\frac{T}{N}}(\boldsymbol{\theta}^{(\widetilde{\sigma}^{(per)})} - \boldsymbol{\theta}_0^{(\sigma^{(per)})})^\top, (\boldsymbol{\phi}_0^{(\sigma^{(per)})} - \boldsymbol{\phi}^{(\widetilde{\sigma}^{(per)})})^\top, \mu_{*0} - \mu_*, \boldsymbol{\phi}_0^{(\sigma^{(per)})\top}\right)$$
$$\times \operatorname{diag}\left(\boldsymbol{D}_\phi(\widetilde{\sigma}^{(per)}), \mathbb{I}_{2G+1}\right)\operatorname{diag}\left(\sqrt{\frac{N}{T}}\mathbb{I}_{G\times d_x}, \mathbb{I}_{2G+1}\right)\frac{1}{T}\sum_{t=1}^{T}\boldsymbol{F}_t(\boldsymbol{U},\boldsymbol{U}_0)\epsilon_{0,*,t}^{(w)}$$
$$+\left(\sqrt{\frac{T}{N}}(\boldsymbol{\theta}^{(\widetilde{\sigma}^{(per)})} - \boldsymbol{\theta}_0^{(\sigma^{(per)})})^\top, (\boldsymbol{\phi}_0^{(\sigma^{(per)})} - \boldsymbol{\phi}^{(\widetilde{\sigma}^{(per)})})^\top, \mu_{*0} - \mu_*, \boldsymbol{\phi}_0^{(\sigma^{(per)})\top}\right)$$
$$\times \operatorname{diag}\left(\boldsymbol{D}_\phi(\widetilde{\sigma}^{(per)}), \mathbb{I}_{2G+1}\right)\boldsymbol{\Lambda}_{N,T}(\boldsymbol{U},\boldsymbol{U}_0)\operatorname{diag}\left(\boldsymbol{D}_\phi(\widetilde{\sigma}^{(per)}), \mathbb{I}_{2G+1}\right)$$

$$\times \left( \sqrt{\frac{T}{N}} (\boldsymbol{\theta}^{(\widetilde{\sigma}^{(per)})} - \boldsymbol{\theta}_0^{(\sigma^{(per)})})^\top, (\boldsymbol{\phi}_0^{(\sigma^{(per)})} - \boldsymbol{\phi}^{(\widetilde{\sigma}^{(per)})})^\top, \mu_{*0} - \mu_*, \boldsymbol{\phi}_0^{(\sigma^{(per)})\top} \right)^\top . \tag{B.50}$$

Eqs. (B.45) and (B.46) imply that

$$\frac{1}{T} \sum_{t=1}^{T} \boldsymbol{F}_t(\boldsymbol{U}, \boldsymbol{U}_0) \epsilon_{0,*,t}^{(w)} = \left( O_p\left(N^{-1}\right) \iota_{G \times d_x}, O_p\left(\frac{1}{\sqrt{NT}}\right) \iota_{2G+1} \right)$$

uniformly in $\boldsymbol{U}$. Therefore the first term in (B.50) is negligible in probability. One can thus has that

$$\begin{aligned}
\frac{1}{T} \sum_{t=1}^{T} \left( \epsilon_{*,t}^2(\boldsymbol{\psi}, \boldsymbol{U}) - \epsilon_{0,*,t}^{(w)2} \right) = & \left( \sqrt{\frac{T}{N}} (\boldsymbol{\theta}^{(\widetilde{\sigma}^{(per)})} - \boldsymbol{\theta}_0^{(\sigma^{(per)})})^\top, (\boldsymbol{\phi}_0^{(\sigma^{(per)})} - \boldsymbol{\phi}^{(\widetilde{\sigma}^{(per)})})^\top, \mu_{*0} - \mu_*, \boldsymbol{\phi}_0^{(\sigma^{(per)})\top} \right) \\
& \times \operatorname{diag}\left( \boldsymbol{D}_\phi(\widetilde{\sigma}^{(per)}), \mathbb{I}_{2G+1} \right) \boldsymbol{\Lambda}_{N,T}(\boldsymbol{U}, \boldsymbol{U}_0) \operatorname{diag}\left( \boldsymbol{D}_\phi(\widetilde{\sigma}^{(per)}), \mathbb{I}_{2G+1} \right) \\
& \times \left( \sqrt{\frac{T}{N}} (\boldsymbol{\theta}^{(\widetilde{\sigma}^{(per)})} - \boldsymbol{\theta}_0^{(\sigma^{(per)})})^\top, (\boldsymbol{\phi}_0^{(\sigma^{(per)})} - \boldsymbol{\phi}^{(\widetilde{\sigma}^{(per)})})^\top, \mu_{*0} - \mu_*, \boldsymbol{\phi}_0^{(\sigma^{(per)})\top} \right)^\top \\
& + o_p(1).
\end{aligned} \tag{B.51}$$

STEP 3: Define the following open balls: $\mathcal{B}(\boldsymbol{\theta}_0, \eta_\theta) \coloneqq \{\boldsymbol{\theta} \in \Theta_\theta : \sqrt{\frac{T}{N}} H(\boldsymbol{\theta}, \boldsymbol{\theta}_0) < \eta_\theta\}$, $\mathcal{B}(\boldsymbol{\phi}_0, \eta_\phi) \coloneqq \{\boldsymbol{\phi} \in \Theta_\phi : H(\boldsymbol{\phi}, \boldsymbol{\phi}_0) < \eta_\phi\}$, $\mathcal{B}(\mu_{*0}, \eta_\mu) \coloneqq \{|\mu - \mu_{*0}| < \eta_\mu\}$, and $\mathcal{B}_N(\boldsymbol{U}_0, \eta_u) \coloneqq \{\boldsymbol{U} \in \Delta_S^N : H(\boldsymbol{U}, \boldsymbol{U}_0) < \eta_u\}$. Let's denote by $\mathcal{A}(\boldsymbol{\psi}_0, \eta_\psi) \coloneqq \bigcup_{\substack{\eta_\theta, \eta_\phi, \eta_\mu \\ (\eta_\theta^2 + \eta_\phi^2 + \eta_\mu^2)^{1/2} = \eta_\psi}} \mathcal{B}^c(\boldsymbol{\theta}_0, \eta_\theta) \times \mathcal{B}^c(\boldsymbol{\phi}_0, \eta_\phi) \times \mathcal{B}^c(\mu_{*0}, \eta_\mu)$ a union of the complements of the above-defined open balls. It then follows that, for some $\eta_\phi \in (0, 1)$,

$$P\left( \widehat{\boldsymbol{\psi}} \in \mathcal{A}(\boldsymbol{\psi}_0, \eta_\phi), \widehat{\boldsymbol{U}} \in \mathcal{B}_N^c(\boldsymbol{U}_0, \eta_u) \right) \le P\left( \inf_{\substack{\boldsymbol{\psi} \in \mathcal{A}(\boldsymbol{\psi}_0, \eta_\phi) \\ \boldsymbol{U} \in \mathcal{B}_N^c(\boldsymbol{U}_0, \eta_u)}} \frac{1}{T} \sum_{t=1}^{T} \{ \epsilon_{*,t}^2(\boldsymbol{\psi}, \boldsymbol{U}) - \epsilon_{0,*,t}^{(w)2} \} \le 0 \right). \tag{B.52}$$

By the eigenvalue inequality, it follows from (B.51) that

$$\begin{aligned}
\inf_{\substack{\boldsymbol{\psi} \in \mathcal{A}(\boldsymbol{\psi}_0, \eta_\phi) \\ \boldsymbol{U} \in \mathcal{B}_N^c(\boldsymbol{U}_0, \eta_u)}} \frac{1}{T} \sum_{t=1}^{T} \left( \epsilon_{*,t}^2(\boldsymbol{\psi}, \boldsymbol{U}) - \epsilon_{0,*,t}^{(w)2} \right) \ge C_0 & \inf_{\substack{\boldsymbol{\psi} \in \mathcal{A}(\boldsymbol{\psi}_0, \eta_\phi) \\ \boldsymbol{U} \in \mathcal{B}_N^c(\boldsymbol{U}_0, \eta_u)}} \Bigg\{ \lambda_{\min}\left( \boldsymbol{\Lambda}_{N,T}(\boldsymbol{U}, \boldsymbol{U}_0) \right) \\
& \times \left| H\left( \left( \sqrt{\frac{T}{N}} \boldsymbol{\theta}, \boldsymbol{\phi}, \mu_* \right), \left( \sqrt{\frac{T}{N}} \boldsymbol{\theta}_0, \boldsymbol{\phi}_0, \mu_{*0} \right) \right) \right|^2 \Bigg\}.
\end{aligned}$$

By sending $N$ and $T$ to infinity, one obtains from Assumption 8 that

$$\lim_{N\uparrow\infty,T\uparrow\infty} \inf_{\substack{\boldsymbol{\psi}\in\mathcal{A}(\boldsymbol{\psi}_0,\eta_\phi)\\ \boldsymbol{U}\in\mathcal{B}_N^c(\boldsymbol{U}_0,\eta_u)}} \frac{1}{T}\sum_{t=1}^{T}\{\epsilon_{*,t}^2(\boldsymbol{\psi},\boldsymbol{U})-\epsilon_{0,*,t}^{(w)2}\} > C_0\eta_\psi^2 \text{ w.p.1.} \tag{B.53}$$

Eqs. (B.52) and (B.53) imply that

$$P\left(\widehat{\boldsymbol{\psi}}\in\mathcal{A}(\boldsymbol{\psi}_0,\eta_\phi),\widehat{\boldsymbol{U}}\in\mathcal{B}_N^c(\boldsymbol{U}_0,\eta_u)\right)\downarrow 0.$$

Thus, one has shown that $\sqrt{\frac{T}{N}}H(\widehat{\boldsymbol{\theta}},\boldsymbol{\theta}_0)=o_p(1)$, $H\left((\widehat{\boldsymbol{\phi}},\widehat{\mu}_*),(\boldsymbol{\phi}_0,\mu_{*0})\right)=o_p(1)$, and $H(\widehat{\boldsymbol{U}},\boldsymbol{U}_0)=o_p(1)$.

STEP 4: To refine the convergence rates of $\widehat{\boldsymbol{\psi}}$, first define open balls:

$$\mathcal{N}_T(\boldsymbol{\theta}_0,\eta'_\theta)\coloneqq\{\boldsymbol{\theta}\in\mathcal{B}(\boldsymbol{\theta}_0,\eta_\theta):\ \sqrt{T}H(\boldsymbol{\theta},\boldsymbol{\theta}_0)<\eta'_\theta\},$$

$$\mathcal{N}_N(\boldsymbol{\phi}_0,\eta'_\phi)\coloneqq\{\boldsymbol{\phi}\in\mathcal{B}(\boldsymbol{\phi}_0,\eta_\phi):\ \sqrt{N}H(\boldsymbol{\phi},\boldsymbol{\phi}_0)<\eta'_\phi\},$$

$$\mathcal{N}_N(\mu_*,\eta'_\mu)\coloneqq\{\mu_*\in\mathcal{B}(\mu_{*0},\eta_\mu):\ \sqrt{N}|\mu_*-\mu_{*0}|<\eta'_\mu\},$$

and

$$\mathcal{B}(\sigma_{\epsilon,0}^2,\eta_\sigma)\coloneqq\{\sigma_\epsilon^2\in\Theta_\sigma:\ |\sigma_\epsilon^2-\sigma_{\epsilon,0}^2|<\eta_\sigma\}.$$

Let denote by $\mathfrak{A}(\boldsymbol{\psi}_0,\eta'_\psi)\coloneqq\bigcup_{\substack{\eta'_\theta,\eta'_\phi,\eta'_\mu\\(\eta'^2_\theta+\eta'^2_\phi+\eta'^2_\mu)^{\frac{1}{2}}=\eta'_\psi}}\mathcal{N}_T^c(\boldsymbol{\theta}_0,\eta'_\theta)\times\mathcal{N}_N^c(\boldsymbol{\phi}_0,\eta'_\phi)\times\mathcal{N}_N^c(\mu_*,\eta'_\mu)$ a union of the complements of these open balls. One obtains that

$$P\left(\widehat{\boldsymbol{\psi}}\in\mathfrak{A}(\boldsymbol{\psi}_0,\eta'_\psi),\widehat{\boldsymbol{U}}\in\mathcal{B}_N^c(\boldsymbol{U}_0,\eta_u),\widehat{\sigma}_\epsilon^2\in\mathcal{B}^c(\sigma_{\epsilon,0}^2,\eta_\sigma)\right)$$
$$\le P\left(\inf_{\substack{\boldsymbol{\psi}\in\mathfrak{A}(\boldsymbol{\psi}_0,\eta'_\psi)\\ \boldsymbol{U}\in\mathcal{B}_N^c(\boldsymbol{U}_0,\eta_u)\\ \sigma_\epsilon^2\in\mathcal{B}^c(\sigma_{\epsilon,0}^2,\eta_\sigma)}}\left\{\overline{Q}_{N,T}(\boldsymbol{\psi}_0,\sigma_{\epsilon,0}^2,\boldsymbol{U}_0)-\overline{Q}_{N,T}(\boldsymbol{\psi},\sigma_\epsilon^2,\boldsymbol{U})\right\}\le 0\right). \tag{B.54}$$

Similar to the argument in STEP 2, notice that

$$\frac{N}{T}\sum_{t=1}^{T}\left(\epsilon_{*,t}^2(\boldsymbol{\psi},\boldsymbol{U})-\epsilon_{0,*,t}^{(w)2}\right)=2\left((\boldsymbol{\theta}^{(\widetilde{\sigma}^{(per)})}-\boldsymbol{\theta}_0^{(\sigma^{(per)})})^\top,(\boldsymbol{\phi}_0^{(\sigma^{(per)})}-\boldsymbol{\phi}^{(\widetilde{\sigma}^{(per)})})^\top,\mu_{*0}-\mu_*,\boldsymbol{\phi}_0^{(\sigma^{(per)})\top}\right)$$
$$\times\operatorname{diag}\left(\boldsymbol{D}_\phi(\widetilde{\sigma}^{(per)}),\mathbb{I}_{2G+1}\right)\frac{N}{T}\sum_{t=1}^{T}\boldsymbol{F}_t(\boldsymbol{U},\boldsymbol{U}_0)\epsilon_{0,*,t}^{(w)}$$

$$+\Big(\sqrt{T}(\boldsymbol{\theta}^{(\widetilde{\sigma}^{(per)})}-\boldsymbol{\theta}_0^{(\sigma^{(per)})})^\top,\sqrt{N}(\boldsymbol{\phi}_0^{(\sigma^{(per)})}-\boldsymbol{\phi}^{(\widetilde{\sigma}^{(per)})})^\top,\sqrt{N}(\mu_{*0}-\mu_*),$$
$$\sqrt{N}\boldsymbol{\phi}_0^{(\sigma^{(per)})\top}\Big)\operatorname{diag}\left(\boldsymbol{D}_\phi(\widetilde{\sigma}^{(per)}),\mathbb{I}_{2G+1}\right)\boldsymbol{\Lambda}_{N,T}(\boldsymbol{U},\boldsymbol{U}_0)$$
$$\times\operatorname{diag}\left(\boldsymbol{D}_\phi(\widetilde{\sigma}^{(per)}),\mathbb{I}_{2G+1}\right)\Big(\sqrt{T}(\boldsymbol{\theta}^{(\widetilde{\sigma}^{(per)})}-\boldsymbol{\theta}_0^{(\sigma^{(per)})})^\top,$$
$$\sqrt{N}(\boldsymbol{\phi}_0^{(\sigma^{(per)})}-\boldsymbol{\phi}^{(\widetilde{\sigma}^{(per)})})^\top,\sqrt{N}(\mu_{*0}-\mu_*),\sqrt{N}\boldsymbol{\phi}_0^{(\sigma^{(per)})\top}\Big)^\top. \tag{B.55}$$

In view of (B.50), we have that

$$\frac{N}{T}\sum_{t=1}^{T}\boldsymbol{F}_t(\boldsymbol{U},\boldsymbol{U}_0)\epsilon_{0,*,t}^{(w)}=\left(O_p\left(1\right)\iota_{G\times d_x},O_p\left(\sqrt{\frac{N}{T}}\right)\iota_{2G+1}\right)$$

uniformly in $\boldsymbol{U}$. Therefore, it is immediate to see that, if $N/T\to const.\in(0,\infty)$, the first term on the right-hand side of (B.55) is probabilistically negligible for every $\boldsymbol{\theta}\in\mathcal{B}(\boldsymbol{\theta}_0,\eta_\theta)$, $\boldsymbol{\phi}\in\mathcal{B}(\boldsymbol{\phi}_0,\eta_\phi)$,, $\mu_*\in\mathcal{B}(\mu_{*0},\eta_\mu)$, and $\boldsymbol{U}\in\mathcal{B}(\boldsymbol{U}_0,\eta_u)$, i.e.,

$$\Bigg(\max_{\widetilde{\sigma}^{(per)}\in\sigma(\mathcal{P})}\min_{\sigma^{(per)}\in\sigma(\mathcal{P})}\Big\{\Big((\boldsymbol{\theta}^{(\widetilde{\sigma}^{(per)})}-\boldsymbol{\theta}_0^{(\sigma^{(per)})})^\top,(\boldsymbol{\phi}_0^{(\sigma^{(per)})}-\boldsymbol{\phi}^{(\widetilde{\sigma}^{(per)})})^\top,\mu_{*0}-\mu_*,\boldsymbol{\phi}_0^{(\sigma^{(per)})\top}\Big)$$
$$\operatorname{diag}\left(\boldsymbol{D}_\phi(\widetilde{\sigma}^{(per)}),\mathbb{I}_{2G+1}\right)\frac{N}{T}\sum_{t=1}^{T}\boldsymbol{F}_t(\boldsymbol{U},\boldsymbol{U}_0)\epsilon_{0,*,t}^{(w)}\Big\},$$
$$\max_{\sigma^{(per)}\in\sigma(\mathcal{P})}\min_{\widetilde{\sigma}^{(per)}\in\sigma(\mathcal{P})}\Big\{\Big((\boldsymbol{\theta}^{(\widetilde{\sigma}^{(per)})}-\boldsymbol{\theta}_0^{(\sigma^{(per)})})^\top,(\boldsymbol{\phi}_0^{(\sigma^{(per)})}-\boldsymbol{\phi}^{(\widetilde{\sigma}^{(per)})})^\top,\mu_{*0}-\mu_*,\boldsymbol{\phi}_0^{(\sigma^{(per)})\top}\Big)$$
$$\operatorname{diag}\left(\boldsymbol{D}_\phi(\widetilde{\sigma}^{(per)}),\mathbb{I}_{2G+1}\right)\frac{N}{T}\sum_{t=1}^{T}\boldsymbol{F}_t(\boldsymbol{U},\boldsymbol{U}_0)\epsilon_{0,*,t}^{(w)}\Big\}\Bigg)^+=o(1)O_p(1)+o_p\left(\sqrt{\frac{N}{T}}\right).$$

By the eigenvalue inequality, one can show from (B.55) that

$$\inf_{\substack{\boldsymbol{\psi}\in\mathfrak{A}(\boldsymbol{\psi}_0,\eta'_\psi)\\ \boldsymbol{U}\in\mathcal{B}_N^c(\boldsymbol{U}_0,\eta_u)}}\frac{N}{T}\sum_{t=1}^{T}(\epsilon_{*,t}^2(\boldsymbol{\psi},\boldsymbol{U})-\epsilon_{0,*,t}^{(w)2})>o_p(1)+C_0\inf_{\boldsymbol{U}\in\mathcal{B}(\boldsymbol{U}_0,\eta_u)}\lambda_{\min}\left(\boldsymbol{\Lambda}_{N,T}(\boldsymbol{U},\boldsymbol{U}_0)\right)$$
$$\times\left|\inf_{\boldsymbol{\psi}\in\mathfrak{A}(\boldsymbol{\psi}_0,\eta'_\psi)}H\left((\sqrt{T}\boldsymbol{\theta},\sqrt{N}\boldsymbol{\phi},\sqrt{N}\mu_*),(\sqrt{T}\boldsymbol{\theta}_0,\sqrt{N}\boldsymbol{\phi}_0,\sqrt{N}\mu_{*0})\right)\right|^2$$
$$>\eta_\psi'^2$$

in view of Assumption 8. The rest of the proof is immediate by following the same line as the proof of Theorem 7. Hence, in view of (B.54) we have

$$P\left(\widehat{\boldsymbol{\psi}}\in\mathfrak{A}(\boldsymbol{\psi}_0,\eta'_\psi),\widehat{\boldsymbol{U}}\in\mathcal{B}_N^c(\boldsymbol{U}_0,\eta_u),\widehat{\sigma}_\epsilon^2\in\mathcal{B}^c(\sigma_{\epsilon,0}^2,\eta_\sigma)\right)\downarrow 0.$$

The main theorem then follows. ■

**Proof of Theorem 9.** The proof of this theorem follows along the same lines as the proof of Theorem 6; some of the arguments need to be modified due to the nonstationarity of the vector of covariates $\boldsymbol{x}_{i,t}$. First, by Lemma 5, we have $\gamma = o_p\left(N^{-3/2}\right) + O_p\left(N^{-1}\right)$, where $\gamma$ is defined in the proof of Theorem 6. It is then sufficient to derive the convergence rates for $\mathcal{T}_{1,N,T}^{(a)}$, $\mathcal{T}_{1,N,T}^{(b)}$, $\mathcal{T}_{1,N,T}^{(c)}$, and $\mathcal{T}_{1,N,T}^{(d)}$. Recall some notations defined in the proof of Theorem 8: $N(\boldsymbol{u}_c) := \sum_{i=1}^{N} u_{i,c}$ and $g_c \equiv g(\boldsymbol{u}_c) = \frac{N(\boldsymbol{u}_c)}{N}$.

To bound $\mathcal{T}_{1,N,T}^{(a)}$, an application of Lemma 9 yields

$$\frac{1}{\sqrt{N}T^{\frac{3}{2}}}\sum_{t=1}^{T}\sum_{i=1}^{N} u_{i,c}\boldsymbol{x}_{i,t} = \sum_{t=1}^{T}\int_{\frac{t}{T}}^{\frac{t+1}{T}} \frac{1}{\sqrt{NT}}\sum_{s=1}^{\lfloor T\tau\rfloor}\sum_{i=1}^{N} u_{i,c}\boldsymbol{\eta}_{i,s}\frac{1}{T}$$
$$\xrightarrow{w} \sqrt{g_c}\boldsymbol{\Sigma}_{\eta}^{(c,c)\frac{1}{2}}\int_0^1 \boldsymbol{W}_{\eta}^{(c)}(\tau)d\tau \tag{B.56}$$

for $c = 1, \ldots, G$, where $\boldsymbol{\Sigma}_{\eta}^{(c,c)} = \operatorname{plim}_{N,T\uparrow\infty}\frac{1}{TN(\boldsymbol{u}_c)}E\left[\boldsymbol{S}_{\eta}(N(\boldsymbol{u}_c),T)\boldsymbol{S}_{\eta}(N(\boldsymbol{u}_c),T)^{\top}\right]$ with $\boldsymbol{S}_{\eta}(N(\boldsymbol{u}_c),t) = \sum_{s=1}^{t}\sum_{i=1}^{N} u_{i,c}\boldsymbol{\eta}_{i,s}$ and $\boldsymbol{W}_{\eta}^{(c)}(\tau)$ is a $d_x \times 1$ vector of Brownian motions with the covariance kernel $E[\boldsymbol{W}_{\eta}^{(c)}(\tau)\boldsymbol{W}_{\eta}^{(c)}(\kappa)^{\top}] = \min(\tau,\kappa)\mathbb{I}_{d_x}$. It then follows that, for a given $\boldsymbol{\psi} \in \mathcal{N}_{N,T}(\boldsymbol{\psi}_0, \eta_{\psi})$,

$$\min_{\sigma^{(per)}\in\sigma(\mathcal{P})}\sum_{c=1}^{G}\left|\frac{1}{NT}\sum_{i=1}^{N}\sum_{t=1}^{T}\phi_{\sigma^{(per)}(c)}(u_{0,i,c}-u_{i,c})(\mu_{*0}-\mu_{*})(\boldsymbol{\theta}_{\sigma^{(per)}(c)}-\boldsymbol{\theta}_{0,c})^{\top}\boldsymbol{x}_{i,t}^{(w)}\right|$$
$$< C_0\frac{1}{N}\left(\sqrt{N}|\mu_{*0}-\mu_{*}|\right)\min_{\sigma^{(per)}\in\sigma(\mathcal{P})}\sum_{c=1}^{G}\left(\sqrt{T}(\boldsymbol{\theta}_{\sigma^{(per)}(c)}-\boldsymbol{\theta}_{0,c})^{\top}\right)\left|\frac{1}{\sqrt{N}T^{\frac{3}{2}}}\sum_{t=1}^{T}\sum_{i=1}^{N}(u_{0,i,c}-u_{i,c})\boldsymbol{x}_{i,t}\right|.$$

Moreover the weak convergence to a Gaussian process in (B.56) implies that

$$\lim_{N\uparrow\infty,T\uparrow\infty} E\left[\left|\frac{1}{\sqrt{N}T^{\frac{3}{2}}}\sum_{t=1}^{T}\sum_{i=1}^{N}(u_{0,i,c}-u_{i,c})\boldsymbol{x}_{i,t}\right|^{C_{\alpha}}\right] < \infty$$

for every $C_{\alpha} \geq 1$. Therefore, by the consistency of $\widehat{\boldsymbol{\psi}}$ as demonstrated in Theorem 8, one obtains that the first term in (B.34) $\mathcal{T}_{1,N,T}^{(a*)} = O\left(N^{-C_{\alpha}}\right)$; and the convergence rates of the remaining terms $\mathcal{T}_{1,N,T}^{(a**)}$ and $\mathcal{T}_{1,N,T}^{(a***)}$ remain the same. Therefore, it follows that

$$\mathcal{T}_{1,N,T}^{(a)} < C_0\Big(T^{-C_{\alpha}} + N^{-C_{\alpha}} + N^{\gamma_M}\log(T)T^{\gamma_M-\frac{3}{4}\theta_{\alpha}}$$
$$+\max\left(\exp\left(-C_{\epsilon_{\eta}}NT\right), \exp\left(-C_{\epsilon_{\eta}}\frac{T^{1/4}}{\log(T)}\right)\right)\Big). \tag{B.57}$$

To bound $\mathcal{T}^{(b)}_{1,N,T}$, notice that

$$\begin{aligned}
&\sum_{c=1}^{G}\frac{1}{NT}\sum_{t=1}^{T}\sum_{i=1}^{N}(\phi_{0,c}-\phi_{\sigma^{(per)}(c)})(u_{0,i,c}-u_{i,\sigma^{(per)}(c)})\left\{(\boldsymbol{\theta}_{\sigma^{(per)}(c)}-\boldsymbol{\theta}_{0,c})^{\top}\boldsymbol{x}^{(w)}_{i,t}-\xi^{(w)}_{i,t}(\boldsymbol{\theta}_{0,c})\right\}\xi^{(w)}_{*,t}(\boldsymbol{\theta}_{0,c},\boldsymbol{u}_{0,c})\\
&=\sum_{c=1}^{G}\left(g_{*,c}E[\xi_{*,t}(\boldsymbol{\theta}_{0,c})]+o_p(1)\right)\left\{(\phi_{0,c}-\phi_{\sigma^{(per)}(c)})(\boldsymbol{\theta}_{\sigma^{(per)}(c)}-\boldsymbol{\theta}_{0,c})^{\top}\frac{1}{NT}\sum_{t=1}^{T}\sum_{i=1}^{N}(u_{0,i,c}-u_{i,\sigma^{(per)}(c)})\boldsymbol{x}^{(w)}_{i,t}\right\}\\
&\quad+\sum_{c=1}^{G}\left(g_{*,c}E[\xi_{*,t}(\boldsymbol{\theta}_{0,c})]+o_p(1)\right)(\phi_{0,c}-\phi_{\sigma^{(per)}(c)})\frac{1}{NT}\sum_{t=1}^{T}\sum_{i=1}^{N}(u_{0,i,c}-u_{i,\sigma^{(per)}(c)})\left(\xi_{i,t}(\boldsymbol{\theta}_{0,c})-E[\xi_{i,t}(\boldsymbol{\theta}_{0,c})]\right)\\
&\quad+\sum_{c=1}^{G}\left(g_{*,c}E[\xi_{*,t}(\boldsymbol{\theta}_{0,c})]+o_p(1)\right)(\phi_{0,c}-\phi_{\sigma^{(per)}(c)})\frac{1}{NT}\sum_{t=1}^{T}\sum_{i=1}^{N}(u_{0,i,c}-u_{i,\sigma^{(per)}(c)})\Big(E[\xi_{i,t}(\boldsymbol{\theta}_{0,\sigma^{(per)}(c)})]\\
&\quad+o_p(1)\Big)\\
&=:\widetilde{A}_{N,T}(\boldsymbol{U},\sigma^{(per)})+\widetilde{B}_{N,T}(\boldsymbol{U},\sigma^{(per)})+\widetilde{C}_{N,T}(\boldsymbol{U},\sigma^{(per)}).
\end{aligned}$$

Because

$$\begin{aligned}
\min_{\sigma^{(per)}\in\sigma(\mathcal{P})}\left|\widetilde{A}_{N,T}(\boldsymbol{U},\sigma^{(per)})+\widetilde{B}_{N,T}(\boldsymbol{U},\sigma^{(per)})+\widetilde{C}_{N,T}(\boldsymbol{U},\sigma^{(per)})\right| &\approx \min_{\sigma^{(per)}\in\sigma(\mathcal{P})}\left|\widetilde{A}_{N,T}(\boldsymbol{U},\sigma^{(per)})\right|\\
&+\min_{\sigma^{(per)}\in\sigma(\mathcal{P})}\left|\widetilde{B}_{N,T}(\boldsymbol{U},\sigma^{(per)})\right|\\
&+\min_{\sigma^{(per)}\in\sigma(\mathcal{P})}\left|\widetilde{C}_{N,T}(\boldsymbol{U},\sigma^{(per)})\right|
\end{aligned}$$

for every $\boldsymbol{\psi}\in\mathcal{N}_{N,T}(\boldsymbol{\psi}_0,\eta_\psi)$, an application of Boole's inequality yields

$$\begin{aligned}
\mathcal{T}^{(b)}_{1,N,T} &\le P\left(\min_{\sigma^{(per)}\in\sigma(\mathcal{P})}\widetilde{A}_{N,T}(\sigma^{(per)})>\frac{\epsilon_\eta}{12}\right)\\
&\quad+P\left(\min_{\sigma^{(per)}\in\sigma(\mathcal{P})}\widetilde{B}_{N,T}(\sigma^{(per)})>\frac{\epsilon_\eta}{12}\right)\\
&\quad+P\left(\min_{\sigma^{(per)}\in\sigma(\mathcal{P})}\widetilde{C}_{N,T}(\sigma^{(per)})>\frac{\epsilon_\eta}{12}\right)\\
&=:\mathcal{T}^{(b*)}_{1,N,T}+\mathcal{T}^{(b**)}_{1,N,T}+\mathcal{T}^{(b***)}_{1,N,T}.
\end{aligned}$$

In view of (B.56), we have $\mathcal{T}^{(b*)}_{1,N,T}=O\left(N^{-C_\alpha}\right)$ for every $C_\alpha>1$. The other terms $\mathcal{T}^{(b**)}_{1,N,T}$ and $\mathcal{T}^{(b***)}_{1,N,T}$ have the same convergence rates as $\mathcal{T}^{(a**)}_{1,N,T}$ and $\mathcal{T}^{(a***)}_{1,N,T}$. It then follows that

$$\mathcal{T}^{(b)}_{1,N,T}<C_0\Bigg(T^{-C_\alpha}+N^{-C_\alpha}+N^{\gamma_M}\log(T)T^{\gamma_M-\frac{3}{4}\theta_\alpha}+\max\left(\exp\left(-C_{\epsilon_\eta}NT\right),\exp\left(-C_{\epsilon_\eta}\frac{T^{1/4}}{\log(T)}\right)\right)\Bigg). \tag{B.58}$$

To bound $\mathcal{T}_{1,N,T}^{(c)}$, some simple calculations yield

$$\sum_{c=1}^{G} \frac{1}{NT} \sum_{i=1}^{N} \sum_{t=1}^{T} \phi_{\sigma^{(per)}(c)} (u_{0,i,c} - u_{i,\sigma^{(per)}(c)}) \left(\boldsymbol{\theta}_{\sigma^{(per)}(c)} - \boldsymbol{\theta}_{0,c}\right)^{\top} \boldsymbol{x}_{*,t}^{(w)} (\boldsymbol{u}_{0,c})$$
$$\times \left\{ (\boldsymbol{\theta}_{\sigma^{(per)}(c)} - \boldsymbol{\theta}_{0,c})^{\top} \boldsymbol{x}_{i,t}^{(w)} - \xi_{i,t}^{(w)} (\boldsymbol{\theta}_{0,c}) \right\}$$
$$= \mathfrak{T}_{1,N,T}(\sigma^{(per)}, \boldsymbol{U}) + \mathfrak{T}_{2,N,T}(\sigma^{(per)}, \boldsymbol{U}) + \mathfrak{T}_{3,N,T}(\sigma^{(per)}, \boldsymbol{U}),$$

where

$$\mathfrak{T}_{1,N,T}(\sigma^{(per)}, \boldsymbol{U}) \coloneqq \frac{1}{N} \sum_{c=1}^{G} \phi_{\sigma^{(per)}(c)} \sqrt{T} \left(\boldsymbol{\theta}_{\sigma^{(per)}(c)} - \boldsymbol{\theta}_{0,c}\right)^{\top}$$
$$\times \left\{ \frac{1}{T^2} \sum_{i=1}^{N} \sum_{t=1}^{T} (u_{0,i,c} - u_{i,\sigma^{(per)}(c)}) \boldsymbol{x}_{i,t}^{(w)} \boldsymbol{x}_{*,t}^{(w)\top} \right\} \sqrt{T} \left(\boldsymbol{\theta}_{\sigma^{(per)}(c)} - \boldsymbol{\theta}_{0,c}\right),$$

$$\mathfrak{T}_{2,N,T}(\sigma^{(per)}, \boldsymbol{U}) \coloneqq \frac{1}{NT} \sum_{c=1}^{G} \phi_{\sigma^{(per)}(c)} \sqrt{T} \left(\boldsymbol{\theta}_{\sigma^{(per)}(c)} - \boldsymbol{\theta}_{0,c}\right)^{\top}$$
$$\times \frac{1}{\sqrt{T}} \sum_{i=1}^{N} \sum_{t=1}^{T} (u_{0,i,c} - u_{i,\sigma^{(per)}(c)}) \left( \xi_{i,t}^{(w)}(\boldsymbol{\theta}_{0,c}) - E[\xi_{i,t}^{(w)}(\boldsymbol{\theta}_{0,c})] \right) \boldsymbol{x}_{*,t}^{(w)},$$

$$\mathfrak{T}_{3,N,T}(\sigma^{(per)}, \boldsymbol{U}) \coloneqq \frac{1}{\sqrt{N}} \sum_{c=1}^{G} \phi_{\sigma^{(per)}(c)} \sqrt{T} \left(\boldsymbol{\theta}_{\sigma^{(per)}(c)} - \boldsymbol{\theta}_{0,c}\right)^{\top}$$
$$\times E[\xi_{i,t}^{(w)}(\boldsymbol{\theta}_{0,c})] \frac{N^{\frac{1}{2}}}{T^{\frac{3}{2}}} \sum_{t=1}^{T} \boldsymbol{x}_{*,t}^{(w)} \frac{1}{N} \sum_{i=1}^{N} (u_{0,i,c} - u_{i,\sigma^{(per)}(c)}).$$

By (B.9) in Lemma 5, one can show that

$$\frac{N}{T^2} \sum_{t=1}^{T} u_{i,c} \boldsymbol{x}_{*,t}^{(w)} \boldsymbol{x}_{*,t}^{(w)\top} \xrightarrow{w} g_c \boldsymbol{\Sigma}_{\eta}^{(c,c)} \int_0^1 \boldsymbol{W}_{\eta}^{(c)}(\tau) \boldsymbol{W}_{\eta}^{(c)}(\tau)^{\top} d\tau.$$

Therefore, an application of the dominated convergence theorem and the Tchebyshev inequality yields

$$P\left( \sup_{\boldsymbol{U} \in \Delta_S^N} \min_{\sigma^{(per)} \in \sigma(\mathcal{P})} \left| \mathfrak{T}_{1,N,T}(\sigma^{(per)}, \boldsymbol{U}) \right| > \frac{\epsilon_\eta}{8} \right) = O\left(N^{-C_\alpha}\right) \text{ for every } C_\alpha > 1.$$

Moreover, by Lemma 9 and the continuous mapping theorem, one can show that

$$\frac{1}{\sqrt{T}} \sum_{i=1}^{N} \sum_{t=1}^{T} u_{i,c} \left( \xi_{i,t}^{(w)}(\boldsymbol{\theta}_{0,c}) - E[\xi_{i,t}^{(w)}(\boldsymbol{\theta}_{0,c})] \right) \boldsymbol{x}_{*,t}^{(w)} \xrightarrow{w} g_c \sigma_{\xi}^{(c)} \boldsymbol{\Sigma}_{\eta}^{(c,c)} \int_0^1 \boldsymbol{W}_{\eta}^{(c)}(\tau) dW_{\xi}(d\tau),$$

where $\sigma_{\xi}^{(c)2} = \lim_{N\uparrow\infty, T\uparrow\infty} \frac{1}{N(\boldsymbol{u}_c)T} Var\left(\sum_{t=1}^{T}\sum_{i=1}^{N} u_{i,c}\xi_{i,t}(\boldsymbol{\theta}_{0,c})\right)$. Thus, we have

$$P\left(\sup_{\boldsymbol{U}\in\Delta_S^N}\min_{\sigma^{(per)}\in\sigma(\mathcal{P})}\left|\mathfrak{T}_{2,N,T}(\sigma^{(per)},\boldsymbol{U})\right| > \frac{\epsilon_\eta}{8}\right) = O\left((NT)^{-C_\alpha}\right) \text{ for every } C_\alpha > 1.$$

By the same argument, we also obtain

$$P\left(\sup_{\boldsymbol{U}\in\Delta_S^N}\min_{\sigma^{(per)}\in\sigma(\mathcal{P})}\left|\mathfrak{T}_{3,N,T}(\sigma^{(per)},\boldsymbol{U})\right| > \frac{\epsilon_\eta}{8}\right) = O\left(N^{-C_\alpha/2}\right) \text{ for every } C_\alpha > 1.$$

It then follows that

$$\mathcal{T}_{1,N,T}^{(c)} < C_0\left(N^{-C_\alpha/2} + (NT)^{-C_\alpha}\right). \tag{B.59}$$

Finally, to bound $\mathcal{T}_{4,N,T}^{(d)}$, notice that

$$\begin{aligned}
&\frac{1}{NT}\sum_{i=1}^{N}\sum_{t=1}^{T}\phi_{\sigma^{(per)}(c)}(u_{0,i,c} - u_{i,\sigma^{(per)}(c)})\left\{(\boldsymbol{\theta}_{\sigma^{(per)}(c)} - \boldsymbol{\theta}_{0,c})^\top\boldsymbol{x}_{i,t}^{(w)} - \xi_{i,t}^{(w)}(\boldsymbol{\theta}_{0,c})\right\}\epsilon_{0,*,t}^{(w)} \\
&\quad = \frac{1}{N\sqrt{T}}\phi_{\sigma^{(per)}(c)}\sqrt{T}(\boldsymbol{\theta}_{\sigma^{(per)}(c)} - \boldsymbol{\theta}_{0,c})^\top\frac{N}{T}\sum_{t=1}^{T}\left(\boldsymbol{x}_{*,t}^{(w)}(\boldsymbol{u}_{0,c})\epsilon_{0,*,t} - \boldsymbol{x}_{*,t}^{(w)}(\boldsymbol{u}_{\sigma^{(per)}(c)})\epsilon_{0,*,t}\right) \\
&\qquad + \phi_{\sigma^{(per)}(c)}\frac{1}{T}\sum_{t=1}^{T}\left(\xi_{*,t}^{(w)}(\boldsymbol{u}_{0,c})\epsilon_{0,*,t} - \xi_{*,t}^{(w)}(\boldsymbol{u}_{\sigma^{(per)}(c)})\epsilon_{0,*,t}\right) \\
&\quad =: \mathfrak{L}_{1,N,T}(\sigma^{(per)},\boldsymbol{U}) + \mathfrak{L}_{2,N,T}(\sigma^{(per)},\boldsymbol{U}).
\end{aligned}$$

An application of Lemma 3 yields that $\frac{N}{T}\sum_{t=1}^{T}\boldsymbol{x}_{*,t}^{(w)}(\boldsymbol{u}_c)\epsilon_{0,*,t} \xrightarrow{w} \sqrt{g_c}\sigma_\epsilon\boldsymbol{\Sigma}_\eta^{(c,c)}\int_0^1\boldsymbol{W}_\eta^{(c,c)}(\tau)dW_\epsilon(\tau)$, $c = 1,\ldots,G$. Thus, by the dominated convergence theorem and the Tchebyshev inequality, one readily obtains that

$$P\left(\sup_{\boldsymbol{U}\in\Delta_S^N}\min_{\sigma^{(per)}\in\sigma(\mathcal{P})}\left|\mathfrak{L}_{1,N,T}(\sigma^{(per)},\boldsymbol{U})\right| > \frac{\epsilon_\eta}{8}\right) = O\left((N\sqrt{T})^{-C_\alpha}\right) \text{ for every } C_\alpha > 1.$$

Also, by Lemma 12, we can show that

$$\begin{aligned}
P\left(\sup_{\boldsymbol{U}\in\Delta_S^N}\min_{\sigma^{(per)}\in\sigma(\mathcal{P})}\left|\mathfrak{L}_{2,N,T}(\sigma^{(per)},\boldsymbol{U})\right| > \frac{\epsilon_\eta}{8}\right) = O\Bigg(&T^{-C_\alpha} + N^{2\gamma_M}\log^2(T)T^{\gamma_M - \frac{3}{4}\theta_\alpha} \\
&+ \max\left\{\exp\left(-C_\sigma N^2\log^2(T)T^{7/4}\right),\right. \\
&\left.\exp\left(-C_M\frac{T^{1/4}}{\log^2(T)}\right)\right\}\Bigg).
\end{aligned}$$

It then follows that

$$\mathcal{T}^{(d)}_{4,N,T} < C_0 \Big( T^{-C_\alpha} + N^{2\gamma_M} \log^2(T) T^{\gamma_M - \frac{3}{4}\theta_\alpha} + \max\left\{ \exp\left(-C_\sigma N^2 \log^2(T) T^{7/4}\right), \exp\left(-C_M \frac{T^{1/4}}{\log^2(T)}\right) \right\} \Big). \tag{B.60}$$

Collecting all the terms derived in (B.57)-(B.60), we obtain the main theorem. ■

**Proof of Theorem 10.** Recall some commonly-used notations: $\epsilon_{*,t} = (\boldsymbol{\psi} - \widetilde{\boldsymbol{\psi}})^\top \boldsymbol{D}_\phi \boldsymbol{D}_g \boldsymbol{X}_{N,T,t}(\boldsymbol{\theta}) + \epsilon_{*,t}(\widetilde{\boldsymbol{\psi}}, \boldsymbol{U}_0)$, where

$$\boldsymbol{X}_{N,T,t}(\boldsymbol{\theta}) := \left( \boldsymbol{x}^{(1)\top}_{*,t}, \ldots, \boldsymbol{x}^{(G)\top}_{*,t}, -\xi^{(1)}_{*,t}(\boldsymbol{\theta}_1), \ldots, -\xi^{(G)}_{*,t}(\boldsymbol{\theta}_G), -1 \right)^\top$$

with $\boldsymbol{D}_\phi := \operatorname{diag}\left(\boldsymbol{\phi} \otimes \mathbb{I}_{d_x}, \mathbb{I}_{G+1}\right)$ and $\boldsymbol{D}_g := \operatorname{diag}\left(\boldsymbol{g} \otimes \mathbb{I}_{d_x}, \boldsymbol{g}, 1\right)$; and $\widetilde{\mathcal{Q}}_{N,T}(\boldsymbol{\psi}) := \frac{N}{T} \sum_{t=1}^T \epsilon^2_{*,t}(\boldsymbol{\psi}, \boldsymbol{U}_0)$. This proof proceeds along the lines of the proof of Theorem 7. As in the Step 1, it can be shown that

$$\begin{aligned}\widetilde{\mathcal{Q}}_{N,T}(\widehat{\boldsymbol{\psi}}) - \widetilde{\mathcal{Q}}_{N,T}(\widetilde{\boldsymbol{\psi}}) \geq \lambda_{\lim} \Bigg( &\operatorname{diag}\left(T^{-1/2}\mathbb{I}_{G\times d_x}, N^{-1/2}\mathbb{I}_{G+1}\right) \left( \frac{N}{T} \sum_{t=1}^T \boldsymbol{X}_{N,T,t}(\widehat{\boldsymbol{\theta}}) \boldsymbol{X}_{N,T,t}(\widehat{\boldsymbol{\theta}})^\top \right) \\ &\times \operatorname{diag}\left(T^{-1/2}\mathbb{I}_{G\times d_x}, N^{-1/2}\mathbb{I}_{G+1}\right) \Bigg) \\ &\times H\left( \operatorname{diag}\left(\sqrt{T}\mathbb{I}_{G\times d_x}, \sqrt{N}\mathbb{I}_{G+1}\right) \widehat{\boldsymbol{\psi}}, \operatorname{diag}\left(\sqrt{T}\mathbb{I}_{G\times d_x}, \sqrt{N}\mathbb{I}_{G+1}\right) \widetilde{\boldsymbol{\psi}} \right).\end{aligned}$$

Since $\widehat{\boldsymbol{\theta}}$ is consistent by Theorem 8, it then follows from Assumption 6 that

$$H\left( \operatorname{diag}\left(\sqrt{T}\mathbb{I}_{G\times d_x}, \sqrt{N}\mathbb{I}_{G+1}\right) \widehat{\boldsymbol{\psi}}, \operatorname{diag}\left(\sqrt{T}\mathbb{I}_{G\times d_x}, \sqrt{N}\mathbb{I}_{G+1}\right) \widetilde{\boldsymbol{\psi}} \right) \leq \widetilde{\mathcal{Q}}_{N,T}(\widehat{\boldsymbol{\psi}}) - \widetilde{\mathcal{Q}}_{N,T}(\widetilde{\boldsymbol{\psi}}). \tag{B.61}$$

Next, by applying Lemmas 4 and 5, it can be shown that

$$\frac{N^{1/2}}{T^{3/2}} \sum_{t=1}^T \xi^{(w)}_{i,t} \boldsymbol{x}^{(w)}_{i,t} = O_p(1), \tag{B.62}$$

$$\frac{1}{T} \sum_{t=1}^T \xi^{(w)}_{i,t}(\boldsymbol{\theta}_{0,c}) \xi^{(w)}_{i,t}(\boldsymbol{\theta}_{0,d}) = O_p(1), \ \forall\ c, d \in [1, G], \tag{B.63}$$

$$\sqrt{\frac{N}{T}} \sum_{t=1}^T \xi^{(w)}_{i,t}(\boldsymbol{\theta}_{0,c}) \epsilon_{0,*,t} = O_p(1), \tag{B.64}$$

$$\frac{N}{T^2} \sum_{t=1}^T \boldsymbol{x}^{(w)}_{i,t} \boldsymbol{x}^{(w)\top}_{i,t} = O_p(1), \tag{B.65}$$

$$\frac{N}{T}\sum_{t=1}^{T}\boldsymbol{x}_{i,t}^{(w)}\epsilon_{0,*,t} \quad = \quad O_p(1), \tag{B.66}$$

$$\frac{N^{1/2}}{T^{3/2}}\sum_{t=1}^{T}\xi_{i,t}^{(w)}(\boldsymbol{\theta}_{0,c})\boldsymbol{x}_{i,t}^{(w)} \quad = \quad O_p(1). \tag{B.67}$$

It then follows from (4.13), (B.62)-(B.64) that

$$\frac{N}{T}\sum_{t=1}^{T}\xi_{i,t}^{(w)}(\boldsymbol{\theta}_{0,c})\left(\epsilon_{*,t}(\boldsymbol{\psi},\widehat{\boldsymbol{U}})+\epsilon_{*,t}(\boldsymbol{\psi},\boldsymbol{U}_0)\right)=o_p\left(\sqrt{N}\right); \tag{B.68}$$

and also from (4.13), (B.65)-(B.67), one obtains that

$$\sqrt{T}\min_{\sigma\in\sigma(\mathcal{P})}\sum_{c=1}^{G}\left|\left(\boldsymbol{\theta}_{\sigma^{(per)}(c)}-\boldsymbol{\theta}_{0,c}\right)^{\top}\right|\left|\frac{N}{T^{3/2}}\sum_{t=1}^{T}\boldsymbol{x}_{i,t}^{(w)}\left(\epsilon_{*,t}(\boldsymbol{\psi},\widehat{\boldsymbol{U}})+\epsilon_{*,t}(\boldsymbol{\psi},\boldsymbol{U}_0)\right)\right| \quad = \quad o_p\left(\sqrt{N}\right) \tag{B.69}$$

for some $\boldsymbol{\psi}\in\mathcal{B}_T(\boldsymbol{\theta}_0,\eta_\theta)\times\mathcal{B}_N(\boldsymbol{\phi}_0,\eta_\phi)\times\mathcal{B}_N\left(\mu_{*0},\eta_\mu\right)$, where these shrinking balls are defined in Theorem 9. Recall the representation (B.41) established in the proof of Theorem 7. By Theorem 9 together with (B.68) and (B.69), one then has that

$$\sup_{\substack{\boldsymbol{\psi}=(\boldsymbol{\theta}^\top,\boldsymbol{\phi}^\top,\mu_*)^\top,\\ \boldsymbol{\theta}\in\mathcal{B}_T(\boldsymbol{\theta}_0,\eta_\theta),\\ \boldsymbol{\phi}\in\mathcal{B}_N(\boldsymbol{\phi}_0,\eta_\phi),\\ \mu_*\in\mathcal{B}_N(\mu_{*0},\eta_\mu).}}\left|\widehat{\mathcal{Q}}_{N,T}(\boldsymbol{\psi})-\widetilde{\mathcal{Q}}_{N,T}(\boldsymbol{\psi})\right|=O_p\Bigg(N^{\frac{1}{2}-\frac{C_\alpha}{2}}+N^{\frac{1}{2}}T^{-\frac{C_\alpha}{2}}+N^{\gamma_M+\frac{1}{2}}\log(T)T^{\frac{\gamma_M}{2}-\frac{3}{8}\theta_\alpha}$$
$$+N^{\frac{1}{2}}\exp\left(-C_{\epsilon_\eta}\frac{T^{\frac{1}{4}}}{2\log^2(T)}\right)\Bigg).$$

The rest of this proof follows exactly the same argument in Step 3 of the proof of Theorem 7. ■

**Proof of Theorem 11.** We shall consider two scenarios: 1) $\boldsymbol{x}_{i,t}$ is stationary under Assumption 3, and 2) $x_{i,t}$ is nonstationary under Assumption 5. We start with the **stationary case**. Recall $\boldsymbol{x}_{*,t}^{(w)}(\boldsymbol{u}_c):=\frac{1}{N}\sum_{i=1}^{N}u_{i,c}\boldsymbol{x}_{i,t}^{(w)}$ with $\boldsymbol{u}_c:=(u_{1,c},\ldots,u_{N,c})^\top$, $\xi_{*,t}^{(w)}(\boldsymbol{\theta}_c,\boldsymbol{u}_c):=\frac{1}{N}\sum_{i=1}^{N}u_{i,c}\xi_{i,t}^{(w)}(\boldsymbol{\theta}_c)$, and $\xi_{0,*,t}(\boldsymbol{\theta}_{0,c})\equiv\xi_{*,t}(\boldsymbol{\theta}_{0,c},\boldsymbol{u}_{0,c})$.

⋄ For $G<G_0$, the model is underfit. Write

$$\epsilon_{*,G,t}(\boldsymbol{\psi},\boldsymbol{U}):=(\mu_{*0}-\mu_*)1_t^{(w)}+\sum_{c=1}^{G}(\phi_{0,c}-\phi_{\sigma^{(per)}(c)})\xi_{0,*,t}(\boldsymbol{\theta}_{0,c})+\sum_{c=G+1}^{G_0}\phi_{0,c}\xi_{0,*,t}^{(w)}(\boldsymbol{\theta}_{0,c})$$
$$+\sum_{c=1}^{G}\phi_{\sigma^{(per)}(c)}\left(\boldsymbol{\theta}_{\sigma^{(per)}(c)}-\boldsymbol{\theta}_{0,c}\right)^{\top}\boldsymbol{x}_{*,t}^{(w)}(\boldsymbol{u}_{0,c})$$

$$+\sum_{c=1}^{G}\phi_{\sigma^{(per)}(c)}\frac{1}{N}\sum_{i=1}^{N}\left(u_{0,i,c}-u_{i,\sigma^{(per)}(c)}\right)\xi_{i,t}^{(w)}(\boldsymbol{\theta}_{\sigma^{(per)}(c)})+\epsilon_{0,*,t}.$$

Since the permutation operator $\sigma^{(per)}$ is a strict one-to-one mapping, we let $\sigma^{(per)}(c)=c$ for simplicity without compromising the main results. It then follows that

$$\begin{aligned}
ASSE_{G,N,T}(\boldsymbol{\psi},\boldsymbol{U}) &\coloneqq \frac{N}{T}\sum_{t=1}^{T}\epsilon_{*,G,t}^{2}(\boldsymbol{\psi},\boldsymbol{U})=\frac{N}{T}\sum_{t=1}^{T}\Bigg|(\mu_{*0}-\mu_{*})1_{t}^{(w)}+\sum_{c=1}^{G}(\phi_{0,c}-\phi_{c})\xi_{0,*,t}(\boldsymbol{\theta}_{0,c}) \\
&+\sum_{c=G+1}^{G_0}\phi_{0,c}\xi_{0,*,t}^{(w)}(\boldsymbol{\theta}_{0,c})+\sum_{c=1}^{G}\phi_{c}\left(\boldsymbol{\theta}_{c}-\boldsymbol{\theta}_{0,c}\right)^{\top}\boldsymbol{x}_{*,t}^{(w)}(\boldsymbol{u}_{0,c}) \\
&+\sum_{c=1}^{G}\phi_{c}\frac{1}{N}\sum_{i=1}^{N}\left(u_{0,i,c}-u_{i,c}\right)\xi_{i,t}^{(w)}(\boldsymbol{\theta}_{c})\Bigg|^{2} \\
&+2\frac{N}{T}\sum_{t=1}^{T}\epsilon_{0,*,t}\left((\mu_{*0}-\mu_{*})1_{t}^{(w)}+\sum_{c=1}^{G_0}\phi_{0,c}\xi_{0,*,t}^{(w)}(\boldsymbol{\theta}_{0,c})-\sum_{c=1}^{G}\phi_{c}\xi_{*,t}^{(w)}(\boldsymbol{\theta}_{c},\boldsymbol{u}_{c})\right) \\
&+\frac{N}{T}\sum_{t=1}^{T}|\epsilon_{0,*,t}^{(w)}|^{2} \\
&\eqqcolon \mathcal{T}_{1}(\boldsymbol{\psi},\boldsymbol{U})+\mathcal{T}_{2}(\boldsymbol{\psi},\boldsymbol{U})+\mathcal{T}_{3}.
\end{aligned}\tag{B.70}$$

Let $\boldsymbol{\lambda}_N\coloneqq\sqrt{N}\left(\mu_{*0}-\mu_{*},\phi_{0,1}-\phi_{1},\ldots,\phi_{0,G}-\phi_{G},\phi_{0,G+1},\ldots,\phi_{0,G_0},(\boldsymbol{\theta}_{1}-\boldsymbol{\theta}_{0,1})^{\top},\ldots,(\boldsymbol{\theta}_{0,G}-\boldsymbol{\theta}_{G})^{\top},\frac{1}{N}(\boldsymbol{u}_{0,1}-\boldsymbol{u}_{1})^{\top},\frac{1}{N}(\boldsymbol{u}_{0,G}-\boldsymbol{u}_{G})^{\top}\right)^{\top}$ and $\boldsymbol{D}_N\coloneqq\operatorname{diag}\left(\iota_{G_0+1}^{\top},\boldsymbol{\phi}^{\top}\otimes\iota_{d_x}^{\top},\boldsymbol{\phi}^{\top}\otimes\iota_{N}^{\top}\right)$. We have that

$$\begin{aligned}
\mathcal{T}_{1}(\boldsymbol{\psi},\boldsymbol{U}) &= \boldsymbol{\lambda}_N^{\top}\boldsymbol{D}_N\boldsymbol{\mathcal{X}}_{N,T}^{*}(\boldsymbol{\theta})\boldsymbol{D}_N\boldsymbol{\lambda}_N \\
&\geq C_0\lambda_{\min}\left(\boldsymbol{\mathcal{X}}_{N,T}^{*}(\boldsymbol{\theta})\right)\Bigg(N(\mu_{*0}-\mu_{*})^{2}+N\sum_{g=1}^{G}(\phi_{0,g}-\phi_{g})^{2}+N\sum_{g=1}^{G}\|\boldsymbol{\theta}_{g}-\boldsymbol{\theta}_{0,g}\|^{2} \\
&\quad +N\sum_{g=G+1}^{G_0}\phi_{0,g}^{2}+\frac{1}{N}\sum_{g=1}^{G}\sum_{i=1}^{N}|u_{i,g}-u_{0,i,g}|\Bigg) \\
&> C_0N\sum_{g=G+1}^{G_0}\phi_{0,g}^{2}.
\end{aligned}\tag{B.71}$$

An application of Lemma 1 yields $\mathcal{T}_{2}(\boldsymbol{\psi},\boldsymbol{U})=O_p\left(\sqrt{\frac{N}{T}}\right)$. Since $\mathcal{T}_{3}=O_p(1)$ due to the weak law of large numbers, one can then obtain from (B.70) and (B.71) that

$$ASSE_{G,N,T}(\widehat{\boldsymbol{\psi}},\widehat{\boldsymbol{U}})>C_0N\sum_{g=G+1}^{G_0}\phi_{0,g}^{2}.$$

Since $\frac{N}{T}\sum_{t=1}^{T}\epsilon_{*,G_0,t}\left(\widetilde{\boldsymbol{\psi}},\widetilde{\boldsymbol{U}}\right) \leq \frac{N}{T}\sum_{t=1}^{T}\epsilon_{*,G_0,t}\left(\boldsymbol{\psi}_0,\boldsymbol{U}_0\right) = \frac{N}{T}\sum_{t=1}^{T}|\epsilon_{0,*,t}^{(w)}|^2 = O_p(1)$, where $(\widetilde{\boldsymbol{\psi}},\widetilde{\boldsymbol{U}}) \coloneqq \text{argmin}_{\boldsymbol{\psi},\boldsymbol{U}} ASSE_{G_0,N,T}\left(\boldsymbol{\psi},\boldsymbol{U}\right)$. Therefore,

$$ASSE_{G,N,T}\left(\widehat{\boldsymbol{\psi}},\widehat{\boldsymbol{U}}\right) - ASSE_{G_0,N,T}\left(\widetilde{\boldsymbol{\psi}},\widetilde{\boldsymbol{U}}\right) > C_0 N,$$

where $C_0$ is some positive constant. It then follows that

$$IC(G) - IC(G_0) \geq C_0 N + (G-G_0)\omega_N \uparrow \infty. \tag{B.72}$$

$\diamond$ For $G > G_0$, the model is overfit. $ASSE_{G,N,G}(\boldsymbol{\psi},\boldsymbol{U}) = \frac{N}{T}\sum_{t=1}^{T}\left|\epsilon_{0,*,t}^{(w)}\right|^2 = O_p(1)$ when $\mu_* = \mu_{*0}$; $\phi_c = \phi_{0,c}$, $\boldsymbol{\theta}_c = \boldsymbol{\theta}_{0,c}$, and $\boldsymbol{u}_c = \boldsymbol{u}_{0,c}$ for $c = 1,\ldots,G_0$; and $\phi_c = 0$ for $c = G_0+1,\ldots,G$. Since $ASSE_{G,N,G}(\widehat{\boldsymbol{\psi}},\widehat{\boldsymbol{U}}) \leq \frac{N}{T}\sum_{t=1}^{T}\left|\epsilon_{0,*,t}^{(w)}\right|^2$, It must be the case that $ASSE_{G,N,G}(\widehat{\boldsymbol{\psi}},\widehat{\boldsymbol{U}}) = O_p(1)$. Write

$$\begin{aligned}\epsilon_{*,G,t}(\boldsymbol{\psi},\boldsymbol{U}) \coloneqq (\mu_{*0}-\mu_*)1_t^{(w)} &+ \sum_{c=1}^{G_0}\left(\phi_{0,c}-\phi_c\right)\xi_{0,*,t}^{(w)}(\boldsymbol{\theta}_{0,c}) + \sum_{c=1}^{G_0}\phi_c\left(\boldsymbol{\theta}_c-\boldsymbol{\theta}_{0,c}\right)^\top \boldsymbol{x}_{*,t}^{(w)}(\boldsymbol{u}_{0,c}) \\ &+ \sum_{c=1}^{G_0}\phi_c\frac{1}{N}\sum_{i=1}^{N}\left(u_{0,i,c}-u_{i,c}\right)\xi_{i,t}^{(w)}(\boldsymbol{\theta}_c) - \sum_{G_0+1}^{G}\phi_c\xi_{*,t}^{(w)}(\boldsymbol{\theta}_c,\boldsymbol{u}_c) + \epsilon_{0,*,t}\end{aligned}$$

It is immediate to show that

$$\begin{aligned}ASSE_{G,N,T}(\boldsymbol{\psi},\boldsymbol{U}) \coloneqq \frac{N}{T}\sum_{t=1}^{T}\epsilon_{*,G,t}^2(\boldsymbol{\psi},\boldsymbol{U}) = \frac{N}{T}\sum_{t=1}^{T}\Bigg(&(\mu_{*0}-\mu_*)1_t^{(w)} + \sum_{c=1}^{G_0}\left(\phi_{0,c}-\phi_c\right)\xi_{0,*,t}^{(w)}(\boldsymbol{\theta}_{0,c}) \\ &+ \sum_{c=1}^{G_0}\phi_c\left(\boldsymbol{\theta}_c-\boldsymbol{\theta}_{0,c}\right)^\top \boldsymbol{x}_{*,t}^{(w)}(\boldsymbol{u}_{0,c}) + \sum_{c=1}^{G_0}\phi_c\frac{1}{N}\sum_{i=1}^{N}\left(u_{0,i,c}-u_{i,c}\right)\xi_{i,t}^{(w)}(\boldsymbol{\theta}_c) \\ &- \sum_{G_0+1}^{G}\phi_c\xi_{*,t}^{(w)}(\boldsymbol{\theta}_c,\boldsymbol{u}_c)\Bigg)^2 \\ + 2\frac{N}{T}\sum_{t=1}^{T}\epsilon_{0,*,t}\Bigg(&(\mu_{*0}-\mu_*)1_t^{(w)} + \sum_{c=1}^{G_0}\left(\phi_{0,c}-\phi_c\right)\xi_{0,*,t}^{(w)}(\boldsymbol{\theta}_{0,c}) \\ &+ \sum_{c=1}^{G_0}\phi_c\left(\boldsymbol{\theta}_c-\boldsymbol{\theta}_{0,c}\right)^\top \boldsymbol{x}_{*,t}^{(w)}(\boldsymbol{u}_{0,c}) + \sum_{c=1}^{G_0}\phi_c\frac{1}{N}\sum_{i=1}^{N}\left(u_{0,i,c}-u_{i,c}\right)\xi_{i,t}^{(w)}(\boldsymbol{\theta}_c) \\ &- \sum_{G_0+1}^{G}\phi_c\xi_{*,t}^{(w)}(\boldsymbol{\theta}_c,\boldsymbol{u}_c)\Bigg) + \frac{N}{T}\sum_{t=1}^{T}\epsilon_{0,*,t}^2 \\ =: \mathfrak{T}_1(\boldsymbol{\psi},\boldsymbol{U}) + \mathfrak{T}_2(\boldsymbol{\psi},\boldsymbol{U}) + \mathfrak{T}_3,&\end{aligned} \tag{B.73}$$

To bound $\mathfrak{T}_1(\boldsymbol{\psi}, \boldsymbol{U})$ and $\mathfrak{T}_2(\boldsymbol{\psi}, \boldsymbol{U})$, let

$$\begin{aligned}\boldsymbol{\kappa}_N &:= \sqrt{N}\Big(\mu_{*0}-\mu_*, \phi_{0,1}-\phi_1, \ldots, \phi_{0,G_0}-\phi_{G_0}, (\boldsymbol{\theta}_1-\boldsymbol{\theta}_{0,1})^\top, \ldots, (\boldsymbol{\theta}_{G_0}-\boldsymbol{\theta}_{0,G_0})^\top, \\ &\qquad \frac{1}{N}(\boldsymbol{u}_{0,1}-\boldsymbol{u}_1)^\top, \ldots, \frac{1}{N}(\boldsymbol{u}_{0,G_0}-\boldsymbol{u}_{G_0})^\top, \phi_{G_0+1}\frac{1}{N}\boldsymbol{u}_{G_0+1}^\top, \ldots, \phi_G\frac{1}{N}\boldsymbol{u}_G^\top\Big)^\top, \\ \boldsymbol{J}_N &:= \operatorname{diag}\left(\iota_{G_0+1}^\top, \boldsymbol{\phi}^\top \otimes \iota_{d_x}^\top, \boldsymbol{\phi}^\top \otimes \iota_N^\top, \iota_{(G-G_0)N}^\top\right),\end{aligned}$$

and

$$\boldsymbol{X}_{G,N,T}^{**}(\boldsymbol{\theta}) := \frac{1}{T}\sum_{t=1}^{T}\boldsymbol{X}_{G_0,G_0,G,t}(\boldsymbol{\theta})\boldsymbol{X}_{G_0,G_0,G,t}(\boldsymbol{\theta})^\top.$$

We then have that

$$\begin{aligned}\mathfrak{T}_1(\boldsymbol{\psi}, \boldsymbol{U}) &= \boldsymbol{\kappa}_N^\top \boldsymbol{J}_N \boldsymbol{X}_{G,N,T}^{**}(\boldsymbol{\theta}) \boldsymbol{J}_N \boldsymbol{\kappa}_N \\ &\geq C_0 \lambda_{\min}\left(\boldsymbol{X}_{G,N,T}^{**}(\boldsymbol{\theta})\right)\Bigg\{N(\mu_{*0}-\mu_*)^2 + N\sum_{g=1}^{G_0}(\phi_{0,g}-\phi_g)^2 + N\sum_{g=1}^{G_0}\|\boldsymbol{\theta}_g-\boldsymbol{\theta}_{0,g}\|^2 \\ &\quad + \frac{1}{N}\sum_{g=1}^{G_0}\sum_{i=1}^{N}|u_{0,i,c}-u_{i,c}| + \sum_{c=G_0+1}^{G}\phi_c^2\frac{1}{N}\sum_{i=1}^{N}|u_{i,c}|\Bigg\} \qquad \text{(B.74)}\end{aligned}$$

Moreover, an application of Lemma 1 yields

$$\begin{aligned}\mathfrak{T}_2(\boldsymbol{\psi}, \boldsymbol{U}) = o_p\Bigg(&\sqrt{N}|\mu_{*0}-\mu_*| + \sqrt{N}\sum_{g=1}^{G_0}|\phi_{0,g}-\phi_g| + \sqrt{N}\sum_{g=1}^{G_0}\|\boldsymbol{\theta}_g-\boldsymbol{\theta}_{0,g}\| \\ &+ \frac{1}{\sqrt{N}}\sum_{g=1}^{G_0}\sum_{i=1}^{N}|u_{i,c}-u_{0,i,c}| + \frac{1}{\sqrt{N}}\sum_{c=G_0+1}^{G}|\phi_c|\sum_{i=1}^{N}|u_{i,c}|\Bigg). \qquad \text{(B.75)}\end{aligned}$$

Because $ASSE_{G,N,T}\left(\widehat{\boldsymbol{\psi}}, \widehat{\boldsymbol{U}}\right) - \mathfrak{T}_3 \leq 0$, it then follows from (B.73) - (B.75) that

$$\begin{aligned}&N(\mu_{*0}-\widehat{\mu}_*)^2 + N\sum_{g=1}^{G_0}(\phi_{0,g}-\widehat{\phi}_g)^2 + N\sum_{g=1}^{G_0}\|\widehat{\boldsymbol{\theta}}_g-\boldsymbol{\theta}_{0,g}\|^2 + \frac{1}{N}\sum_{g=1}^{G_0}\sum_{i=1}^{N}|u_{0,i,c}-\widehat{u}_{i,c}| \\ &+ \sum_{c=G_0+1}^{G}\widehat{\phi}_c^2\frac{1}{N}\sum_{i=1}^{N}|\widehat{u}_{i,c}| = o_p\Bigg(\sqrt{N}|\mu_{*0}-\widehat{\mu}_*| + \sqrt{N}\sum_{g=1}^{G_0}|\phi_{0,g}-\widehat{\phi}_g| + \sqrt{N}\sum_{g=1}^{G_0}\|\widehat{\boldsymbol{\theta}}_g-\boldsymbol{\theta}_{0,g}\| \\ &+ \frac{1}{\sqrt{N}}\sum_{g=1}^{G_0}\sum_{i=1}^{N}|\widehat{u}_{i,c}-u_{0,i,c}| + \frac{1}{\sqrt{N}}\sum_{c=G_0+1}^{G}|\widehat{\phi}_c|\sum_{i=1}^{N}|\widehat{u}_{i,c}|\Bigg).\end{aligned}$$

This implies that $\sqrt{N}|\mu_{*0}-\widehat{\mu}_*|+\sqrt{N}\sum_{g=1}^{G_0}|\phi_{0,g}-\widehat{\phi}_g|+\sqrt{N}\sum_{g=1}^{G_0}\|\widehat{\boldsymbol{\theta}}_g-\boldsymbol{\theta}_{0,g}\|+\frac{1}{\sqrt{N}}\sum_{g=1}^{G_0}\sum_{i=1}^{N}|\widehat{u}_{i,c}-u_{0,i,c}|+\frac{1}{\sqrt{N}}\sum_{c=G_0+1}^{G}|\widehat{\phi}_c|\sum_{i=1}^{N}|\widehat{u}_{i,c}|=O_p(1)$. Therefore, we have that

$$\frac{1}{\sqrt{N}}\sum_{c=G_0+1}^{G}|\widehat{\phi}_c|\sum_{i=1}^{N}|\widehat{u}_{i,c}|=O_p(1). \tag{B.76}$$

Notice that

$$\begin{aligned} ASSE_{G,N,T}(\widehat{\boldsymbol{\psi}},\widehat{\boldsymbol{U}}) = & \underbrace{\frac{N}{T}\sum_{t=1}^{T}\left((\mu_{*0}-\widehat{\mu}_*)1_t^{(w)}+\sum_{c=1}^{G_0}\left(\phi_{0,c}\xi_{0,*,t}^{(w)}(\boldsymbol{\theta}_{0,c})-\widehat{\phi}_c\xi_{*,t}^{(w)}(\widehat{\boldsymbol{\theta}}_c,\widehat{\boldsymbol{u}}_c)\right)+\epsilon_{0,*,t}^{(w)}\right)^2}_{=ASSE_{G_0,N,T}(\widehat{\boldsymbol{\psi}},\widehat{\boldsymbol{U}})} \\ & -\frac{N}{T}\sum_{t=1}^{T}\left|\sum_{c=G_0+1}^{G}\widehat{\phi}_c\xi_{*,t}^{(w)}(\widehat{\boldsymbol{\theta}}_c,\widehat{\boldsymbol{u}}_c)\right|^2. \end{aligned}$$

Therefore,

$$\begin{aligned} ASSE_{G,N,T}(\widehat{\boldsymbol{\psi}},\widehat{\boldsymbol{U}})-ASSE_{G_0,N,T}(\widetilde{\boldsymbol{\psi}},\widetilde{\boldsymbol{U}}) = & \underbrace{ASSE_{G_0,N,T}(\widehat{\boldsymbol{\psi}},\widehat{\boldsymbol{U}})-ASSE_{G_0,N,T}(\widetilde{\boldsymbol{\psi}},\widetilde{\boldsymbol{U}})}_{\geq 0} \\ & -\frac{N}{T}\sum_{t=1}^{T}\left|\sum_{c=G_0+1}^{G}\widehat{\phi}_c\xi_{*,t}^{(w)}(\widehat{\boldsymbol{\theta}}_c,\widehat{\boldsymbol{u}}_c)\right|^2 \\ \geq & -\frac{N}{T}\sum_{t=1}^{T}\left|\sum_{c=G_0+1}^{G}\widehat{\phi}_c\xi_{*,t}^{(w)}(\widehat{\boldsymbol{\theta}}_c,\widehat{\boldsymbol{u}}_c)\right|^2 \\ \geq & -(G-G_0)\frac{N}{T}\sum_{t=1}^{T}\sum_{c=G_0+1}^{G}\left|\widehat{\phi}_c\xi_{*,t}^{(w)}(\widehat{\boldsymbol{\theta}}_c,\widehat{\boldsymbol{u}}_c)\right|^2 \\ \geq & -C_0(G-G_0), \end{aligned}$$

where the last inequality is obtained from (B.76) and the fact that $\sup_{\boldsymbol{\theta}_c,\boldsymbol{u}_c}\xi_{*,t}^{(w)}(\boldsymbol{\theta}_c,\boldsymbol{u}_c)=O_p(1)$. It then immediately follows that

$$IC(G)-IC(G_0)\geq -C_0(G-G_0)+(G-G_0)\omega_N\uparrow\infty \text{ w.p.1.} \tag{B.77}$$

In view of (B.72) and (B.77), this theorem has been proved for the stationary case.

We next proceed to the **nonstationary case**. The proof here follows along the same line as in the stationary case with some modifications to accommodate persistent covariates.

$\diamond$ For $G < G_0$, the model is underfit. Write

$$\epsilon_{*,G,t}(\boldsymbol{\psi}, \boldsymbol{U}) \coloneqq (\mu_{*0} - \mu_*)1_t^{(w)} + \sum_{c=1}^{G} (\phi_{0,c} - \phi_c)\, \xi_{*,t}^{(w)}(\boldsymbol{\theta}_{0,c}) + \sum_{c=G+1}^{G_0} \phi_{0,c}\xi_{*,t}^{(w)}(\boldsymbol{\theta}_{0,c})$$
$$+ \sum_{c=1}^{G} \phi_c (\boldsymbol{\theta}_c - \boldsymbol{\theta}_{0,c})^\top \boldsymbol{x}_{*,t}^{(w)}(\boldsymbol{u}_c) + \sum_{c=1}^{G} \phi_c \frac{1}{N} \sum_{i=1}^{N} (u_{0,i,c} - u_{i,c})\, \xi_{i,t}^{(w)}(\boldsymbol{\theta}_{0,c}) + \epsilon_{0,*,t}.$$

It then follows that

$$\begin{aligned}
ASSE_{G,N,T}(\boldsymbol{\psi}, \boldsymbol{U}) \coloneqq \frac{N}{T}\sum_{t=1}^{T} \epsilon_{*,G,t}^2(\boldsymbol{\psi}, \boldsymbol{U}) &= \frac{N}{T}\sum_{t=1}^{T} \Bigg( (\mu_{*0} - \mu_*)1_t^{(w)} + \sum_{c=1}^{G}(\phi_{0,c} - \phi_c)\xi_{*,t}^{(w)}(\boldsymbol{\theta}_{0,c}) \\
&\quad + \sum_{c=G+1}^{G_0} \phi_{0,c}\xi_{*,t}^{(w)}(\boldsymbol{\theta}_{0,c}) + \sum_{c=1}^{G} \phi_c(\boldsymbol{\theta}_c - \boldsymbol{\theta}_{0,c})^\top \boldsymbol{x}_{*,t}^{(w)}(\boldsymbol{u}_c) \\
&\quad + \sum_{c=1}^{G} \phi_c \frac{1}{N} \sum_{i=1}^{N}(u_{0,i,c} - u_{i,c})\xi_{i,t}^{(w)}(\boldsymbol{\theta}_{0,c}) \Bigg)^2 \\
&+ 2\frac{N}{T}\sum_{t=1}^{T} \epsilon_{0,*,t}^{(w)} \Bigg( (\mu_{*0} - \mu_*)1_t^{(w)} + \sum_{c=1}^{G} (\phi_{0,c} - \phi_c)\, \xi_{*,t}^{(w)}(\boldsymbol{\theta}_{0,c}) \\
&\quad + \sum_{c=G+1}^{G_0} \phi_{0,c}\xi_{*,t}^{(w)}(\boldsymbol{\theta}_{0,c}) + \sum_{c=1}^{G} \phi_c (\boldsymbol{\theta}_c - \boldsymbol{\theta}_{0,c})^\top \boldsymbol{x}_{*,t}^{(w)}(\boldsymbol{u}_c) \\
&\quad + \sum_{c=1}^{G} \phi_c \frac{1}{N} \sum_{i=1}^{N} (u_{0,i,c} - u_{i,c})\, \xi_{i,t}^{(w)}(\boldsymbol{\theta}_{0,c}) \Bigg) \\
&+ \frac{N}{T}\sum_{t=1}^{T} |\epsilon_{0,*,t}^{(w)}|^2 \\
&\eqqcolon \mathcal{T}_1(\boldsymbol{\psi}, \boldsymbol{U}) + \mathcal{T}_2(\boldsymbol{\psi}, \boldsymbol{U}) + \mathcal{T}_3.
\end{aligned} \tag{B.78}$$

To bound $\mathcal{T}_1(\boldsymbol{\psi}, \boldsymbol{U})$, let $\boldsymbol{\lambda}_N \coloneqq \sqrt{N}\big(\mu_{*0}-\mu_*, \phi_{0,1}-\phi_1, \dots, \phi_{0,G}-\phi_G, \phi_{0,G+1}, \dots, \phi_{0,G_0}, (\boldsymbol{\theta}_1-\boldsymbol{\theta}_{0,1})^\top, \dots, (\boldsymbol{\theta}_G - \boldsymbol{\theta}_{0,G})^\top, \frac{1}{N}(\boldsymbol{u}_{0,1} - \boldsymbol{u}_1)^\top, \dots, \frac{1}{N}(\boldsymbol{u}_{0,G} - \boldsymbol{u}_G)\big)$ and $\boldsymbol{D}_N \coloneqq \operatorname{diag}\left( \iota_{G_0+1}^\top, \underbrace{\boldsymbol{\phi}^\top}_{1\times G} \otimes \iota_{d_x}^\top, \boldsymbol{\phi}^\top \otimes \iota_N^\top \right)$. We then have that

$$\mathcal{T}_1(\boldsymbol{\psi}, \boldsymbol{U}) = \boldsymbol{\lambda}_N^\top \boldsymbol{\ell}_{G,G,N,T}^{-1} \boldsymbol{D}_N \boldsymbol{\ell}_{G,G,N,T} \mathcal{X}_{G,G,N,T}^{**}(\boldsymbol{U}) \boldsymbol{\ell}_{G,G,N,T} \boldsymbol{D}_N \boldsymbol{\ell}_{G,G,N,T}^{-1} \boldsymbol{\lambda}_N,$$

where each element of $\boldsymbol{\ell}_{G,G,N,T}\mathcal{X}_{G,G,N,T}^{**}(\boldsymbol{U})\boldsymbol{\ell}_{G,G,N,T}$ has non-zero probability limit according to

Lemma 5. By the minimum eigenvalue inequality, it holds that

$$\begin{aligned}\mathcal{T}_1(\boldsymbol{\psi},\boldsymbol{U}) &\geq C_0\lambda_{\min}\left(\boldsymbol{\ell}_{G,G,N,T}\mathcal{X}^{**}_{G,G,N,T}(\boldsymbol{U})\boldsymbol{\ell}_{G,G,N,T}\right)\Bigg(N(\mu_{*0}-\mu_*)^2+N\sum_{g=1}^{G}(\phi_{0,g}-\phi_g)^2\\ &\quad +N\sum_{g=G+1}^{G_0}\phi_{0,g}^2+T\sum_{g=1}^{G}\|\boldsymbol{\theta}_g-\boldsymbol{\theta}_{0,g}\|^2+\frac{1}{N}\sum_{g=1}^{G}\sum_{i=1}^{N}|u_{0,i,c}-u_{i,c}|\Bigg)\\ &> N\sum_{g=G+1}^{G_0}\phi_{0,g}^2\\ &> C_0(G_0-G)N.\end{aligned}$$

Moreover, an application of Lemma 4 yields $\mathcal{T}_2(\boldsymbol{\psi},\boldsymbol{U})=O_p\left(1+\sqrt{\frac{N}{T}}\right)$. Since $\mathcal{T}_3=O_p(1)$ as usual, we immediately obtain that

$$ASSE_{G,N,T}(\widehat{\boldsymbol{\psi}},\widehat{\boldsymbol{U}})>C_0(G_0-G)N.$$

Since $ASSE_{G_0,N,T}(\widetilde{\boldsymbol{\psi}},\widetilde{\boldsymbol{U}})\leq\mathcal{T}_3=O_p(1)$, it must be the case that

$$IC(G)-IC(G_0)\geq C_0(G_0-G)N+(G-G_0)\omega_N\uparrow\infty \text{ w.p.1.} \tag{B.79}$$

⋄ For $G>G_0$, the model is overfit. $ASSE_{G,N,T}(\boldsymbol{\psi},\boldsymbol{U})=\frac{N}{T}\sum_{t=1}^{T}|\epsilon_{0,*,t}^{(w)}|^2=O_p(1)$ when $\mu_*=\mu_{*0}$, $\phi_c=\phi_{0,c}$, $\boldsymbol{\theta}_c=\boldsymbol{\theta}_{0,c}$, and $\boldsymbol{u}_c=\boldsymbol{u}_{0,c}$ for $c=1,\ldots,G_0$; and $\phi_c=0$ for $c=G_0+1,\ldots,G$. Since $ASSE_{G,N,T}(\widehat{\boldsymbol{\psi}},\widehat{\boldsymbol{U}})\leq\frac{N}{T}\sum_{t=1}^{T}|\epsilon_{0,*,t}^{(w)}|^2$, it must be the case that $ASSE_{G,N,T}(\widehat{\boldsymbol{\psi}},\widehat{\boldsymbol{U}})=O_p(1)$. Write

$$\begin{aligned}\epsilon_{*,G,t}(\boldsymbol{\psi},\boldsymbol{U}) &:= (\mu_{*0}-\mu_*)1_t^{(w)}+\sum_{c=1}^{G_0}(\phi_{0,c}-\phi_c)\xi_{0,*,t}^{(w)}(\boldsymbol{\theta}_{0,c})+\sum_{c=1}^{G_0}\phi_c\left(\boldsymbol{\theta}_c-\boldsymbol{\theta}_{0,c}\right)^\top\boldsymbol{x}_{*,t}^{(w)}(\boldsymbol{u}_c)\\ &\quad+\sum_{c=1}^{G_0}\phi_c\frac{1}{N}\sum_{i=1}^{N}(u_{0,i,c}-u_{i,c})\xi_{i,t}^{(w)}(\boldsymbol{\theta}_{0,c})-\sum_{c=G_0+1}^{G}\phi_c\frac{1}{N}\sum_{i=1}^{N}u_{i,c}\xi_{i,t}^{(w)}(\boldsymbol{\theta}_c)+\epsilon_{0,*,t}^{(w)}.\end{aligned}$$

Some calculations yields

$$\begin{aligned}ASSE_{G,N,T}(\boldsymbol{\psi},\boldsymbol{U}) &:= \frac{N}{T}\sum_{t=1}^{T}|\epsilon_{*,t}^{(w)}|^2=\frac{N}{T}\sum_{t=1}^{T}\Bigg((\mu_{*0}-\mu_*)1_t^{(w)}+\sum_{c=1}^{G_0}(\phi_{0,c}-\phi_c)\xi_{0,*,t}^{(w)}(\boldsymbol{\theta}_{0,c})\\ &\quad+\sum_{c=1}^{G_0}\phi_c\left(\boldsymbol{\theta}_c-\boldsymbol{\theta}_{0,c}\right)^\top\boldsymbol{x}_{*,t}^{(w)}(\boldsymbol{u}_c)+\sum_{c=1}^{G_0}\phi_c\frac{1}{N}\sum_{i=1}^{N}(u_{0,i,c}-u_{i,c})\xi_{i,t}^{(w)}(\boldsymbol{\theta}_{0,c})\\ &\quad-\sum_{c=G_0+1}^{G}\phi_c\frac{1}{N}\sum_{i=1}^{N}u_{i,c}\xi_{i,t}^{(w)}(\boldsymbol{\theta}_c)\Bigg)^2\end{aligned}$$

$$
\begin{aligned}
&+ 2\frac{N}{T}\sum_{t=1}^{T}\epsilon_{0,*,t}^{(w)}\Bigg((\mu_{*0}-\mu_*)1_t^{(w)} + \sum_{c=1}^{G_0}(\phi_{0,c}-\phi_c)\xi_{0,*,t}^{(w)}(\boldsymbol{\theta}_{0,c}) \\
&\quad + \sum_{c=1}^{G_0}\phi_c\left(\boldsymbol{\theta}_c-\boldsymbol{\theta}_{0,c}\right)^\top \boldsymbol{x}_{*,t}^{(w)}(\boldsymbol{u}_c) + \sum_{c=1}^{G_0}\phi_c\frac{1}{N}\sum_{i=1}^{N}(u_{0,i,c}-u_{i,c})\xi_{i,t}^{(w)}(\boldsymbol{\theta}_{0,c}) \\
&\quad - \sum_{c=G_0+1}^{G}\phi_c\frac{1}{N}\sum_{i=1}^{N}u_{i,c}\xi_{i,t}^{(w)}(\boldsymbol{\theta}_c)\Bigg) + \frac{N}{T}\sum_{t=1}^{T}|\epsilon_{0,*,t}^{(w)}|^2 \\
&=: \mathfrak{T}_1(\boldsymbol{\psi},\boldsymbol{U}) + \mathfrak{T}_2(\boldsymbol{\psi},\boldsymbol{U}) + \mathfrak{T}_3, \qquad \text{(B.80)}
\end{aligned}
$$

where $\mathfrak{T}_3 = O_p(1)$ as usual.

To bound $\mathfrak{T}_1(\boldsymbol{\psi},\boldsymbol{U})$, define $\boldsymbol{\kappa}_N \coloneqq \sqrt{N}\Big(\mu_{*0}-\mu_*, \phi_{0,1}-\phi_1,\ldots,\phi_{0,G_0}-\phi_{G_0},(\boldsymbol{\theta}_1-\boldsymbol{\theta}_{0,1})^\top,\ldots,(\boldsymbol{\theta}_{G_0}-\boldsymbol{\theta}_{0,G_0})^\top, \frac{1}{N}(\boldsymbol{u}_{0,1}-\boldsymbol{u}_1)^\top,\ldots,\frac{1}{N}(\boldsymbol{u}_{0,G_0}-\boldsymbol{u}_{G_0})^\top, \phi_{G_0+1}\frac{1}{N}\boldsymbol{u}_{G_0+1}^\top,\ldots,\phi_G\frac{1}{N}\boldsymbol{u}_G^\top\Big)$ and $\boldsymbol{D}_N \coloneqq \mathrm{diag}\Bigg(\iota_{G_0+1}^\top, \underbrace{\boldsymbol{\phi}^\top}_{1\times G_0}\otimes \iota_{d_x}^\top, \boldsymbol{\phi}^\top\otimes\iota_N^\top, \iota_{(G-G_0)\times N}^\top\Bigg)$. It then follows from the minimum eigenvalue inequality that

$$
\begin{aligned}
\mathfrak{T}_1(\boldsymbol{\psi},\boldsymbol{U}) &= \boldsymbol{\kappa}_N^\top \boldsymbol{\ell}_{G_0,G,N,T}^{-1}\boldsymbol{D}_N\boldsymbol{\ell}_{G_0,G,N,T}\mathcal{X}_{G_0,G,N,T}(\boldsymbol{\theta},\boldsymbol{U})\boldsymbol{\ell}_{G_0,G,N,T}\boldsymbol{D}_N\boldsymbol{\ell}_{G_0,G,N,T}^{-1}\boldsymbol{\kappa}_N \\
&\geq C_0\lambda_{\min}\left(\boldsymbol{\ell}_{G_0,G,N,T}\mathcal{X}_{G_0,G,N,T}(\boldsymbol{\theta},\boldsymbol{U})\boldsymbol{\ell}_{G_0,G,N,T}\right)\Bigg(N(\mu_{*0}-\mu_*)^2 + N\sum_{g=1}^{G_0}(\phi_{0,g}-\phi_g)^2 \\
&\quad + T\sum_{g=1}^{G_0}\|\boldsymbol{\theta}_g-\boldsymbol{\theta}_{0,g}\|^2 + \frac{1}{N}\sum_{g=1}^{G_0}\sum_{i=1}^{N}|u_{0,i,c}-u_{i,c}| + T^2\sum_{g=G_0+1}^{G}\phi_g^2\frac{1}{N}\sum_{i=1}^{N}u_{i,g}\Bigg), \qquad \text{(B.81)}
\end{aligned}
$$

where each element of $\boldsymbol{\ell}_{G_0,G,N,T}\mathcal{X}_{G_0,G,N,T}(\boldsymbol{\theta},\boldsymbol{U})\boldsymbol{\ell}_{G_0,G,N,T}$ has non-zero probability limit according to Lemma 5 and the fact that, for each $i\in[1,N]$, $y_{i,t}$ is a unit-root process with non-zero intercept, thus $y_{i,t} = O_p(T)$. By the sam argument, we can also verify that

$$
\mathfrak{T}_2(\boldsymbol{\psi},\boldsymbol{U}) = O_p\left(T\sqrt{N}\sum_{g=G_0+1}^{G}|\phi_c|\frac{1}{N}\sum_{i=1}^{N}u_{i,g}\right). \qquad \text{(B.82)}
$$

Since $ASSE_{G,N,T}(\widehat{\boldsymbol{\psi}},\widehat{\boldsymbol{U}}) = O_1(1)$, it follows that

$$
\sum_{g=G_0+1}^{G}\widehat{\phi}_g^2 = O_p\left(\frac{1}{T^2}\right) \qquad \text{(B.83)}
$$

and

$$\frac{1}{\sqrt{N}}\sum_{g=G_0+1}^{G}\sum_{i=1}^{N}\widehat{u}_{i,g}=O_p(1). \tag{B.84}$$

By the same argument leading to (B.77), one can also verify that

$$\begin{aligned} ASSE_{G,N,T}\left(\widehat{\boldsymbol{\psi}},\widehat{\boldsymbol{U}}\right)-ASSE_{G_0,N,T}\left(\widetilde{\boldsymbol{\psi}},\widetilde{\boldsymbol{U}}\right) &\geq -(G-G_0)\frac{N}{T}\sum_{g=G_0+1}^{G}\sum_{t=1}^{T}|\widehat{\phi}_g|^2\left|\xi_{*,t}^{(w)}(\widehat{\boldsymbol{\theta}}_g,\widehat{\boldsymbol{u}}_g)\right|^2 \\ &\geq -(G-G_0)\sum_{g=G_0+1}^{G}\widehat{\phi}_g^2\left(\frac{1}{\sqrt{N}}\sum_{i=1}^{N}\widehat{u}_{i,g}\right)^2 \\ &\quad\times\max_{i\in[1,N]}\sup_{\boldsymbol{\theta}}\frac{1}{T}\sum_{t=1}^{T}\left|\xi_{i,t}^{(w)}(\boldsymbol{\theta})\right|^2 \\ &> -C_0(G-G_0), \end{aligned}$$

where the last inequality follows from (B.83)-(B.84) and the fact that $\xi_{i,t}^{(w)}(\boldsymbol{\theta})=O_p(T)$ for every $i\in[1,N]$ and $\|\boldsymbol{\theta}\|<\infty$. Therefore, we have

$$IC(G)-IC(G_0)>-C_0(G-G_0)+(G-G_0)\omega_N\uparrow\infty. \tag{B.85}$$

In view of (B.79) and (B.85) the main result also holds for the nonstationary case. ■

## B.3 Auxiliary Lemmata

**Lemma 6** *Let $\{\eta_{\boldsymbol{s}},\ \boldsymbol{s}\in V_N\}$ represent a centered mixing random field. Suppose that $\eta_{\boldsymbol{s}}$, $\boldsymbol{s}\in V_N$ are identically distributed across locations such that $E[|\eta_{\boldsymbol{s}}|^{\gamma_\eta}]<\infty$ for some $\gamma_\eta>2$; and $\sum_{r=1}^{diam(V_N)}r^{d_v-1}\alpha(r)^{1-\frac{2}{\gamma_\eta}}<\infty$. Then,*

$$E\left|\sum_{\boldsymbol{s}\in V_N}\eta_{\boldsymbol{s}}\right|^2\leq C_0|V_N|.$$

**Proof of Lemma 6.** For brevity, define $S(V_N):=\sum_{\boldsymbol{s}\in V_N}\eta_{\boldsymbol{s}}$. One has

$$\begin{aligned} \frac{E|S(V_N)|^2}{|V_N|} &= E[\eta_{\boldsymbol{s}}^2]+\frac{1}{|V_N|}\sum_{\boldsymbol{s},\boldsymbol{w}\in V_N,\boldsymbol{s}\neq\boldsymbol{w}}E[\eta_{\boldsymbol{s}}\eta_{\boldsymbol{w}}] \\ &= E[\eta_{\boldsymbol{s}}^2]+\frac{1}{|V_N|}\sum_{\boldsymbol{s}\in V_N}\sum_{r=1}^{\mathrm{diam}(V_N)}\sum_{\boldsymbol{w}\in V_N,\|\boldsymbol{w}-\boldsymbol{s}\|=r}E[\eta_{\boldsymbol{s}}\eta_{\boldsymbol{w}}] \end{aligned}$$

$$=: E[\eta_{\boldsymbol{s}}^2] + \mathcal{A}_N.$$

By Lemma 17, one gets $\mathcal{A}_N \le C_0 \frac{1}{|V_N|} \sum_{\boldsymbol{s}\in V_N} \sum_{r=1}^{\text{diam}(V_N)} |\{\boldsymbol{w} \in V_N : \ \|\boldsymbol{w}-\boldsymbol{s}\| = r\}| \, \alpha(r)^{1-2/\gamma_\eta}$. Invoking Lemma 16, it then follows that $\mathcal{A}_N \le C_0 \sum_{r=1}^{\text{diam}(V_N)} r^{d_v-1}\alpha(r)^{1-2/\gamma_\eta} < \infty$. The lemma is proved. ■

**Lemma 7** *Let $\{\eta_{\boldsymbol{s}}, \ \boldsymbol{s} \in V_N\}$ be defined as in Lemma 6 and $S(V_N) \coloneqq \sum_{\boldsymbol{s}\in V_N} \eta_{\boldsymbol{s}}$. Suppose that $E[|\eta_{\boldsymbol{s}}|^{2\gamma_\eta}] < \infty$ for some $\gamma_\eta > 2$. If*

$$\sum_{r=1}^{diam(V_N)} r^{d_v-1}\alpha(r)^{1-\frac{2}{\gamma_\eta}} < \infty$$

*and*

$$|V_N|^{1/2} \sum_{r=|V_N|^{\frac{1}{2d_v}}}^{diam(V_N)} r^{d_v-1}\alpha(r)^{1-2/\gamma_\eta} < \infty$$

*hold, then*

$$E[S(V_N)^3] \le C_0 |V_N|^{3/2}. \tag{B.86}$$

**Proof of Lemma 7.** One can immediately obtain

$$E[S(V)^2] = |V_N| E[\eta_{\boldsymbol{s}}^3] + \sum_{\boldsymbol{s},\boldsymbol{w}\in V_N, \boldsymbol{s}\ne\boldsymbol{w}} E[\eta_{\boldsymbol{s}}^2\eta_{\boldsymbol{w}}] + \sum_{\substack{\boldsymbol{s},\boldsymbol{w},\boldsymbol{z}\in V_N\\ \boldsymbol{s}\ne\boldsymbol{w}\\ \boldsymbol{s}\ne\boldsymbol{z}\\ \boldsymbol{w}\ne\boldsymbol{z}}} E[\eta_{\boldsymbol{s}}\eta_{\boldsymbol{w}}\eta_{\boldsymbol{z}}] =: |V_N| E[\eta_{\boldsymbol{s}}^3] + \mathcal{A}_N + \mathcal{B}_N. \tag{B.87}$$

(Note that the symbols $\mathcal{A}_N$ and $\mathcal{B}_N$ are meant specifically in this proof and different from those defined elsewhere.) By Lemma 17, one has $|E[\eta_{\boldsymbol{s}}^2\eta_{\boldsymbol{w}}]| \le C_0 \|\eta_{\boldsymbol{s}}^2\|_{\gamma_\eta} \|\eta_{\boldsymbol{w}}\|_{\gamma_\eta} M_\alpha(1,1)^{1-2/\gamma_\eta} \alpha(\|\boldsymbol{s}-\boldsymbol{w}\|)^{1-2/\gamma_\eta}$. Thus, in view of Lemmas 16 and 17,

$$\begin{aligned}\mathcal{A}_N \le C_0 \sum_{\boldsymbol{s}\in V_N} \sum_{r=1}^{\text{diam}(V_N)} \sum_{\|\boldsymbol{s}-\boldsymbol{w}\|=r, \boldsymbol{w}\in V_N} \alpha(r)^{1-2/\gamma_\eta} &\le 2d_v C_0 |V_N| \sum_{r=1}^{\text{diam}(V_N)} (2r+1)^{d_v-1}\alpha(r)^{1-2/\gamma_\eta} \\ &\le C_0 |V_N|. \end{aligned} \tag{B.88}$$

Next, to bound $\mathcal{B}_N$, a decomposition of the summation indices yields

$$\mathcal{B}_N = \sum_{\boldsymbol{s}\in V_N} \left( \sum_{\substack{\boldsymbol{w}\in V_N\\ \|\boldsymbol{w}-\boldsymbol{s}\|\le c_N}} + \sum_{\substack{\boldsymbol{w}\in V_N\\ \|\boldsymbol{w}-\boldsymbol{s}\|> c_N}} \right) \left( \sum_{\substack{\boldsymbol{z}\in V_N\\ \|\boldsymbol{z}-\boldsymbol{s}\|\le c_N}} + \sum_{\substack{\boldsymbol{z}\in V_N\\ \|\boldsymbol{z}-\boldsymbol{s}\|> c_N}} \right) E[\eta_{\boldsymbol{s}}\eta_{\boldsymbol{w}}\eta_{\boldsymbol{z}}]$$

$$
\begin{aligned}
&= \sum_{\boldsymbol{s}\in V_N}\left(\sum_{\substack{\boldsymbol{w}\in V_N\\ \|\boldsymbol{w}-\boldsymbol{s}\|\le c_N}}\sum_{\substack{\boldsymbol{z}\in V_N\\ \|\boldsymbol{z}-\boldsymbol{s}\|\le c_N}} + \sum_{\substack{\boldsymbol{w}\in V_N\\ \|\boldsymbol{w}-\boldsymbol{s}\|\le c_N}}\sum_{\substack{\boldsymbol{z}\in V_N\\ \|\boldsymbol{z}-\boldsymbol{s}\|> c_N}} + \sum_{\substack{\boldsymbol{w}\in V_N\\ \|\boldsymbol{w}-\boldsymbol{s}\|> c_N}}\sum_{\substack{\boldsymbol{z}\in V_N\\ \|\boldsymbol{z}-\boldsymbol{s}\|\le c_N}}\right.\\
&\quad \left. + \sum_{\substack{\boldsymbol{w}\in V_N\\ \|\boldsymbol{w}-\boldsymbol{s}\|> c_N}}\sum_{\substack{\boldsymbol{z}\in V_N\\ \|\boldsymbol{z}-\boldsymbol{s}\|> c_N}}\right) E[\eta_{\boldsymbol{s}}\eta_{\boldsymbol{w}}\eta_{\boldsymbol{z}}]\\
&=: \mathcal{B}_{N,1} + \mathcal{B}_{N,2} + \mathcal{B}_{N,3} + \mathcal{B}_{N,4}. \qquad \text{(B.89)}
\end{aligned}
$$

Notice that, by Lemma 16, for a given $\boldsymbol{s}\in V_N$, $\sum_{\substack{\boldsymbol{s}\in V_N\\ 1\le\|\boldsymbol{w}-\boldsymbol{s}\|\le c_N}} = \sum_{r=1}^{c_N}\sum_{\substack{\boldsymbol{w}\in V_N\\ \|\boldsymbol{w}-\boldsymbol{s}\|=r}} \le 2d_v\sum_{r=1}^{c_N}(2r+1)^{d_v-1} \le 2d_v 2^{d_v-2}(2^{d_v-1}+1)\sum_{r=1}^{c_N} r^{d_v-1} < C_0 c_N^{d_v}$, where the last inequality holds by the formula: $\sum_{k=1}^{N} k^p \approx \frac{N^{p+1}}{p+1}$; and, by Lemma 17, $|E[\eta_{\boldsymbol{s}}\eta_{\boldsymbol{w}}\eta_{\boldsymbol{z}}]| \le C_0\alpha\left(\min(\|\boldsymbol{w}-\boldsymbol{s}\|, \|\boldsymbol{z}-\boldsymbol{s}\|)\right)^{1-2/\gamma_\eta}$. It immediately follows that

$$
\mathcal{B}_{N,1} \le C_0|V_N|c_N^{d_v}\sum_{r=1}^{c_N} r^{d_v-1}\alpha(r)^{1-2/\gamma_\eta}. \qquad \text{(B.90)}
$$

Since $\mathcal{B}_{N,2} = \sum_{\boldsymbol{s}\in V_N}\sum_{\substack{\boldsymbol{w}\in V_N\\ \|\boldsymbol{w}-\boldsymbol{s}\|\le c_N}}\left(\sum_{\substack{\boldsymbol{z}\in V_N\\ \|\boldsymbol{z}-\boldsymbol{s}\|>c_N\\ \|\boldsymbol{z}-\boldsymbol{w}\|\le c_N}} + \sum_{\substack{\boldsymbol{z}\in V_N\\ \|\boldsymbol{z}-\boldsymbol{s}\|>c_N\\ \|\boldsymbol{z}-\boldsymbol{w}\|> c_N}}\right) E[\eta_{\boldsymbol{s}}\eta_{\boldsymbol{w}}\eta_{\boldsymbol{z}}] =: \mathcal{B}_{N,2,a} + \mathcal{B}_{N,2,b}$, where $|E[\eta_{\boldsymbol{s}}\eta_{\boldsymbol{w}}\eta_{\boldsymbol{z}}]| \le C_0\alpha\left(\min(\|\boldsymbol{z}-\boldsymbol{s}\|, \|\boldsymbol{z}-\boldsymbol{w}\|)\right)^{1-2/\gamma_\eta}$, one has that

$$
\begin{aligned}
\mathcal{B}_{N,2,a} &\le C_0\sum_{\boldsymbol{s}\in V_N}\sum_{\substack{\boldsymbol{w}\in V_N\\ \|\boldsymbol{w}-\boldsymbol{s}\|\le c_N}}\sum_{\substack{\boldsymbol{z}\in V_N\\ \|\boldsymbol{z}-\boldsymbol{s}\|\le c_N}} \alpha\left(\|\boldsymbol{z}-\boldsymbol{w}\|\right)^{1-2/\gamma_\eta}\\
&\le C_0|V_N|c_N^{d_v}\sum_{r=1}^{c_N} r^{d_v-1}\alpha(r)^{1-2/\gamma_\eta}
\end{aligned}
$$

and

$$
\begin{aligned}
\mathcal{B}_{N,2,b} &\le C_0\sum_{\boldsymbol{s}\in V_N}\sum_{\substack{\boldsymbol{w}\in V_N\\ \|\boldsymbol{w}-\boldsymbol{s}\|\le c_N}}\sum_{r=c_N}^{\mathrm{diam}(V_N)} r^{d_v-1}\alpha(r)^{1-2/\gamma_\eta}\\
&\le C_0|V_N|c_N^{d_v}\sum_{r=c_N}^{\mathrm{diam}(V_N)} r^{d_v-1}\alpha(r)^{1-2/\gamma_\eta}.
\end{aligned}
$$

Therefore,

$$
\mathcal{B}_{N,2} \le C_0|V_N|c_N^{d_v}\sum_{r=1}^{\mathrm{diam}(V_N)} r^{d_v-1}\alpha(r)^{1-2/\gamma_\eta}; \qquad \text{(B.91)}
$$

and similarly, one also has

$$\mathcal{B}_{N,3} \leq C_0 |V_N| c_N^{d_v} \sum_{r=1}^{\mathrm{diam}(V_N)} r^{d_v - 1} \alpha(r)^{1-2/\gamma_\eta}. \tag{B.92}$$

By the same argument, we can also show that

$$\begin{aligned} \mathcal{B}_{N,4} &\leq C_0 \sum_{\boldsymbol{s}\in V_N} \sum_{\substack{\boldsymbol{w}\in V_N \\ \|\boldsymbol{w}-\boldsymbol{s}\|>c_N}} \sum_{r=c_N}^{\mathrm{diam}(V_N)} r^{d_v-1}\alpha(r)^{1-2/\gamma_\eta} \\ &\leq C_0 |V_N|^2 \sum_{r=c_N}^{\mathrm{diam}(V_N)} r^{d_v-1}\alpha(r)^{1-2/\gamma_\eta}. \end{aligned} \tag{B.93}$$

The lemma readily follows from (B.87)-(B.93) by choosing $c_N = |V_N|^{\frac{1}{2d_v}}$. ∎

**Lemma 8** *Again let* $S(V_N) := \sum_{\boldsymbol{s}\in V_N} \eta_{\boldsymbol{s}}$ *be defined as in Lemma 6 above. Suppose that* $\alpha(\tau) \leq C_\theta \tau^{-\theta_\alpha}$ *for some* $\theta_\alpha \geq \max\left(\frac{p d_v \gamma_\eta}{(p-q)(\gamma_\eta - 2)} + d_v\gamma_M, \frac{d_v}{1-\frac{2}{\gamma_\eta}}, \frac{1}{1-\frac{\delta}{\delta+2}-\frac{2}{p}} - \gamma_M\right)$*, where* $\gamma_\eta > 2$*,* $p > \delta + 2$*,* $q = \frac{p(2+\delta)}{2p-2-\delta}$ *for some* $\delta > 0$*, and* $\gamma_M$ *is defined in Definition B.1. Moreover, assume that*

$$\max_{\boldsymbol{i}} \left(E|\eta_{\boldsymbol{i}}|^p, E|\eta_{\boldsymbol{i}}|^{\gamma_\eta}, E|\eta_{\boldsymbol{i}}|^{\delta+2}\right) < \infty.$$

*Then,*

$$E[|S(V_N)|^{2+\delta}] < C_* |V_N|^{1+\frac{\delta}{2}},$$

*where* $C_*$ *is some sufficiently large generic constant such that* $C_* > \frac{4c_\delta A_\delta C_u}{1-\tau_0-4c_\delta \xi A_\delta}$*,* $c_\delta = \begin{cases} 1 & \text{if } \delta<1, \\ 2^{\delta-1} & \text{if } \delta\geq 1 \end{cases}$*,* $C_u$ *is the generic constant chosen in Lemma 6,* $A_\delta > 0$ *is to ensure that (B.94) holds,* $\xi \in \left(0, \frac{1-\tau_0}{4c_\delta A_\delta}\right)$*, and* $\tau_0$ *is some generic constant chosen less than 1.*

**Proof of Lemma 8.** As the argument based on the decomposition of summation indices (used in the proof of Lemma 7) is rather cumbersome to apply in this current context, especially when $\delta$ is greater than 3, we shall here base the proof on an inductive argument, reminiscent of the one used in Bulinski and Shashkin (2007). First, note that, for a given $\delta$, one can always choose an $A_\delta > 0$ such that

$$(x+y)^2(1+x+y)^\delta \leq x^{2+\delta} + y^{2+\delta} + A_\delta\left((1+x)^\delta y^2 + x^2(1+y)^\delta\right) \text{ for any } x, y \geq 0. \tag{B.94}$$

Let $h(N) := \min\{k \in \mathbb{Z}_+ : \ 2^k \geq N\}$, $N \in \mathbb{N}$. For any sublattice, $V_N \subset \mathbb{Z}^{d_v}$, having edges of lengths, at most equal to $\ell_1, \ldots, \ell_{d_v}$, we define $h(V_N) := \sum_{i=1}^{d_v} h(\ell_i)$. We need to show that, for some $C*$ large enough and all sublattices, $V_N \subset \mathbb{Z}^{d_v}$,

$$E\left[S^2(V_N)\left(1+S(V_N)\right)^\delta\right] \le C_*|V_N|^{1+\frac{\delta}{2}}. \tag{B.95}$$

When $h(V_N) = 0$ (i.e., $|V_N| = 1$), (B.95) is obviously true. Suppose that (B.95) holds for every $U_N$ such that $h(U_N) \le h_0$. One needs to verify that it also holds for any $V_N$ such that $h(V_N) = h_0+1$, say. Let $\ell_+(V_N)$ represent the maximum length of the longest edge of $V_N$. Draw a hyperplane orthogonal to this longest edge, cutting this edge into two intervals of lengths, $\lfloor\frac{\ell_+(V_N)}{2}\rfloor$ and $\ell_+(V_N) - \lfloor\frac{\ell_+(V_N)}{2}\rfloor$. The hyperplane then divides $V_N$ into two non-overlapping sublattices, say $V_{1,N}$ and $V_{2,N}$, with $h(V_{1,N}), h(V_{2,N}) \le h_0$.

Let $Q_{1,N} = S(V_{1,N})$ and $Q_{2,N} = S(V_{2,N})$. By (B.94) and Lemma 18, one obtains that, for some $\tau_0 < 1$,

$$\begin{aligned}
E\left[S^2(V_N)\left(1+S(V_N)\right)^\delta\right] &= E\left[(Q_{1,N}+Q_{2,N})^2(1+Q_{1,N}+Q_{2,N})^\delta\right] \\
&\le C_*(|V_{1,N}|^{1+\frac{\delta}{2}} + |V_{2,N}|^{1+\frac{\delta}{2}}) \\
&\quad + A_\delta\left(E\left[(1+|Q_{1,N}|)^\delta Q_{2,N}^2\right] + E\left[(1+|Q_{2,N}|)^\delta Q_{1,N}^2\right]\right) \\
&\le C_*\tau_0|V_N|^{1+\frac{\delta}{2}} + A_\delta E\left[|1+Q_{1,N}|^\delta Q_{2,N}^2\right] \\
&\quad + A_\delta E\left[|1+Q_{2,N}|^\delta Q_{1,N}^2\right].
\end{aligned} \tag{B.96}$$

We still need to bound $E\left[|1+Q_{1,N}|^\delta Q_{2,N}^2\right]$ and $E\left[|1+Q_{2,N}|^\delta Q_{1,N}^2\right]$. We shall now proceed with the former as the latter is quite similar. Introduce the subset $\mathcal{U}_N := \left\{\boldsymbol{s} \in V_{2,N}:\ d(\boldsymbol{s}, V_{1,N}) \le \xi|V_N|^{1/d_v}\right\}$ for some $\xi \in \left(0, \frac{1-\tau_0}{4c_\delta A_\delta}\right)$, where $c_\delta$ is defined in (B.98). An application of the elementary inequality $((a+b)^r \le c_r(a^r+b^r)$, $c_r = 1$ if $r<1$ and $c_r = 2^{r-1}$ if $r \ge 1)$ yields

$$\begin{aligned}
E\left[|1+Q_{1,N}|^\delta Q_{2,N}^2\right] &= E\left[|1+Q_{1,N}|^\delta\left(S(\mathcal{U}_N)+S(V_{2,N}\backslash\mathcal{U}_N)\right)^2\right] \\
&\le 2E\left[|1+Q_{1,N}|^\delta S^2(\mathcal{U}_N)\right] + 2E\left[|1+Q_{1,N}|^\delta S^2(V_{2,N}\backslash\mathcal{U}_N)\right] \\
&\le 2c_\delta E[S^2(\mathcal{U}_N)] + 2c_\delta E\left[|Q_{1,N}|^\delta S^2(\mathcal{U}_N)\right] + 2E\left[|1+Q_{1,N}|^\delta S^2(V_{2,N}\backslash\mathcal{U}_N)\right] \\
&\le 2c_\delta E[S^2(\mathcal{U}_N)] + 2c_\delta\left(E\left[|Q_{1,N}|^{2+\delta}\right]\right)^{\frac{\delta}{\delta+2}}\left(E[S^{2+\delta}(\mathcal{U}_N)]\right)^{\frac{2}{\delta+2}} \\
&\quad + 2E\left[|1+Q_{1,N}|^\delta S^2(V_{2,N}\backslash\mathcal{U}_N)\right] \\
&\le 2c_\delta E[S^2(\mathcal{U}_N)] + 2c_\delta\xi C_*|V_N|^{1+\frac{\delta}{2}} + 2E\left[|1+Q_{1,N}|^\delta S^2(V_{2,N}\backslash\mathcal{U}_N)\right].
\end{aligned} \tag{B.97}$$

Since $E[S^2(\mathcal{U}_N)] \le C_u|V_N|$, where $C_u$ is some given constant, by Lemma 6, one then has

$$E\left[|1+Q_{1,N}|^\delta Q_{2,N}^2\right] \le 2c_\delta C_u|V_N| + 2c_\delta\xi C_*|V_N|^{1+\frac{\delta}{2}} + 2E\left[|1+Q_{1,N}|^\delta S^2(V_{2,N}\backslash\mathcal{U}_N)\right]. \tag{B.98}$$

Moreover, notice that

$$E\left[|1+Q_{1,N}|^{\delta}S^2(V_{2,N}\backslash\mathcal{U}_N)\right] \le \left|\sum_{\boldsymbol{i},\boldsymbol{j}\in V_{2,N}\backslash\mathcal{U}_N,\boldsymbol{i}\neq\boldsymbol{j}} E\left[|1+Q_{1,N}|^{\delta}\eta_{\boldsymbol{i}}\eta_{\boldsymbol{j}}\right]\right| + \left|\sum_{\boldsymbol{i}\in V_{2,N}\backslash\mathcal{U}_N} E\left[|1+Q_{1,N}|^{\delta}\eta_{\boldsymbol{i}}^2\right]\right|$$
$$=: \mathcal{A}_N + \mathcal{B}_N. \tag{B.99}$$

To bound $\mathcal{A}_N$, note that, for each $\boldsymbol{i}\in V_{2,N}\backslash\mathcal{U}_N$,

$$\left|\sum_{\boldsymbol{j}\in V_{2,N}\backslash\mathcal{U}_N,\boldsymbol{i}\neq\boldsymbol{j}} E\left[|1+Q_{1,N}|^{\delta}\eta_{\boldsymbol{i}}\eta_{\boldsymbol{j}}\right]\right| \le \left|\sum_{\boldsymbol{j}\in V_{2,N}^{(1)}} E\left[|1+Q_{1,N}|^{\delta}\eta_{\boldsymbol{i}}\eta_{\boldsymbol{j}}\right]\right| + \left|\sum_{\boldsymbol{j}\in V_{2,N}^{(2)},\boldsymbol{j}\neq\boldsymbol{i}} E\left[|1+Q_{1,N}|^{\delta}\eta_{\boldsymbol{i}}\eta_{\boldsymbol{j}}\right]\right|$$
$$=: \mathcal{A}_{1,N} + \mathcal{A}_{2,N},$$

where $V_{2,N}^{(1)} := \{\boldsymbol{j}\in V_{2,N}\backslash\mathcal{U}_N:\ \|\boldsymbol{j}-\boldsymbol{i}\|\ge\xi|V_N|^{1/d_v}\}$ and $V_{2,N}^{(2)} := \{\boldsymbol{j}\in V_{2,N}\backslash\mathcal{U}_N:\ \|\boldsymbol{j}-\boldsymbol{i}\|<\xi|V_N|^{1/d_v}\}$. For each pair, $\boldsymbol{i}\neq\boldsymbol{j}\in V_{2,N}\backslash\mathcal{U}_N$, define truncated random variables, $\eta_{1,\boldsymbol{i}} := \eta_{\boldsymbol{i}}\mathbf{1}(|\eta_{\boldsymbol{i}}|\le M(\boldsymbol{i},\boldsymbol{j}))$ and $\eta_{2,\boldsymbol{i}} := \eta_{\boldsymbol{i}}-\eta_{1,\boldsymbol{i}}$, where $M(\boldsymbol{i},\boldsymbol{j}) := (d(\{\boldsymbol{j}\},\{\boldsymbol{i}\}\bigcup V_{1,N})))^{\theta_\eta}$ with $\frac{qd_v}{p-q}\le\theta_\eta\le(\theta_\alpha-d_v\gamma_M)(1-2/\gamma_\eta)-d_v$. Also let $|1+Q_{1,N}|^{\delta} = |1+Q_{1,N}|^{\delta}\mathbf{1}(|Q_{1,N}|\le L_N)+|1+Q_{1,N}|^{\delta}\mathbf{1}(|Q_{1,N}|>L_N)$, where $L_N := |V_N|^{1/2}$, one obtains that

$$E\left[|1+Q_{1,N}|^{\delta}\eta_{\boldsymbol{i}}\eta_{\boldsymbol{j}}\right] = Cov\left(|1+Q_{1,N}|^{\delta}\mathbf{1}(|Q_{1,N}|\le L_N)\eta_{1,\boldsymbol{i}},\eta_{\boldsymbol{j}}\right)$$
$$+\,Cov\left(|1+Q_{1,N}|^{\delta}\mathbf{1}(|Q_{1,N}|>L_N)\eta_{1,\boldsymbol{i}},\eta_{\boldsymbol{j}}\right)$$
$$+\,Cov\left(|1+Q_{1,N}|^{\delta}\eta_{2,\boldsymbol{i}},\eta_{\boldsymbol{j}}\right)$$
$$=: I+II+III, \tag{B.100}$$

where, by Lemma 17,

$$I \le C_\alpha\|\eta_{\boldsymbol{i}}\|_{\gamma_\eta}\left\|\eta_{1,\boldsymbol{i}}|1+Q_{1,N}|^{\delta}\mathbf{1}(|Q_{1,N}|\le L_N)\right\|_{\gamma_\eta}\left\{M_\alpha(1,|V_{1,N}|+1)\alpha\left(d(\{\boldsymbol{j}\},\{\boldsymbol{i}\}\bigcup V_{1,N})\right)\right\}^{1-2/\gamma_\eta}$$
$$\le C_\alpha C_{\gamma_\eta}L_N^{\delta}M(\boldsymbol{i},\boldsymbol{j})\left\{M_\alpha(1,|V_{1,N}|+1)\alpha\left(d(\{\boldsymbol{j}\},\{\boldsymbol{i}\}\bigcup V_{1,N})\right)\right\}^{1-2/\gamma_\eta},$$

where $C_{\gamma_\eta} := \|\eta_{\boldsymbol{i}}\|_{\gamma_\eta}$ and $C_\alpha$ is the generic constant defined by Lemma 17. By the same argument, one can prove that

$$II \le C_\alpha C_{\gamma_\eta}M(\boldsymbol{i},\boldsymbol{j})\left(\frac{E\left[|1+Q_{1,N}|^{\delta\gamma_\eta}Q_{1,N}^2\right]}{L_N^2}\right)^{1/\gamma_\eta}\left\{M_\alpha(1,|V_{1,N}|+1)\alpha\left(d(\{\boldsymbol{j}\},\{\boldsymbol{i}\}\bigcup V_{1,N})\right)\right\}^{1-2/\gamma_\eta}$$
$$\le C_\alpha C_{\gamma_\eta}C_*^{1/\gamma_\eta}M(\boldsymbol{i},\boldsymbol{j})L_N^{-2/\gamma_\eta}|V_{1,N}|^{\frac{\delta}{2}+\frac{1}{\gamma_\eta}}\left\{M_\alpha(1,|V_{1,N}|+1)\alpha\left(d(\{\boldsymbol{j}\},\{\boldsymbol{i}\}\bigcup V_{1,N})\right)\right\}^{1-2/\gamma_\eta}.$$

An application of Hölder's inequality, one has

$$III \le \left(E[|1+Q_{1,N}|^{2+\delta}]\right)^{\frac{\delta}{2+\delta}} \left(E|\eta_{2,\boldsymbol{i}}|^q\right)^{1/q} \|\eta_{\boldsymbol{i}}\|_p \le C_p^{1+p/q} C_*^{\frac{\delta}{2+\delta}} \frac{|V_{1,N}|^{\frac{\delta}{2}}}{M(\boldsymbol{i},\boldsymbol{j})^{p/q-1}},$$

where $C_p \coloneqq \|\eta_{\boldsymbol{i}}\|_p$. Therefore, in view of (B.100),

$$\begin{aligned} E\left[|1+Q_{1,N}|^\delta \eta_{\boldsymbol{i}}\eta_{\boldsymbol{j}}\right] \le C_\alpha C_{\gamma_\eta} M(\boldsymbol{i},\boldsymbol{j}) &\left\{L_N^\delta + C_*^{1/\gamma_\eta} L_N^{-2/\gamma_\eta} |V_{1,N}|^{\frac{\delta}{2}+\frac{1}{\gamma_\eta}}\right\} \\ &\times \left\{M_\alpha(1,|V_{1,N}|+1)\alpha\left(d(\{\boldsymbol{j}\},\{\boldsymbol{i}\}\bigcup V_{1,N})\right)\right\}^{1-2/\gamma_\eta} \\ &+ C_p^{1+p/q} C_*^{\frac{\delta}{2+\delta}} \frac{|V_{1,N}|^{\frac{\delta}{2}}}{M(\boldsymbol{i},\boldsymbol{j})^{p/q-1}}. \end{aligned}$$

It then follows that

$$\begin{aligned} \mathcal{A}_{1,N} \le C_\alpha C_\theta^{1-2/\gamma_\eta} C_{\gamma_\eta} &\left\{L_N^\delta + C_*^{1/\gamma_\eta} L_N^{-2/\gamma_\eta} |V_{1,N}|^{\frac{\delta}{2}+\frac{1}{\gamma_\eta}}\right\} \left\{M_\alpha(1,|V_N|+1)\right\}^{1-2/\gamma_\theta} \\ &\times \sum_{\boldsymbol{j}\in V_{2,N}^{(1)},\boldsymbol{j}\neq\boldsymbol{i}} d(\{\boldsymbol{j}\},\{\boldsymbol{i}\}\bigcup V_{1,N})^{\theta_\eta-\theta_\alpha(1-2/\gamma_\eta)} \\ &+ C_p^{1+p/q} C_*^{\frac{\delta}{2+\delta}} |V_{1,N}|^{\frac{\delta}{2}} \sum_{\boldsymbol{j}\in V_{2,N}^{(1)},\boldsymbol{j}\neq\boldsymbol{i}} d(\{\boldsymbol{j}\},\{\boldsymbol{i}\}\bigcup V_{1,N})^{-\theta_\eta\frac{p-q}{q}}. \end{aligned}$$

Notice that, by Lemma 16, one can effectively show that

$$\begin{aligned} \sum_{\boldsymbol{j}\in V_{2,N}^{(1)},\boldsymbol{j}\neq\boldsymbol{i}} d(\{\boldsymbol{j}\},\{\boldsymbol{i}\}\bigcup V_{1,N})^{\theta_\eta-\theta_\alpha(1-2/\gamma_\eta)} &\le \sum_{m\ge\xi|V_N|^{1/d_v}} |\{\boldsymbol{j}\in V_{2,N}\backslash\mathcal{U}_N:\ \|\boldsymbol{j}-\boldsymbol{i}\|=m\}|\, m^{\theta_\eta-\theta_\alpha(1-2/\gamma_\eta)} \\ &\le 4d_v 2^{2d_v-3} \sum_{m\ge\xi|V_N|^{1/d_v}} m^{\theta_\eta+d_v-1-\theta_\alpha(1-2/\gamma_\eta)} \\ &\approx \frac{4d_v 2^{2d_v-3}}{\theta_\alpha(1-2/\gamma_\eta)-\theta_\eta-d_v}\left(\xi|V_N|^{\frac{1}{d_v}}\right)^{\theta_\eta+d_v-\theta_\alpha\left(1-\frac{2}{\gamma_\eta}\right)} \end{aligned}$$

and

$$\begin{aligned} \sum_{\boldsymbol{j}\in V_{2,N}^{(1)},\boldsymbol{j}\neq\boldsymbol{i}} d(\{\boldsymbol{j}\},\{\boldsymbol{i}\}\bigcup V_{1,N})^{-\theta_\eta\frac{p-q}{q}} &\le 4d_v 2^{2d_v-3} \sum_{m\ge\xi|V_N|^{1/d_v}} m^{d_v-1-\theta_\eta\frac{p-q}{q}} \\ &\approx \frac{4d_v 2^{2d_v-3}}{\theta_\eta\frac{p-q}{q}-d_v}\left(\xi|V_N|^{1/d_v}\right)^{d_v-\theta_\eta\frac{p-q}{q}}. \end{aligned}$$

It then follows that

$$\mathcal{A}_{1,N} \quad \leq \quad 4d_v 2^{2d_v-3}\left\{\frac{\xi^{\theta_\eta+d_v-\theta_\alpha\left(1-\frac{2}{\gamma_\eta}\right)}C_\alpha C_\theta^{1-2/\gamma_\eta}C_{\gamma_\eta}(C_*^{1/\gamma_\eta}+1)}{\theta_\alpha\left(1-\frac{2}{\gamma_\eta}\right)-\theta_\eta-d_v}+\frac{\xi^{d_v-\theta_\eta\frac{p-q}{q}}C_p^{1+p/q}C_*^{\frac{\delta}{2+\delta}}}{\theta_\eta\frac{p-q}{q}-d_v}\right\}|V_N|^{\delta/2}. \tag{B.101}$$

Next, to derive the upper bound for $\mathcal{A}_{2,N}$, note that, for every pair, $\boldsymbol{i},\boldsymbol{j}\in V_{2,N}^{(2)}$, an application of the triangle inequality yields $\left|E\left[|1+Q_{1,N}|^\delta\eta_{\boldsymbol{i}}\eta_{\boldsymbol{j}}\right]\right|\leq\left|E\left[|1+Q_{1,N}|^\delta\eta_{\boldsymbol{i}}\eta_{\boldsymbol{j}}\right]-E\left[|1+Q_{1,N}|^\delta\right]E\left[\eta_{\boldsymbol{i}}\eta_{\boldsymbol{j}}\right]\right|+\left|E\left[|1+Q_{1,N}|^\delta\right]E\left[\eta_{\boldsymbol{i}}\eta_{\boldsymbol{j}}\right]\right|=:\mathcal{A}_{2,a,N}+\mathcal{A}_{2,b,N}$. First, by Lemma 17 and Hölder's inequality, one can show that

$$\begin{aligned}\mathcal{A}_{2,a,N}&\leq C_\alpha\left(E[|1+Q_{1,N}|^{2+\delta}]\right)^{\frac{\delta}{2+\delta}}\|\eta_{\boldsymbol{i}}\eta_{\boldsymbol{j}}\|_{\frac{p}{2}}\left\{M_\alpha(2,|V_{1,N}|)\alpha(d(\{\boldsymbol{i},\boldsymbol{j}\},V_{1,N}))\right\}^{1-\frac{\delta}{2+\delta}-\frac{1}{p}}\\&\leq C_\alpha C_p^2C_*^{\frac{\delta}{\delta+2}}\xi^{-\theta_\alpha\left(1-\frac{\delta}{\delta+2}-\frac{2}{p}\right)}|V_N|^{\frac{\delta}{2}+(\gamma_M-\theta_\alpha)\left(1-\frac{\delta}{\delta+2}-\frac{2}{p}\right)}\end{aligned}$$

and

$$\mathcal{A}_{2,b,N}\leq C_\alpha C_{\gamma_\eta}^2C_*^{\frac{\delta}{2+\delta}}|V_N|^{\frac{\delta}{2}}\alpha(\|\boldsymbol{i}-\boldsymbol{j}\|)^{1-\frac{2}{\gamma_\eta}}.$$

Therefore, one has that

$$\begin{aligned}\mathcal{A}_{2,N}&\leq C_\alpha C_p^2C_*^{\frac{\delta}{\delta+2}}\xi^{-\theta_\alpha\left(1-\frac{\delta}{\delta+2}-\frac{2}{p}\right)}|V_N|^{\frac{\delta}{2}+(\gamma_M-\theta_\alpha)\left(1-\frac{\delta}{\delta+2}-\frac{2}{p}\right)}\left|\{\boldsymbol{j}\in V_{2,N}^{(2)},\ \boldsymbol{j}\neq\boldsymbol{i}\}\right|\\&\quad+C_\alpha C_{\gamma_\eta}^2C_*^{\frac{\delta}{2+\delta}}|V_N|^{\frac{\delta}{2}}\sum_{\boldsymbol{j}\in V_{2,N}^{(2)},\boldsymbol{j}\neq\boldsymbol{i}}\|\boldsymbol{i}-\boldsymbol{j}\|^{-\theta_\alpha\left(1-\frac{2}{\gamma_\eta}\right)}.\end{aligned}$$

Moreover, an application of Lemma 17 and an elementary inequality (i.e., $\sum_{k=1}^N k^p\approx\frac{N^{p+1}}{p+1},\ p>-1$ as $N\uparrow\infty$) yields

$$\begin{aligned}\left|\{\boldsymbol{j}\in V_{2,N}^{(2)},\ \boldsymbol{j}\neq\boldsymbol{i}\}\right|&=\left|\{\boldsymbol{j}\in V_{2,N}\backslash\mathcal{U}_N:\ \|\boldsymbol{j}-\boldsymbol{i}\|<\xi|V_N|^{1/d_v}\}\right|\\&\leq\left|\{\boldsymbol{j}\in\mathbb{Z}^{d_v}:\ \|\boldsymbol{j}-\boldsymbol{i}\|<\xi|V_N|^{1/d_v}\}\right|\\&\leq 2^{2(d_v-1)}\xi^{d_v}|V_N|\end{aligned}$$

and

$$\begin{aligned}\sum_{\boldsymbol{j}\in V_{2,N}^{(2)},\boldsymbol{j}\neq\boldsymbol{i}}\|\boldsymbol{i}-\boldsymbol{j}\|^{-\theta_\alpha\left(1-\frac{2}{\gamma_\eta}\right)}&=\sum_{r=1}^{\xi|V_N|^{\frac{1}{d_v}}}\sum_{\boldsymbol{j}\in V_{2,N}^{(2)},\ \|\boldsymbol{j}-\boldsymbol{i}\|=r}r^{-\theta_\alpha\left(1-\frac{2}{\gamma_\eta}\right)}\\&=\sum_{r=1}^{\xi|V_N|^{1/d_v}}\left|\{\boldsymbol{j}\in V_{2,N}^{(2)}:\ \|\boldsymbol{j}-\boldsymbol{i}\|=r\}\right|r^{-\theta_\alpha\left(1-\frac{2}{\gamma_\eta}\right)}\end{aligned}$$

$$
\begin{aligned}
&\leq 2d_v 2^{2d_v-3} \sum_{r=1}^{\xi|V_N|^{\frac{1}{d_v}}} r^{d_v-1-\theta_\alpha\left(1-\frac{2}{\gamma_\eta}\right)} \\
&\leq 2d_v 2^{2d_v-3} \max\left(1, (\xi|V_N|)^{d_v-\theta_\alpha\left(1-\frac{2}{\gamma_\eta}\right)}\right).
\end{aligned}
$$

It then follows that

$$
\mathcal{A}_{2,N} \leq C_\alpha C_*^{\frac{\delta}{\delta+2}} \left\{2^{2(d_v-1)} C_p^2 \xi^{d_v-\theta_\alpha\left(1-\frac{\delta}{\delta+2}-\frac{2}{p}\right)} + 2d_v 2^{2d_v-3} C_{\gamma_\eta}^2\right\} |V_N|^{\frac{\delta}{2}}. \tag{B.102}
$$

Finally, an application of Hölder's inequality yields

$$
\begin{aligned}
E[|1+Q_{1,N}|^\delta \eta_{\boldsymbol{i}}^2] &\leq \left(E|1+Q_{1,N}|^{2+\delta}\right)^{\frac{\delta}{\delta+2}} \|\eta_{\boldsymbol{i}}\|_{2+\delta}^2 \\
&\leq C_*^{\frac{\delta}{\delta+2}} C_{\delta+2}^2 |V_N|^{\frac{\delta}{2}},
\end{aligned}
$$

where $C_{\delta+2} := \|\eta_{\boldsymbol{i}}\|_{\delta+2}$. Therefore, one has that

$$
\mathcal{B}_N \leq C_*^{\frac{\delta}{\delta+2}} C_{\delta+2}^2 |V_N|^{\frac{1+\delta}{2}}. \tag{B.103}
$$

Collecting all the results derived in (B.98)-(B.103), we have that

$$
\begin{aligned}
E\left[|1+Q_{1,N}|^\delta Q_{2,N}^2\right] \leq 2c_\delta C_u |V_N| + 2 \Bigg\{ & c_\delta \xi C_* + 4d_v 2^{2d_v-3} \Bigg\{ \frac{\xi^{\theta_\eta+d_v-\theta_\alpha\left(1-\frac{2}{\gamma_\eta}\right)} C_\alpha C_\theta^{1-2/\gamma_\eta} C_{\gamma_\eta}(C_*^{1/\gamma_\eta}+1)}{\theta_\alpha\left(1-\frac{2}{\gamma_\eta}\right)-\theta_\eta-d_v} \\
& + \frac{\xi^{d_v-\theta_\eta\frac{p-q}{q}} C_p^{1+p/q} C_*^{\frac{\delta}{2+\delta}}}{\theta_\eta\frac{p-q}{q}-d_v} \Bigg\} + C_\alpha C_*^{\frac{\delta}{\delta+2}} \left\{2^{2(d_v-1)} C_p^2 \xi^{d_v-\theta_\alpha\left(1-\frac{\delta}{\delta+2}-\frac{2}{p}\right)} + 2d_v 2^{2d_v-3} C_{\gamma_\eta}^2\right\} \\
& + C_*^{\frac{\delta}{\delta+2}} C_{\delta+2}^2 \Bigg\} |V_N|^{\frac{1+\delta}{2}}.
\end{aligned}
$$

In view of (B.96), some algebraic manipulations yield

$$
\begin{aligned}
E\left[S^2(V_N)\left(1+S(V_N)\right)^\delta\right] \leq & \left((\tau_0 + 4c_\delta \xi A_\delta) C_* + B_1 C_*^{\frac{\delta}{\delta+2}} + C_1 C_*^{\frac{1}{\gamma_\eta}} + D_1\right) |V_N|^{\frac{1+\delta}{2}} \\
& + 4c_\delta C_u A_\delta |V_N|,
\end{aligned} \tag{B.104}
$$

where

$$
B_1 := 4A_\delta \left\{4d_v 2^{2d_v-3} \frac{\xi^{d_v} - \theta_\eta \frac{p-q}{q} C_p^{1+\frac{p}{q}}}{\theta_\eta \frac{p-q}{q} - d_v} + C_\alpha \left\{2^{2(d_v-1)} C_p^2 \xi^{d_v-\theta_\alpha\left(1-\frac{\delta}{\delta+2}-\frac{2}{p}\right)} + 2d_v 2^{2d_v-3} C_{\gamma_\eta}^2\right\} + C_{\delta+2}^2\right\}
$$

and $C_1 = D_1 := 16A_\delta d_v 2^{2d_v-3}\frac{\xi^{\theta_\eta+d_v-\theta_\alpha\left(1-\frac{2}{\gamma_\eta}\right)}C_\alpha C_\theta^{1-\frac{2}{\gamma_\eta}}C_{\gamma_\eta}}{\theta_\alpha\left(1-\frac{2}{\gamma_\eta}\right)-\theta_\eta-d_v}$. The right-hand side of (B.104) is a root-polynomial function of $C_*$, thus will become less than $C_*$ if $C_*$ is large. The inductive argument has been proved. ■

**Lemma 9 (FCLT for Mixing Spatio-Temporal Data)** *Let $S(V_N, \lfloor T\tau\rfloor) := \sum_{t=1}^{\lfloor T\tau\rfloor}\sum_{\boldsymbol{i}\in V_N}\eta_{\boldsymbol{i},t}$ represent a partial-sum process of mixing centered spatio-temporal random fields, $\{\eta_{\boldsymbol{i},t},\ \boldsymbol{i}\in V_N, t\in[1,T]\}$. Suppose that $\{\eta_{\boldsymbol{i},t},\ \boldsymbol{i}\in V_N, t\in[1,T]\}$ are identically distributed across both space and time. Moreover, let the following conditions holds: (a) $\alpha(\tau)\le C_\theta\tau^{-\theta_\alpha}$ for some $\theta_\alpha \ge \max\left(\frac{p(d_v+1)\gamma_\eta}{(p-q)(\gamma_\eta-2)} + (d_v+1)\gamma_M, \frac{d_v+1}{1-\frac{2}{\gamma_\eta}}, \frac{1}{1-\frac{\delta}{\delta+2}-\frac{2}{p}} - \gamma_M\right)$, where $\gamma_\eta > 2$, $p > \delta+2$, $q = \frac{p(2+\delta)}{2p-2-\delta}$ for some $\delta > 0$, $d_v$ is the dimension of $V_N$, and $\gamma_M$ is given in Definition B.1; (b) $\max_{\boldsymbol{i}}\left(E|\eta_{\boldsymbol{i}}|^p, E|\eta_{\boldsymbol{i}}|^{\gamma_\eta}, E|\eta_{\boldsymbol{i}}|^{\delta+2}\right) < \infty$; (c) $|V_N|^{\gamma_M}T^{\gamma_M+1-\theta_\alpha}\downarrow 0$. Then,*

$$\frac{1}{\sigma\sqrt{T}|V_N|^{1/2}}S(V_N,\lfloor T\tau\rfloor)\xrightarrow{w} W(\tau),$$

*where $\sigma^2 := \lim_{N,T\uparrow\infty}\frac{1}{T|V_N|}E[S^2(V_N,T)] < \infty$ and $W(\tau)$ is the Brownian motion.*

**Proof of Lemma 9.** Let $D[0,1]$ denote the Skorohod space of càdlàg functions on $[0,1]$. (All the properties that we need can be found in Billingsley (1968).) The partial-sum process $S(V_N,\lfloor T\tau\rfloor)$ can be considered as a random function in $D[0,1]$. Therefore, the FCLT for $S(V_N,\lfloor T\tau\rfloor)$ is reminiscent of Billingsley (1968, Theorem20.1). As in Deo (1975) the weak convergence of $S(V_N,\lfloor T\tau\rfloor)$ to the Brownian motion requires the following conditions: Define $\overline{S}(V_N,\lfloor T\tau\rfloor) := \frac{1}{\sigma\sqrt{T}|V_N|^{1/2}}S(V_N,\lfloor T\tau\rfloor)$,

**(i)** $\lim_{N,T\uparrow\infty}E[\overline{S}^2(V_N,\lfloor T\tau\rfloor)] = \tau$ for each $\tau\in(0,1]$,

**(ii)** $\overline{S}^2(V_N,\lfloor T\tau\rfloor)$ is uniformly integrable for each $\tau\in(0,1]$,

**(iii)** $\overline{S}(V_N,\lfloor T\tau\rfloor)$ has asymptotically independent increments,

**(iv)** $\overline{S}(V_N,\lfloor T\tau\rfloor)$ is tight in $D[0,1]$ (see Billingsley (1968, Theorem 19.2)).

*Verification of* **(i)***:* $\frac{1}{T|V_N|}Var\left(S(V_N,\lfloor T\tau\rfloor)\right) = \frac{\lfloor T\tau\rfloor}{T}E[\eta_{\boldsymbol{i},t}^2] + \frac{1}{T|V_N|}\sum_{s=1}^{\lfloor T\tau\rfloor}\sum_{t=1,t\neq s}^{\lfloor T\tau\rfloor}\sum_{\boldsymbol{i},\boldsymbol{j}\in V_N}E[\eta_{\boldsymbol{i},s}\eta_{\boldsymbol{j},t}] + \frac{1}{T|V_N|}\sum_{s=1}^{\lfloor T\tau\rfloor}\sum_{\boldsymbol{i},\boldsymbol{j}\in V_N,\boldsymbol{i}\neq\boldsymbol{j}}E[\eta_{\boldsymbol{i},s}\eta_{\boldsymbol{j},s}] =: \frac{\lfloor T\tau\rfloor}{T}E[\eta_{\boldsymbol{i},t}^2] + \mathcal{A}_{N,T} + \mathcal{B}_{N,T}$. We now proceed to bound $\mathcal{A}_{N,T}$ and $\mathcal{B}_{N,T}$.

$$\begin{aligned}
\mathcal{A}_{N,T} &= \frac{\lfloor T\tau\rfloor}{T}\frac{1}{|V_N|}\sum_{t=1}^{\lfloor T\tau\rfloor}\sum_{\boldsymbol{i},\boldsymbol{j}\in V_N}E[\eta_{\boldsymbol{i},0}\eta_{\boldsymbol{j},t}] \\
&= \frac{\lfloor T\tau\rfloor}{T}\left(\sum_{t=1}^{\lfloor T\tau\rfloor}\frac{1}{|V_N|}\sum_{\boldsymbol{i}\in V_N}E[\eta_{\boldsymbol{i},0}\eta_{\boldsymbol{i},t}] + \sum_{t=1}^{\lfloor T\tau\rfloor}\frac{1}{|V_N|}\sum_{\boldsymbol{i},\boldsymbol{j}\in V_N,\boldsymbol{i}\neq\boldsymbol{j}}E[\eta_{\boldsymbol{i},0}\eta_{\boldsymbol{j},t}]\right)
\end{aligned}$$

$$=: \frac{\lfloor T\tau \rfloor}{T}(\mathcal{A}_{1,N,T} + \mathcal{A}_{2,N,T}), \tag{B.105}$$

where $\mathcal{A}_{1,N,T} \leq C_0 \sum_{\tau=1}^{\infty} \|\eta_{\boldsymbol{i},0}\|_{\gamma_\eta} \|\eta_{\boldsymbol{i},t}\|_{\gamma_\eta} \alpha(\tau)^{1-2/\gamma_\eta} < \infty$ by Lemma 17 and Conditions (a) and (b); and, by Lemma 16,

$$\begin{aligned}
\mathcal{A}_{2,N,T} &= \sum_{m=1}^{\operatorname{diam}(V_N \times [1,\lfloor T\tau \rfloor])} \frac{1}{|V_N|} \sum_{\boldsymbol{i} \in V_N} \sum_{\substack{t \in [1,\lfloor T\tau \rfloor] \\ \boldsymbol{j} \in V_N \\ d(\{\boldsymbol{i},0\},\{\boldsymbol{j},t\})=m}} E[\eta_{\boldsymbol{i},0}\eta_{\boldsymbol{j},t}] \\
&\leq C_0 \sum_{m=1}^{\operatorname{diam}(V_N \times [1,\lfloor T\tau \rfloor])} \frac{1}{|V_N|} \sum_{\boldsymbol{i} \in V_N} |\{\{\boldsymbol{j},t\} \in V_N \times [1,\lfloor T\tau \rfloor] : \; d(\{\boldsymbol{i},0\},\{\boldsymbol{j},t\}) = m\}| \\
&\quad \times \|\eta_{\boldsymbol{i},0}\|_{\gamma_\eta} \|\eta_{\boldsymbol{j},t}\|_{\gamma_\eta} \alpha(m)^{1-2/\gamma_\eta} \\
&\leq C_0 \sum_{m=1}^{\infty} m^{d_v} \alpha(m)^{1-2/\gamma_\eta} \\
&< \infty.
\end{aligned}$$

By using the same argument, one can also verify that

$$\mathcal{B}_{N,T} = \frac{\lfloor T\tau \rfloor}{T} \frac{1}{|V_N|} \sum_{\substack{\boldsymbol{i},\boldsymbol{j} \in V_N \\ \boldsymbol{i} \neq \boldsymbol{j}}} E[\eta_{\boldsymbol{i},s}\eta_{\boldsymbol{j},s}] =: \frac{\lfloor T\tau \rfloor}{T} \mathcal{B}_{1,N,T}, \tag{B.106}$$

where $\mathcal{B}_{1,N,T} \leq C_0 \sum_{m=1}^{\infty} m^{d_v - 1} \alpha(m)^{1-2/\gamma_\eta} < \infty$.

Notice that $\frac{\lfloor T\tau \rfloor}{T} \to \tau$ and $\sigma^2 := E[\eta_{\boldsymbol{i},t}^2] + \lim_{N,T\uparrow\infty}(\mathcal{A}_{1,N,T} + \mathcal{A}_{2,N,T} + \mathcal{B}_{1,N,T}) < \infty$, Condition (i) has been verified.

*Verification of* **(ii)***:* Is is sufficient to show that $\overline{S}^2(V_N, T)$ is uniformly integrable. An application of the Tchebyshev inequality and Lemma 8 yields that, for some $\delta > 0$,

$$\begin{aligned}
E\left[\overline{S}^2(V_N,T)\mathbf{1}(|\overline{S}(V_N,T)| \geq C)\right] &= \frac{1}{\sigma^2 T|V_N|} E\left[S^2(V_N,T)\mathbf{1}\left(|S(V_N,T)| \geq \sigma\sqrt{T|V_N|}C\right)\right] \\
&\leq \frac{1}{\sigma^{2+\delta} C^\delta (T|V_N|)^{1+\delta/2}} E\left[|S(V_N,T)|^{2+\delta}\right] \\
&\leq \frac{C_*}{\sigma^{2+\delta} C^\delta} \to 0 \text{ as } C \to \infty.
\end{aligned}$$

*Verification of* **(iii)***:* Let $0 = s_1 \leq t_1 < s_2 \leq t_2 < \cdots < s_m \leq t_m = 1$ denote a partition of the unit interval $[0,1]$. For all Borel sets, $H_1, \ldots, H_m$, of $\mathbb{R}$, one needs to show that

$$\lim_{N,T\uparrow\infty}\left|P\left(\overline{S}(V_N,\lfloor Tt_i\rfloor)-\overline{S}(V_N,\lfloor Ts_i\rfloor)\in H_i,\ i=1,\dots m\right)-\prod_{i=1}^m P\left(\overline{S}(V_N,\lfloor Tt_i\rfloor)-\overline{S}(V_N,\lfloor Ts_i\rfloor)\in H_i\right)\right|=0. \tag{B.107}$$

Note that, as the event $\{\overline{S}(V_N,\lfloor Tt_i\rfloor)-\overline{S}(V_N,\lfloor Ts_i\rfloor)\in H_i\}$ belongs to the $\sigma$-algebra $\mathcal{B}_i$ generated by the sequence $\{\eta_{\boldsymbol{s}}:\ \boldsymbol{s}\in V_N\times[\lfloor Ts_i\rfloor+1,\lfloor Tt_i\rfloor]\}$, the random element $\xi:=\mathbf{1}\big(\overline{S}(V_N,\lfloor Tt_i\rfloor)-\overline{S}(V_N,\lfloor Ts_i\rfloor)\in H_i\big)$ is $\mathcal{B}_i$-measurable. By Lemma 19, one obtains that

$$\left|E\left[\prod_{s=1}^m\xi_s\right]-\prod_{s=1}^m E[\xi_s]\right|\le\sum_{i=1}^{m-1}\sum_{j=i+1}^N\left|Cov\left(\xi_i-1,(\xi_j-1)\prod_{s=j+1}^m\xi_s\right)\right|. \tag{B.108}$$

Since $(\xi_j-1)\prod_{s=j+1}^m\xi_s$ is $\bigcup_{i=j}^m\mathcal{B}_i$-measurable and $d\big(V_N\times[\lfloor Ts_i\rfloor+1,\lfloor Tt_i\rfloor],V_N\times\bigcup_{\ell=j}^m[\lfloor Ts_\ell\rfloor+1,\lfloor Tt_\ell\rfloor]\big)=\lfloor T(s_j-t_i)\rfloor>\lfloor Tb\rfloor>0$ for every $j>i$, where $b$ is some positive number, by Lemma 17, one obtains that

$$\begin{aligned}Cov\left(\xi_i-1,(\xi_j-1)\prod_{s=j+1}^m\xi_s\right)&\le C_0M_\alpha\left(|V_N\times[\lfloor Ts_i\rfloor+1,\lfloor Tt_i\rfloor]|,\left|V_N\times\bigcup_{\ell=j}^m[\lfloor Ts_\ell\rfloor+1,\lfloor Tt_\ell\rfloor]\right|\right)\\&\quad\times\alpha(\lfloor Tb\rfloor)\\&\le C_0\,(T|V_N|)^{\gamma_M}\,\alpha(\lfloor Tb\rfloor)\\&\approx(T|V_N|)^{\gamma_M}\,(\lfloor Tb\rfloor)^{-\theta_\alpha}\downarrow 0\text{ by Condition }(c).\end{aligned}$$

Therefore, in view of (B.108), (B.107) has been proved.

*Verification of* **(iv)***:* In view of Billingsley (1968, Theorem 8.4) (adapted to $D[0,1]$), the tightness condition will follow if one can prove that, for each positive $\epsilon$, there exist a positive $\lambda$ and integers, $N_0$ and $T_0$, such that $N\ge N_0$ and $T\ge T_0$ together imply

$$P\left(\max_{1\le t\le T}|S(V_N,t)|\ge\sigma\lambda\sqrt{T|V_N|}\right)\le\frac{\epsilon}{\lambda^2\sigma^2}. \tag{B.109}$$

First, introduce the events $E_1:=\left\{|S(V_N,1)|\ge\sigma\lambda\sqrt{T|V_N|}\right\}$ and $E_j:=\left\{\max_{1\le i<j}|S(V_N,i)|<\sigma\lambda\sqrt{T|V_N|}\right.$ $\left.<|S(V_N,j)|\right\}$ for every $j>1$. It then follows that

$$\begin{aligned}P\left(\max_{1\le t\le T}|S(V_N,t)|\ge\sigma\lambda\sqrt{T|V_N|}\right)&\le P\left(|S(V_N,T)|\ge(\lambda-\lambda_1)\sigma\sqrt{T|V_N|}\right)\\&\quad+\sum_{j=1}^{T-1}P\left(E_j\bigcap\left\{|S(V_N,T)-S(V_N,j)|\ge\sigma\lambda\sqrt{T|V_N|}\right\}\right)\end{aligned} \tag{B.110}$$

for any $\lambda_1 < \lambda$. Note that

$$P\left(E_j \bigcap \left\{|S(V_N,T) - S(V_N,j)| \geq \sigma\lambda\sqrt{T|V_N|}\right\}\right) \leq P\left(E_j \bigcap \{|S(V_N,T) - S(V_N,j+k)| \geq \sigma(\lambda_1 - \lambda_2) \times \sqrt{T|V_N|}\}\right) + P\left(|S(V_N,j+k) - S(V_N,j)| \geq \sigma\lambda_2\sqrt{T|V_N|}\right) =: \mathcal{A}_{N,T} + \mathcal{B}_{N,T},$$

where $\lambda_2 \in (0, \lambda_1)$ and $k$ takes some value less than $T$. To bound the right-hand side of (B.110), one first needs to bound $\mathcal{A}_{N,T}$ and $\mathcal{B}_{N,T}$. Let $\mathcal{B}_i^j$ be the $\sigma$-algebra generated by $\{\eta_{\boldsymbol{s}} : \boldsymbol{s} \in V_N \times [i,j]\}$. Thus the Bernoulli random variables $\mathbf{1}(E_j)$ is $\mathcal{B}_1^j$-measurable and $\mathbf{1}\big(|S(V_N,T) - S(V_N,j+k)| \geq \sigma(\lambda_1 - \lambda_2)\sqrt{T|V_N|}\big)$ is $\mathcal{B}_{j+k+1}^T$-measurable. Invoking Lemma 17, one can obtain that

$$\begin{aligned}
&\Big|P\left(E_j \bigcap |S(V_N,T) - S(V_N,j+k)| \geq \sigma(\lambda_1 - \lambda_2)\sqrt{T|V_N|}\right) \\
&\quad - P(E_j)P\left(|S(V_N,T) - S(V_N,j+k)| \geq \sigma(\lambda_1 - \lambda_2)\sqrt{T|V_N|}\right)\Big| \\
&\leq C_0 M_\alpha\left(|V_N|j, |V_N|(T-j-k)\right)\alpha\left(d\left(V_N \times [1,j], V_N \times [j+k+1,T]\right)\right) \\
&\leq C_0(T|V_N|)^{\gamma_M}\alpha(k).
\end{aligned}$$

In addition,

$$\begin{aligned}
P\left(|S(V_N,T) - S(V_N,j+k)| \geq \sigma(\lambda_1 - \lambda_2)\sqrt{T|V_N|}\right) &\overset{(a)}{\leq} \frac{E[|S(V_N,T) - S(V_N,j+k)|^2]}{\sigma^2(\lambda_1 - \lambda_2)^2 T|V_N|} \\
&\leq \frac{1}{\sigma^2(\lambda_1 - \lambda_2)^2 T|V_N|}\left(\sum_{\boldsymbol{s} \in V_N \times [1,j+k]} E[|\eta_{\boldsymbol{s}}|^2]\right. \\
&\quad \left. + \sum_{\boldsymbol{s},\boldsymbol{w} \in V_N \times [1,j+k],\ \boldsymbol{s} \neq \boldsymbol{w}} E[\eta_{\boldsymbol{s}}\eta_{\boldsymbol{w}}]\right) \\
&\overset{(b)}{\leq} |V_N|T\left(\|\eta_{\boldsymbol{s}}\|^2 + 2C_0 d_v 3^{\gamma_M(1-2/\gamma_\eta)}\|\eta_{\boldsymbol{s}}\|_{\gamma_\eta}\right. \\
&\quad \left. \times \sum_{r=1}^{\infty}(2r+1)^{d_v}\alpha(r)^{1-2/\gamma_\eta}\right) \\
&= |V_N|T\Theta_\eta,
\end{aligned}$$

where Conditions (a) and (b) ensure that $\Theta_\eta < \infty$; (a) follows from the Tchebyshev inequality; and (b) follows from Lemma 17. It then follows that

$$\mathcal{A}_{N,T} \leq P(E_j)\frac{\Theta_\eta}{\sigma^2(\lambda_1-\lambda_2)^2} + C_0(T|V_N|)^{\gamma_M}\alpha(k). \tag{B.111}$$

By the Tchebyshev inequality and Lemma 8, one also has that

$$\begin{aligned}
\mathcal{B}_{N,T} &\leq \sum_{s=j+1}^{j+k} P\left(\left|\sum_{\boldsymbol{i}\in V_N}\eta_{\boldsymbol{i},s}\right| \geq \frac{\sigma\lambda_2\sqrt{|V_N|T}}{k}\right) \\
&\leq \frac{k^{3+\delta}}{\sigma^{2+\delta}\lambda_2^{2+\delta}(|V_N|T)^{1+\delta/2}} E\left[\left|\sum_{\boldsymbol{i}\in V_N}\eta_{\boldsymbol{i},s}\right|^{2+\delta}\right] \\
&\leq C_*\frac{k^{3+\delta}}{\sigma^{2+\delta}\lambda_2^{2+\delta}T^{1+\frac{\delta}{2}}} \qquad \text{(B.112)}
\end{aligned}$$

for some $\delta > 0$. In view of (B.110), (B.111), and (B.112), we have that

$$\begin{aligned}
P\left(\max_{1\leq t\leq T}|S(V_N,t)| \geq \sigma\lambda\sqrt{T|V_N|}\right) &\leq P\left(|S(V_N,T)| \geq (\lambda-\lambda_1)\sigma\sqrt{T|V_N|}\right) \\
&\quad + \frac{\Theta_\eta}{\sigma^2(\lambda_1-\lambda_2)^2}\sum_{j=1}^{T-1}P(E_j) + C_0T^{\gamma_M+1}|V_N|^{\gamma_M}\alpha(k) \\
&\quad + C_*\frac{k^{3+\delta}}{\sigma^{2+\delta}\lambda_2^{2+\delta}T^{\frac{\delta}{2}}}.
\end{aligned}$$

Because the events $E_j$, $j = 1, \ldots, T-1$ are disjoint and $\bigcup_{j=1}^{T-1} E_j \subset \{\max_{1\leq j\leq T}|S(V_N,t)| \geq \sigma\lambda\sqrt{T|V_N|}\}$, one can immediately show that

$$\begin{aligned}
P\left(\max_{1\leq t\leq T}|S(V_N,t)| \geq \sigma\lambda\sqrt{T|V_N|}\right) &\leq \left(1-\frac{\Theta_\eta}{\sigma^2(\lambda_1-\lambda_2)^2}\right)^{-1}\left(P\left(|S(V_N,T)| \geq (\lambda-\lambda_1)\sigma\sqrt{T|V_N|}\right)\right. \\
&\quad \left. + C_0T^{\gamma_M+1}|V_N|^{\gamma_M}\alpha(k) + C_*\frac{k^{3+\delta}}{\sigma^{2+\delta}\lambda_2^{2+\delta}T^{\frac{\delta}{2}}}\right). \qquad \text{(B.113)}
\end{aligned}$$

Now, let $\lambda_1 = \lambda/2$. For a given $\epsilon > 0$, one can choose $\lambda$ sufficiently large so that

$$P\left(|S(V_N,T)| \geq \frac{1}{2}\lambda\sigma\sqrt{T|V_N|}\right) \leq \frac{\epsilon}{9\lambda^2\sigma^2},$$

which is possible because of the uniform integrability condition (ii). One can also choose $\lambda_2 < \lambda_1$ so that $\frac{\Theta_\eta}{\sigma^2(\lambda_1-\lambda_2)^2} < \frac{2}{3}$. Therefore, (B.113) leads to

$$P\left(\max_{1\le t\le T}|S(V_N,t)| \ge \sigma\lambda\sqrt{T|V_N|}\right) \le 3\left(\frac{\epsilon}{9\sigma^2\lambda^2} + C_0|V_N|^{\gamma_M}T^{\gamma_M+1-\theta_\alpha} + C_*\frac{k^{3+\delta}}{\sigma^{2+\delta}\lambda_2^{2+\delta}T^{\frac{\delta}{2}}}\right).$$

If one chooses $k < T$ such that $k^{3+\delta}/T^{\delta/2}$ is arbitrarily small, Condition (c) then implies that

$$P\left(\max_{1\le t\le T}|S(V_N,t)| \ge \sigma\lambda\sqrt{T|V_N|}\right) \le \frac{\epsilon}{\sigma^2\lambda^2}.$$

The tightness condition was verified. ■

**Lemma 10** *Let $S(U_N,V_N,T) := \sum_{t=1}^T\sum_{\boldsymbol{i}\in U_N}\sum_{\boldsymbol{j}\in V_N} w_{\boldsymbol{i},t}\epsilon_{\boldsymbol{j},t}$, where $\{w_{\boldsymbol{i},t},\ \boldsymbol{i}\in U_N\}$ and $\{\epsilon_{\boldsymbol{j},t},\ \boldsymbol{j}\in V_N\}$ are contemporaneously independent centered spatio-temporal processes; and for given $\boldsymbol{i}\in U_N$ and $\boldsymbol{j}\in V_N$, $w_{\boldsymbol{i},t}$ is a causal process and $\{\epsilon_{\boldsymbol{j},t},\ t=1,\dots,T\}$ are serially independent. In addition, suppose that (a) the processes are identically distributed across both space and time, (b) both $w_{\boldsymbol{i},t}$ and $\epsilon_{\boldsymbol{i},t}$ are mixing with the mixing coefficient satisfying $\alpha(\tau) \le C_\theta\tau^{-\theta_\alpha}$ for some $\theta_\alpha \ge \max\left(\frac{pd_v\gamma_\eta}{(p-q)(\gamma_\eta-2)} + d_v\gamma_M, \frac{d_v}{1-\frac{2}{\gamma_\eta}}, \frac{2p}{p-4}-\gamma_M\right)$, where $\gamma_\eta > 2$, $p > 4$, $q = \frac{4p}{2p-4}$, $d_v$ is the dimension of $V_N$, and $\gamma_M$ is given in Definition B.1, (c) $\max\Big((|U_N|+|V_N|)^{\gamma_M(1-2/\gamma_\eta)}T^{\epsilon-1/2}, T^{(\gamma_M+\theta_\alpha-1)\epsilon-\frac{1}{2}(\theta_\alpha-\gamma_M-1)}$ $\max(|U_N|,|V_N|)^{\gamma_M}\Big) \downarrow 0$ for some $\epsilon \in \left(0, \min\left(\frac{1}{2}, \frac{\theta_\alpha-\gamma_M-1}{2(\gamma_M+\theta_\alpha-1)}\right)\right)$. Then,*

$$\frac{1}{\sqrt{T|U_N||V_N|}}S(U_N,V_N,T) \xrightarrow{d} N(0,\sigma^2),$$

*where $\sigma^2 := \lim_{N\uparrow\infty}\frac{1}{|U_N||V_N|}E\left|\sum_{\boldsymbol{i}\in U_N}w_{\boldsymbol{i},t}\right|^2 E\left|\sum_{\boldsymbol{i}\in V_N}\epsilon_{\boldsymbol{i},t}\right|^2$.*

**Proof of Lemma 10.** Let $S^*(U_N,V_N,T) := \sum_{t=1}^T w_{*,t}\epsilon_{*,t}$, where $w_{*,t} := \frac{1}{|U_N|}\sum_{\boldsymbol{i}\in U_N}w_{\boldsymbol{i},t}$ and $\epsilon_{*,t} := \frac{1}{|V_N|}\sum_{\boldsymbol{i}\in V_N}\epsilon_{\boldsymbol{i},t}$.

*Step I:* Divide the time-period index set $[1,T]$ into $k_T$ big blocks, $\{\eta^{(b)}_{N,T,i},\ i=0,\dots,k_T-1\}$, of size $p_T = \lfloor T^{1/2+\epsilon}\rfloor$ and $k_T+1$ small blocks, $\{\eta^{(s)}_{N,T,i},\ i=0,\dots,k_T\}$, of size $q_T = \lfloor T^{1/2-\epsilon}\rfloor$ for some small $0 < \epsilon < \frac{\theta_\alpha-\gamma_M-1}{2(\gamma_M+\theta_\alpha-1)}$. Put

$$\eta^{(b)}_{N,T,i} := \sum_{j=1}^{p_T} w_{*,i(p_T+q_T)+j}\epsilon_{*,i(p_T+q_T)+j},\ i=0,\dots,k_T-1,$$
$$\eta^{(s)}_{N,T,i} := \sum_{j=1}^{p_T} w_{*,i(p_T+q_T)+p_T+j}\epsilon_{*,i(p_T+q_T)+p_T+j},\ i=0,\dots,k_T-1,$$

and the small remaining block $\eta^{(s)}_{N,T,k_T} := \sum_{j=k_T(p_T+q_T)+1}^{T} w_{*,j}\epsilon_{*,j}$. It then follows that

$$S^*(U_N, V_N, T) = \sum_{i=0}^{k_T-1} \eta^{(b)}_{N,T,i} + \sum_{i=0}^{k_T} \eta^{(s)}_{N,T,i} =: \mathcal{B}_{N,T} + \mathcal{S}_{N,T}. \tag{B.114}$$

*Step II:* Derive the asymptotic variance for $S^*(U_N, V_N, T)$: Notice that $E\left|S^*(U_N, V_N, T)\right|^2 = \sum_{t=1}^{T} E\left|w_{*,t}\epsilon_{*,t}\right|^2 + \sum_{t\neq s} E[w_{*,t}\epsilon_{*,t}w_{*,s}\epsilon_{*,s}] =: \mathcal{T}_{1,N,T} + \mathcal{T}_{2,N,T}$, where $\frac{|U_N||V_N|}{T}\mathcal{T}_{1,N,T} = \frac{1}{|U_N|}E\left|\sum_{\boldsymbol{i}\in U_N} w_{\boldsymbol{i},t}\right|^2 \frac{1}{|V_N|}E\left|\sum_{\boldsymbol{i}\in V_N}\epsilon_{\boldsymbol{i},t}\right|^2 < \infty$ as $N\to\infty$ by Lemma 6 and $\mathcal{T}_{2,N,T} = 2\sum_{s<t} E[\epsilon_{*,t}]E[w_{*,t}w_{*,s}\epsilon_{*,s}] = 0$. One can then obtain that

$$\sigma^2 := \lim_{N\uparrow\infty}\left(\frac{1}{|U_N|}E\left|\sum_{\boldsymbol{i}\in U_N} w_{\boldsymbol{i},t}\right|^2\right)\left(\frac{1}{|V_N|}E\left|\sum_{\boldsymbol{i}\in V_N}\epsilon_{\boldsymbol{i},t}\right|^2\right).$$

*Step III:* Let $\mathcal{S}^*_{N,T} := \sqrt{\frac{|U_N||V_N|}{T}}\frac{1}{\sigma}\mathcal{S}_{N,T}$. Then,

$$\begin{aligned} E\left|\mathcal{S}^*_{N,T}\right|^2 &= \frac{|U_N||V_N|}{T\sigma^2}E\left|\sum_{i=0}^{k_T-1}\eta^{(s)}_{N,T,i} + \eta_{N,T,k_T}\right|^2 \\ &= \frac{|U_N||V_N|}{T\sigma^2}\left(E|\eta^{(s)}_{N,T,k_T}|^2 + k_T E\left|\eta^{(s)}_{N,T,0}\right|^2 + 2E\left[\sum_{i=0}^{k_T-1}\eta^{(s)}_{N,T,i}\eta^{(s)}_{N,T,k_T}\right]\right), \end{aligned}$$

where $E|\eta^{(s)}_{N,T,k_T}|^2 = O\left(\frac{\lfloor T-k_T(p_T+q_T)\rfloor}{|U_N||V_N|}\right)$ and $E\left|\eta^{(s)}_{N,T,0}\right|^2 = O\left(\frac{q_T}{|U_N||V_N|}\right)$ by Lemma 6. In addition, $E\left[\sum_{i=0}^{k_T-1}\eta^{(s)}_{N,T,i}\eta^{(s)}_{N,T,k_T}\right] = \sum_{j=1}^{q_T}\sum_{\ell=k_T(p_T+q_T)+1}^{T} E[w_{*,i(p_T+q_T)+p_T+j}w_{*,\ell}\epsilon_{*,i(p_T+q_T)+p_T+j}\epsilon_{*,\ell}] = 0$. It then follows that $E\left|\mathcal{S}^*_{N,T}\right|^2 = O\left(\frac{\lfloor T-k_T(p_T+q_T)\rfloor}{T} + \frac{q_T k_T}{T}\right) = o(1)$.

*Step IV:* Let $\mathcal{B}^*_{N,T} := \sqrt{\frac{|U_N||V_N|}{T}}\frac{1}{\sigma}\mathcal{B}_{N,T}$. Show that

$$Q_1 = \left|E\exp\left(i\theta\mathcal{B}^*_{N,T}\right) - \prod_{i=0}^{k_T-1}E\exp\left(i\theta\sqrt{\frac{|U_N||V_N|}{T}}\frac{1}{\sigma}\eta^{(b)}_{N,T,i}\right)\right| = o(1), \text{ where } i = \sqrt{-1}. \tag{B.115}$$

To do so, invoking Lemma 17 yields $Q_1 \le C_0\sum_{j=0}^{k_T-2} M_\alpha\left(|V_N|p_T, (k_T-j-1)p_T|V_N|\right)\alpha(q_T)$ $\le C_0 k_T\left(|V_N|p_T\right)^{\gamma_M}\alpha(q_T) \overset{(a)}{\le} C_0 T^{1/2-\epsilon}|V_N|^{\gamma_M}T^{\gamma_M(1/2+\epsilon)}\alpha(q_T) \overset{(b)}{\le} C_0|V_N|^{\gamma_M}T^{(\gamma_M+\theta_\alpha-1)\epsilon-\frac{1}{2}(\theta_\alpha-\gamma_M-1)}$, where (a) follows because $k_T\lfloor\frac{T}{p_T+q_T}\rfloor \le q_T$, and (b) follows because of Condition (b). Now, invoking Condition (c), we obtain $Q_1 = o(1)$.

*Step V:* Show that $\sum_{i=0}^{k_T-1} E\left|\sqrt{\frac{|U_N||V_N|}{T}}\frac{1}{\sigma}\eta^{(b)}_{N,T,i}\right|^2 \to 1$. An application of Lemma 6 yields

$$\sum_{i=0}^{k_T-1} E\left|\sqrt{\frac{|U_N||V_N|}{T}}\frac{1}{\sigma}\eta^{(b)}_{N,T,i}\right|^2 = \frac{|U_N||V_N|}{T}\frac{1}{\sigma^2}\sum_{i=0}^{k_T} p_T E[w^2_{*,t}\epsilon^2_{*,t}] = \frac{p_T k_T}{T} \to 1.$$

*Step VI:* Finally, we need to verify the following uniform integrability condition

$$\frac{|U_N||V_N|}{T}\sum_{i=0}^{k_T-1} E\left[|\eta^{(b)}_{N,T,i}|^2 \mathbf{1}\left(|\eta^{(b)}_{N,T,i}| > \lambda\sqrt{\frac{T}{|U_N||V_N|}}\sigma\right)\right] \to 0 \tag{B.116}$$

for every $\lambda > 0$. Invoking the Tchebyshev inequality, the left-hand side of (B.116) is dominated by $\frac{1}{\lambda^2\sigma^2}\left(\frac{|U_N||V_N|}{T}\right)^2 \sum_{i=0}^{k_T} E\left|\eta^{(b)}_{N,T,i}\right|^4$. To study this upper bound, one needs to bound $E\left|\eta^{(b)}_{N,T,i}\right|^4$. For ease of notation, we shall write $\widetilde{w}_{*,j} = w_{*,i(p_T+q_T)+j}$ and $\widetilde{\epsilon}_{*,j} = \epsilon_{*,i(p_T+q_T)+j}$. Therefore,

$$\begin{aligned}
E\left|\eta^{(b)}_{N,T,i}\right|^4 &= \sum_{j=1}^{p_T} E\left|\widetilde{w}_{*,j}\widetilde{\epsilon}_{*,j}\right|^4 + \sum_{j_1\neq j_2}^{p_T} E[\widetilde{w}^2_{*,j_1}\widetilde{\epsilon}^2_{*,j_1}\widetilde{w}^2_{*,j_2}\widetilde{\epsilon}^2_{*,j_2}] + \sum_{j_1\neq j_2}^{p_T} E[\widetilde{w}^3_{*,j_1}\widetilde{\epsilon}^3_{*,j_1}\widetilde{w}_{*,j_2}\widetilde{\epsilon}_{*,j_2}] \\
&\quad + \sum_{j_1\neq j_2\neq j_3 (\text{pairwise})}^{p_T} E[\widetilde{w}^2_{*,j_1}\widetilde{\epsilon}^2_{*,j_1}\widetilde{w}_{*,j_2}\widetilde{\epsilon}_{*,j_2}\widetilde{w}_{*,j_3}\widetilde{\epsilon}_{*,j_3}] \\
&\quad + \sum_{j_1\neq j_2\neq j_3\neq j_4 (\text{pairwise})}^{p_T} E[\widetilde{w}_{*,j_1}\widetilde{\epsilon}_{*,j_1}\widetilde{w}_{*,j_2}\widetilde{\epsilon}_{*,j_2}\widetilde{w}_{*,j_3}\widetilde{\epsilon}_{*,j_3}\widetilde{w}_{*,j_4}\widetilde{\epsilon}_{*,j_4}] \\
&=: \mathcal{A}_{N,T} + \mathcal{B}_{N,T} + \mathcal{C}_{N,T} + \mathcal{D}_{N,T} + \mathcal{E}_{N,T}.
\end{aligned} \tag{B.117}$$

An application of Hölder's inequality and Lemma 8 under Condition (b) yields

$$\mathcal{A}_{N,T} < C_*\frac{p_T}{|U_N|^2|V_N|^2} \tag{B.118}$$

and

$$\begin{aligned}
\mathcal{B}_{N,T} &< 2\sum_{j_1<j_2}^{p_T} E[\widetilde{\epsilon}^2_{*,j_2}]E[\widetilde{w}^2_{*,j_1}\widetilde{\epsilon}^2_{*,j_1}\widetilde{w}^2_{*,j_2}] \\
&\leq 2p_T^2 E[\widetilde{\epsilon}^2_{*,j_1}]E^{1/3}[\widetilde{w}^6_{*,1}]E^{1/3}[\widetilde{\epsilon}^6_{*,j_1}]E^{1/3}[\widetilde{w}^6_{*,1}] \\
&\leq C_0\frac{p_T^2}{|U_N|^2|V_N|^2}.
\end{aligned} \tag{B.119}$$

Moreover, by the covariance inequality (Lemma 17),

$$\begin{aligned}\mathcal{C}_{N,T} &= \sum_{j_1>j_2} E[\widetilde{\epsilon}^3_{*,j_1}]E[\widetilde{w}^3_{*,j_1}\widetilde{w}_{*,j_2}\widetilde{\epsilon}_{*,j_2}] \\ &< C_0 \left|E[\widetilde{\epsilon}^3_{*,j_1}]\right| \|\widetilde{w}^3_{*,j_1}\|_{\gamma_\eta}\|\widetilde{w}_{*,j_2}\|_{\gamma_\eta}\|\widetilde{\epsilon}_{*,j_2}\|_{\gamma_\eta} M_\alpha(|U_N|,|V_N|)^{1-2/\gamma_\eta}\sum_{j_1>j_2}\alpha(|j_1-j_2|)^{1-2/\gamma_\eta}.\end{aligned}$$

It is immediate to verify that all the conditions set out in Lemma 7 hold, therefore, one has $E[\widetilde{\epsilon}^3_{*,j_1}] < C_*\frac{1}{|V_N|^{3/2}}$. Also, by invoking Lemma 8, one obtains $\|\widetilde{w}^3_{*,j_1}\|_{\gamma_\eta} < C_*\frac{1}{|U_N|^{3/2}}$, $\|\widetilde{w}_{*,j_2}\|_{\gamma_\eta} < C_*\frac{1}{|U_N|^{1/2}}$, and $\|\widetilde{\epsilon}_{*,j_2}\|_{\gamma_\eta} < C_*\frac{1}{|V_N|^{1/2}}$. It then follows that

$$\mathcal{C}_{N,T} < C_*\frac{p_T(|U_N|+|V_N|)^{\gamma_M(1-2/\gamma_\eta)}}{|U_N|^2|V_N|^2}\sum_{\tau=1}^{\infty}\alpha(\tau)^{1-2/\gamma_\eta}. \tag{B.120}$$

Notice that

$$\begin{aligned}\mathcal{D}_{N,T} &= \underbrace{\sum_{\substack{j_1<\min(j_2,j_3)\\ j_2\neq j_3}}^{p_T} E[\widetilde{w}^2_{*,j_1}\widetilde{\epsilon}^2_{*,j_1}\widetilde{w}_{*,j_2}\widetilde{\epsilon}_{*,j_2}\widetilde{w}_{*,j_3}\widetilde{\epsilon}_{*,j_3}]}_{=0} + \sum_{\substack{j_1>\min(j_2,j_3)\\ j_2\neq j_3}} E[\widetilde{\epsilon}^2_{*,j_1}]E[\widetilde{w}^2_{*,j_1}\widetilde{w}_{*,j_2}\widetilde{\epsilon}_{*,j_2}\widetilde{w}_{*,j_3}\widetilde{\epsilon}_{*,j_3}] \\ &\leq C_0 E[\widetilde{\epsilon}^2_{*,j_1}]\sum_{j_1>j_2>j_3}^{p_T}\left|E[\widetilde{w}^2_{*,j_1}\widetilde{w}_{*,j_2}\widetilde{\epsilon}_{*,j_2}\widetilde{w}_{*,j_3}\widetilde{\epsilon}_{*,j_3}]\right|,\end{aligned}$$

where - by the same argument as above - one has $E[\widetilde{\epsilon}^2_{*,j_1}] < C\frac{1}{|V_N|}$ and

$$\left|E[\widetilde{w}^2_{*,j_1}\widetilde{w}_{*,j_2}\widetilde{\epsilon}_{*,j_2}\widetilde{w}_{*,j_3}\widetilde{\epsilon}_{*,j_3}]\right| \leq C_0\|\widetilde{w}^2_{*,j_1}\|_{\gamma_\eta}\left\|\widetilde{w}_{*,j_2}\widetilde{\epsilon}_{*,j_2}\widetilde{w}_{*,j_3}\widetilde{\epsilon}_{*,j_3}\right\|_{\gamma_\eta} M_\alpha(|U_N|,|V_N|)^{1-2/\gamma_\eta}\alpha(|j_1-j_2|)^{1-2/\gamma_\eta}$$

with $\|\widetilde{w}^2_{*,j_1}\|_{\gamma_\eta} < C_0\frac{1}{|U_N|}$ and $\left\|\widetilde{w}_{*,j_2}\widetilde{\epsilon}_{*,j_2}\widetilde{w}_{*,j_3}\widetilde{\epsilon}_{*,j_3}\right\|_{\gamma_\eta} < C_0\frac{1}{|U_N||V_N|}$. It then follows that

$$\mathcal{D}_{N,T} < C_0\frac{p_T^2(|U_N|+|V_N|)^{\gamma_M(1-2/\gamma_\eta)}}{|U_N|^2|V_N|^2}\sum_{\tau=1}^{\infty}\alpha(\tau)^{1-2/\gamma_\eta}. \tag{B.121}$$

Finally, it is not difficult to see that

$$\mathcal{E}_{N,T} < C_0\sum_{j_1<j_2<j_3<j_4}^{p_T} E[\widetilde{w}_{*,j_1}\widetilde{\epsilon}_{*,j_1}\widetilde{w}_{*,j_2}\widetilde{\epsilon}_{*,j_2}\widetilde{w}_{*,j_3}\widetilde{\epsilon}_{*,j_3}\widetilde{w}_{*,j_4}\widetilde{\epsilon}_{*,j_4}] = 0. \tag{B.122}$$

Therefore, in view of (B.117)-(B.122), we have

$$E\left|\eta_{N,T,i}^{(b)}\right|^4 < C_0\left\{\frac{p_T^2}{|U_N|^2|V_N|^2} + \frac{p_T^2(|U_N|+|V_N|)^{\gamma_M(1-2/\gamma_\eta)}}{|U_N|^2|V_N|^2}\sum_{\tau=1}^{\infty}\alpha(\tau)^{1-2/\gamma_\eta}\right\}.$$

Invoking Condition (c), (B.116) has been verified. The main theorem then follows in view of Steps *I-VI* above. ■

**Lemma 11** *Let* $\{X_{\boldsymbol{i},t}: \boldsymbol{i}\in V_N,\ t\in[1,T]\}$ *be a mixing spatio-temporal process. Suppose that (a)* $X_{\boldsymbol{i},t}$, $\boldsymbol{i}\in V_N$ *and* $t\in[1,T]$, *are identically distributed over time and space; (b)* $\max_{\boldsymbol{i}} E[\exp(\ell\|X_{\boldsymbol{i},t}\|)] \le C_\ell$ *for a constant* $C_\ell > 0$ *and* $\ell > 0$ *small enough; (c)* $\max_{\boldsymbol{i}}\|X_{\boldsymbol{i},t}\|_{\delta_\alpha} < \infty$ *for some* $\delta_\alpha > 2$; *(d)* $\alpha(\tau)\le C_0\tau^{-\theta_\alpha}$ *for some* $\theta_\alpha > \left(\frac{4\gamma_M}{3}, \frac{d_v+1}{1-2/\delta_\alpha}\right)^+$. *Then,*

$$P\left(\left|\frac{1}{T}\sum_{t=1}^{T}X_{*,t}\right|\ge M\right)\le \frac{2}{M}T^{-C_\alpha} + C_0N^{\gamma_M}\log(T)T^{\gamma_M-\frac{3}{4}\theta_\alpha}$$
$$+4\max\left\{\exp\left(-\frac{1}{256}\left(\frac{M}{C_\ell}\right)^2\frac{NT}{2C_\sigma}\right),\ \exp\left(-\frac{1}{32}\frac{N}{C_\ell}\frac{T^{1/4}}{\log(T)}\right)\right\}^+,$$

*where* $C_\sigma$ *and* $C_\alpha$ *are sufficiently large constants.*

**Proof of Lemma 11.** First, we employ the following truncation: $X_{\boldsymbol{i},t} = X_{\boldsymbol{i},t}^{(<)} + X_{\boldsymbol{i},t}^{(>)}$ with $X_{\boldsymbol{i},t}^{(<)} := X_{\boldsymbol{i},t}\mathbf{1}(|X_{\boldsymbol{i},t}|\le C_x\log(T))$ and $X_{\boldsymbol{i},t}^{(>)} := X_{\boldsymbol{i},t} - X_{\boldsymbol{i},t}^{(<)}$. It then follows that

$$\begin{aligned}P\left(\left|\frac{1}{T}\sum_{t=1}^{T}X_{*,t}\right|\ge M\right) &\le P\left(\left|\frac{1}{T}\sum_{t=1}^{T}X_{*,t}^{(<)} - E[X_{*,t}^{(<)}]\right|\ge M/2\right)\\ &\quad + P\left(\left|\frac{1}{T}\sum_{t=1}^{T}X_{*,t}^{(>)} - E[X_{*,t}^{(>)}]\right|\ge M/2\right)\\ &=: \mathcal{T}_{<,N,T} + \mathcal{T}_{>,N,T}. \end{aligned} \tag{B.123}$$

By the Tchebyshev inequality and Conditions (a) and (b), one could show that

$$\mathcal{T}_{>,N,T}\le \frac{2}{M}E[|X_{\boldsymbol{i},t}|\mathbf{1}(|X_{\boldsymbol{i},t}|>C_x\log(T))]\le\frac{2}{M}T^{-C_\alpha}, \tag{B.124}$$

where $C_x$ can always be chosen to make $C_\alpha$ large enough. To bound $\mathcal{T}_{<,N,T}$, let $\mu_T$ and $b_T$ denote two divergent sequences so that $T-b_T < 2\mu_Tb_T\le T$, divide $\{X_{*,1},\dots,X_{*,T}\}$ into $2\mu_T$ blocks of size $b_T$. We can always choose $b_T$ and $\mu_T$ in such a way that the remainder $\{X_{*,T-2\mu_Tb_T},\dots,X_{*,T}\}$ can be ignored. Let $(\xi_{\boldsymbol{i},1},\dots,\xi_{\boldsymbol{i},b_T}), (\xi_{\boldsymbol{i},b_T+1},\dots,\xi_{\boldsymbol{i},2b_T}),\dots,(\xi_{\boldsymbol{i},(2\mu_T-1)b_T+1},\dots,\xi_{\boldsymbol{i},2\mu_Tb_T})$ be independent blocks of random elements such that $(\xi_{\boldsymbol{i},jb_T+1},\dots,\xi_{\boldsymbol{i},(j+1)b_T})$ and $(X_{\boldsymbol{i},jb_T+1}^{(<)},\dots,X_{\boldsymbol{i},(j+1)b_T}^{(<)})$, $j=1,\dots,2\mu_T$, have the same distribution. Moreover, define

$$Z_{*,j} := \sum_{t=(2j-1)b_T+1}^{2jb_T} \xi_{*,t}, \; j = 1, 2 \ldots, \mu_T.$$

It then follows that

$$\begin{aligned}
\mathcal{T}_{<,N,T} &\le 2P\left(\left|\frac{1}{T}\sum_{j=1}^{\mu_T}\sum_{t=(2j-1)b_T+1}^{2jb_T} X_{*,t}^{(<)} - E[X_{*,t}^{(<)}]\right| \ge \frac{M}{4}\right) \\
&\le 2P\left(\left|\frac{1}{T}\sum_{j=1}^{\mu_T}\left\{\sum_{t=(2j-1)b_T+1}^{2jb_T} X_{*,t}^{(<)} - Z_{*,j}\right\}\right| \ge \frac{M}{8}\right) \\
&\quad + 2P\left(\left|\frac{1}{T}\sum_{j=1}^{\mu_T} Z_{*,j} - E[Z_{*,j}]\right| \ge \frac{M}{8}\right) \\
&=: \mathcal{T}_{<,N,T}^{(a)} + \mathcal{T}_{<,N,T}^{(b)}.
\end{aligned}$$

Let $S_{N,j} := [(2j-1)b_T + 1, 2jb_T] \times V_N$, $j = 1, \ldots, \mu_T$, then $d(S_{N,i}, S_{N,j}) \ge b_T$ for every $i \ne j$. Since $Z_{*,j}$ is $\mathcal{B}(S_{N,j})$-measurable, $|Z_{*,j}| \le C_x b_T \log(T)$, and $S_{N,j}$ contains $Nb_T$ sites, an application of Rio's coupling inequality (Lemma 20) yields

$$\mathcal{T}_{<,N,T}^{(a)} \le 2C_x \mu_T b_T \log(T) M_\alpha(Nb_T\mu_T, Nb_T)\alpha(b_T) \le C_0 N^{\gamma_M} \log(T) T^{\gamma_M - \frac{3}{4}\theta_\alpha}. \tag{B.125}$$

In addition, as $|Z_{*,j} - E[Z_{*,j}]| \le 2C_x b_T \log(T)$, thus $|\widetilde{Z}_{*,j}| = \frac{|Z_{*,j} - E[Z_{*,j}]|}{2C_x b_T \log(T)} \le 1$ and

$$Var(\widetilde{Z}_{*,j}) = \frac{Var(Z_{*,j})}{4C_x^2 b_T^2 \log^2(T)} \le \frac{1}{2C_x^2 b_T^2 \log^2(T)} Var\left(\sum_{t=1}^{b_T} X_{*,t}\right). \tag{B.126}$$

Some combinatorics arguments yield

$$\begin{aligned}
Var\left(\sum_{t=1}^{b_T} X_{*,t}\right) &= \frac{1}{N^2}\sum_{t=1}^{b_T}\sum_{\boldsymbol{i}\in V_N} Var(X_{\boldsymbol{i},t}) + \frac{1}{N^2}\sum_{r=1}^{\text{diam}(V_N\times[1,b_T])}\sum_{(\boldsymbol{i},t)\in V_N\times[1,b_T]}\sum_{\substack{(\boldsymbol{w},\tau)\in V_N\times[1,b_T] \\ \|(\boldsymbol{i},t)-(\boldsymbol{w},\tau)\|=r}} Cov(X_{\boldsymbol{i},t}, X_{\boldsymbol{w},\tau}) \\
&\le \frac{1}{N^2}\sum_{t=1}^{b_T}\sum_{\boldsymbol{i}\in V_N} Var(X_{\boldsymbol{i},t}) \\
&\quad + C_0\frac{1}{N^2}\sum_{(\boldsymbol{w},\tau)\in V_N\times[1,b_T]}\sum_{r=1}^{\text{diam}(V_N\times[1,b_T])} |\{(\boldsymbol{i},t)\in V_N\times[1,b_T] : \; \|(\boldsymbol{i},t)-(\boldsymbol{w},\tau)\| = r\}|
\end{aligned}$$

$$\times \|X_{\boldsymbol{i},t}\|_{\delta_\alpha}^2 \{M_\alpha(1,1)\alpha(r)\}^{1-\frac{2}{\delta_\alpha}}$$
$$< C_\sigma \frac{b_T}{N}.$$

It then follows from (B.126) that

$$Var(\widetilde{Z}_{*,j}) \leq C_\sigma \frac{1}{N b_T \log^2(T)}.$$

Invoking Lemma 24, one readily obtains that

$$\mathcal{T}^{(b)}_{<,N,T} = P\left(\left|\sum_{j=1}^{\mu_T} \widetilde{Z}_{*,j}\right| \geq \frac{M}{16C_x}\frac{T}{b_T\log(T)}\right)$$
$$\leq 2\max\left\{\exp\left(-\frac{1}{4}\left(\frac{M}{8C_x}\right)^2 \frac{Nb_T}{C_\sigma}\right), \exp\left(-\frac{M}{32C_x}\frac{T^{1/4}}{\log(T)}\right)\right\}^+. \tag{B.127}$$

The main lemma then follows from (B.123)-(B.127). ■

**Lemma 12** *Let* $\{(X_{\boldsymbol{i},t}, \epsilon_{\boldsymbol{i},t}) : \boldsymbol{i} \in V_N,\ t \in [1,T]\}$ *represent a bivariate spatio-temporal process. Suppose that (a)* $\{X_{*,t}, \epsilon_{\boldsymbol{i},t}\}$, $\boldsymbol{i} \in V_N$ *and* $t \in [1,T]$ *are mixing and identically distributed over time and space; (b)* $\alpha(\tau) < C_0\tau^{-\theta_\alpha}$, $\theta_\alpha > \left(\frac{4\gamma_M}{3}, \frac{2d_v+1}{1-2/\delta_\alpha}\right)^+$ *for some* $\delta_\alpha > 2$*; (c)* $\max_{\boldsymbol{i}} \|X_{\boldsymbol{i},t}\epsilon_{\boldsymbol{i},t}\|_{\delta_\alpha} < \infty$*; (d)* $\max_{\boldsymbol{i}} E[\exp(\ell\|X_{\boldsymbol{i},t}\|)] \leq C_\ell$ *for a constant* $C_\ell > 0$ *and* $\ell > 0$ *small enough. Then,*

$$P\left(\frac{1}{T}\left|\sum_{t=1}^{T}\{X_{*,t}\epsilon_{*,t} - E[X_{*,t}\epsilon_{*,t}]\}\right| \geq M\right) \leq C_0\Bigg(T^{-C_\alpha} + N^{2\gamma_M}\log^2(T)T^{\gamma_M - \frac{3}{4}\theta_\alpha}$$
$$+ \max\left\{\exp\left(-C_\sigma N^2\log^2(T)T^{7/4}\right), \exp\left(-C_M\frac{T^{1/4}}{\log^2(T)}\right)\right\}\Bigg),$$

*where* $C_\sigma$ *and* $C_M$ *are some sufficiently large constants.*

**Proof of Lemma 12.** Define the following truncated random variables: $X_{\boldsymbol{i},t} = X^{(<)}_{\boldsymbol{i},t} + X^{(>)}_{\boldsymbol{i},t}$ with $X^{(<)}_{\boldsymbol{i},t} := X_{\boldsymbol{i},t}\mathbf{1}(|X_{\boldsymbol{i},t}| \leq C_x\log(T))$ and $X^{(>)}_{\boldsymbol{i},t} := X_{\boldsymbol{i},t}\mathbf{1}(|X_{\boldsymbol{i},t}| > C_x\log(T))$; $\epsilon_{\boldsymbol{i},t} = \epsilon^{(<)}_{\boldsymbol{i},t} + \epsilon^{(>)}_{\boldsymbol{i},t}$ with $\epsilon^{(<)}_{\boldsymbol{i},t} := \epsilon_{\boldsymbol{i},t}\mathbf{1}(|\epsilon_{\boldsymbol{i},t}| \leq C_\epsilon\log(T))$ and $\epsilon^{(>)}_{\boldsymbol{i},t} := \epsilon_{\boldsymbol{i},t}\mathbf{1}(|\epsilon_{\boldsymbol{i},t}| > C_\epsilon\log(T))$. Thus, $X\epsilon = X^{(<)}\epsilon^{(<)} + X^{(<)}\epsilon^{(>)} + X^{(>)}\epsilon^{(<)} + X^{(>)}\epsilon^{(>)}$. It then follows that

$$P\left(\frac{1}{T}\left|\sum_{t=1}^{T}\{X_{*,t}\epsilon_{*,t} - E[X_{*,t}\epsilon_{*,t}]\}\right| \geq M\right) \leq P\left(\frac{1}{T}\left|\sum_{t=1}^{T}\left\{X^{(<)}_{*,t}\epsilon^{(<)}_{*,t} - E\left[X^{(<)}_{*,t}\epsilon^{(<)}_{*,t}\right]\right\}\right| \geq \frac{M}{4}\right)$$
$$+ P\left(\frac{1}{T}\left|\sum_{t=1}^{T}\left\{X^{(<)}_{*,t}\epsilon^{(>)}_{*,t} - E\left[X^{(<)}_{*,t}\epsilon^{(>)}_{*,t}\right]\right\}\right| \geq \frac{M}{4}\right)$$

$$
\begin{aligned}
&+ P\left(\frac{1}{T}\left|\sum_{t=1}^{T}\left\{X_{*,t}^{(>)}\epsilon_{*,t}^{(>)} - E\left[X_{*,t}^{(>)}\epsilon_{*,t}^{(>)}\right]\right\}\right| \geq \frac{M}{4}\right) \\
&+ P\left(\frac{1}{T}\left|\sum_{t=1}^{T}\left\{X_{*,t}^{(>)}\epsilon_{*,t}^{(>)} - E\left[X_{*,t}^{(>)}\epsilon_{*,t}^{(>)}\right]\right\}\right| \geq \frac{M}{4}\right) \\
&=: \mathcal{T}_{1,N,T} + \mathcal{T}_{2,N,T} + \mathcal{T}_{3,N,T} + \mathcal{T}_{4,N,T}. \qquad \text{(B.128)}
\end{aligned}
$$

To bound $\mathcal{T}_{1,N,T}$, let $w_{*,t} := X_{*,t}^{(<)}\epsilon_{*,t}^{(<)}$. Divide $\{w_{*,1}, \dots, w_{*,T}\}$ into $2\mu_T$ blocks, $\{w_{*,(j-1)b_T+1}, \dots, w_{*,jb_T}\}$, $j = 1, \dots, 2\mu_T$, of size $b_T$ and a smaller remaining block. One can always choose $\mu_T$ and $b_T$ such that the remaining block is negligible so that it can be ignored. Define $2\mu_T$ contemporaneously independent blocks, $\{(\xi_{\boldsymbol{i},1}, \zeta_{\boldsymbol{i},1}), \dots (\xi_{\boldsymbol{i},b_T}, \zeta_{\boldsymbol{i},b_T})\}, \{(\xi_{\boldsymbol{i},b_T+1}, \zeta_{\boldsymbol{i},b_T+1}), \dots, (\xi_{\boldsymbol{i},2b_T}, \zeta_{\boldsymbol{i},2b_T})\}, \dots, \{(\xi_{\boldsymbol{i},(2\mu_T-1)b_T}, \zeta_{\boldsymbol{i},(2\mu_T-1)b_T}), \dots, (\xi_{\boldsymbol{i},2\mu_T b_T}, \zeta_{\boldsymbol{i},2\mu_T b_T})\}$, such that $(\xi_{\boldsymbol{i},(j-1)b_T+1}, \dots, \xi_{\boldsymbol{i},jb_T})$ and $\{X_{\boldsymbol{i},(j-1)b_T+1}, \dots, X_{\boldsymbol{i},jb_T}\}$ are identically distributed; and $(\zeta_{\boldsymbol{i},(j-1)b_T+1}, \dots, \zeta_{\boldsymbol{i},jb_T})$ and $\{\epsilon_{\boldsymbol{i},(j-1)b_T+1}, \dots, \epsilon_{\boldsymbol{i},jb_T}\}$ are identically distributed. Let $Z_{*,j} := \sum_{t=(2j-1)b_T+1}^{2jb_T} \xi_{*,t}\zeta_{*,t}$, $j = 1, \dots, \mu_T$. One can obtain that

$$
\begin{aligned}
\mathcal{T}_{1,N,T} &\leq 2P\left(\left|\frac{1}{T}\sum_{j=1}^{\mu_T}\sum_{t=(2j-1)b_T+1}^{2jb_T} w_{*,t} - E[w_{*,t}]\right| \geq \frac{M}{8}\right) \\
&\leq 2P\left(\left|\frac{1}{T}\sum_{j=1}^{\mu_T}\left(\sum_{t=(2j-1)b_T+1}^{2jb_T} w_{*,t} - Z_{*,j}\right)\right| \geq \frac{M}{16}\right) \\
&\quad + 2P\left(\left|\frac{1}{T}\sum_{j=1}^{\mu_T}(Z_{*,j} - E[Z_{*,j}])\right| \geq \frac{M}{16}\right) \\
&=: \mathcal{T}_{1,N,T}^{(a)} + \mathcal{T}_{1,N,T}^{(b)}. \qquad \text{(B.129)}
\end{aligned}
$$

Define $S_{N,j} := [(2j-1)b_T+1, 2jb_T] \times V_N \times V_N$. Then, $d(S_{N,j}, S_{N,k}) \geq b_T$ for $j \neq k$ and $\sum_{t=(2j-1)b_T+1}^{2jb_T} w_{*,t}$ is $\mathcal{B}(S_{N,j})$-measurable. Since $|Z_{*,j}| \leq C_x C_\epsilon b_T \log^2(T)$, an application of Lemma 20 yields

$$
\mathcal{T}_{1,N,T}^{(a)} \leq C_0 N^{2\gamma_M} \log^2(T) T^{\gamma_M - \frac{3}{4}\theta_\alpha}.
$$

To bound $\mathcal{T}_{1,N,T}^{(b)}$, notice that

$$
\begin{aligned}
Var\left(Z_{*,1}\right) &= \frac{1}{N^4} Var\left(\sum_{(\boldsymbol{i},\boldsymbol{j},t) \in V_{N,T}} X_{\boldsymbol{i},t}^{(<)}\epsilon_{\boldsymbol{j},t}^{(<)}\right) \\
&= \frac{1}{N^4}\sum_{(\boldsymbol{i},\boldsymbol{j},t) \in V_{N,T}} Var(X_{\boldsymbol{i},t}^{(<)}\epsilon_{\boldsymbol{j},t}^{(<)})
\end{aligned}
$$

$$+\frac{1}{N^4}\sum_{\substack{(\boldsymbol{i}_1,\boldsymbol{j}_1,s)\in V_{N,T}\\(\boldsymbol{i}_2,\boldsymbol{j}_2,t)\in V_{N,T}\\\|(\boldsymbol{i}_1,\boldsymbol{j}_1,s)-(\boldsymbol{i}_2,\boldsymbol{j}_2,t)\|\neq 0}} Cov(X^{(<)}_{\boldsymbol{i}_1,s}\epsilon^{(<)}_{\boldsymbol{j}_1,s},X^{(<)}_{\boldsymbol{i}_2,t}\epsilon^{(<)}_{\boldsymbol{j}_2,t}),$$

where the second term is bounded by
$C_0\frac{1}{N^4}\sum_{(\boldsymbol{i}_1,\boldsymbol{j}_1,s)\in V_{N,T}}\sum_{r=1}^{\text{diam}(V_{N,T})}|\{(\boldsymbol{i}_2,\boldsymbol{j}_2,t)\in V_{N,T}:\ \|(\boldsymbol{i}_2,\boldsymbol{j}_2,t)-(\boldsymbol{i}_1,\boldsymbol{j}_1,s)\|=4\}|\,\|X_{\boldsymbol{i}_1,s}\epsilon_{\boldsymbol{j}_1,s}\|^2_{\delta_\alpha}$ $\{M_\alpha(1,1)\alpha(r)\}^{1-2/\delta_\alpha}\leq C_0\frac{b_T}{N^2}\sum_{r=1}^{\infty}r^{2d_v}\alpha(r)^{1-\frac{2}{\delta_\alpha}}$ in view of Lemma 17. Conditions (b) and (c) imply that

$$Var\,(Z_{*,1})\leq C_\sigma\frac{b_T}{N^2}.$$

Let $\widetilde{Z}_{*,j}=\frac{1}{C_xC_\epsilon b_T\log^2(T)}|Z_{*,j}|\leq 1$. Invoking Lemma 24, one can show that

$$\begin{aligned}\mathcal{T}^{(b)}_{1,N,T}&\leq 2P\left(\left|\frac{1}{T}\sum_{j=1}^{\mu_T}(Z_{*,j}-E[Z_{*,j}])\right|\geq C_\sigma\mu_T\frac{1}{b_T\log^4(T)N^2}\frac{M}{32C_\epsilon C_xC_\sigma}\frac{N^2\log^2(T)T}{\mu_T}\right)\\&\leq 4\max\left\{\exp\left(-C_\sigma\frac{N^2\log^2(T)T^2}{\mu_T}\right),\exp\left(-C_M\frac{\mu_T}{\log^2(T)}\right)\right\}.\end{aligned}$$

By choosing $\mu_T=O(T^{1/4})$ and $b_T=O\left(\lfloor\frac{T}{2\mu_T}\rfloor\right)$, we obtain that

$$\mathcal{T}^{(b)}_{1,N,T}\leq 4\max\left\{\exp\left(-C_\sigma N^2\log^2(T)T^{7/4}\right),\exp\left(-C_M\frac{T^{1/4}}{\log^2(T)}\right)\right\}.$$

It then follows from (B.129) that

$$\begin{aligned}\mathcal{T}_{1,N,T}&\leq C_0N^{2\gamma_M}\log^2(T)T^{\gamma_M-\frac{3}{4}\theta_\alpha}\\&\quad+4\max\left\{\exp\left(-C_\sigma N^2\log^2(T)T^{7/4}\right),\exp\left(-C_M\frac{T^{1/4}}{\log^2(T)}\right)\right\}.\end{aligned}\tag{B.130}$$

To bound the remaining terms in (B.128), Condition (d) implies that one can choose $C_\epsilon$ such that, for $C_\alpha$ large enough, $E\left[|\epsilon_{\boldsymbol{i},t}|^2\mathbf{1}(|\epsilon_{\boldsymbol{i},t}|>C_\epsilon\log(T))\right]\leq T^{-C_\alpha}$ and $E\left[|X_{\boldsymbol{i},t}|^2\mathbf{1}(|X_{\boldsymbol{i},t}|>C_\epsilon\log(T))\right]\leq T^{-C_\alpha}$. Therefore, we have

$$\mathcal{T}_{2,N,T}\leq\frac{8}{M}\frac{1}{N^2}\sum_{\boldsymbol{i},\boldsymbol{j}\in V_N}E[|X^{(<)}_{\boldsymbol{i},t}\epsilon^{(>)}_{\boldsymbol{j},t}|]\leq\frac{8}{M}\frac{1}{N^2}\sum_{\boldsymbol{i},\boldsymbol{j}\in V_N}\|X_{\boldsymbol{i},t}\|_2\left\|\epsilon^{(>)}_{\boldsymbol{j},t}\right\|_2\leq C_0T^{-C_\alpha}.\tag{B.131}$$

Similarly, we also obtain $\mathcal{T}_{3,N,T}\leq C_0T^{-C_\alpha}$ and $\mathcal{T}_{4,N,T}\leq C_0T^{-2C_\alpha}$. The main lemma then follows from (B.128)-(B.131). ■

**Lemma 13** *Let all the symbols be defined as in Section C.1. The function*

$$\mathcal{H}_{1,N,T}(\boldsymbol{\gamma},\boldsymbol{U}) \coloneqq \frac{1}{N^2}\sum_{c=1}^{G}\sum_{i=1}^{N}\frac{1}{2}\left(\rho_u u_{i,c}^2 + \rho_\phi \frac{\phi_c^2}{N} + \rho_\theta \frac{\boldsymbol{\theta}_c^\top \boldsymbol{\theta}_c}{N}\right) - \mathcal{E}_{1,N,T}(\boldsymbol{\gamma},\boldsymbol{U}) + \mathcal{A}_0$$

*is convex for every* $\boldsymbol{\rho} \doteq (\rho_u, \rho_\phi, \rho_\theta)$ *satisfying* (B.132)-(B.133), (B.138)-(B.140).

**Proof of Lemma 13.** Write $\mathcal{H}_{1,N,T}(\boldsymbol{\gamma},\boldsymbol{U}) = \frac{1}{N^2}\sum_{c=1}^{G}\sum_{i=1}^{N}\mathfrak{f}_{1,N,T}^{(c,i)}(\boldsymbol{\gamma},\boldsymbol{U})$, where $\mathfrak{f}_{1,N,T}^{(c,i)}(\boldsymbol{\gamma},\boldsymbol{U}) \coloneqq \frac{1}{2}\left(\rho_u u_{i,c}^2 + \rho_\phi \frac{\phi_c^2}{N} + \rho_\theta \frac{\boldsymbol{\theta}_c^\top \boldsymbol{\theta}_c}{N}\right) - N^2\mathcal{A}_0 - u_{i,c}^2\phi_c^2\mathcal{A}_{1,i} - u_{i,c}^2\phi_c^2\boldsymbol{\theta}_c^\top\mathcal{B}_{1,i}\boldsymbol{\theta}_c + 2u_{i,c}^2\phi_c^2\boldsymbol{\theta}_c^\top\mathcal{C}_{1,i} + 2Nu_{i,c}\phi_c\mathcal{D}_{1,i} - 2Nu_{i,c}\phi_c\boldsymbol{\theta}_c^\top\mathcal{F}_{1,i}$. One needs to verify that $\mathfrak{f}_{1,N,T}^{(c,i)}(\boldsymbol{\gamma},\boldsymbol{U})$ is convex for each $i \in [1,N]$ and $c \in [1,G]$. It is equivalent to showing that the minimum eigenvalue of the Hessian matrix is strictly positive. The positivity of the minimum eigenvalue of a matrix can be verified by the positive definiteness of all the sub-matrices. Some simple calculations yield the second-order derivatives of $\mathfrak{f}_{1,N,T}^{(c,i)}(\boldsymbol{\gamma},\boldsymbol{U})$ :

$$\begin{aligned}
a_{2,2}^{(1)} &\coloneqq \partial_{\phi_c}^2 \mathfrak{f}_{1,N,T}^{(c,i)}(\boldsymbol{\gamma},\boldsymbol{U}) = \frac{\rho_\phi}{N} - 2u_{i,c}^2\mathcal{A}_{1,i} - 2u_{i,c}^2\boldsymbol{\theta}_c^\top\mathcal{B}_{1,i}\boldsymbol{\theta}_c + 4u_{i,c}^2\boldsymbol{\theta}_c^\top\mathcal{C}_{1,i},\\
\boldsymbol{a}_{2,3}^{(1)} &\coloneqq \partial_{\phi_c,\boldsymbol{\theta}_c}^2 \mathfrak{f}_{1,N,T}^{(c,i)}(\boldsymbol{\gamma},\boldsymbol{U}) = -4u_{i,c}^2\phi_c\mathcal{B}_{1,i}\boldsymbol{\theta}_c + 4u_{i,c}^2\phi_c\mathcal{C}_{1,i} - 2Nu_{i,c}\mathcal{F}_{1,i} \eqqcolon \boldsymbol{a}_{3,2}^{(1)},\\
\boldsymbol{a}_{3,3}^{(1)} &\coloneqq \partial_{\boldsymbol{\theta}_c}^2 \mathfrak{f}_{1,N,T}^{(c,i)}(\boldsymbol{\gamma},\boldsymbol{U}) = \frac{\rho_\theta}{N}\mathbb{I}_{d_x} - 2u_{i,c}^2\phi_c^2\mathcal{B}_{1,i},\\
a_{1,1}^{(1)} &\coloneqq \partial_{u_{i,c}}^2 \mathfrak{f}_{1,N,T}^{(c,i)}(\boldsymbol{\gamma},\boldsymbol{U}) = \rho_u - 2\phi_c^2\mathcal{A}_{1,i} - 2\phi_c^2\boldsymbol{\theta}_c^\top\mathcal{B}_{1,i}\boldsymbol{\theta}_c + 4\phi_c^2\boldsymbol{\theta}_c^\top\mathcal{C}_{1,i},\\
a_{1,2}^{(1)} &\coloneqq \partial_{u_{i,c}\phi_c}^2 \mathfrak{f}_{1,N,T}^{(c,i)}(\boldsymbol{\gamma},\boldsymbol{U}) = -4u_{i,c}\phi_c\mathcal{A}_{1,i} - 4u_{i,c}\phi_c\boldsymbol{\theta}_c^\top\mathcal{B}_{1,i}\boldsymbol{\theta}_c + 8u_{i,c}\phi_c\boldsymbol{\theta}_c^\top\mathcal{C}_{1,i} + 2N\mathcal{D}_{1,i} - 2N\boldsymbol{\theta}_c^\top\mathcal{F}_{1,i}\\
&\eqqcolon a_{2,1}^{(1)},\\
\boldsymbol{a}_{1,3}^{(1)} &\coloneqq \partial_{u_{i,c}\boldsymbol{\theta}_c}^2 \mathfrak{f}_{1,N,T}^{(c,i)}(\boldsymbol{\gamma},\boldsymbol{U}) = -4u_{i,c}\phi_c^2\mathcal{B}_{1,i}\boldsymbol{\theta}_c + 4u_{i,c}\phi_c^2\mathcal{C}_{1,i} - 2N\phi_c\mathcal{F}_{1,i}\\
&\eqqcolon \boldsymbol{a}_{3,1}^{(1)}.
\end{aligned}$$

Let $\mathcal{H}_1 \coloneqq \begin{pmatrix} a_{1,1}^{(1)} & a_{1,2}^{(1)} & \boldsymbol{a}_{1,3}^{(1)\top} \\ a_{2,1}^{(1)} & a_{2,2}^{(1)} & \boldsymbol{a}_{2,3}^{(1)\top} \\ \boldsymbol{a}_{3,1}^{(1)} & \boldsymbol{a}_{3,2}^{(1)} & \boldsymbol{a}_{3,3}^{(1)} \end{pmatrix}$ denote the Hessian matrix of $\mathfrak{f}_{1,N,T}^{(c,i)}(\boldsymbol{\gamma},\boldsymbol{U})$. The positive definiteness of the first Hessian sub-matrix is warranted by

$$\rho_u \geq \max_{i,c}\left\{2\ell_{\phi,c}^2|\mathcal{A}_{1,i}| + 2\ell_{\phi,c}^2\|\boldsymbol{\ell}_{\theta,c}\|^2\lambda_{\max}(\mathcal{B}_{1,i}) + 4\ell_{\phi,c}^2|\boldsymbol{\ell}_{\theta,c}^\top||\mathcal{C}_{1,i}|\right\}, \tag{B.132}$$

where $\boldsymbol{\ell}_{\theta,c} \coloneqq (\ell_{\theta,c,1},\dots,\ell_{\theta,c,d_x})^\top$. The positive definiteness of the second Hessian sub-matrix is warranted by

$$\begin{aligned}\rho_u \geq \max_{i,c}\Big\{&(2\ell_{\phi,c}^2 + 4\ell_{\phi,c})|\mathcal{A}_{1,i}| + 2\|\boldsymbol{\ell}_{\theta,c}\|^2\lambda_{\max}(\mathcal{B}_{1,i})\left(\ell_{\phi,c}^2 + 2\ell_{\phi,c}\right) + 4(\ell_{\phi,c}^2 + 2\ell_{\phi,c})|\boldsymbol{\ell}_{\theta,c}^\top||\mathcal{C}_{1,i}|\\
&+2N|\mathcal{D}_{1,i}| + 2N|\boldsymbol{\ell}_{\theta,c}^\top||\mathcal{F}_{1,i}|\Big\}\end{aligned} \tag{B.133}$$

and

$$\rho_\phi \geq N \max_{i,c} \left\{2|\mathcal{A}_{1,i}|(1+2\ell_{\phi,c}) + 2(1+2\ell_{\phi,c})\|\boldsymbol{\ell}_{\theta,c}\|^2\lambda_{\max}(\mathcal{B}_{1,i}) + 4(1+2\ell_{\phi,c})|\boldsymbol{\ell}_{\theta,c}^\top||\mathcal{C}_{1,i}| \right.$$
$$\left. +2N|\mathcal{D}_{1,i}| + 2N|\boldsymbol{\ell}_{\theta,c}^\top||\mathcal{F}_{1,i}|\right\}. \tag{B.134}$$

In view of Lemma 25 the positive definiteness of the third Hessian sub-matrices is warranted by

$$\frac{1}{d+2}\left(a_{1,1}^{(1)} + a_{1,2}^{(1)} + \boldsymbol{a}_{1,3}^{(1)\top}\imath_{d,d_x}\right) > \max(0, a_{1,2}^{(1)}, \boldsymbol{a}_{1,3}^{(1)\top}), \tag{B.135}$$
$$\frac{1}{d+2}\left(a_{1,2}^{(1)} + a_{2,2}^{(1)} + \boldsymbol{a}_{2,3}^{(1)\top}\imath_{d,d_x}\right) > \max(0, a_{1,2}^{(1)}, \boldsymbol{a}_{2,3}^{(1)\top}), \tag{B.136}$$
$$\frac{1}{d+2}\left(\boldsymbol{a}_{1,3}^{(1)} + \boldsymbol{a}_{2,3}^{(1)} + \boldsymbol{a}_{3,3}^{(1)}\imath_{d,d_x}\right) > \max\left(0, \boldsymbol{a}_{1,3}^{(1)}, \boldsymbol{a}_{2,3}^{(1)}, -2\phi_c^2 u_{i,c}^2\mathcal{B}_{1,i}\right) \tag{B.137}$$

for all $d = 1, \ldots, d_x$, where $\imath_{d,d_x} := \big(\underbrace{\underbrace{1, \ldots, 1}_{d}, 0, \ldots, 0}_{d_x}\big)^\top$. We immediately verify that $|a_{1,2}^{(1)}| \leq \ell_{a_{12}}^{(1)} :=$ $\max_{c,i}\left\{4\ell_{\phi,c}|\mathcal{A}_{1,i}| + 4\ell_{\phi,c}\|\boldsymbol{\ell}_{\theta,c}\|^2\lambda_{\max}(\mathcal{B}_{1,i}) + 8\ell_{\phi,c}|\boldsymbol{\ell}_{\theta,c}^\top||\mathcal{C}_{1,i}| + 2N|\mathcal{D}_{1,i}| - 2N|\boldsymbol{\ell}_{\theta,c}^\top||\mathcal{F}_{1,i}|\right\}$, $|\boldsymbol{a}_{2,3}^{(1)}| \leq \boldsymbol{\ell}_{a_{23}}^{(1)}$ $:= \max_{c,i}\left\{4\ell_{\phi,c}(|\mathcal{B}_{1,i}|\boldsymbol{\ell}_{\theta,c} + \mathcal{C}_{1,i}) + 2N\mathcal{F}_{1,i}\right\}$, and $|\boldsymbol{a}_{1,3}^{(1)}| \leq \boldsymbol{\ell}_{a_{13}}^{(1)} := \max_{c,i}\big\{4\ell_{\phi,c}^2|\mathcal{B}_{1,i}||\boldsymbol{\ell}_{\theta,c}| + 4\ell_{\phi,c}^2|\mathcal{C}_{1,i}| + 2N\ell_{\phi,c}|\mathcal{F}_{1,i}|\big\}$. Therefore, Eq. (B.135) is implied by

$$\rho_u \geq \max_{d,c,i}\left\{2\ell_{\phi,c}^2|\mathcal{A}_{1,i}| + 2\ell_{\phi,c}^2\|\boldsymbol{\ell}_{\theta,c}\|^2\lambda_{\max}(\mathcal{B}_{1,i}) + 4\ell_{\phi,c}^2\left|\boldsymbol{\ell}_{\theta,c}^\top\right||\mathcal{C}_{1,i}| + \ell_{a_{12}}^{(1)} + \boldsymbol{\ell}_{a_{13}}^\top\imath_{d,d_x}\right.$$
$$\left. +(d+2)\max\left(\ell_{a_{12}}^{(1)}, \boldsymbol{\ell}_{a_{13}}^{(1)\top}\right)\right\}. \tag{B.138}$$

Eq. (B.136) is implied by

$$\rho_\phi \geq N\max_{d,c,i}\left\{2|\mathcal{A}_{1,i}| + 2\|\boldsymbol{\ell}_{\theta,c}\|^2\lambda_{\max}(\mathcal{B}_{1,i}) + 4\boldsymbol{\ell}_{\theta,c}^\top|\mathcal{C}_{1,i}| + \ell_{a_{12}}^{(1)} + \boldsymbol{\ell}_{a_{23}}^{(1)\top}\imath_{d,d_x}\right.$$
$$\left. +(d+2)\max(\ell_{a_{12}}^{(1)}, \boldsymbol{\ell}_{a_{23}}^{(1)\top})\right\}. \tag{B.139}$$

Eq. (B.137) is implied by

$$\rho_\theta\iota_{d_x} \geq N\max_{c,i}\left\{2u_{i,c}^2\ell_{\phi,c}^2|\mathcal{B}_{1,i}|\iota_{d_x} + \boldsymbol{\ell}_{a_{13}}^{(1)} + \boldsymbol{\ell}_{a_{23}}^{(1)} + (d_x+2)\max\left(\boldsymbol{\ell}_{a_{13}}^{(1)}, \boldsymbol{\ell}_{a_{23}}^{(1)}, 2\ell_{\phi,c}^2|\mathcal{B}_{1,i}|\right)\right\}. \tag{B.140}$$

■

**Lemma 14** *The function*

$$\mathcal{H}_{2,N,T}(\boldsymbol{\gamma}, \boldsymbol{U}) := \frac{1}{N^2}\sum_{c=1}^{G}\sum_{i\neq j}^{N}\frac{1}{2}\left(\rho_u\frac{(u_{i,c}^2 + u_{j,c}^2)}{N-1} + \rho_\phi\frac{\phi_c^2}{N(N-1)} + \rho_\theta\frac{\boldsymbol{\theta}_c^\top\boldsymbol{\theta}_c}{N(N-1)}\right) - \mathcal{E}_{2,N,T}(\boldsymbol{\gamma}, \boldsymbol{U})$$

*is convex for every* $\boldsymbol{\rho} := (\rho_u, \rho_\phi, \rho_\theta)$ *satisfying* (B.146)-(B.150).

**Proof of Lemma 14.** Write $\mathcal{H}_{2,N,T}(\boldsymbol{\gamma}, \boldsymbol{U}) = \frac{1}{N^2}\sum_{c=1}^{G}\sum_{i\neq j}^{N} \mathfrak{f}_{2,N,T}^{(c,i,j)}(\boldsymbol{\gamma}, \boldsymbol{U})$, where

$$\mathfrak{f}_{2,N,T}^{(c,i,j)}(\boldsymbol{\gamma}, \boldsymbol{U}) := \frac{1}{2}\left(\rho_u \frac{(u_{i,c}^2 + u_{j,c}^2)}{N-1} + \rho_\phi \frac{\phi_c^2}{N(N-1)} + \rho_\theta \frac{\boldsymbol{\theta}_c^\top \boldsymbol{\theta}_c}{N(N-1)}\right) - u_{i,c}u_{j,c}\phi_c^2 \mathcal{A}_{2,i,j} + u_{i,c}u_{j,c}\phi_c^2 \boldsymbol{\theta}_c^\top \mathcal{B}_{2,i,j} - u_{i,c}u_{j,c}\phi_c^2 \boldsymbol{\theta}_c^\top \mathcal{C}_{2,i,j}\boldsymbol{\theta}_c.$$

One then needs to prove that $\mathfrak{f}_{2,N,T}^{(c,i,j)}(\boldsymbol{\gamma}, \boldsymbol{U})$ is a convex function for some sufficiently large $\boldsymbol{\rho}$. The second-order partial derivatives of $\mathfrak{f}_{2,N,T}^{(c,i,j)}(\boldsymbol{\gamma}, \boldsymbol{U})$ can be derived in a straight-forward manner.

$$\begin{aligned}
a_{1,1}^{(2)} &:= D_{u_{i,c}}^2 \mathfrak{f}_{2,N,T}^{(c,i,j)}(\boldsymbol{\gamma}, \boldsymbol{U}) = \frac{\rho_u}{N-1},\\
a_{1,2}^{(2)} &:= D_{u_{i,c}u_{j,c}}^2 \mathfrak{f}_{2,N,T}^{(c,i,j)}(\boldsymbol{\gamma}, \boldsymbol{U}) = \phi_c^2\left(\boldsymbol{\theta}_c^\top \mathcal{B}_{2,i,j} - \mathcal{A}_{2,i,j} - \boldsymbol{\theta}_c^\top \mathcal{C}_{2,i,j}\boldsymbol{\theta}_c\right) =: a_{2,1}^{(2)},\\
a_{1,3}^{(2)} &:= D_{u_{i,c}\phi_c}^2 \mathfrak{f}_{2,N,T}^{(c,i,j)}(\boldsymbol{\gamma}, \boldsymbol{U}) = 2u_{j,c}\phi_c\left(\boldsymbol{\theta}_c^\top \mathcal{B}_{2,i,j} - \mathcal{A}_{2,i,j} - \boldsymbol{\theta}_c^\top \mathcal{C}_{2,i,j}\boldsymbol{\theta}_c\right) =: a_{3,1}^{(2)},\\
\boldsymbol{a}_{1,4}^{(2)} &:= D_{u_{i,c}\boldsymbol{\theta}_c}^2 \mathfrak{f}_{2,N,T}^{(c,i,j)}(\boldsymbol{\gamma}, \boldsymbol{U}) = u_{j,c}\phi_c^2(\mathcal{B}_{2,i,j} - 2\mathcal{C}_{2,i,j}\boldsymbol{\theta}_c) =: \boldsymbol{a}_{4,1}^{(2)},\\
a_{2,2}^{(2)} &:= D_{u_{j,c}}^2 \mathfrak{f}_{2,N,T}^{(c,i,j)}(\boldsymbol{\gamma}, \boldsymbol{U}) = \frac{\rho_u}{N-1},\\
a_{2,3}^{(2)} &:= D_{u_{j,c}\phi_c}^2 \mathfrak{f}_{2,N,T}^{(c,i,j)}(\boldsymbol{\gamma}, \boldsymbol{U}) = 2u_{i,c}\phi_c\left(\boldsymbol{\theta}_c^\top \mathcal{B}_{2,i,j} - \mathcal{A}_{2,i,j} - \boldsymbol{\theta}_c^\top \mathcal{C}_{2,i,j}\boldsymbol{\theta}_c\right) =: a_{3,2}^{(2)},\\
\boldsymbol{a}_{2,4}^{(2)} &:= D_{u_{j,c}\boldsymbol{\theta}_c}^2 \mathfrak{f}_{2,N,T}^{(c,i,j)}(\boldsymbol{\gamma}, \boldsymbol{U}) = u_{i,c}\phi_c^2\left(\mathcal{B}_{2,i,j} - 2\mathcal{C}_{2,i,j}\boldsymbol{\theta}_c\right) =: \boldsymbol{a}_{4,2}^{(2)},\\
a_{3,3}^{(2)} &:= D_{\phi_c}^2 \mathfrak{f}_{2,N,T}^{(c,i,j)}(\boldsymbol{\gamma}, \boldsymbol{U}) = \frac{\rho_\phi}{N(N-1)},\\
\boldsymbol{a}_{3,4}^{(2)} &:= D_{\phi_c\boldsymbol{\theta}_c}^2 \mathfrak{f}_{2,N,T}^{(c,i,j)}(\boldsymbol{\gamma}, \boldsymbol{U}) = 2u_{i,c}u_{j,c}\phi_c\left(\mathcal{B}_{2,i,j} - 2\mathcal{C}_{2,i,j}\boldsymbol{\theta}_c\right) =: \boldsymbol{a}_{4,3}^{(2)},\\
\boldsymbol{a}_{4,4}^{(2)} &:= D_{\boldsymbol{\theta}_c}^2 \mathfrak{f}_{2,N,T}^{(c,i,j)}(\boldsymbol{\gamma}, \boldsymbol{U}) = \frac{\rho_\theta}{N(N-1)}\mathbb{I}_{d_x} - 2u_{i,c}u_{j,c}\phi_c^2 \mathcal{C}_{2,i,j}.
\end{aligned}$$

Let $\mathcal{H}_2 := \begin{pmatrix} a_{1,1}^{(2)} & a_{1,2}^{(2)} & a_{1,3}^{(2)} & \boldsymbol{a}_{1,4}^{(2)\top} \\ a_{2,1}^{(2)} & a_{2,2}^{(2)} & a_{2,3}^{(2)} & \boldsymbol{a}_{2,3}^{(2)\top} \\ a_{3,1}^{(2)} & a_{3,2}^{(2)} & a_{3,3}^{(2)} & \boldsymbol{a}_{3,4}^{(2)\top} \\ \boldsymbol{a}_{4,1}^{(2)} & \boldsymbol{a}_{4,2}^{(2)} & \boldsymbol{a}_{4,3}^{(2)} & \boldsymbol{a}_{4,4}^{(2)} \end{pmatrix}$ be the Hessian matrix of $\mathfrak{f}_{2,N,T}^{(c,i,j)}(\boldsymbol{\gamma}, \boldsymbol{U})$. Similar to the proof of Lemma 13, all the sub-matrices of the Hessian matrix $\mathcal{H}_2$ are positively definite if the following conditions hold: for every $d = 0, \ldots, d_x$,

$$a_{1,1}^{(2)}a_{2,2}^{(2)} - a_{1,2}^{(2)}a_{2,1}^{(2)} > 0, \tag{B.141}$$

$$\frac{1}{d+3}(a_{1,1}^{(2)} + a_{1,2}^{(2)} + a_{1,3}^{(3)} + \boldsymbol{a}_{1,4}^{(2)\top}\imath_{d,d_x}) > \max\left(0, a_{1,2}^{(2)}, a_{1,3}^{(2)}, \boldsymbol{a}_{1,4}^{(2)\top}\right), \tag{B.142}$$

$$\frac{1}{d+3}(a_{2,1}^{(2)} + a_{2,2}^{(2)} + a_{2,3}^{(2)} + \boldsymbol{a}_{2,4}^{(2)\top}\imath_{d,d_x}) > \max\left(0, a_{2,1}^{(2)}, a_{2,3}^{(2)}, \boldsymbol{a}_{2,4}^{(2)\top}\right), \tag{B.143}$$

$$\frac{1}{d+3}(a_{3,1}^{(2)} + a_{3,2}^{(2)} + a_{3,3}^{(2)} + \boldsymbol{a}_{3,4}^{(2)\top}\imath_{d,d_x}) > \max\left(0, a_{3,1}^{(2)}, a_{3,2}^{(2)}, \boldsymbol{a}_{3,4}^{(2)\top}\right), \tag{B.144}$$

$$\frac{1}{d+3}\left(\mathbb{S}_{d,d_x}\boldsymbol{a}_{4,1}^{(2)}+\mathbb{S}_{d,d_x}\boldsymbol{a}_{4,2}^{(2)}+\mathbb{S}_{d,d_x}\boldsymbol{a}_{4,3}^{(2)}+\mathbb{S}_{d,d_x}\boldsymbol{a}_{4,4}^{(2)}\imath_{d,d_x}\right)>\max\left(0,\boldsymbol{a}_{4,1}^{(2)},\boldsymbol{a}_{4,2}^{(2)},\boldsymbol{a}_{4,3}^{(2)},\boldsymbol{a}_{4,4}^{*(2)}\right), \quad \text{(B.145)}$$

where $\boldsymbol{a}_{4,4}^{*(2)}$ is the matrix with zero diagonal elements and the off-diagonal elements are the same as those of $\boldsymbol{a}_{4,4}^{(2)}$; and $\underbrace{\mathbb{S}_{d,d_x}}_{d\times d_x}:=\begin{pmatrix}1&0&0&0&\underbrace{\cdots}_{d_x-5}&0\\0&1&0&0&\cdots&0\\0&0&1&0&\cdots&0\\\vdots&\vdots&\vdots&\vdots&\ddots&\vdots\\0&0&0&0&\cdots&1\end{pmatrix}$ is the selection matrix.

The relation (B.141) is valid when

$$\rho_u\geq(N-1)\max_{c,i,j}\left(\ell_{\phi,c}^2|\mathcal{A}_{2,i,j}|+\ell_{\phi,c}^2|\boldsymbol{\ell}_{\theta,c}^\top|\mathcal{B}_{2,i,j}|+\ell_{\phi,c}^2\boldsymbol{\ell}_{\theta,c}^\top|\mathcal{C}_{2,i,j}|\boldsymbol{\ell}_{\theta,c}\right). \quad \text{(B.146)}$$

Moreover, notice that

$$\begin{aligned}
|a_{1,2}^{(2)}| &\leq \ell_{a_{12}}^{(2)}:=\max_{c,i,j}\left\{\ell_{\phi,c}^2\left(\mathcal{A}_{2,i,j}+|\boldsymbol{\ell}_{\theta,c}^\top||\mathcal{B}_{2,i,j}|+\boldsymbol{\ell}_{\theta,c}^\top|\mathcal{C}_{3,i,j}|\boldsymbol{\ell}_{\theta,c}\right)\right\};\\
|a_{1,3}^{(2)}| &\leq \ell_{a_{13}}^{(2)}:=2\max_{c,i,j}\left\{\ell_{\phi,c}\left(\mathcal{A}_{2,i,j}+\boldsymbol{\ell}_{\theta,c}^\top|\mathcal{B}_{2,i,j}|+\boldsymbol{\ell}_{\theta,c}^\top|\mathcal{C}_{2,i,j}|\boldsymbol{\ell}_{\theta,c}\right)\right\};\\
|a_{2,3}^{(2)}| &\leq \ell_{a_{23}}^{(2)}:=2\max_{c,i,j}\left\{\ell_{\phi,c}\left(|\mathcal{A}_{2,i,j}|+\boldsymbol{\ell}_{\theta,c}^\top|\mathcal{B}_{2,i,j}|+\boldsymbol{\ell}_{\theta,c}^\top|\mathcal{C}_{2,i,j}|\boldsymbol{\ell}_{\theta,c}\right)\right\};\\
|\boldsymbol{a}_{1,4}^{(2)}| &\leq \boldsymbol{\ell}_{a_{14}}^{(2)}:=\max_{c,i,j}\{\ell_{\phi,c}^2(|\mathcal{A}_{2,i,j}|+2|\mathcal{C}_{2,i,j}|\boldsymbol{\ell}_{\theta,c})\};\\
|\boldsymbol{a}_{2,4}^{(2)}| &\leq \boldsymbol{\ell}_{a_{24}}^{(2)}=\boldsymbol{\ell}_{a_{14}}^{(2)};\\
|\boldsymbol{a}_{3,4}^{(2)}| &\leq \boldsymbol{\ell}_{a_{34}}^{(2)}:=2\max_{c,i,j}\left\{\ell_{\phi,c}\left(|\mathcal{B}_{2,i,j}|+2|\mathcal{C}_{2,i,j}|\boldsymbol{\ell}_{\theta,c}\right)\right\};\\
|\boldsymbol{a}_{4,4}^{*(2)}| &\leq 2\max_{c,i,j}\{\ell_{\phi,c}^2|\mathcal{C}_{2,i,j}^*|\}\text{ with }\mathcal{C}_{2,i,j}^*\text{ is }\mathcal{C}_{2,i,j}\text{ except for zero diagonal elements.}
\end{aligned}$$

Therefore, it follows that (B.142) holds if

$$\rho_u\geq(N-1)\left\{\ell_{a_{12}}^{(2)}+\ell_{a_{13}}^{(2)}+\boldsymbol{\ell}_{a_{14}}^{(2)}\imath_{d,d_x}+(d+3)\max(\ell_{a_{12}}^{(2)},\ell_{a_{13}}^{(2)},\boldsymbol{\ell}_{a_{14}}^{(2)\top})\right\}; \quad \text{(B.147)}$$

(B.143) holds if

$$\rho_u\geq(N-1)\left\{\ell_{a_{12}}^{(2)}+\ell_{a_{23}}^{(2)}+\boldsymbol{\ell}_{a_{14}}^{(2)\top}\imath_{d,d_x}+(d+3)\max\left(\ell_{a_{21}}^{(2)},\ell_{a_{23}}^{(2)},\boldsymbol{\ell}_{a_{14}}^{(2)\top}\right)\right\}; \quad \text{(B.148)}$$

(B.144) holds if

$$\rho_\phi\geq N(N-1)\left\{\ell_{a_{13}}^{(2)}+\ell_{a_{23}}^{(2)}+\boldsymbol{\ell}_{a_{34}}^{(2)\top}\imath_{d,d_x}+(d+3)\max\left(\ell_{a_{13}}^{(2)},\ell_{a_{23}}^{(2)},\boldsymbol{\ell}_{a_{34}}^{(2)\top}\right)\right\}; \quad \text{(B.149)}$$

and, finally, (B.145) holds if

$$\rho_\theta\geq N(N-1)\max_{c,i,j}\|2\ell_{\phi,c}^2\mathbb{S}_{d,d_x}|\mathcal{C}_{2,i,j}|\imath_{d,d_x}+\mathbb{S}_{d,d_x}\boldsymbol{\ell}_{a_{14}}^{(2)}+\mathbb{S}_{d,d_x}\boldsymbol{\ell}_{a_{24}}^{(2)}+\mathbb{S}_{d,d_x}\boldsymbol{\ell}_{a_{34}}^{(2)}$$

$$+ (d+3)\max\left(0, \boldsymbol{\ell}^{(2)}_{a_{14}}, \boldsymbol{\ell}^{(2)}_{a_{24}}, \boldsymbol{\ell}^{(2)}_{a_{34}}, 2\ell^2_{\phi,c}|\mathcal{C}^*_{2,i,j}|\right)\|_\infty, \tag{B.150}$$

where $\|\cdot\|_\infty$ is the matrix sup-norm. ■

**Lemma 15** *The function*

$$\begin{aligned}\mathcal{H}_{3,N,T}(\boldsymbol{\gamma},\boldsymbol{U}) \coloneqq \frac{1}{N^2}\sum_{c\neq\ell}^{G}\sum_{i\neq j}^{N}\frac{1}{2}\Bigg(&\rho_u\left(\frac{u^2_{i,c}}{(G-1)(N-1)}+\frac{u^2_{j,\ell}}{(G-1)(N-1)}\right)+\rho_\phi\left(\frac{\phi^2_c}{(G-1)N(N-1)}\right.\\ &\left.+\frac{\phi^2_\ell}{(G-1)N(N-1)}\right)+\rho_\theta\left(\frac{\boldsymbol{\theta}_c^\top\boldsymbol{\theta}_c}{(G-1)N(N-1)}+\frac{\boldsymbol{\theta}_\ell^\top\boldsymbol{\theta}_\ell}{(G-1)N(N-1)}\right)\Bigg)\\ &-\mathcal{E}_{3,N,T}(\boldsymbol{\gamma},\boldsymbol{U})\end{aligned}$$

*is convex for every* $\boldsymbol{\rho} \coloneqq (\rho_u, \rho_\phi, \rho_\theta)$ *satisfying* (B.151), (B.165)-(B.177).

**Proof of Lemma 15.** Write $\mathcal{H}_{3,N,T}(\boldsymbol{\gamma},\boldsymbol{U}) = \frac{1}{N^2}\sum_{c\neq\ell}^{G}\sum_{i\neq j}^{N}\mathfrak{f}^{(c,\ell,i,j)}_{3,N,T}(\boldsymbol{\gamma},\boldsymbol{U})$, where

$$\begin{aligned}\mathfrak{f}^{(c,\ell,i,j)}_{3,N,T}(\boldsymbol{\gamma},\boldsymbol{U}) \coloneqq \frac{1}{2}\Bigg\{&\rho_u\left(\frac{u^2_{i,c}}{(G-1)(N-1)}+\frac{u^2_{j,\ell}}{(G-1)(N-1)}\right)\\ &+\rho_\phi\left(\frac{\phi^2_c}{(G-1)N(N-1)}+\frac{\phi^2_\ell}{(G-1)N(N-1)}\right)\\ &+\rho_\theta\left(\frac{\boldsymbol{\theta}_c^\top\boldsymbol{\theta}_c}{(G-1)N(N-1)}+\frac{\boldsymbol{\theta}_\ell^\top\boldsymbol{\theta}_\ell}{(G-1)N(N-1)}\right)\Bigg\}\\ &-u_{i,c}u_{j,\ell}\phi_c\phi_\ell\mathcal{A}_{2,i,j}+u_{i,c}u_{j,\ell}\phi_c\phi_\ell\boldsymbol{\theta}_\ell^\top\mathcal{B}_{3,i,j}+u_{i,c}u_{j,\ell}\phi_c\phi_\ell\boldsymbol{\theta}_c^\top\mathcal{C}_{3,i,j}\\ &-u_{i,c}u_{j,\ell}\phi_c\phi_\ell\boldsymbol{\theta}_c^\top\mathcal{C}_{2,i,j}\boldsymbol{\theta}_\ell.\end{aligned}$$

We then need to find conditions to ensure the convexity of $\mathfrak{f}^{(c,\ell,i,j)}_{3,N,T}(\boldsymbol{\gamma},\boldsymbol{U})$. To start with, let's compute the second-order partial derivatives:

$$\begin{aligned}
a^{(3)}_{1,1} &\coloneqq D^2_{u_{i,c}}\mathfrak{f}^{(c,\ell,i,j)}_{3,N,T}(\boldsymbol{\gamma},\boldsymbol{U}) = \frac{\rho_u}{(G-1)(N-1)},\\
a^{(3)}_{1,3} &\coloneqq D^2_{u_{i,c}\phi_c}\mathfrak{f}^{(c,\ell,i,j)}_{3,N,T}(\boldsymbol{\gamma},\boldsymbol{U}) = -u_{j,\ell}\phi_\ell\mathcal{A}_{2,i,j}+u_{j,\ell}\phi_\ell\left(\boldsymbol{\theta}_\ell^\top\mathcal{B}_{3,i,j}+\boldsymbol{\theta}_c^\top\mathcal{C}_{3,i,j}\right)-u_{j,\ell}\phi_\ell\boldsymbol{\theta}_c^\top\mathcal{C}_{2,i,j}\boldsymbol{\theta}_\ell \eqqcolon a^{(3)}_{3,2},\\
a^{(3)}_{1,4} &\coloneqq D^2_{u_{i,c}\phi_\ell}\mathfrak{f}^{(c,\ell,i,j)}_{3,N,T}(\boldsymbol{\gamma},\boldsymbol{U}) = -u_{j,\ell}\phi_c\mathcal{A}_{2,i,j}+u_{j,\ell}\phi_c\left(\boldsymbol{\theta}_\ell^\top\mathcal{B}_{3,i,j}+\boldsymbol{\theta}_c^\top\mathcal{C}_{3,i,j}\right)-u_{j,\ell}\phi_c\boldsymbol{\theta}_c^\top\mathcal{C}_{2,i,j}\boldsymbol{\theta}_\ell \eqqcolon a^{(3)}_{4,1},\\
\boldsymbol{a}^{(3)}_{1,5} &\coloneqq D^2_{u_{i,c}\boldsymbol{\theta}_c}\mathfrak{f}^{(c,\ell,i,j)}_{3,N,T}(\boldsymbol{\gamma},\boldsymbol{U}) = u_{j,\ell}\phi_c\phi_\ell\mathcal{C}_{3,i,j}-u_{j,\ell}\phi_c\phi_\ell\mathcal{C}_{2,i,j}\boldsymbol{\theta}_\ell \eqqcolon \boldsymbol{a}^{(3)}_{5,1},\\
\boldsymbol{a}^{(3)}_{1,6} &\coloneqq D^2_{u_{i,c}\boldsymbol{\theta}_\ell}\mathfrak{f}^{(c,\ell,i,j)}_{3,N,T}(\boldsymbol{\gamma},\boldsymbol{U}) = u_{j,\ell}\phi_c\phi_\ell\left(\mathcal{B}_{3,i,j}-\mathcal{C}_{2,i,j}\boldsymbol{\theta}_c\right) \eqqcolon \boldsymbol{a}^{(3)}_{6,1},\\
a^{(3)}_{2,2} &\coloneqq D^2_{u_{j,\ell}}\mathfrak{f}^{(c,\ell,i,j)}_{3,N,T}(\boldsymbol{\gamma},\boldsymbol{U}) = \frac{\rho_u}{(G-1)(N-1)},\\
a^{(3)}_{1,2} &\coloneqq D^2_{u_{j,\ell},u_{i,c}}\mathfrak{f}^{(c,\ell,i,j)}_{3,N,T}(\boldsymbol{\gamma},\boldsymbol{U}) = -\phi_c\phi_\ell\left(\mathcal{A}_{2,i,j}-\boldsymbol{\theta}_\ell^\top\mathcal{B}_{3,i,j}-\boldsymbol{\theta}_c^\top\mathcal{C}_{3,i,j}+\boldsymbol{\theta}_c^\top\mathcal{C}_{2,i,j}\boldsymbol{\theta}_\ell\right) \eqqcolon a^{(3)}_{2,1},\\
a^{(3)}_{2,3} &\coloneqq D^2_{u_{j,\ell},\phi_c}\mathfrak{f}^{(c,\ell,i,j)}_{3,N,T}(\boldsymbol{\gamma},\boldsymbol{U}) = -u_{i,c}\phi_\ell\mathcal{A}_{2,i,j}+u_{i,c}\phi_\ell\left(\boldsymbol{\theta}_\ell^\top\mathcal{B}_{3,i,j}+\boldsymbol{\theta}_c^\top\mathcal{C}_{3,i,j}\right)-u_{i,c}\phi_\ell\boldsymbol{\theta}_c^\top\mathcal{C}_{2,i,j}\boldsymbol{\theta}_\ell \eqqcolon a^{(3)}_{3,2},
\end{aligned}$$

$$
\begin{aligned}
a_{2,4}^{(3)} &\coloneqq D^2_{u_{j,\ell},\phi_\ell}\mathfrak{f}^{(c,\ell,i,j)}_{3,N,T}(\boldsymbol{\gamma},\boldsymbol{U}) = -u_{i,c}\phi_c\mathcal{A}_{2,i,j} + u_{i,c}\phi_c\left(\boldsymbol{\theta}_\ell^\top\mathcal{B}_{3,i,j} + \boldsymbol{\theta}_c^\top\mathcal{C}_{3,i,j}\right) - u_{i,c}\phi_c\boldsymbol{\theta}_c^\top\mathcal{C}_{2,i,j}\boldsymbol{\theta}_\ell \eqqcolon a_{4,2}^{(3)},\\
\boldsymbol{a}_{2,5}^{(3)} &\coloneqq D^2_{u_{j,\ell},\boldsymbol{\theta}_c}\mathfrak{f}^{(c,\ell,i,j)}_{3,N,T}(\boldsymbol{\gamma},\boldsymbol{U}) = u_{i,c}\phi_c\phi_\ell\left(\mathcal{C}_{3,i,j} - \mathcal{C}_{2,i,j}\boldsymbol{\theta}_\ell\right) \eqqcolon \boldsymbol{a}_{5,2}^{(3)},\\
\boldsymbol{a}_{2,6}^{(3)} &\coloneqq D^2_{u_{j,\ell},\boldsymbol{\theta}_\ell}\mathfrak{f}^{(c,\ell,i,j)}_{3,N,T}(\boldsymbol{\gamma},\boldsymbol{U}) = u_{i,c}\phi_c\phi_\ell\left(\mathcal{B}_{3,i,j} - \mathcal{C}_{2,i,j}\boldsymbol{\theta}_c\right) \eqqcolon \boldsymbol{a}_{6,2}^{(3)},\\
a_{3,4}^{(3)} &\coloneqq D^2_{\phi_c,\phi_\ell}\mathfrak{f}^{(c,\ell,i,j)}_{3,N,T}(\boldsymbol{\gamma},\boldsymbol{U}) = -u_{i,c}u_{j,\ell}\left(\mathcal{A}_{2,i,j} - \boldsymbol{\theta}_\ell^\top\mathcal{B}_{3,i,j} - \boldsymbol{\theta}_c^\top\mathcal{C}_{3,i,j} + \boldsymbol{\theta}_c^\top\mathcal{C}_{2,i,j}\boldsymbol{\theta}_\ell\right),\\
a_{3,3}^{(3)} &\coloneqq D^2_{\phi_c}\mathfrak{f}^{(c,\ell,i,j)}_{3,N,T}(\boldsymbol{\gamma},\boldsymbol{U}) = \frac{\rho_\phi}{(G-1)N(N-1)},\\
\boldsymbol{a}_{3,5}^{(3)} &\coloneqq D^2_{\phi_c,\boldsymbol{\theta}_c}\mathfrak{f}^{(c,\ell,i,j)}_{3,N,T}(\boldsymbol{\gamma},\boldsymbol{U}) = u_{i,c}u_{j,\ell}\phi_\ell\left(\mathcal{C}_{3,i,j} - \mathcal{C}_{2,i,j}\boldsymbol{\theta}_\ell\right) \eqqcolon \boldsymbol{a}_{5,3}^{(3)},\\
\boldsymbol{a}_{3,6}^{(3)} &\coloneqq D^2_{\phi_c,\boldsymbol{\theta}_\ell}\mathfrak{f}^{(c,\ell,i,j)}_{3,N,T}(\boldsymbol{\gamma},\boldsymbol{U}) = u_{i,c}u_{j,\ell}\phi_\ell\left(\mathcal{B}_{3,i,j} - \mathcal{C}_{2,i,j}\boldsymbol{\theta}_c\right) \eqqcolon \boldsymbol{a}_{6,3}^{(3)},\\
\boldsymbol{a}_{4,5}^{(3)} &\coloneqq D^2_{\phi_\ell,\boldsymbol{\theta}_c}\mathfrak{f}^{(c,\ell,i,j)}_{3,N,T}(\boldsymbol{\gamma},\boldsymbol{U}) = u_{i,c}u_{j,\ell}\phi_c\left(\mathcal{C}_{3,i,j} - \mathcal{C}_{2,i,j}\boldsymbol{\theta}_\ell\right) \eqqcolon \boldsymbol{a}_{5,4}^{(3)},\\
\boldsymbol{a}_{4,4}^{(3)} &\coloneqq D^2_{\phi_\ell}\mathfrak{f}^{(c,\ell,i,j)}_{3,N,T}(\boldsymbol{\gamma},\boldsymbol{U}) = \frac{\rho_\phi}{(G-1)N(N-1)},\\
\boldsymbol{a}_{4,6}^{(3)} &\coloneqq D^2_{\phi_\ell,\boldsymbol{\theta}_\ell}\mathfrak{f}^{(c,\ell,i,j)}_{3,N,T}(\boldsymbol{\gamma},\boldsymbol{U}) = u_{i,c}u_{j,\ell}\phi_c\left(\mathcal{B}_{3,i,j} - \mathcal{C}_{2,i,j}\boldsymbol{\theta}_c\right) \eqqcolon \boldsymbol{a}_{6,4}^{(3)},\\
\boldsymbol{a}_{5,5}^{(3)} &\coloneqq D^2_{\boldsymbol{\theta}_c,\boldsymbol{\theta}_c}\mathfrak{f}^{(c,\ell,i,j)}_{3,N,T}(\boldsymbol{\gamma},\boldsymbol{U}) = \frac{\rho_\theta}{(G-1)N(N-1)}\mathbb{I}_{d_x},\\
\boldsymbol{a}_{5,6}^{(3)} &\coloneqq D^2_{\boldsymbol{\theta}_c,\boldsymbol{\theta}_\ell}\mathfrak{f}^{(c,\ell,i,j)}_{3,N,T}(\boldsymbol{\gamma},\boldsymbol{U}) = u_{i,c}u_{j,\ell}\phi_c\phi_\ell\mathcal{C}_{2,i,j} \eqqcolon \boldsymbol{a}_{6,5}^{(3)},\\
\boldsymbol{a}_{6,6}^{(3)} &\coloneqq D^2_{\boldsymbol{\theta}_\ell,\boldsymbol{\theta}_\ell}\mathfrak{f}^{(c,\ell,i,j)}_{3,N,T}(\boldsymbol{\gamma},\boldsymbol{U}) = \frac{\rho_\theta}{(G-1)N(N-1)}\mathbb{I}_{d_x}.
\end{aligned}
$$

Let's denote by $\mathcal{H}_3 \coloneqq \begin{pmatrix} a_{1,1}^{(3)} & a_{1,2}^{(3)} & a_{1,3}^{(3)} & a_{1,4}^{(3)} & \boldsymbol{a}_{1,5}^{(3)\top} & \boldsymbol{a}_{1,6}^{(3)\top} \\ a_{2,1}^{(3)} & a_{2,2}^{(3)} & a_{2,3}^{(3)} & a_{2,4}^{(3)} & \boldsymbol{a}_{2,5}^{(3)\top} & \boldsymbol{a}_{2,6}^{(3)\top} \\ a_{3,1}^{(3)} & a_{3,2}^{(3)} & a_{3,3}^{(3)} & a_{3,4}^{(3)} & \boldsymbol{a}_{3,5}^{(3)\top} & \boldsymbol{a}_{3,6}^{(3)\top} \\ a_{4,1}^{(3)} & a_{4,2}^{(3)} & a_{4,3}^{(3)} & a_{4,4}^{(3)} & \boldsymbol{a}_{4,5}^{(3)\top} & \boldsymbol{a}_{4,6}^{(3)\top} \\ \boldsymbol{a}_{1,5}^{(3)} & \boldsymbol{a}_{2,5}^{(3)} & \boldsymbol{a}_{3,5}^{(3)} & \boldsymbol{a}_{4,5}^{(3)} & \boldsymbol{a}_{5,5}^{(3)} & \boldsymbol{a}_{6,5}^{(3)\top} \\ \boldsymbol{a}_{1,6}^{(3)} & \boldsymbol{a}_{2,6}^{(3)} & \boldsymbol{a}_{3,6}^{(3)} & \boldsymbol{a}_{4,6}^{(3)} & \boldsymbol{a}_{5,6}^{(3)} & \boldsymbol{a}_{6,6}^{(3)} \end{pmatrix}$ the Hessian matrix of $\mathfrak{f}^{(c,\ell,i,j)}_{3,N,T}(\boldsymbol{\gamma},\boldsymbol{U})$. Since $\boldsymbol{\gamma}$ satisfies Assumption 1, one obtains that

$$
|a_{1,2}^{(3)}| \le \ell_{a_{12}}^{(3)} \coloneqq \max_{i,j,c,\ell}\left\{\ell_{\phi,c}\ell_{\phi,\ell}\left(|\mathcal{A}_{2,i,j}| + \boldsymbol{\ell}_{\theta,\ell}^\top|\mathcal{B}_{3,i,j}| + \boldsymbol{\ell}_{\theta,c}^\top|\mathcal{C}_{3,i,j}| + \boldsymbol{\ell}_{\theta,c}^\top|\mathcal{C}_{2,i,j}|\boldsymbol{\ell}_{\theta,\ell}\right)\right\};
$$

$$
|a_{1,3}^{(3)}| \le \ell_{a_{13}}^{(3)} \coloneqq \max_{i,j,c,\ell}\left\{\ell_{\phi,\ell}\left(|\mathcal{A}_{2,i,j}| + \boldsymbol{\ell}_{\theta,\ell}^\top|\mathcal{B}_{3,i,j}| + \boldsymbol{\ell}_{\theta,c}^\top|\mathcal{C}_{3,i,j}| + \boldsymbol{\ell}_{\theta,c}^\top|\mathcal{C}_{2,i,j}|\boldsymbol{\ell}_{\theta,\ell}\right)\right\};
$$

$$
|a_{2,3}^{(3)}| \le \ell_{a_{23}}^{(3)} \coloneqq \max_{i,j,c,\ell}\left\{\ell_{\phi,\ell}\left(|\mathcal{A}_{2,i,j}| + \boldsymbol{\ell}_{\theta,\ell}^\top|\mathcal{B}_{3,i,j}| + \boldsymbol{\ell}_{\theta,c}^\top|\mathcal{C}_{3,i,j}| + \boldsymbol{\ell}_{\theta,c}^\top|\mathcal{C}_{2,i,j}|\boldsymbol{\ell}_{\theta,\ell}\right)\right\};
$$

$$
|a_{1,4}^{(3)}| \le \ell_{a_{14}}^{(3)} \coloneqq \max_{i,j,c,\ell}\left\{\ell_{\phi,c}\left(|\mathcal{A}_{2,i,j}| + \boldsymbol{\ell}_{\theta,\ell}^\top|\mathcal{B}_{3,i,j}| + \boldsymbol{\ell}_{\theta,c}^\top|\mathcal{C}_{3,i,j}| + \boldsymbol{\ell}_{\theta,c}^\top|\mathcal{C}_{2,i,j}|\boldsymbol{\ell}_{\theta,\ell}\right)\right\};
$$

$$
|a_{2,4}^{(3)}| \le \ell_{a_{24}}^{(3)} \coloneqq \ell_{a_{14}}^{(3)};
$$

$$
|a_{3,4}^{(3)}| \le \ell_{a_{34}}^{(3)} \coloneqq \max_{i,j,c,\ell}\left\{|\mathcal{A}_{2,i,j}| + \boldsymbol{\ell}_{\theta,\ell}^\top|\mathcal{B}_{3,i,j}| + \boldsymbol{\ell}_{\theta,c}^\top|\mathcal{C}_{3,i,j}| + \boldsymbol{\ell}_{\theta,c}^\top|\mathcal{C}_{2,i,j}|\boldsymbol{\ell}_{\theta,\ell}\right\};
$$

$$
|\boldsymbol{a}_{1,5}^{(3)}| \le \boldsymbol{\ell}_{a_{15}}^{(3)} \coloneqq \max_{i,j,c,\ell}\left\{\ell_{\phi,c}\ell_{\phi,\ell}\left(|\mathcal{C}_{3,i,j}| + |\mathcal{C}_{2,i,j}|\boldsymbol{\ell}_{\theta,\ell}\right)\right\};
$$

$$|\boldsymbol{a}_{1,6}^{(3)}| \leq \boldsymbol{\ell}_{a16}^{(3)} := \max_{i,j,c,\ell} \{\ell_{\phi,c}\ell_{\phi,\ell} (|\mathcal{B}_{3,i,j}| + |\mathcal{C}_{2,i,j}|\boldsymbol{\ell}_{\theta,c})\};$$

$$|\boldsymbol{a}_{2,5}^{(3)}| \leq \boldsymbol{\ell}_{a25}^{(3)} := \boldsymbol{\ell}_{a15}^{(3)};\ |\boldsymbol{a}_{2,6}^{(3)}| \leq \boldsymbol{\ell}_{a26}^{(3)} := \boldsymbol{\ell}_{a16}^{(3)};\ |\boldsymbol{a}_{3,5}^{(3)}| \leq \boldsymbol{\ell}_{a35}^{(3)} := \frac{1}{\ell_{\phi,c}}\boldsymbol{\ell}_{a15}^{(3)}; |\boldsymbol{a}_{3,6}^{(3)}| \leq \boldsymbol{\ell}_{a36}^{(3)} := \frac{1}{\ell_{\phi,c}}\boldsymbol{\ell}_{a16}^{(3)};$$

$$|\boldsymbol{a}_{4,5}^{(3)}| \leq \boldsymbol{\ell}_{a45}^{(3)} := \frac{1}{\ell_{\phi,\ell}}\boldsymbol{\ell}_{a15}^{(3)};\ |\boldsymbol{a}_{4,6}^{(3)}| \leq \boldsymbol{\ell}_{a46}^{(3)} := \frac{1}{\ell_{\phi,\ell}}\boldsymbol{\ell}_{a16}^{(3)};\ |\boldsymbol{a}_{5,6}^{(3)}| \leq \boldsymbol{\ell}_{a56}^{(3)} := \max_{i,j,c,\ell}\{\ell_{\phi,c}\ell_{\phi,\ell}|\mathcal{C}_{2,i,j}|\}.$$

Invoking Lemma 25 the minimum eigenvalue of $\mathcal{H}_3$ is positive is implied by the following inequality constraints:

$$\rho_u \geq (G-1)(N-1) \max_{i,j,c,\ell} \{\phi_c\phi_\ell \left(|\mathcal{A}_{2,i,j}| + \boldsymbol{\ell}_{\theta,\ell}^\top|\mathcal{B}_{3,i,j}| + \boldsymbol{\ell}_{\theta,c}^\top|\mathcal{C}_{3,i,j}| + \boldsymbol{\ell}_{\theta,c}^\top|\mathcal{C}_{2,i,j}|\boldsymbol{\ell}_{\theta,\ell}\right)\}, \tag{B.151}$$

$$\frac{1}{3}(a_{1,1}^{(3)} + a_{1,2}^{(3)} + a_{1,3}^{(3)}) > \max(0, a_{1,2}^{(3)}, a_{1,3}^{(3)}), \tag{B.152}$$

$$\frac{1}{3}(a_{2,1}^{(3)} + a_{2,2}^{(3)} + a_{2,3}^{(3)}) > \max(0, a_{2,1}^{(3)}, a_{2,3}^{(3)}), \tag{B.153}$$

$$\frac{1}{3}(a_{3,1}^{(3)} + a_{3,2}^{(3)} + a_{3,3}^{(3)}) > \max(0, a_{3,1}^{(3)}, a_{3,2}^{(3)}), \tag{B.154}$$

$$\frac{1}{4}(a_{1,1}^{(3)} + a_{1,2}^{(3)} + a_{1,3}^{(3)} + a_{1,4}^{(3)}) > \max(0, a_{1,2}^{(3)}, a_{1,3}^{(3)}, a_{1,4}^{(3)}), \tag{B.155}$$

$$\frac{1}{4}(a_{2,1}^{(3)} + a_{2,2}^{(3)} + a_{2,3}^{(3)} + a_{2,4}^{(3)}) > \max(0, a_{2,1}^{(3)}, a_{2,3}^{(3)}, a_{2,4}^{(3)}), \tag{B.156}$$

$$\frac{1}{4}(a_{1,3}^{(3)} + a_{3,2}^{(3)} + a_{3,3}^{(3)} + a_{3,4}^{(4)}) > \max(0, a_{3,1}^{(3)}, a_{3,2}^{(3)}, a_{3,4}^{(3)}), \tag{B.157}$$

$$\frac{1}{4}(a_{4,1}^{(3)} + a_{4,2}^{(3)} + a_{4,3}^{(3)} + a_{4,4}^{(4)}) > \max(0, a_{4,1}^{(3)}, a_{4,2}^{(3)}, a_{4,3}^{(3)}), \tag{B.158}$$

$$\frac{1}{4+d+e}\left(a_{1,1}^{(3)} + a_{1,2}^{(3)} + a_{1,3}^{(3)} + a_{1,4}^{(3)} + \boldsymbol{a}_{1,5}^{(3)\top}\imath_{d,d_x} + \boldsymbol{a}_{1,6}^{(3)\top}\imath_{e,d_x}\right) > \max(0, a_{1,2}^{(3)}, a_{1,3}^{(3)}, a_{1,4}^{(3)}, \boldsymbol{a}_{1,5}^{(3)\top}, \boldsymbol{a}_{1,6}^{(3)\top}) \tag{B.159}$$

for all $d = 1, \ldots, d_x$ and $e = 1, \ldots, d_x$,

$$\frac{1}{4+d+e}\left(a_{2,1}^{(3)} + a_{2,2}^{(3)} + a_{2,3}^{(3)} + a_{2,4}^{(3)} + \boldsymbol{a}_{2,5}^{(3)\top}\imath_{d,d_x} + \boldsymbol{a}_{2,6}^{(3)\top}\imath_{e,d_x}\right) > \max\left(0, a_{2,1}^{(3)}, a_{2,3}^{(3)}, a_{2,4}^{(3)}, \boldsymbol{a}_{2,5}^{(3)\top}, \boldsymbol{a}_{2,6}^{(3)\top}\right) \tag{B.160}$$

for all $d = 1, \ldots, d_x$ and $e = 1, \ldots, d_x$,

$$\frac{1}{4+d+e}\left(a_{3,1}^{(3)} + a_{3,2}^{(3)} + a_{3,3}^{(3)} + a_{3,4}^{(3)} + \boldsymbol{a}_{3,5}^{(3)\top}\imath_{d,d_x} + \boldsymbol{a}_{3,6}^{(3)\top}\imath_{e,d_x}\right)$$

$$> \max\left(0, a_{3,1}^{(3)}, a_{3,2}^{(3)}, a_{3,4}^{(3)}, \boldsymbol{a}_{3,5}^{(3)\top}, \boldsymbol{a}_{3,6}^{(3)\top}\right) \quad \text{(B.161)}$$

for all $d = 1, \ldots, d_x$ and $e = 1, \ldots, d_x$,

$$\frac{1}{4+d+e}\left(a_{4,1}^{(3)} + a_{4,2}^{(3)} + a_{4,3}^{(3)} + a_{4,4}^{(3)} + \boldsymbol{a}_{4,5}^{(3)\top}\imath_{d,d_x} + \boldsymbol{a}_{4,6}^{(3)\top}\imath_{e,d_x}\right)$$
$$> \max\left(0, a_{4,1}^{(3)}, a_{4,2}^{(3)}, a_{4,3}^{(3)}, \boldsymbol{a}_{4,5}^{(3)\top}, \boldsymbol{a}_{4,6}^{(3)\top}\right) \quad \text{(B.162)}$$

for all $d = 1, \ldots, d_x$ and $e = 1, \ldots, d_x$,

$$\frac{1}{4+d+3}\left(\mathbb{S}_{d,d_x}\boldsymbol{a}_{1,5}^{(3)} + \mathbb{S}_{d,d_x}\boldsymbol{a}_{2,5}^{(3)} + \mathbb{S}_{d,d_x}\boldsymbol{a}_{3,5}^{(3)} + \mathbb{S}_{d,d_x}\boldsymbol{a}_{4,5}^{(3)} + \mathbb{S}_{d,d_x}\boldsymbol{a}_{5,5}^{(3)}\imath_{d,d_x} + \mathbb{S}_{d,d_x}\boldsymbol{a}_{6,5}^{(3)}\imath_{e,d_x}\right)$$
$$> \max\left(0, \mathbb{S}_{d,d_x}\boldsymbol{a}_{1,5}^{3}, \mathbb{S}_{d,d_x}\boldsymbol{a}_{2,5}^{(3)}, \mathbb{S}_{d,d_x}\boldsymbol{a}_{3,5}^{(3)}, \mathbb{S}_{d,d_x}\boldsymbol{a}_{4,5}^{(3)}, \mathbb{S}_{d,d_x}\boldsymbol{a}_{6,5}^{(3)}\right) \quad \text{(B.163)}$$

for all $d = 1, \ldots, d_x$ and $e = 1, \ldots, d_x$, and

$$\frac{1}{4+d+e}\left(\mathbb{S}_{e,d_x}\boldsymbol{a}_{1,6}^{(3)} + \mathbb{S}_{e,d_x}\boldsymbol{a}_{2,6}^{(3)} + \mathbb{S}_{e,d_x}\boldsymbol{a}_{3,6}^{(3)} + \mathbb{S}_{e,d_x}\boldsymbol{a}_{4,6}^{(3)\top} + \mathbb{S}_{e,d_x}\boldsymbol{a}_{6,5}^{(3)}\imath_{d,d_x} + \mathbb{S}_{e,d_x}\boldsymbol{a}_{6,6}^{(3)}\imath_{e,d_x}\right)$$
$$> \max\left(0, \mathbb{S}_{e,d_x}\boldsymbol{a}_{1,6}^{(3)}, \mathbb{S}_{e,d_x}\boldsymbol{a}_{2,6}^{(3)}, \mathbb{S}_{e,d_x}\boldsymbol{a}_{3,6}^{(3)}, \mathbb{S}_{e,d_x}\boldsymbol{a}_{4,6}^{(3)}, \mathbb{S}_{e,d_x}\boldsymbol{a}_{6,5}^{(3)}\right) \quad \text{(B.164)}$$

for all $d = 1, \ldots, d_x$ and $e = 1, \ldots, d_x$.

By some simple calculations, Eqs. (B.152)-(B.164) hold if the following conditions hold:

$$\rho_u \geq (G-1)(N-1)\left(\ell_{a_{12}}^{(3)} + \ell_{a_{13}}^{(3)} + 3\max(0, \ell_{a_{12}}^{(3)}, \ell_{a_{13}}^{(3)})\right); \quad \text{(B.165)}$$
$$\rho_u \geq (G-1)(N-1)\left(\ell_{a_{12}}^{(3)} + \ell_{a_{23}}^{(3)} + 3\max\left(\ell_{a_{12}}^{(3)}, \ell_{a_{23}}^{(3)}\right)\right); \quad \text{(B.166)}$$
$$\rho_\phi \geq (G-1)N(N-1)\left(\ell_{a_{13}}^{(3)} + \ell_{a_{23}}^{(3)} + 3\max\left(\ell_{a_{13}}^{(3)}, \ell_{a_{23}}^{(3)}\right)\right); \quad \text{(B.167)}$$
$$\rho_u \geq (G-1)(N-1)\left(\ell_{a_{12}}^{(3)} + \ell_{a_{13}}^{(3)} + 4\max(\ell_{a_{12}}^{(3)}, \ell_{a_{13}}^{(3)}, \ell_{a_{14}}^{(3)})\right); \quad \text{(B.168)}$$
$$\rho_u \geq (G-1)(N-1)\left(\ell_{a_{12}}^{(3)} + \ell_{a_{23}}^{(3)} + \ell_{a_{24}}^{(3)} + 4\max(\ell_{a_{12}}^{(3)}, \ell_{a_{23}}^{(3)}, \ell_{a_{24}}^{(3)})\right); \quad \text{(B.169)}$$
$$\rho_\phi \geq (G-1)N(N-1)\left(\ell_{a_{13}}^{(3)} + \ell_{a_{23}}^{(3)} + \ell_{a_{34}}^{(3)} + \max\left(\ell_{a_{13}}^{(3)}, \ell_{a_{23}}^{(3)}, \ell_{a_{34}}^{(3)}\right)\right); \quad \text{(B.170)}$$
$$\rho_\phi \geq (G-1)N(N-1)\left(\ell_{a_{14}}^{(3)} + \ell_{a_{24}}^{(3)} + \ell_{a_{34}}^{(3)} + \max\left(\ell_{a_{14}}^{(3)}, \ell_{a_{24}}^{(3)}, \ell_{a_{34}}^{(3)}\right)\right); \quad \text{(B.171)}$$
$$\rho_u \geq (G-1)(N-1)\left(\ell_{a_{12}}^{(3)} + \ell_{a_{13}}^{(3)} + \ell_{a_{14}}^{(3)} + \boldsymbol{\ell}_{a_{15}}^{(3)\top}\imath_{d,d_x} + \boldsymbol{\ell}_{a_{16}}^{(3)\top}\imath_{e,d_x}\right.$$
$$\left. + (4+d+e)\max\left(\ell_{a_{12}}^{(3)}, \ell_{a_{13}}^{(3)}, \ell_{a_{14}}^{(3)}, \boldsymbol{\ell}_{a_{15}}^{(3)\top}, \boldsymbol{\ell}_{a_{16}}^{(3)\top}\right)\right); \quad \text{(B.172)}$$
$$\rho_u \geq (G-1)(N-1)\left(\ell_{a_{12}}^{(3)} + \ell_{a_{13}}^{(3)} + \ell_{a_{14}}^{(3)} + \boldsymbol{\ell}_{a_{15}}^{(3)\top}\imath_{d,d_x} + \boldsymbol{\ell}_{a_{16}}^{(3)\top}\imath_{e,d_x}\right.$$
$$\left. + (4+d+e)\max\left(\ell_{a_{12}}^{(3)}, \ell_{a_{13}}^{(3)}, \ell_{a_{14}}^{(3)}, \boldsymbol{\ell}_{a_{15}}^{(3)\top}, \boldsymbol{\ell}_{a_{16}}^{(3)\top}\right)\right); \quad \text{(B.173)}$$
$$\rho_\phi \geq (G-1)N(N-1)\left(\ell_{a_{13}}^{(3)} + \ell_{a_{23}}^{(3)} + \ell_{a_{34}}^{(3)} + \boldsymbol{\ell}_{a_{35}}^{(3)\top}\imath_{d,d_x} + \boldsymbol{\ell}_{a_{36}}^{(3)\top}\imath_{e,d_x}\right.$$
$$\left. + (4+d+e)\max\left(\ell_{a_{13}}^{(3)}, \ell_{a_{23}}^{(3)}, \ell_{a_{34}}^{(3)}, \boldsymbol{\ell}_{a_{35}}^{(3)\top}, \boldsymbol{\ell}_{a_{36}}^{(3)\top}\right)\right); \quad \text{(B.174)}$$

$$\rho_\phi \geq (G-1)N(N-1)\left(\ell^{(3)}_{a_{14}} + \ell^{(3)}_{a_{24}} + \ell^{(3)}_{a_{34}} + \boldsymbol{\ell}^{(3)\top}_{a_{45}}\imath_{d,d_x} + \boldsymbol{\ell}^{(3)\top}_{a_{46}}\imath_{e,d_x}\right.$$
$$\left. +(4+d+e)\max\left(\ell^{(3)}_{a_{14}}, \ell^{(3)}_{a_{24}}, \ell^{(3)}_{a_{34}}, \boldsymbol{\ell}^{(3)\top}_{a_{45}}, \boldsymbol{\ell}^{(3)\top}_{a_{46}}\right)\right); \tag{B.175}$$
$$\rho_\theta \iota_d \geq (G-1)N(N-1)\left(\mathbb{S}_{d,d_x}\boldsymbol{\ell}^{(3)}_{a_{15}} + \mathbb{S}_{d,d_x}\boldsymbol{\ell}^{(3)}_{a_{25}} + \mathbb{S}_{d,d_x}\boldsymbol{\ell}^{(3)}_{a_{35}} + \mathbb{S}_{d,d_x}\boldsymbol{\ell}^{(3)}_{a_{45}} + \mathbb{S}_{d,d_x}\boldsymbol{\ell}^{(3)\top}_{a_{65}}\imath_{e,d_x}\right.$$
$$\left. +(4+d+e)\max\left(\mathbb{S}_{d,d_x}\boldsymbol{\ell}^{(3)}_{a_{15}}, \mathbb{S}_{d,d_x}\boldsymbol{\ell}^{(3)}_{a_{25}}, \mathbb{S}_{d,d_x}\boldsymbol{\ell}^{(3)}_{a_{35}}, \mathbb{S}_{d,d_x}\boldsymbol{\ell}^{(3)}_{a_{45}}, \mathbb{S}_{d,d_x}\boldsymbol{\ell}^{(3)\top}_{a_{65}}\right)\right); \tag{B.176}$$
$$\rho_\theta \iota_e \geq (G-1)N(N-1)\left(\mathbb{S}_{e,d_x}\boldsymbol{\ell}^{(3)}_{a_{16}} + \mathbb{S}_{e,d_x}\boldsymbol{\ell}^{(3)}_{a_{26}} + \mathbb{S}_{e,d_x}\boldsymbol{\ell}^{(3)}_{a_{36}} + \mathbb{S}_{e,d_x}\boldsymbol{\ell}^{(3)}_{a_{46}} + \mathbb{S}_{e,d_x}\boldsymbol{\ell}^{(3)\top}_{a_{65}}\imath_{d,d_x}\right.$$
$$\left. +(4+d+e)\max\left(\mathbb{S}_{e,d_x}\boldsymbol{\ell}^{(3)}_{a_{16}}, \mathbb{S}_{e,d_x}\boldsymbol{\ell}^{(3)}_{a_{26}}, \mathbb{S}_{e,d_x}\boldsymbol{\ell}^{(3)}_{a_{36}}, \mathbb{S}_{e,d_x}\boldsymbol{\ell}^{(3)}_{a_{46}}, \mathbb{S}_{e,d_x}\boldsymbol{\ell}^{(3)}_{a_{65}}\right)\right). \tag{B.177}$$

■

## B.4 Known Results

**Definition B.1** *A random field, $\{X_{\boldsymbol{j}},\ \boldsymbol{j} \in V_N\}$, on a sublattice indexed by $N$, say $V_N$, in the standard integer lattice $\mathbb{Z}^{d_v}$ is mixing with the mixing coefficient $\alpha_s(\cdot)$ if there exists a function $\alpha(\tau) \downarrow 0$ as $\tau \uparrow \infty$ such that, for any pair of subsets, $S$ and $S'$ in $V_N$,*

$$\alpha_s(\mathcal{B}(S), \mathcal{B}(S')) := \sup\left\{\left|P(A \bigcap B) - P(A)P(B)\right|,\ A \subset \mathcal{B}(S) \text{ and } B \subset \mathcal{B}(S')\right\}$$
$$\leq M_\alpha(|S|, |S'|)\alpha(d(S, S')),$$

*where $\mathcal{B}(S)$ is the Borel $\sigma$-field generated by the random elements $\{X_{\boldsymbol{j}},\ \boldsymbol{j} \in S\}$ and $M_\alpha(\cdot,\cdot)$ is a symmetric positive function non-decreasing in its arguments. This cardinality-based mixing condition has been widely used to characterize weak dependence for random fields in the literature (see, e.g., Bradley (2007, Chap. 29)). Throughout this paper, we assume that $M_\alpha(\cdot,\cdot)$ satisfies one of the following conditions:*

$$M_\alpha(n, m) \leq C_0 \min(n, m), \tag{B.178}$$
$$M_\alpha(n, m) \leq C_0 (n + m_1)^{\gamma_M} \text{ for some } \gamma_M \geq 1. \tag{B.179}$$

*Conditions (B.178) and (B.179) correspond to the ones used by Neaderhouser (1980) and Takahata (1983) respectively. They are satisfied by many spatial models (see, e.g., Rosenblatt (1985) or Guyon (1995)). It is important to note that, if $M_\alpha(n, m) = 1$ for every $n, m \geq 1$, then we call $\{X_{\boldsymbol{j}},\ \boldsymbol{j} \in V_N\}$ a strongly mixing random field. There are many random fields which do not satisfy the strong-mixing condition, but they do satisfy the mixing condition (see, e.g., Neaderhouser (1980)).*

**Lemma 16** *For any fixed $\boldsymbol{a} \in \mathbb{Z}^d$ with $d \geq 1$,*

$$\left|\{\boldsymbol{b} \in \mathbb{Z}^d :\ \|\boldsymbol{a} - \boldsymbol{b}\| = r\}\right| \leq 2d(2r+1)^{d-1}.$$

**Proof.** See, e.g., Sunklodas (2008). ∎

**Lemma 17** *Suppose that the random field $\{\eta_{\boldsymbol{s}} : \ \boldsymbol{s} \in V_N\}$ is mixing. Let $\mathcal{L}_r(\mathcal{F})$ denote the class of $\mathcal{F}$-measurable random functions, say $f(X)$, satisfying $\|f(X)\|_r := \{E|f(X)|^r\}^{1/r} < \infty$. Let $U := u(\eta_{\boldsymbol{s}}) \in \mathcal{L}\left(\mathcal{B}(S)\right)$ and $V := v(\eta_{\boldsymbol{s}}) \in \mathcal{L}\left(\mathcal{B}(S')\right)$ be measurable functions of $\eta_{\boldsymbol{s}}$. If $\max(\|U\|_r, \|V\|_s) < \infty$ for some $r, s > 2$, one then has, for some $r > 1$ and $1/s + 1/r < 1$,*

$$|Cov(U, V)| \leq C_0 \|U\|_r \|V\|_s \left(M_\alpha(|S|, |S'|)\alpha(d(S, S'))\right)^{1-1/r-1/s}.$$

*In case where $U < C_1$ and $V < C_2$ almost surely, one has*

$$Cov(U, v) \leq C_0 C_1 C_2 M_\alpha(|S|, |S'|)\alpha(d(S, S')).$$

**Proof.** This lemma is a variant of Davydov's inequality (see, e.g., Truong and Stone (1992)). ∎

**Lemma 18** *There exists a value $\tau_0 = \tau_0(\delta) < 1$ such that, for any subset $U \subset \mathbb{Z}^{d_v}$ with $|U| > 1$, one has that*

$$|U_1|^{1+\frac{\delta}{2}} + |U_2|^{1+\frac{\delta}{2}} \leq \tau_0 |U|^{1+\frac{\delta}{2}}, \text{ where } U = U_1 \bigcup U_2 \text{ and } U_1 \bigcap U_2 \neq \emptyset.$$

**Proof.** See Bulinski and Shashkin (2006). ∎

**Lemma 19** *Let $(\xi_1, \ldots, \xi_N)$ be a random vector such that $\max_{i=1,\ldots,N-1} \left| E\left[\prod_{s=i}^N \xi_s\right]\right| < \infty$ and $|C_0 \xi_i| \leq 1, \ i = 1, \ldots, N$. Then,*

$$\left| E\left[\prod_{s=1}^N \xi_s\right] - \prod_{s=1}^N E[\xi_s]\right| \leq \sum_{i=1}^{N-1} \sum_{j=i+1}^N \left| E\left[(\xi_i - 1)(\xi_j - 1) \prod_{s=j+1}^N \xi_s\right] - E[\xi_i - 1] E\left[(\xi_j - 1) \prod_{s=j+1}^N \xi_s\right]\right|.$$

**Proof.** See Nakhapetyan (1988). ∎

**Lemma 20** *Suppose $S_1, S_2, \ldots, S_r$ be sets, each containing $m$ sites with $dist(S_i, S_j) := \inf_{\boldsymbol{u} \in S_i, \boldsymbol{v} \in S_j} \|\boldsymbol{u} - \boldsymbol{v}\| \geq \delta$ for all $i \neq j$, where $1 \leq i \leq j$ and $1 \leq j \leq r$. Suppose that $Y_1, Y_2, \ldots, Y_r$ be a sequence of real-valued random variables measurable with respect to Borel fields, $\mathcal{B}(S_1), \mathcal{B}(S_2), \ldots, \mathcal{B}(S_r)$, respectively; and $Y_i$ takes values in $[a, b]$. Then, there exists a sequence of independent random variables, $Y_1^*, Y_2^*, \ldots, Y_r^*$, independent from $Y_1, Y_2, \ldots, Y_r$ such that $Y_i^*$ has the same distribution as $Y_i$ and satisfies*

$$\sum_{i=1}^r E|Y_i - Y_i^*| \leq 2r(b - a) M_\alpha((r - 1)m, m)\alpha(\delta).$$

**Proof.** The proof based on Rio (1995) can be found in Carbon, Tran, and Wu (1997). ∎

**Lemma 21 (CLT for Double Arrays of Martingale Difference Sequences (M.D.S.))** *Let $u_{N,t}$ be a double arrays of m.d.s. with respect to some sequence, $\mathcal{F}_{N,t}$, $t = 1, \dots, T$, of $\sigma$-fields such that $E[u_{N,t}|\mathcal{F}_t] = 0$, and let $\boldsymbol{z}_{N,t}$ be a sequence of $G$-dimensional random vectors measurable with respect to $\mathcal{F}_{N,t}$. Suppose that (i) $\lim_{N,T\uparrow\infty} \sum_{t=1}^{T} \boldsymbol{z}_{N,t}\boldsymbol{z}_{N,t}^{\top} \xrightarrow{p} \boldsymbol{\eta}$, where $\boldsymbol{\eta}$ is possibly a stochastic matrix, and (ii) $\lim_{N,T\uparrow\infty} \sum_{t=1}^{T} E\left[\|\boldsymbol{z}_{N,t}\|^{2+\delta}\right] < \infty$ for some $\delta > 0$. Then,*

$$\sum_{t=1}^{T} \boldsymbol{z}_{N,t} u_{N,t} \xrightarrow{w} \sigma_u \boldsymbol{\eta}^{1/2} N(\mathbf{0}, \mathbb{I}_G),$$

*where $\sigma_u^2 := \lim_{N\uparrow\infty} E[u_{N,t}^2|\mathcal{F}_{N,t}]$, and $\boldsymbol{\eta}$ and $N(\mathbf{0}, \mathbb{I}_G)$ are independent.*

**Proof.** See Rao (1987, p. 50). ■

**Lemma 22** *If a sequence of random variables $\{X_i,\ i \in \mathbb{N}\}$ satisfies $\sum_{i=1}^{\infty} E|X_i| < \infty$, then the summation $\sum_{i=1}^{\infty} X_i$ almost surely converges to a random variable $X = O_p(1)$.*

**Proof.** See Taniguchi, Hirukawa, and Tamaki (2008, Theorem A.2). ■

**Lemma 23** *Let $C$ be a nonempty bounded polyhedral convex set, $f$ be a d.c. function on $C$, and $g$ be a nonnegative concave function on $C$. Then, there exists $\gamma_0 \geq 0$ such that, for all $\gamma > \gamma_0$, the following problems have the same optimal value and the same solution set:*

$$\textit{(P)}\ \inf\{f(x) :\ x \in C,\ g(x) \leq 0\}$$

$$\textit{(P')}\ \inf\{f(x) + \gamma g(x) :\ x \in C\}.$$

**Proof.** See Le Thi Hoai An, Huynh Van Ngai, and Pham Dinh Tao (2012). ■

**Lemma 24 (Chernoff-type inequality)** *Let $X_i$, $i = 1, \dots, N$, represent jointly independent centered random variables. Let $S_N := \sum_{i=1}^{N} X_i$, where $\max_{1\leq i\leq N} |X_i| \leq 1$ and $\max_{1\leq i\leq N} Var(X_i) \leq \sigma^2 < \infty$. Then,*

$$P\left(|S_N| \geq N\lambda\sigma\right) \leq 2\max\left(\exp\left(-N\frac{\lambda^2}{4}\right), \exp\left(-\frac{N\lambda\sigma}{2}\right)\right).$$

**Proof of Lemma 24.** Notice that, by the independence of $X_i$, $i = 1, \dots, N$, one immediately has $E[\exp(\theta S_N)] = \prod_{i=1}^{N} E[\exp(\theta X_i)]$. Using an elementary inequality, $\exp(\theta X_i) \leq 1 + \theta X_i + \theta^2 X_i^2$ for $|\theta| \leq 1$, we obtain that $E[\exp(\theta X_i)] \leq 1 + \theta^2 Var(X_i) \leq \exp\left(\theta^2 Var(X_i)\right)$, thus, $E[\exp(\theta S_N)] \leq \exp(N\theta^2\sigma^2)$. Invoking Chernoff's inequality, one can immediately show that

$$P\left(|S_N| \geq N\lambda\sigma\right) \leq 2P\left(S_N \geq N\lambda\sigma\right) \leq 2\exp\left(\min_{0<\theta\leq 1}\{-\theta\lambda\sigma + N\theta^2\sigma^2\}\right) = \begin{cases} 2\exp\left(-\frac{N\lambda^2}{4}\right) & \text{if } \lambda\leq 2\sigma \\ 2\exp\left(-\frac{N\lambda\sigma}{2}\right) & \text{if } \lambda>2\sigma \end{cases}.$$

■

**Lemma 25** *Let* $\boldsymbol{A} := (a_{i,j})_{1\le i,j\le N}$ *be a matrix satisfying, for* $i = 1,\ldots,N$, $\sum_{k=1}^{N} a_{i,k} > 0$, *and* $a_{i,j} < \frac{1}{N}\sum_{k=1}^{N} a_{i,k}$ *for every* $j \neq i$. *Then, det*$(\boldsymbol{A}) > 0$.

**Proof.** See Carnicer, Goodman, and Pena (1999, Corollary 4.5). ■

# C Computational Considerations

## C.1 Computation: A New VNS-DCA Algorithm

Some background material on the gist of the DC programming and DCA is provided in C.3. First, recall some notations defined earlier: $\Delta y_{*,t}^{(w)} := \Delta y_{*,t} - \left(\sum_{t=1}^{T} \Delta y_{*,t}\boldsymbol{w}_{*,t}^{\top}\right)\left(\sum_{t=1}^{T} \boldsymbol{w}_{*,t}\boldsymbol{w}_{*,t}^{\top}\right)^{-1}\boldsymbol{w}_{*,t}$, $y_{i,t-1}^{(w)} := y_{i,t-1} - \left(\sum_{t=1}^{T} y_{i,t}\right)\left(\sum_{t=1}^{T} \boldsymbol{w}_{*,t}\boldsymbol{w}_{*,t}\right)^{-1}\boldsymbol{w}_{*,t}$, $\boldsymbol{x}_{i,t}^{(w)} := \boldsymbol{x}_{i,t} - \left(\sum_{t=1}^{T} \boldsymbol{x}_{i,t}\boldsymbol{w}_{*,t}^{\top}\right)\left(\sum_{t=1}^{T} \boldsymbol{w}_{*,t}\boldsymbol{w}_{*,t}^{\top}\right)^{-1}$ $\times\, \boldsymbol{w}_{*,t}$, and $1_t^{(w)} := 1 - \left(\sum_{t=1}^{T} \boldsymbol{w}_{*,t}^{\top}\right)\left(\sum_{t=1}^{T} \boldsymbol{w}_{*,t}\boldsymbol{w}_{*,t}^{\top}\right)^{-1}\boldsymbol{w}_{*,t}$. The *concentrated* composite errors are defined as

$$\epsilon_{*,t}(\boldsymbol{\psi},\boldsymbol{U}) := \Delta y_{*,t}^{(w)} - \sum_{c=1}^{G}\frac{1}{N}\sum_{i=1}^{N} u_{i,c}\phi_c\left(y_{i,t-1}^{(w)} - \boldsymbol{\theta}_c^{\top}\boldsymbol{x}_{i,t}^{(w)}\right) - \mu_* 1_t^{(w)}. \tag{C.1}$$

For a given $\boldsymbol{U} \in \Delta_S^N$, local minimum values, $\widehat{\boldsymbol{\psi}}(\boldsymbol{U})$, of $\mathcal{E}_{N,T}(\boldsymbol{\psi},\boldsymbol{U}) := \frac{1}{T}\sum_{t=1}^{T}\epsilon_{*,t}^2(\boldsymbol{\psi},\boldsymbol{U})$ satisfy the Karush-Kuhn-Tucker (KKT) conditions. Since $\frac{\partial}{\partial\mu_*}\epsilon_{*,t}(\boldsymbol{\psi},\boldsymbol{U}) = -1_t^{(w)}$, it then follows that

$$\widehat{\mu}_* = \left(\sum_{t=1}^{T} 1_t^{(w)2}\right)^{-1}\left\{\sum_{t=1}^{T}\Delta y_{*,t}^{(w)}1_t^{(w)} - \sum_{c=1}^{G}\frac{1}{N}\sum_{i=1}^{N} u_{i,c}\widehat{\phi}_c\left(\sum_{t=1}^{T} y_{i,t-1}^{(w)}1_t^{(w)} - \widehat{\boldsymbol{\theta}}_c^{\top}\sum_{t=1}^{T}\boldsymbol{x}_{i,t}^{(w)}1_t^{(w)}\right)\right\}. \tag{C.2}$$

Thus, $\widehat{\boldsymbol{\gamma}}(\boldsymbol{U}) = \left(\widehat{\boldsymbol{\theta}}(\boldsymbol{U}), \widehat{\boldsymbol{\phi}}(\boldsymbol{U})\right)$ are the minimum values of

$$\mathcal{E}_{N,T}(\boldsymbol{\gamma},\boldsymbol{U}) := \frac{1}{T}\sum_{t=1}^{T}\epsilon_{*,t}^2(\boldsymbol{\gamma},\boldsymbol{U}), \tag{C.3}$$

where

$$\epsilon_{*,t}(\boldsymbol{\gamma},\boldsymbol{U}) := A_t - \sum_{c=1}^{G}\frac{1}{N}\sum_{i=1}^{N} u_{i,c}\phi_c\left\{B_{i,t} - \boldsymbol{\theta}_c^{\top}\boldsymbol{C}_{i,t}\right\}$$

with

$$A_t \equiv A_{N,T,t} := \Delta y_{*,t}^{(w)} - \left(\sum_{t=1}^{T} 1_t^{(w)2}\right)^{-1}\left\{\sum_{t=1}^{T}\Delta y_{*,t}^{(w)}1_t^{(w)}\right\}1_t^{(w)},$$

$$
\begin{aligned}
B_{i,t} &\equiv B_{N,T,i,t} \coloneqq y_{i,t-1}^{(w)} - \left(\sum_{t=1}^{T} 1_t^{(w)2}\right)^{-1} \left\{\sum_{t=1}^{T} y_{i,t-1}^{(w)} 1_t^{(w)}\right\} 1_t^{(w)}, \\
\boldsymbol{C}_{i,t} &\equiv \boldsymbol{C}_{N,T,i,t} \coloneqq \boldsymbol{x}_{i,t}^{(w)} - \left(\sum_{t=1}^{T} 1_t^{(w)2}\right)^{-1} \left\{\sum_{t=1}^{T} \boldsymbol{x}_{i,t}^{(w)} 1_t^{(w)}\right\} 1_t^{(w)}.
\end{aligned}
$$

Let

$$
\begin{aligned}
\mathcal{E}_{1,N,T}(\boldsymbol{\gamma}, \boldsymbol{U}) \coloneqq \mathcal{A}_0 + \frac{1}{N^2} \sum_{c=1}^{G} \sum_{i=1}^{N} \big(& u_{i,c}^2 \phi_c^2 \mathcal{A}_{1,i} + u_{i,c}^2 \phi_c^2 \boldsymbol{\theta}_c^\top \mathcal{B}_{1,i} \boldsymbol{\theta}_c - 2u_{i,c}^2 \phi_c^2 \boldsymbol{\theta}_c^\top \mathcal{C}_{1,i} - 2N u_{i,c} \phi_c \mathcal{D}_{1,i} \\
&+ 2N u_{i,c} \phi_c \boldsymbol{\theta}_c^\top \mathcal{F}_{1,i} \big), \qquad \text{(C.4)}
\end{aligned}
$$

where $\mathcal{A}_0 \coloneqq \frac{1}{T}\sum_{t=1}^{T} A_t^2$; $\mathcal{A}_{1,i} \equiv \mathcal{A}_{1,N,T,i} \coloneqq \frac{1}{T}\sum_{t=1}^{T} B_{i,t}^2$; $\mathcal{B}_{1,i} \equiv \mathcal{B}_{1,N,T,i} \coloneqq \frac{1}{T}\sum_{t=1}^{T} \boldsymbol{C}_{i,t}\boldsymbol{C}_{i,t}^\top$; $\mathcal{C}_{1,i} \equiv \mathcal{C}_{1,N,T,i} \coloneqq \frac{1}{T}\sum_{t=1}^{T} B_{i,t}\boldsymbol{C}_{i,t}$; $\mathcal{D}_{1,i} \equiv \mathcal{D}_{1,N,T,i} \coloneqq \frac{1}{T}\sum_{t=1}^{T} A_t B_{i,t}$; and $\mathcal{F}_{1,i} \equiv \mathcal{F}_{N,T,i} \coloneqq \frac{1}{T}\sum_{t=1}^{T} A_t \boldsymbol{C}_{i,t}$.

$$
\mathcal{E}_{2,N,T}(\boldsymbol{\gamma}, \boldsymbol{U}) \coloneqq \frac{1}{N^2} \sum_{c=1}^{G} \sum_{i \neq j}^{N} \left(u_{i,c} u_{j,c} \phi_c^2 \mathcal{A}_{2,i,j} - u_{i,c} u_{j,c} \phi_c^2 \boldsymbol{\theta}_c^\top \mathcal{B}_{2,i,j} + u_{i,c} u_{j,c} \phi_c^2 \boldsymbol{\theta}_c^\top \mathcal{C}_{2,i,j} \boldsymbol{\theta}_c\right), \qquad \text{(C.5)}
$$

where $\mathcal{A}_{2,i,j} \equiv \mathcal{A}_{2,N,T,i,j} \coloneqq \frac{1}{T}\sum_{t=1}^{T} B_{i,t}B_{j,t}$; $\mathcal{B}_{2,i,j} \equiv \mathcal{B}_{2,N,T,i,j} \coloneqq \frac{1}{T}\sum_{t=1}^{T} \left(B_{i,t}\boldsymbol{C}_{j,t} + B_{j,t}\boldsymbol{C}_{i,t}\right)$; and $\mathcal{C}_{2,i,j} \equiv \mathcal{D}_{2,N,T,i,j} \coloneqq \frac{1}{T}\sum_{t=1}^{T} \boldsymbol{C}_{i,t}\boldsymbol{C}_{j,t}^\top$.

$$
\begin{aligned}
\mathcal{E}_{3,N,T}(\boldsymbol{\gamma}, \boldsymbol{U}) \coloneqq \frac{1}{N^2} \sum_{c \leq \ell}^{G} \sum_{i \neq j}^{N} \big(& u_{i,c} u_{j,\ell} \phi_c \phi_\ell \mathcal{A}_{2,i,j} - u_{i,c} u_{j,\ell} \phi_c \phi_\ell \boldsymbol{\theta}_c^\top \mathcal{B}_{3,i,j} - u_{i,c} u_{j,\ell} \phi_c \phi_\ell \boldsymbol{\theta}_\ell^\top \mathcal{C}_{3,i,j} \\
&+ u_{i,c} u_{j,\ell} \phi_c \phi_\ell \boldsymbol{\theta}_c^\top \mathcal{C}_{2,i,j} \boldsymbol{\theta}_\ell \big), \qquad \text{(C.6)}
\end{aligned}
$$

where $\mathcal{B}_{3,i,j} \coloneqq \frac{1}{T}\sum_{t=1}^{T} B_{j,t}\boldsymbol{C}_{i,t}$; and $\mathcal{C}_{3,i,j} \equiv \mathcal{C}_{3,N,T,i,j} \coloneqq \mathcal{B}_{2,i,j} - \mathcal{B}_{3,i,j}$.

Using the relation $2g_1g_2 = (g_1+g_2)^2 - (g_1^2 + g_2^2)$, we can immediately verify that $\mathcal{E}_{1,N,T}(\boldsymbol{\gamma}, \boldsymbol{U})$, $\mathcal{E}_{2,N,T}(\boldsymbol{\gamma}, \boldsymbol{U})$, and $\mathcal{E}_{3,N,T}(\boldsymbol{\gamma}, \boldsymbol{U})$ defined in (C.4)-(C.6) are d.c. functions.

**Assumption 1** *For ease of exposition, let the parameters $\boldsymbol{\gamma}$ take values in symmetric boxes, $\boldsymbol{\phi} \in \prod_{c=1}^{G}[-\ell_{\phi,c}, \ell_{\phi,c}]$ and*
$\boldsymbol{\theta} \in \prod_{c=1}^{G}\prod_{i=1}^{d_x}[-\ell_{\theta,c,i} \leq \theta_{i,c} \leq \ell_{\theta,c,i}]$.

Define

$$
\begin{aligned}
\mathcal{H}_{N,T}(\boldsymbol{\gamma}, \boldsymbol{U}) \equiv \mathcal{H}_{\rho,N,T}(\boldsymbol{\gamma}, \boldsymbol{U}) &\coloneqq N^2 \left(\mathcal{H}_{1,N,T}(\boldsymbol{\gamma}, \boldsymbol{U}) + \mathcal{H}_{2,N,T}(\boldsymbol{\gamma}, \boldsymbol{U}) + \mathcal{H}_{3,N,T}(\boldsymbol{\gamma}, \boldsymbol{U})\right), \\
F_{N,T}(\boldsymbol{\gamma}, \boldsymbol{U}) &\coloneqq N^2 \left\{\mathcal{E}_{N,T}(\boldsymbol{\gamma}, \boldsymbol{U}) - \mathcal{A}_0\right\} \\
&= N^2 \left\{\mathcal{E}_{1,N,T}(\boldsymbol{\gamma}, \boldsymbol{U}) + \mathcal{E}_{2,N,T}(\boldsymbol{\gamma}, \boldsymbol{U}) + \mathcal{E}_{3,N,T}(\boldsymbol{\gamma}, \boldsymbol{U}) - \mathcal{A}_0\right\}
\end{aligned}
$$

$$= \widetilde{\mathcal{G}}_{N,T}(\boldsymbol{\gamma}, \boldsymbol{U}) - \mathcal{H}_{N,T}(\boldsymbol{\gamma}, \boldsymbol{U}),$$

where

$$\widetilde{\mathcal{G}}_{N,T}(\boldsymbol{\gamma}, \boldsymbol{U}) \coloneqq \frac{5}{2}\rho \sum_{c=1}^{G}\sum_{i=1}^{N} u_{i,c}^2 + 2\rho \sum_{c=1}^{G} \phi_c^2 + 2\rho \sum_{c=1}^{G} \boldsymbol{\theta}_c^\top \boldsymbol{\theta}_c.$$

Note that the following equivalence between mixed-integer sets and polyhedral sets (see, e.g., Hoang (1995)): $\{\boldsymbol{U} \in \Delta_S^N \bigcap \{0,1\}^{G\times N}\} \equiv \{\boldsymbol{U} \in \Delta_S^N : \ g(\boldsymbol{U}) \le 0\}$, where $g(\boldsymbol{U}) \coloneqq \sum_{c=1}^{G}\sum_{i=1}^{N} u_{i,c}(1-u_{i,c})$ is finite concave on $\mathbb{R}^{G\times N}$ and nonnegative on $\Delta_S^N$. In view of Le Thi Hoai An, Huynh Van Ngai, and Pham Dinh Tao (2012, Theorem 1), one can immediately obtain that

$$\begin{aligned} &\min_{\boldsymbol{U}\in\Delta_S^N \bigcap\{0,1\}^{G\times N}} \min_{\substack{\boldsymbol{\phi}\in\prod_{c=1}^{G}[-\ell_{\phi,c},\ell_{\phi,c}] \\ \boldsymbol{\theta}\in\prod_{c=1}^{G}\prod_{i=1}^{d_x}[-\ell_{\theta,c,i},\ell_{\theta,c,i}]}} F_{N,T}(\boldsymbol{\gamma}, \boldsymbol{U}) \\ &\qquad = \min_{\boldsymbol{U}\in\Delta_S^N} \min_{\substack{\boldsymbol{\phi}\in\prod_{c=1}^{G}[-\ell_{\phi,c},\ell_{\phi,c}] \\ \boldsymbol{\theta}\in\prod_{c=1}^{G}\prod_{i=1}^{d_x}[-\ell_{\theta,c,i},\ell_{\theta,c,i}]}} \left\{ \widetilde{F}_{N,T}(\boldsymbol{\gamma}, \boldsymbol{U}) \coloneqq F_{N,T}(\boldsymbol{\gamma}, \boldsymbol{U}) + \widetilde{\gamma} g(\boldsymbol{U}) \right\} \end{aligned} \tag{C.7}$$

for some $\widetilde{\gamma} > 0$. The function $\widetilde{\mathcal{H}}_{N,T}(\boldsymbol{\gamma}, \boldsymbol{U}) \coloneqq \mathcal{H}_{N,T}(\boldsymbol{\gamma}, \boldsymbol{U}) - \widetilde{\gamma} g(\boldsymbol{U})$ is also convex for some appropriately chosen $\boldsymbol{\rho}$, which is stipulated by Lemmas 13-15. The gradient $\nabla\widetilde{\mathcal{H}}_{N,T}(\boldsymbol{\gamma}, \boldsymbol{U})$ of $\widetilde{\mathcal{H}}_{N,T}(\boldsymbol{\gamma}, \boldsymbol{U})$ is given by

$$\begin{aligned} \frac{\partial}{\partial u_{i,c}} \widetilde{\mathcal{H}}_{N,T}(\boldsymbol{\gamma}, \boldsymbol{U}) &= 5\rho_u u_{i,c} + 2N\phi_c \frac{1}{T}\sum_{t=1}^{T} \epsilon_{*,t}(\boldsymbol{\gamma}, \boldsymbol{U}) \left(B_{i,t} - \boldsymbol{\theta}_c^\top \boldsymbol{C}_{i,t}\right) + \widetilde{\gamma}(2u_{i,c} - 1), \\ \frac{\partial}{\partial \phi_c} \widetilde{\mathcal{H}}_{N,T}(\boldsymbol{\gamma}, \boldsymbol{U}) &= 4\rho_\phi \phi_c + 2\frac{N}{T}\sum_{i=1}^{N} u_{i,c} \sum_{t=1}^{T} \epsilon_{*,t}(\boldsymbol{\gamma}, \boldsymbol{U}) \left(B_{i,t} - \boldsymbol{\theta}_c^\top \boldsymbol{C}_{i,t}\right), \\ \frac{\partial}{\partial \boldsymbol{\theta}_c} \widetilde{\mathcal{H}}_{N,T}(\boldsymbol{\gamma}, \boldsymbol{U}) &= 4\rho_\theta \boldsymbol{\theta}_c - 2\phi_c \frac{N}{T}\sum_{i=1}^{N}\sum_{t=1}^{T} u_{i,c}\epsilon_{*,t}(\boldsymbol{\gamma}, \boldsymbol{U}) \boldsymbol{C}_{i,t} \end{aligned}$$

for all $i = 1, \ldots, N$ and $c = 1, \ldots, G$.

This then leads to the following d.c. program:

$$\begin{aligned} \min \Big\{ &\chi_{\prod_{c=1}^{G}\prod_{i=1}^{d_x}[-\ell_{\theta,c,i},\ell_{\theta,c,i}] \times \prod_{c=1}^{G}[-\ell_{\phi,c},\ell_{\phi,c}] \times \Delta_S^N}(\boldsymbol{\gamma}, \boldsymbol{U}) + \widetilde{\mathcal{G}}_{N,T}(\boldsymbol{\gamma}, \boldsymbol{U}) \\ &- \widetilde{\mathcal{H}}_{N,T}(\boldsymbol{\gamma}, \boldsymbol{U}) : \ \boldsymbol{U} \in \mathbb{R}^{N\times G}, \boldsymbol{\gamma} = (\boldsymbol{\theta}, \boldsymbol{\phi}) \in \mathbb{R}^{G\times d_x \times G} \Big\}. \end{aligned} \tag{C.8}$$

**Remark C.1** *The above DC decomposition uses the concentrated composite errors. This DC decomposition has some merits in terms of the execution speed as the objective function has fewer* *parameters to be optimized than the full composite likelihood function. To minimize the full composite likelihood function the problem (C.8) then needs the convex functions $\widetilde{\mathcal{G}}_{N,T}$ and $\widetilde{\mathcal{H}}_{N,T}$ provided in Section C.2 instead. All the algorithms described below can effectively be employed to minimize the sum of squared composite errors; and the computer program provided is specifically written for this minimization problem using the DC representation in Section C.2.*

The DCA applied to the problem (C.8) is described in Algorithm 1 below.

**Algorithm 1** DCA

```
procedure DC–A
Choose an initial point to start recursion, say {U^(0), θ^(0), φ^(0)}, and an error tolerance level, ε
Set ℓ ← 0
    repeat
        {λ^(ℓ), γ^(ℓ), V^(ℓ)} ∈ ∇H̃_{N,T}(θ^(ℓ), φ^(ℓ), U^(ℓ))
        min { G̃_{N,T}(θ, φ, U) − < {θ, φ, U}, {λ^(ℓ), γ^(ℓ), V^(ℓ)} >: {θ, φ, U} ∈ ∏_{c=1}^{G} ∏_{i=1}^{d_x} [−ℓ_{θ,c,i}, ℓ_{θ,c,i}]
        × ∏_{c=1}^{G} [−ℓ_{φ,c}, ℓ_{φ,c}] × Δ_S^N }
        (i.e., set U^(ℓ+1) = Proj_{Δ_S^N}(V^(ℓ)/(5ρ_u)), θ^(ℓ+1) = Proj_{∏_{c=1}^{G} ∏_{i=1}^{d_x}[−ℓ_{θ,c,i}, ℓ_{θ,c,i}]}(λ^(ℓ)/(4ρ_θ)),
        and φ^(ℓ+1) = Proj_{∏_{c=1}^{G}[−ℓ_{φ,c}, ℓ_{φ,c}]}(γ^(ℓ)/(4ρ_φ)))
        {γ**, U**} = {θ^(ℓ+1), φ^(ℓ+1), U^(ℓ+1)}
        {γ*, U*} = {θ^(ℓ), φ^(ℓ), U^(ℓ)}
        ℓ ← ℓ + 1
    until ‖{γ**, U**} − {γ*, U*}‖ ≤ ε (or ‖{γ**, U**} − {γ*, U*}‖ / ‖{γ*, U*}‖ ≤ ε if ‖{γ*, U*}‖ > 1)
return {θ^(ℓ+1), φ^(ℓ+1), U^(ℓ+1)}
end procedure
```

$\mathrm{Proj}_{\Delta_S^N}(\boldsymbol{v})$ denotes the projection of $\boldsymbol{v}$ onto the Cartesian product of standard unit $G$-dimensional simplices; there and many efficient algorithms to compute this projection, for example, the spectral projected gradient algorithm (Júdice, Raydan, Rosa, and Santos, 2008). Other projections onto rectangles can be straight-forwardly computed.

In Algorithm 1, there are two important implementation issues that warrant discussion. The first issue is how to choose $\rho$ as small as possible so that the function $\mathcal{H}_{N,T}(\boldsymbol{\gamma}, \boldsymbol{U})$ is still convex and the concave part $-\widetilde{\mathcal{H}}_{N,T}(\boldsymbol{\gamma}, \boldsymbol{U})$ of the d.c. decomposition becomes less important so as to enhance the efficiency of the DCA. Algorithm 2 to update $\rho$ is suggested by Le Thi Hoai An, Le Hoai Minh, and Pham Dinh Tao (2014). The second issue is to choose a 'good' starting point. For the DCA to work, a starting point must not be a local optimal point as the DCA is stationary at that point. The variable neighbourhood search (VNS) algorithm proposed by Hansen and Mladenović (1997) can potentially generate good starting points for the DCA. The VNS is an effective heuristic scheme for combinatorial and global optimization, which can easily implemented using any local search algorithm as a subroutine. The main principle of the VNS is to explore pre-determined distant neighborhoods of the current incumbent solution, and jump from there to a new one if there is an improvement found through a local search routine. A typical VNS routine requires a set of neighborhoods to be specified.

The structure of all non-intersecting neighborhoods [of $\boldsymbol{\gamma}$] in the hyper-rectangle $\prod_{c=1}^{G}\prod_{i=1}^{d_x}[-\ell_{\theta,c,i},\ell_{\theta,c,i}]$ $\times\prod_{c=1}^{G}[-\ell_{\phi,c},\ell_{\phi,c}]$ can be defined by $\mathcal{H}_k(\boldsymbol{\gamma}) := H_k(\boldsymbol{\gamma})/H_{k-1}(\boldsymbol{\gamma})$, where $H_k(\boldsymbol{\gamma}) := \prod_{c=1}^{G}\prod_{i=1}^{d_x}[\ell_{\theta,c,i}^{(k)},\omega_{\theta,c,i}^{(k)}]\times$ $\prod_{c=1}^{G}[\ell_{\phi,c}^{(k)},\omega_{\phi,c}^{(k)}]$, $k=1,\ldots,k_{\max}$, with

$$\begin{aligned}
\ell_{\theta,c,i}^{(k)} &\equiv \ell_{\theta,c,i}^{(k)}(\theta_{c,i}) := \theta_{c,i} - \frac{k}{k_{\max}}(\theta_{c,i}+\ell_{\theta,c,i}),\\
\omega_{\theta,c,i}^{(k)} &\equiv \omega_{\theta,c,i}^{(k)}(\theta_{c,i}) := \theta_{c,i} + \frac{k}{k_{\max}}(\ell_{\theta,c,i}-\theta_{c,i}),\\
\ell_{\phi,c}^{(k)} &\equiv \ell_{\phi,c}^{(k)}(\phi_c) := \phi_c - \frac{k}{k_{\max}}(\phi_c+\ell_{\phi,c}),\\
\omega_{\phi,c}^{(k)} &\equiv \omega_{\phi,c}^{(k)}(\phi_c) := \phi_c + \frac{k}{k_{\max}}(\ell_{\phi,c}-\phi_c).
\end{aligned}$$

Let $\kappa(\boldsymbol{U},\boldsymbol{U}')$ denote the Hamming distance between $\boldsymbol{U}$ and $\boldsymbol{U}'$ (i.e., the number of pairwise different columns of these $G\times N$ matrices). The system of all neighborhoods [of $\boldsymbol{U}$] induced by this metric in $\Delta_S^N$ is then given by $\mathcal{N}_\ell(\boldsymbol{U}) := \left\{\boldsymbol{U}'\in\Delta_S^N:\ \kappa(\boldsymbol{U},\boldsymbol{U}')=\ell\right\}$, $\ell=1,\ldots,\ell_{\max}$, $\ell_{\max}:=N$. Therefore, one can choose $\mathcal{N}_{k,\ell}(\boldsymbol{\gamma},\boldsymbol{U}) := \mathcal{H}_k(\boldsymbol{\gamma})\times\mathcal{N}_\ell(\boldsymbol{U})\ \forall\ k=1,\ldots,k_{\max}$ and $\ell=1,\ldots,\ell_{\max}$ as a structure of neighborhoods [of $\boldsymbol{\gamma}\times\boldsymbol{U}$] in $\prod_{c=1}^{G}\prod_{i=1}^{d_x}[-\ell_{\theta,c,i},\ell_{\theta,c,i}]\times\prod_{c=1}^{G}[-\ell_{\phi,c},\ell_{\phi,c}]\times\Delta_S^N$.

The VNS using the defined neighborhood system is reminiscent of the divide-and-conquer strategy used in a branch-and-bound optimization algorithm - breaking the search space into smaller pieces, then optimizing the objective function on these pieces. Unlike branch-and-bound algorithms the VNS also allows the system of neighborhoods to vary at each iteration. The basic VNS procedure is described in Algorithm 3 below. In this algorithm, local searches can be performed by using Simulated Annealing (SA) (see, e.g., Guyon (1995, p. 212)) instead of the K-means algorithm. The K-means - despite of its appealing computational efficiency - has certain shortcomings, such as it is very sensitive to outliers so that the computed clusters are different from actual ones, and it does not often reach global optimum even when being ignited by different initial values (see, e.g., Tan, Steinbach, and Kumar (2005); Wu (2012)). A properly designed SA-based algorithm can be more efficient than the K-means algorithm in obtaining a globally optimal solution to the clustering problem (Brown and Huntley (1992); Klein and Dubes (1989); Selim and Alsultan (1991)). The annealing process, as implemented via the Metropolis algorithm (Metropolis, Rosenbluth, Rosenbluth, Teller, and Teller (1953)), always allows for some possibility of moving out of a local optimum by probably accepting a 'worse' local value of the objective function. Therefore the SA can eventually generate near global optimum after a number of runs required to first "melt" the system being

**Algorithm 2** Update $\boldsymbol{\rho} = (\rho_u, \rho_\phi, \rho_\theta)$

Initialize the routine using $\{\boldsymbol{U}^{(0)}, \boldsymbol{\theta}^{(0)}, \boldsymbol{\phi}^{(0)}\}$, and choose a step size, $\tau_\rho \in (0, 1)$.
Set $\ell \leftarrow 0$ and $\boldsymbol{\rho}^{(0)} \leftarrow \boldsymbol{\rho}_0$, where $\boldsymbol{\rho}_0$ satisfies Lemmas 13-15.
**repeat**
    $\boldsymbol{\rho}^{(\ell+1)} = \tau_\rho \boldsymbol{\rho}^{(\ell)}$
    $\{\boldsymbol{\lambda}^{(\ell)}, \boldsymbol{\gamma}^{(\ell)}, \boldsymbol{V}^{(\ell)}\} \in \nabla \widetilde{\mathcal{H}}_{\rho^{(\ell+1)},N,T}(\boldsymbol{\theta}^{(\ell)}, \boldsymbol{\phi}^{(\ell)}, \boldsymbol{U}^{(\ell)})$
    Set $\boldsymbol{U}^{(\ell+1)} = \mathrm{Proj}_{\Delta_S^N}\left(\frac{\boldsymbol{V}^{(\ell)}}{5\rho_u^{(\ell+1)}}\right)$, $\boldsymbol{\theta}^{(\ell+1)} = \mathrm{Proj}_{\prod_{c=1}^{G}\prod_{i=1}^{d_x}[-\ell_{\theta,c,i}, \ell_{\theta,c,i}]}\left(\frac{\boldsymbol{\lambda}^{(\ell)}}{4\rho_\theta^{(\ell+1)}}\right)$,
    and $\boldsymbol{\phi}^{(\ell+1)} = \mathrm{Proj}_{\prod_{c=1}^{G}[-\ell_{\phi,c}, \ell_{\phi,c}]}\left(\frac{\boldsymbol{\gamma}^{(\ell)}}{4\rho_\phi^{(\ell+1)}}\right)$
    $\{\boldsymbol{\gamma}^{**}, \boldsymbol{U}^{**}\} = \{\boldsymbol{\theta}^{(\ell+1)}, \boldsymbol{\phi}^{(\ell+1)}, \boldsymbol{U}^{(\ell+1)}\}$
    $\{\boldsymbol{\gamma}^{*}, \boldsymbol{U}^{*}\} = \{\boldsymbol{\theta}^{(\ell)}, \boldsymbol{\phi}^{(\ell)}, \boldsymbol{U}^{(\ell)}\}$
    $\ell \leftarrow \ell + 1$
    $\boldsymbol{\rho}^{(\ell)} \leftarrow \boldsymbol{\rho}^{(\ell+1)}$
**until** $F_{N,T}(\boldsymbol{\gamma}^{**}, \boldsymbol{U}^{**}) > F_{N,T}(\boldsymbol{\gamma}^{*}, \boldsymbol{U}^{*})$
**if** $\ell > 1$ **then**
    **return** $\boldsymbol{\rho} \leftarrow \boldsymbol{\rho}^{(\ell)}$ and $\{\boldsymbol{U}^{(0)}, \boldsymbol{\theta}^{(0)}, \boldsymbol{\phi}^{(0)}\} \leftarrow \{\boldsymbol{U}^{(\ell)}, \boldsymbol{\theta}^{(\ell)}, \boldsymbol{\phi}^{(\ell)}\}$
**else**
    **return** $\boldsymbol{\rho}^{(0)}$ and $\{\boldsymbol{U}^{(0)}, \boldsymbol{\theta}^{(0)}, \boldsymbol{\phi}^{(0)}\}$
**end if**

optimized at a high effective temperature, then to lower the temperature gradually until the system "freezes" and no further changes to the system can be found. In fact the DCA merely needs a 'good' starting point, which must not be a local optimum, to proceed; and ideally, this 'good' starting point is a near global optimum. The SA procedure is given in Algorithm 4.

**Algorithm 3** VNS

**procedure** VARIABLE NEIGHBORHOOD SEARCH (VNS) PROCEDURE

Choose initial values, $\{\boldsymbol{\gamma}^{(0)}, \boldsymbol{U}^{(0)}\}$, and an error tolerance level, $\epsilon$

$\ell \leftarrow 0$

$k \leftarrow 0$

  **do**

*loop*:

    Randomly generate a point $\{\boldsymbol{\gamma}'^{(\ell)}, \boldsymbol{U}'^{(\ell)}\} \in \mathcal{N}_{k,\ell}\left(\boldsymbol{\gamma}^{(\ell)}, \boldsymbol{U}^{(\ell)}\right) := \mathcal{H}_k(\boldsymbol{\gamma}^{(\ell)}) \times \mathcal{N}_\ell(\boldsymbol{U}^{(\ell)})$

    Do a local search starting at $\{\boldsymbol{\gamma}'^{(\ell)}, \boldsymbol{U}'^{(\ell)}\}$ in $\mathcal{N}_{k,\ell}\left(\boldsymbol{\gamma}^{(\ell)}, \boldsymbol{U}^{(\ell)}\right)$ and obtain a local optimum,

    $\{\boldsymbol{\gamma}^{(\ell+1)}, \boldsymbol{U}^{(\ell+1)}\}$

    **if** $F_{N,T}(\boldsymbol{\gamma}^{(\ell+1)}, \boldsymbol{U}^{(\ell+1)}) < F_{N,T}(\boldsymbol{\gamma}^{(\ell)}, \boldsymbol{U}^{(\ell)})$ **then**

      $\{\boldsymbol{\gamma}^{(\ell)}, \boldsymbol{U}^{(\ell)}\} \leftarrow \{\boldsymbol{\gamma}^{(\ell+1)}, \boldsymbol{U}^{(\ell+1)}\}$

      **goto** *loop*

    **else**

      $\ell \leftarrow \ell + 1$

      $k \leftarrow k + 1$

    **end if**

  **while** $(\ell \leq \ell_{\max}$ **AND** $k \leq k_{\max})$ **OR** $\|\{\boldsymbol{\gamma}^{(\ell)}, \boldsymbol{U}^{(\ell)}\} - \{\boldsymbol{\gamma}^{(\ell-1)}, \boldsymbol{U}^{(\ell-1)}\}\| \leq \epsilon$

**return** $\{\boldsymbol{\gamma}^{(\ell)}, \boldsymbol{U}^{(\ell)}\}$

**end procedure**

The algorithm proposed by Selim and Alsultan (1991) is employed for randomly generating a neighboring group assignment, $\boldsymbol{U}'^{(\ell)}$, of $\boldsymbol{U}'$.

## C.2 DC Decomposition of the Sum of Squared Composite Errors (SSCE)

In view of (3.3) the composite errors are given by

$$\begin{aligned}\epsilon_{*,t} &\equiv \epsilon_{*,t}(\boldsymbol{\theta}, \boldsymbol{\phi}, \boldsymbol{\Lambda}, \mu_*, \boldsymbol{U}) \\ &:= \Delta y_{*,t} - \mu_* - \sum_{g=1}^{G} \frac{1}{N} \sum_{i=1}^{N} \phi_g u_{i,g} \left(y_{i,t-1} - \boldsymbol{\theta}_g^\top \boldsymbol{x}_{i,t}\right) - \sum_{g=1}^{G} \frac{1}{N} \sum_{i=1}^{N} u_{i,g} \boldsymbol{\lambda}_g^\top \boldsymbol{w}_{i,t},\end{aligned}$$

where $\boldsymbol{\Lambda} := \left(\boldsymbol{\lambda}_1^\top, \ldots, \boldsymbol{\lambda}_G^\top\right)^\top$. We can show using some simple calculations that

$$\frac{N^2}{T} \sum_{t=1}^{T} \epsilon_{*,t}(\boldsymbol{\theta}, \boldsymbol{\phi}, \boldsymbol{\Lambda}, \mu_*, \boldsymbol{U}) + \widetilde{\gamma} g(\boldsymbol{U}) = \widetilde{\mathcal{G}}_{N,T}(\boldsymbol{\theta}, \boldsymbol{\phi}, \boldsymbol{\Lambda}, \mu_*, \boldsymbol{U}) - \widetilde{\mathcal{H}}_{N,T}(\boldsymbol{\theta}, \boldsymbol{\phi}, \boldsymbol{\Lambda}, \mu_*, \boldsymbol{U})$$

**Algorithm 4** SA

**procedure** SIMULATED ANNEALING (SA) PROCEDURE
Initialize the algorithm using $\{\boldsymbol{\gamma}^{(0)}, \boldsymbol{U}^{(0)}\}$
set an initial temperature, $Te$, a temperature length, $TL$, and a cooling speed, $\alpha$
$\ell \leftarrow 0$
**repeat**
  **for** $i = 1$ `to` $\lfloor TL/a \rfloor$ **do**
    apply Selim and Alsultan's (1991) algorithm to randomly draw a neighboring point, $\{\boldsymbol{\gamma}^{*}, \boldsymbol{U}^{*}\}$, of $\{\boldsymbol{\gamma}^{(\ell)}, \boldsymbol{U}^{(\ell)}\}$
    compute $\Delta F_{N,T} = F_{N,T}(\boldsymbol{\gamma}^{*}, \boldsymbol{U}^{*}) - F_{N,T}(\boldsymbol{\gamma}^{(\ell)}, \boldsymbol{U}^{(\ell)})$
    **if** $\Delta F_{N,T} \leq 0$ **then**
      $\boldsymbol{\gamma}^{(\ell)} \leftarrow \boldsymbol{\gamma}^{*}$
      $\boldsymbol{U}^{(\ell)} \leftarrow \boldsymbol{U}^{*}$
    **else**
      randomly draw $q = \text{Uniform}(0, 1)$
      **if** $q < \exp(-\Delta F_{N,T}/Te)$ **then**
        $\boldsymbol{\gamma}^{(\ell)} \leftarrow \boldsymbol{\gamma}^{*}$
        $\boldsymbol{U}^{(\ell)} \leftarrow \boldsymbol{U}^{*}$
      **end if**
    **end if**
  **end for**
  $\ell \leftarrow \ell + 1$
  set a new 'cooling' temperature, $Te = Te \times \alpha$
**until** a stopping criterion is met
**return** the solution corresponding to the minimum function
**end procedure**

with

$$\widetilde{\mathcal{G}}_{N,T}(\boldsymbol{\theta}, \boldsymbol{\phi}, \boldsymbol{\Lambda}, \mu_*, \boldsymbol{U}) \coloneqq \frac{5}{2}\rho_u \sum_{g=1}^{G}\sum_{i=1}^{N} u_{i,g}^2 + 2\rho_\phi \sum_{g=1}^{G} \phi_c^2 + 2\rho_\theta \sum_{g=1}^{G} \boldsymbol{\theta}_g^\top \boldsymbol{\theta}_g + 2\rho_\lambda \sum_{g=1}^{G} \boldsymbol{\lambda}_g^\top \boldsymbol{\lambda}_g + 2\rho_\mu \mu_*^2$$

and

$$\widetilde{\mathcal{H}}_{N,T}(\boldsymbol{\theta}, \boldsymbol{\phi}, \boldsymbol{\Lambda}, \mu_*, \boldsymbol{U}) \coloneqq \widetilde{\mathcal{G}}_{N,T}(\boldsymbol{\theta}, \boldsymbol{\phi}, \boldsymbol{\Lambda}, \mu_*, \boldsymbol{U}) - \frac{N^2}{T}\sum_{t=1}^{T} \epsilon_{*,t}(\boldsymbol{\theta}, \boldsymbol{\phi}, \boldsymbol{\Lambda}, \mu_*, \boldsymbol{U}) - \widetilde{\gamma} g(\boldsymbol{U})$$

for some $\widetilde{\gamma} > 0$.

The function $\widetilde{\mathcal{H}}_{N,T}$ is convex for some choice of $\boldsymbol{\rho} \coloneqq (\rho_u, \rho_\phi, \rho_\theta, \rho_\lambda, \rho_\mu)$ similar to what was asserted in Lemmas 13-15. Some algebraic manipulations yield the following gradient vector of $\widetilde{\mathcal{H}}_{N,T}(\boldsymbol{\theta}, \boldsymbol{\phi}, \boldsymbol{\Lambda}, \mu_*, \boldsymbol{U}):$

$$
\begin{aligned}
\frac{\partial \widetilde{\mathcal{H}}_{N,T}}{\partial u_{i,g}} &= 5\rho_u u_{i,g} + 2\frac{N}{T}\sum_{t=1}^{T}\epsilon_{*,t}\left\{\phi_g(y_{i,t-1} - \boldsymbol{\theta}_g^\top \boldsymbol{x}_{i,t}) + \boldsymbol{\lambda}_g^\top \boldsymbol{w}_{i,t}\right\} + \widetilde{\gamma}(2u_{i,g} - 1), \\
\frac{\partial \widetilde{\mathcal{H}}_{N,T}}{\partial \phi_g} &= 5\rho_\phi \phi_g + 2\frac{N}{T}\sum_{t=1}^{T}\epsilon_{*,t}\left\{\sum_{i=1}^{N} u_{i,g}\left(y_{i,t-1} - \boldsymbol{\theta}_g^\top \boldsymbol{x}_{i,t}\right)\right\}, \\
\frac{\partial \widetilde{\mathcal{H}}_{N,T}}{\partial \boldsymbol{\theta}_g} &= 4\rho_\theta \boldsymbol{\theta}_g - 2\frac{N}{T}\phi_g \sum_{t=1}^{T}\epsilon_{*,t}\sum_{i=1}^{N} u_{i,g}\boldsymbol{x}_{i,t}, \\
\frac{\partial \widetilde{\mathcal{H}}_{N,T}}{\partial \boldsymbol{\lambda}_g} &= 4\rho_\lambda \boldsymbol{\lambda}_g + 2\frac{N}{T}\sum_{i=1}^{N} u_{i,g}\sum_{t=1}^{T}\epsilon_{*,t}\boldsymbol{w}_{i,t}, \\
\frac{\partial \widetilde{\mathcal{H}}_{N,T}}{\partial \mu_*} &= 4\rho_\mu \mu_* + 2\frac{N^2}{T}\sum_{t=1}^{T}\epsilon_{*,t}.
\end{aligned}
$$

## C.3 DC Programming and DCA: A Synopsis

Recall that, in the case when group memberships are known, the CQL function is convex. One can then employ various algorithms in convex optimization (e.g., Newton-Raphson or Simulated Annealing) to compute solutions to the CQL maximization problem. However, these algorithms for convex optimization are not sufficient to deal with a large-scale non-convex optimization problem that arises when one has to incorporate unknown group membership indicators into the CQL function. To surpass the difficulty of optimizing large-scale non-convex (possibly, non-smooth) functions, a theory of optimization for a superclass of convex functions, so-called difference-of-convex (d.c.) functions, has been extensively developed (see, e.g., Hiriart-Urruty (1985, 1988) for precursors to the DC programming). A concise review of DC programming and global optimization of d.c. functions is also provided in Hoang (1995) and Thoai (1999).

A DC program can be defined as

$$(P_{dc}) \ \min\{f(\boldsymbol{x}) \coloneqq g(\boldsymbol{x}) - h(\boldsymbol{x}) : \ \boldsymbol{x} \in \mathbb{R}^d\},$$

where $g(\cdot)$ and $h(\cdot)$ represent lower semi-continuous proper convex functions on $\mathbb{R}^d$. Such a function as $f(\cdot)$ is called a d.c. function. Class of d.c. functions is rather large so that most of functions encountered in econometric applications are d.c. functions. Note that convex constraints of type, $\boldsymbol{x} \in \mathcal{C} \subset \mathbb{R}^d$, can be taken into account using a set characteristic function, $\min\{f(\boldsymbol{x}) \coloneqq g(\boldsymbol{x}) - h(\boldsymbol{x}) : \ \boldsymbol{x} \in \mathcal{C}\} = \min\{\chi_{\mathcal{C}}(\boldsymbol{x}) + f(\boldsymbol{x}) : \ \boldsymbol{x} \in \mathbb{R}^d\}$, where $\chi_{\mathcal{C}}(\boldsymbol{x}) = 0$ if $\boldsymbol{x} \in \mathcal{C}$, and $= +\infty$ otherwise. Let

$$g^*(\boldsymbol{y}) \coloneqq \sup\{< \boldsymbol{x}, \boldsymbol{y} > - g(\boldsymbol{x}) : \ \boldsymbol{x} \in \mathbb{R}^d\}$$

be the conjugate of $g(\boldsymbol{x})$. One then obtains the following dual program of $P_{dc}$ :

$$(D_{dc})\ \min\{h^*(\boldsymbol{y}) - g^*(\boldsymbol{y}) :\ \boldsymbol{y} \in \mathbb{R}^d\}.$$

To see this duality, notice that, since $g(x) = \sup\{< \boldsymbol{x}, \boldsymbol{y} > -g^*(\boldsymbol{y}) :\ \boldsymbol{y} \in \mathbb{R}^d\}$ and $h(x) = \sup\{< \boldsymbol{x}, \boldsymbol{y} > -h^*(\boldsymbol{y}) :\ \boldsymbol{y} \in \mathbb{R}^d\}$, one obtains that

$$\begin{aligned}\inf\{g(\boldsymbol{x}) - h(\boldsymbol{x}) : \boldsymbol{x} \in \mathbb{R}^d\} &= \inf\left\{g(\boldsymbol{x}) - \sup\{< \boldsymbol{x}, \boldsymbol{y} > -h^*(\boldsymbol{y}) :\ \boldsymbol{y} \in \mathbb{R}^d\} : \boldsymbol{x} \in \mathbb{R}^d\right\}\\ &= \inf\left\{\inf\{g(\boldsymbol{x}) - < \boldsymbol{x}, \boldsymbol{y} > +h^*(\boldsymbol{y}) :\ \boldsymbol{x} \in \mathbb{R}^d\} :\ \boldsymbol{y} \in \mathbb{R}^d\right\}\\ &= \inf\left\{h^*(\boldsymbol{y}) - g^*(\boldsymbol{y}) :\ \boldsymbol{y} \in \mathbb{R}^d\right\}.\end{aligned}$$

Therefore the optimal solution to the program $P_{dc}$ is the same as the optimal solution to the program $D_{dc}$. The existence of these optimal solutions is guaranteed by the generalized Kuhn-Tucker global optimality condition (see, e.g., Hoang (2010, Proporsition 3.19)). However, in many large-scale non-convex problems, a number of algorithms searching for a point satisfying the global optimality condition - such as branch-and-bound and cutting cones - do not often compute optimal points efficiently (see Horst and Hoang (1993)). The DCA based on the duality in d.c. optimization - first introduced by Pham Dinh and Souad (1988) - is among a few algorithms which allow to solve large-scale d.c. optimization problems (see, e.g., Le Thi Hoai An and Pham Dinh Tao (2003); Pham Dinh Tao and Le Thi Hoai An (1998)). In the d.c. programming literature the DCAs converge to local solutions due to their local optimality nature; however, they often yield the global optimum, and a number of regularization and initialization methods can be used to facilitate the finding of global optimum from local ones in many different cases. A comprehensive introduction to DCA is provided in Pham Dinh Tao and Le Thi Hoai An (1997); and an incisive outline of DCA is given in Le Thi Hoai An (2014); Pham Dinh Tao and Le Thi Hoai An (2014). DCA has been successfully applied to many large-scale non-convex problems in applied science, especially, in Machine Learning where the use of the DCA often leads to global solutions and proves to be more robust than the standard methods (see, Le Thi Hoai An, Le Hoai Minh, and Pham Dinh Tao (2014) and references therein).

The DCA is an iterative primal-dual subgradient method consisting of two sequences, $\{\boldsymbol{x}^{(\ell)}\}$ and $\{\boldsymbol{y}^{(\ell)}\}$, chosen such that $\{g(\boldsymbol{x}^{(\ell)}) - h(\boldsymbol{x}^{(\ell)})\}$ and $\{h^*(\boldsymbol{y}^{(\ell)}) - g^*(\boldsymbol{y}^{(\ell)})\}$ are decreasing so that $\{\boldsymbol{x}^{(\ell)}\}$ and $\{\boldsymbol{y}^{(\ell)}\}$ converge to a feasible primal solution, $\boldsymbol{x}^*$, and a feasible dual solution, $\boldsymbol{y}^*$, respectively. These feasible solutions are shown to satisfy local optimality conditions and

$$\boldsymbol{x}^* \in \partial g^*(\boldsymbol{y}^*) \text{ and } \boldsymbol{y}^* \in \partial h(\boldsymbol{x}^*), \tag{C.9}$$

where $\partial h(\boldsymbol{x}^*)$ is the subdifferential of $h(\boldsymbol{x})$ at $\boldsymbol{x}^*$. Replacing $h(\cdot)$ in the program $P_{dc}$ with its affine minorization, $h_\ell(\boldsymbol{x}) \doteq h(\boldsymbol{x}^{(\ell)}) + < \boldsymbol{x} - \boldsymbol{x}^{(\ell)}, \boldsymbol{y}^{(\ell)} >$ with $\boldsymbol{y}^{(\ell)} \in \partial h(\boldsymbol{x}^{(\ell)})$, one can then obtain a convex approximation to the primal d.c. program $P_{dc}$:

$$(P_\ell)\ \min\{g(\boldsymbol{x}) - h_\ell(\boldsymbol{x})\}.$$

By the following property of subdifferentials of convex functions: $\boldsymbol{y} \in \partial g(\boldsymbol{x}) \equiv \boldsymbol{x} \in \partial g^*(\boldsymbol{y}) \equiv < \boldsymbol{x}, \boldsymbol{y} >= g(\boldsymbol{x}) + g^*(\boldsymbol{y})$, the optimal solution $\boldsymbol{x}^{(\ell+1)}$ to the program $P_\ell$ satisfies $\boldsymbol{x}^{(\ell+1)} \in \partial g^*(\boldsymbol{y}^{(\ell)})$. This gives rise to the following generic DCA scheme:

**Algorithm 5** Generic DCA Scheme

given an initial guess, $\boldsymbol{x}^{(0)} \in \mathbb{R}^d$, and an error tolerance level, $\epsilon$
$\ell \leftarrow 0$
**do**
    calculate $\boldsymbol{y}^{(\ell)} \in \partial h(\boldsymbol{x}^{(0)})$
    calculate $\boldsymbol{x}^{(\ell+1)} \in \partial g^*(\boldsymbol{y}^{(\ell)})$, which is equivalent to $\boldsymbol{x}^{(\ell+1)} \in \operatorname{argmin}\left\{g(\boldsymbol{x}) - h(\boldsymbol{x}^{(\ell)}) - < \boldsymbol{x} - \boldsymbol{x}^{(\ell)}, \boldsymbol{y}^{(\ell)} >: \ \boldsymbol{x} \in \mathbb{R}^d\right\}$.
    $\ell \leftarrow \ell + 1$
**while** $\|\boldsymbol{x}^{(\ell+1)} - \boldsymbol{x}^{(\ell)}\| < \epsilon$

The DCA has a linear convergence rate so that every limiting point of the sequence $\{\boldsymbol{x}^{(\ell)}\}$ or $\{\boldsymbol{y}^{(\ell)}\}$ is a generalized KKT point of $g - h$ or $h^* - g^*$ regardless of chosen starting values. It is worth mentioning that many standard methods of convex and non-convex programming are particular cases of DCA, for example, Expectation-Maximization (EM) of Dempster, Laird, and Rubin (1977), Successive Linear Approximation (SLA) of Bradley and Mangasarian (1998), and Convex-Concave Procedure (CCCP) of Yuille and Rangarajan (2003).

Efficient implementation of DCA involves an appropriate d.c. decomposition of $f(\cdot)$ and a 'good' starting point. If $f(\cdot)$ is such a d.c. function that $\frac{1}{2}\rho\|\boldsymbol{x}\|^2 - f(\boldsymbol{x})$ is convex for some sufficiently large $\rho$, then $f(\cdot) = g(\cdot) - h(\cdot)$, where $g(\boldsymbol{x}) := \frac{1}{2}\rho\|\boldsymbol{x}\|^2$ and $h(\boldsymbol{x}) := \frac{1}{2}\rho\|\boldsymbol{x}\|^2 - f(\boldsymbol{x})$. This special decomposition gives rise to the following algorithm:

**Algorithm 6** Special-Decomposition DCA Scheme

1: given an initial guess, $\boldsymbol{x}^{(0)} \in \mathbb{R}^d$, and an error tolerance level, $\epsilon$
2: $\ell \leftarrow 0$
3: **do**
4:     calculate $\boldsymbol{y}^{(\ell)} \in \partial\left(\frac{1}{2}\rho\|\cdot\|^2 - f(\cdot)\right)(\boldsymbol{x}^{(\ell)})$
5:     $\boldsymbol{x}^{(\ell+1)} \in \operatorname{argmin}\left\{\frac{1}{2}\rho\|\boldsymbol{x}\|^2 - < \boldsymbol{x}, \boldsymbol{y}^{(\ell)} >: \ \boldsymbol{x} \in \mathcal{C} \subset \mathbb{R}^d\right\}$.
6:     $\ell \leftarrow \ell + 1$
7: **while** $\|\boldsymbol{x}^{(\ell+1)} - \boldsymbol{x}^{(\ell)}\| < \epsilon$

Algorithm 6 is practically convenient because the convex minimization subproblem on line 5 can easily be solved by using the orthogonal projection, i.e., $\boldsymbol{x}^{(\ell+1)} = \operatorname{Proj}_{\mathcal{C}}(\frac{\boldsymbol{y}^{(\ell)}}{\rho})$; in fact, there are many algorithms to compute the projection onto convex sets (e.g., box constraints, polyhedron, simplices) [see, e.g., Chen and Ye (2011); Júdice, Raydan, Rosa, and Santos (2008)].

# D Tables

Table 1: Simulated Biases of Estimates for the D.G.P. with Known Group Memberships: Stationary Covariate and Linear SAR Errors with Rook-Contiguity Weights

| $T$ | $\widetilde{Bias}(\widehat{\phi}_1)$ | $\widetilde{Bias}(\widehat{\phi}_2)$ | $\widetilde{Bias}(\widehat{\phi}_3)$ | $\widetilde{Bias}(\widehat{\phi}_4)$ | $\widetilde{Bias}(\widehat{\theta}_1)$ | $\widetilde{Bias}(\widehat{\theta}_2)$ | $\widetilde{Bias}(\widehat{\theta}_3)$ | $\widetilde{Bias}(\widehat{\theta}_4)$ | $\widetilde{Bias}(\widehat{\mu}_*)$ |
|---|---|---|---|---|---|---|---|---|---|
| Experiment 1 ($N_1 = 45$, $N_2 = 30$, $N_3 = 30$, and $N_4 = 70$) | | | | | | | | | |
| 50 | -0.86109 | -0.106376 | 0.068128 | -0.212314 | -0.06528 | 0.040126 | 0.013735 | 0.333403 | 0.207969 |
| 150 | -0.72167 | -0.088056 | 0.045632 | -0.218835 | -0.05005 | 0.042901 | -0.017278 | 0.309425 | 0.197628 |
| 250 | -0.65316 | -0.077994 | 0.001431 | -0.191006 | -0.02688 | 0.045023 | 0.001188 | 0.294455 | 0.188259 |
| 350 | -0.64158 | -0.044944 | 0.065820 | -0.216888 | -0.00593 | 0.042386 | 0.011435 | 0.299321 | 0.178516 |
| 450 | -0.66119 | -0.080777 | 0.038246 | -0.214588 | -0.03850 | 0.048995 | 0.019717 | 0.332659 | 0.143224 |
| 550 | -0.78069 | -0.103053 | -0.004172 | -0.199955 | -0.00401 | 0.061558 | -0.010862 | 0.318524 | 0.217725 |
| 650 | -0.60281 | -0.077536 | -0.008501 | -0.176315 | -0.04533 | 0.046447 | 0.019646 | 0.292884 | 0.178645 |
| 750 | -0.56149 | -0.023177 | 0.036038 | -0.196446 | -0.03419 | 0.074197 | -0.039323 | 0.314156 | 0.138525 |
| Experiment 2 ($N_1 = 100$, $N_2 = 60$, $N_3 = 65$, and $N_4 = 150$) | | | | | | | | | |
| 50 | -0.943347 | 0.523168 | 0.234133 | -0.223378 | -1.54787 | 0.046866 | -0.029766 | -0.70695 | 0.35397 |
| 150 | -0.673831 | -0.111102 | 0.057198 | -0.213776 | -0.06720 | 0.034929 | -0.011663 | 0.32481 | 0.16215 |
| 250 | -0.564914 | -0.101189 | 0.079891 | -0.218222 | -0.01697 | 0.052751 | 0.033071 | 0.32475 | 0.16525 |
| 350 | -0.620736 | -0.073731 | 0.024378 | -0.221785 | -0.05171 | 0.023342 | -0.014370 | 0.31309 | 0.15915 |
| 450 | -0.688307 | -0.095300 | 0.016818 | -0.245358 | 0.02800 | 0.021830 | -0.000344 | 0.31039 | 0.19924 |
| 550 | -0.670902 | -0.093836 | 0.068862 | -0.245341 | -0.04130 | 0.044821 | -0.035420 | 0.32380 | 0.18687 |
| 650 | -0.690446 | -0.093473 | 0.012114 | -0.195285 | -0.05326 | 0.038125 | -0.004501 | 0.30867 | 0.20244 |
| 750 | -0.731355 | -0.075748 | 0.033510 | -0.193993 | -0.07986 | 0.0449219 | -0.029733 | 0.32392 | 0.17823 |
| Experiment 3 ($N_1 = 45$, $N_2 = 30$, $N_3 = 30$, and $N_4 = 70$) | | | | | | | | | |
| 50 | -0.110187 | -0.067890 | -0.078441 | -0.001113 | -0.250219 | -0.535216 | 0.082166 | 0.060488 | 0.015810 |
| 150 | -0.045497 | -0.023332 | -0.024274 | -0.001698 | 0.006884 | -0.194721 | 0.052197 | -0.015488 | 0.020460 |
| 250 | -0.019351 | -0.013274 | -0.016588 | -0.001520 | -0.028486 | -0.158752 | 0.038874 | -0.026079 | 0.021075 |
| 350 | -0.013855 | -0.009533 | -0.012156 | -0.000485 | -0.016486 | -0.087744 | 0.042756 | -0.004560 | 0.021392 |
| 450 | -0.009807 | -0.009852 | -0.010282 | -0.001130 | -0.028529 | -0.047788 | 0.039635 | -0.002043 | 0.021594 |
| 550 | -0.004898 | -0.005949 | -0.007844 | -0.000706 | -0.031241 | -0.039467 | 0.039654 | 0.004700 | 0.021781 |
| 650 | -0.002667 | -0.001809 | -0.006196 | -0.000637 | -0.041310 | -0.065682 | 0.037479 | 0.015756 | 0.021915 |
| 750 | -0.001270 | -0.003121 | -0.006111 | -0.000585 | -0.043350 | -0.022452 | 0.035340 | 0.022582 | 0.021924 |
| Experiment 4 ($N_1 = 100$, $N_2 = 60$, $N_3 = 65$, and $N_4 = 150$) | | | | | | | | | |
| 50 | -0.092588 | -0.078116 | -0.077676 | -0.004888 | -0.544623 | -0.098329 | 0.060390 | 0.106963 | 0.015661 |
| 150 | -0.033139 | -0.021422 | -0.027837 | -0.000733 | -0.059642 | -0.035168 | 0.040841 | -0.005210 | 0.019703 |
| 250 | -0.016082 | -0.011987 | -0.017922 | 0.000818 | -0.049598 | 0.009395 | 0.033612 | 0.020992 | 0.020504 |
| 350 | -0.008162 | -0.005361 | -0.012352 | 0.000812 | -0.066102 | -0.003925 | 0.031797 | 0.026391 | 0.020996 |
| 450 | -0.000513 | -0.004085 | -0.009919 | 0.000457 | -0.082942 | 0.042681 | 0.030141 | 0.027726 | 0.021239 |
| 550 | 0.003646 | -0.004057 | -0.008363 | 0.000633 | -0.084011 | 0.045922 | 0.025307 | 0.044636 | 0.021374 |
| 650 | 0.006919 | -0.003900 | -0.007256 | 0.000152 | -0.094483 | 0.066114 | 0.026684 | 0.035941 | 0.021503 |
| 750 | 0.006309 | -0.001032 | -0.005382 | 7.27E-05 | -0.091688 | 0.040157 | 0.029304 | 0.043090 | 0.021661 |

Table 2: Simulated MSE's of Estimates for the D.G.P. with Known Group Memberships: Stationary Covariate and Linear SAR Errors with Rook-Contiguity Weights

| $T$ | $\widetilde{MSE}(\widehat{\phi}_1)$ | $\widetilde{MSE}(\widehat{\phi}_2)$ | $\widetilde{MSE}(\widehat{\phi}_3)$ | $\widetilde{MSE}(\widehat{\phi}_4)$ | $\widetilde{MSE}(\widehat{\theta}_1)$ | $\widetilde{MSE}(\widehat{\theta}_2)$ | $\widetilde{MSE}(\widehat{\theta}_3)$ | $\widetilde{MSE}(\widehat{\theta}_4)$ | $\widetilde{MSE}(\widehat{\mu}_*)$ | Temp. Ave. Error* |
|---|---|---|---|---|---|---|---|---|---|---|
| | | | | Experiment 1 ($N_1 = 45$, $N_2 = 30$, $N_3 = 30$, and $N_4 = 70$) | | | | | | |
| 50 | 2.061030 | 0.238045 | 0.226894 | 0.339639 | 0.192405 | 0.161238 | 0.216442 | 0.248103 | 0.228033 | 7488.58 |
| 150 | 2.099090 | 0.261776 | 0.150218 | 0.246824 | 0.187279 | 0.168096 | 0.230186 | 0.202591 | 0.162543 | 1.71E+41 |
| 250 | 2.256810 | 0.226814 | 0.179554 | 0.226734 | 0.198282 | 0.159402 | 0.205081 | 0.230600 | 0.243666 | 7.44E+78 |
| 350 | 2.156730 | 0.256681 | 0.173522 | 0.261128 | 0.196061 | 0.149078 | 0.220355 | 0.219168 | 0.192356 | 4.04E+116 |
| 450 | 1.839240 | 0.278468 | 0.169363 | 0.291816 | 0.200510 | 0.173884 | 0.216671 | 0.252268 | 0.171778 | 2.73E+154 |
| 550 | 2.198300 | 0.239358 | 0.216220 | 0.223929 | 0.186552 | 0.185950 | 0.252353 | 0.243869 | 0.209425 | 1.24E+192 |
| 650 | 2.132060 | 0.254277 | 0.205297 | 0.206942 | 0.179464 | 0.196177 | 0.222694 | 0.194514 | 0.192571 | 8.36E+229 |
| 750 | 1.860390 | 0.197142 | 0.172236 | 0.209194 | 0.180146 | 0.174973 | 0.214772 | 0.219093 | 0.178035 | 5.53E+267 |
| | | | | Experiment 2 ($N_1 = 100$, $N_2 = 60$, $N_3 = 65$, and $N_4 = 150$) | | | | | | |
| 50 | 2.384320 | 212.738000 | 13.758200 | 0.269305 | 1152.860000 | 0.160198 | 0.351016 | 549.327000 | 10.862600 | 6559.82 |
| 150 | 1.982950 | 0.253339 | 0.145084 | 0.214627 | 0.182533 | 0.142790 | 0.205465 | 0.222483 | 0.169886 | 1.56E+41 |
| 250 | 2.426260 | 0.271670 | 0.238700 | 0.238971 | 0.200262 | 0.154655 | 0.256025 | 0.233882 | 0.160257 | 6.49E+78 |
| 350 | 2.000270 | 0.223061 | 0.162757 | 0.241096 | 0.201454 | 0.140084 | 0.237468 | 0.210311 | 0.163328 | 3.63E+116 |
| 450 | 2.582510 | 0.252512 | 0.202008 | 0.275982 | 0.256633 | 0.189170 | 0.317486 | 0.239224 | 0.218675 | 2.19E+154 |
| 550 | 2.426350 | 0.238815 | 0.185976 | 0.265754 | 0.200350 | 0.139763 | 0.222850 | 0.226840 | 0.166296 | 1.23E+192 |
| 650 | 2.100940 | 0.273460 | 0.159649 | 0.206538 | 0.211218 | 0.186339 | 0.200136 | 0.211990 | 0.205575 | 8.83E+229 |
| 750 | 2.182040 | 0.256051 | 0.181827 | 0.217218 | 0.189705 | 0.187588 | 0.249155 | 0.228272 | 0.189553 | 5.08E+267 |
| | | | | Experiment 3 ($N_1 = 45$, $N_2 = 30$, $N_3 = 30$, and $N_4 = 70$) | | | | | | |
| 50 | 0.368280 | 0.463023 | 0.049834 | 0.007872 | 6.653660 | 49.719200 | 66.073800 | 2.199690 | 0.059562 | 0.010184 |
| 150 | 0.081101 | 0.100259 | 0.008181 | 0.001855 | 0.542767 | 2.508030 | 6.304430 | 0.468265 | 0.048007 | 0.013849 |
| 250 | 0.044113 | 0.052732 | 0.004153 | 0.001073 | 0.277699 | 1.387700 | 2.670080 | 0.246744 | 0.047652 | 0.014541 |
| 350 | 0.029081 | 0.040395 | 0.002893 | 0.000740 | 0.175204 | 0.778334 | 1.813750 | 0.156959 | 0.048048 | 0.014838 |
| 450 | 0.022085 | 0.030767 | 0.002141 | 0.000578 | 0.133426 | 0.450826 | 1.246650 | 0.106520 | 0.048363 | 0.014997 |
| 550 | 0.017186 | 0.024377 | 0.001691 | 0.000491 | 0.103069 | 0.396044 | 0.961720 | 0.091526 | 0.048825 | 0.01509 |
| 650 | 0.014513 | 0.019639 | 0.001408 | 0.000424 | 0.083507 | 0.269695 | 0.741135 | 0.071040 | 0.049166 | 0.015162 |
| 750 | 0.011942 | 0.016315 | 0.001262 | 0.000364 | 0.075899 | 0.203642 | 0.674555 | 0.061651 | 0.049069 | 0.015225 |
| | | | | Experiment 4 ($N_1 = 100$, $N_2 = 60$, $N_3 = 65$, and $N_4 = 150$) | | | | | | |
| 50 | 0.336490 | 0.499184 | 0.045415 | 0.008392 | 29.993100 | 38.285400 | 73.256500 | 1.995760 | 0.057128 | 0.004828 |
| 150 | 0.066222 | 0.111923 | 0.008543 | 0.002018 | 0.487734 | 3.733990 | 5.550900 | 0.415036 | 0.045350 | 0.006514 |
| 250 | 0.037101 | 0.060085 | 0.004390 | 0.001123 | 0.226384 | 1.019560 | 2.447260 | 0.204707 | 0.045568 | 0.006854 |
| 350 | 0.027267 | 0.039802 | 0.003075 | 0.000786 | 0.150483 | 0.665346 | 1.417690 | 0.129436 | 0.046570 | 0.006975 |
| 450 | 0.020279 | 0.031139 | 0.002171 | 0.000599 | 0.117064 | 0.431270 | 1.045770 | 0.103245 | 0.046913 | 0.007063 |
| 550 | 0.016296 | 0.023378 | 0.001734 | 0.000485 | 0.092761 | 0.320943 | 0.708957 | 0.077289 | 0.047121 | 0.007114 |
| 650 | 0.013217 | 0.019180 | 0.001400 | 0.000396 | 0.080031 | 0.235476 | 0.679439 | 0.061645 | 0.047398 | 0.007150 |
| 750 | 0.011669 | 0.016570 | 0.001143 | 0.000354 | 0.070570 | 0.234721 | 0.565636 | 0.054522 | 0.047895 | 0.007168 |

* abbrev. for the temporal average error defined as $N\frac{1}{T}\sum_{t=1}^{T}\epsilon_{*,t}(\widehat{\boldsymbol{\psi}})$.

Table 3: Simulated Biases of Estimates for the D.G.P. with Known Group Memberships: Stationary Covariate and Linear SAR Errors with Queen-Contiguity Weights

| $T$ | $\widetilde{Bias}(\widehat{\phi}_1)$ | $\widetilde{Bias}(\widehat{\phi}_2)$ | $\widetilde{Bias}(\widehat{\phi}_3)$ | $\widetilde{Bias}(\widehat{\phi}_4)$ | $\widetilde{Bias}(\widehat{\theta}_1)$ | $\widetilde{Bias}(\widehat{\theta}_2)$ | $\widetilde{Bias}(\widehat{\theta}_3)$ | $\widetilde{Bias}(\widehat{\theta}_4)$ | $\widetilde{Bias}(\widehat{\mu}_*)$ |
|---|---|---|---|---|---|---|---|---|---|
| Experiment 1 ($N_1 = 45$, $N_2 = 30$, $N_3 = 30$, and $N_4 = 70$) | | | | | | | | | |
| 50 | -0.858399 | -0.088697 | 0.051561 | -0.251999 | -0.073639 | 0.033987 | -0.002832 | 0.316786 | 0.173743 |
| 150 | -0.701195 | -0.076070 | 0.037866 | -0.228812 | -0.043758 | 0.053390 | 0.016597 | 0.309517 | 0.168948 |
| 250 | -0.663698 | -0.041974 | 0.044686 | -0.213368 | -0.019480 | 0.040885 | -0.004228 | 0.316052 | 0.177749 |
| 350 | -0.698057 | -0.087288 | 0.028046 | -0.225867 | -0.065472 | 0.054591 | 0.005902 | 0.300461 | 0.165818 |
| 450 | -0.612329 | -0.089999 | 0.027263 | -0.213758 | -0.035901 | 0.031392 | 0.006506 | 0.314442 | 0.171976 |
| 550 | -0.699745 | -0.069692 | 0.029242 | -0.201872 | -0.074569 | 0.058946 | -0.011789 | 0.298964 | 0.176924 |
| 650 | -0.623609 | -0.042775 | 0.055336 | -0.213672 | -0.055918 | 0.053703 | 0.005600 | 0.299446 | 0.156117 |
| 750 | -0.626769 | -0.071302 | 0.021131 | -0.195425 | -0.063051 | 0.041090 | 0.016664 | 0.300204 | 0.162929 |
| Experiment 2 ($N_1 = 100$, $N_2 = 60$, $N_3 = 65$, and $N_4 = 150$) | | | | | | | | | |
| 50 | -0.763373 | -0.110235 | 0.081199 | -0.227623 | -0.075237 | 0.037752 | 0.025481 | 0.324663 | 0.185544 |
| 150 | -0.703738 | -0.084416 | 0.015175 | -0.227624 | -0.048532 | 0.039578 | -0.000360 | 0.300304 | 0.163391 |
| 250 | -0.703967 | -0.093266 | 0.062224 | -0.196770 | -0.033552 | 0.062668 | 0.021123 | 0.329863 | 0.175616 |
| 350 | -0.670058 | -0.114683 | 0.024481 | -0.205503 | -0.021432 | 0.044461 | 0.020606 | 0.317632 | 0.176906 |
| 450 | -0.715018 | -0.068980 | 0.058923 | -0.223209 | -0.040791 | 0.056332 | -0.010557 | 0.321255 | 0.172536 |
| 550 | -0.671723 | -0.076533 | 0.050059 | -0.234680 | -0.044193 | 0.044087 | 0.021208 | 0.301685 | 0.164421 |
| 650 | -0.662065 | -0.117619 | 0.062229 | -0.226658 | -0.056909 | 0.030790 | 0.003325 | 0.327446 | 0.181650 |
| 750 | -0.647779 | -0.063958 | 0.049050 | -0.221746 | -0.032152 | 0.012884 | -0.015911 | 0.323810 | 0.177286 |
| Experiment 3 ($N_1 = 45$, $N_2 = 30$, $N_3 = 30$, and $N_4 = 70$) | | | | | | | | | |
| 50 | -0.087895 | -0.073020 | -0.074460 | -0.001997 | -0.241938 | -0.643001 | 0.032662 | 0.025829 | 0.016142 |
| 150 | -0.035869 | -0.029701 | -0.026189 | 0.000696 | -0.024583 | -0.075703 | 0.038675 | 0.020277 | 0.020158 |
| 250 | -0.013167 | -0.015026 | -0.016090 | 0.001048 | -0.034377 | -0.187792 | 0.036391 | 0.029213 | 0.021046 |
| 350 | -0.006550 | -0.005223 | -0.010406 | 0.000806 | -0.047235 | -0.131618 | 0.040012 | 0.008716 | 0.021520 |
| 450 | 0.000266 | -0.002780 | -0.006648 | 0.000274 | -0.065099 | -0.081072 | 0.030614 | 0.033891 | 0.021862 |
| 550 | 0.002312 | 0.001495 | -0.005860 | 0.000100 | -0.068602 | -0.061908 | 0.030486 | 0.031070 | 0.021907 |
| 650 | 0.003062 | 0.000076 | -0.004954 | 0.000158 | -0.056065 | -0.041058 | 0.034975 | 0.029493 | 0.021987 |
| 750 | 0.004951 | 0.005124 | -0.004059 | -0.000061 | -0.070581 | -0.039654 | 0.035146 | 0.043064 | 0.022064 |
| Experiment 4 ($N_1 = 100$, $N_2 = 60$, $N_3 = 65$, and $N_4 = 150$) | | | | | | | | | |
| 50 | -0.089830 | -0.101346 | -0.078642 | -0.006556 | -0.297800 | -0.486428 | 0.032444 | 0.107005 | 0.015765 |
| 150 | -0.018141 | -0.042031 | -0.023972 | -0.002195 | -0.075558 | 0.048009 | 0.031493 | 0.015916 | 0.020249 |
| 250 | -0.014801 | -0.024221 | -0.011953 | -0.001133 | -0.046614 | 0.011725 | 0.030177 | 0.008297 | 0.021163 |
| 350 | -0.006422 | -0.016582 | -0.006240 | -0.000029 | -0.050570 | 0.038305 | 0.038714 | 0.025374 | 0.021597 |
| 450 | -0.000105 | -0.018953 | -0.005170 | -0.000627 | -0.066211 | 0.063712 | 0.030313 | 0.029299 | 0.021733 |
| 550 | -0.002778 | -0.014857 | -0.003442 | 0.000424 | -0.072256 | 0.047690 | 0.030318 | 0.035364 | 0.021815 |
| 650 | 0.000942 | -0.009095 | -0.003085 | 0.000752 | -0.067760 | 0.054388 | 0.030497 | 0.033301 | 0.021825 |
| 750 | 0.004076 | -0.004472 | -0.003347 | 0.000480 | -0.092348 | 0.063148 | 0.027805 | 0.039053 | 0.021818 |

Table 4: Simulated MSE's of Estimates for the D.G.P. with Known Group Memberships: Stationary Covariate and Linear SAR Errors with Queen-Contiguity Weights

| $T$ | $\widetilde{MSE}(\widehat{\phi}_1)$ | $\widetilde{MSE}(\widehat{\phi}_2)$ | $\widetilde{MSE}(\widehat{\phi}_3)$ | $\widetilde{MSE}(\widehat{\phi}_4)$ | $\widetilde{MSE}(\widehat{\theta}_1)$ | $\widetilde{MSE}(\widehat{\theta}_2)$ | $\widetilde{MSE}(\widehat{\theta}_3)$ | $\widetilde{MSE}(\widehat{\theta}_4)$ | $\widetilde{MSE}(\widehat{\mu}_*)$ | Temp. Ave. Error* |
|---|---|---|---|---|---|---|---|---|---|---|
| Experiment 1 ($N_1 = 45$, $N_2 = 30$, $N_3 = 30$, and $N_4 = 70$) | | | | | | | | | | |
| 50 | 1.874520 | 0.295715 | 0.543656 | 0.491988 | 0.233785 | 0.283593 | 0.249145 | 0.313727 | 0.247791 | 7329.74 |
| 150 | 2.117680 | 0.269733 | 0.183599 | 0.228730 | 0.186245 | 0.183630 | 0.235978 | 0.228160 | 0.184288 | 1.62E+41 |
| 250 | 2.001020 | 0.218764 | 0.175765 | 0.242151 | 0.180947 | 0.148920 | 0.198553 | 0.226997 | 0.184359 | 6.98E+78 |
| 350 | 1.957610 | 0.261443 | 0.181741 | 0.233389 | 0.197940 | 0.170263 | 0.251513 | 0.209617 | 0.175732 | 3.67E+116 |
| 450 | 1.946160 | 0.237436 | 0.173029 | 0.232767 | 0.176523 | 0.154068 | 0.224669 | 0.209865 | 0.179699 | 2.17E+154 |
| 550 | 2.210790 | 0.272458 | 0.179004 | 0.249963 | 0.211208 | 0.177857 | 0.214837 | 0.224487 | 0.232034 | 1.35E+192 |
| 650 | 1.853950 | 0.249563 | 0.196796 | 0.253207 | 0.177171 | 0.168365 | 0.235829 | 0.205364 | 0.184621 | 7.57E+229 |
| 750 | 1.784650 | 0.241543 | 0.179591 | 0.231717 | 0.187445 | 0.154874 | 0.214978 | 0.205378 | 0.183574 | 5.62E+267 |
| Experiment 2 ($N_1 = 100$, $N_2 = 60$, $N_3 = 65$, and $N_4 = 150$) | | | | | | | | | | |
| 50 | 1.898220 | 0.545762 | 0.331516 | 0.836769 | 0.231477 | 0.328107 | 0.274496 | 0.303754 | 0.266136 | 5646.86 |
| 150 | 2.000690 | 0.230504 | 0.169907 | 0.229354 | 0.178778 | 0.156767 | 0.217029 | 0.212232 | 0.168470 | 1.39E+41 |
| 250 | 2.360900 | 0.274960 | 0.179689 | 0.229213 | 0.201960 | 0.167710 | 0.242299 | 0.241965 | 0.191065 | 6.53E+78 |
| 350 | 2.139580 | 0.272874 | 0.174808 | 0.208555 | 0.190818 | 0.172576 | 0.217213 | 0.227648 | 0.190945 | 3.30E+116 |
| 450 | 2.308180 | 0.243625 | 0.202452 | 0.249083 | 0.236306 | 0.206432 | 0.214332 | 0.241470 | 0.173112 | 1.90E+154 |
| 550 | 2.224590 | 0.282695 | 0.192194 | 0.250986 | 0.203446 | 0.176593 | 0.270079 | 0.216051 | 0.182502 | 1.27E+192 |
| 650 | 2.356580 | 0.293268 | 0.185516 | 0.240022 | 0.207784 | 0.154649 | 0.290928 | 0.227304 | 0.176041 | 8.12E+229 |
| 750 | 2.223230 | 0.253720 | 0.161511 | 0.243489 | 0.184041 | 0.161936 | 0.246289 | 0.230593 | 0.192008 | 4.80E+267 |
| Experiment 3 ($N_1 = 45$, $N_2 = 30$, $N_3 = 30$, and $N_4 = 70$) | | | | | | | | | | |
| 50 | 0.317611 | 0.447992 | 0.047439 | 0.007840 | 7.398800 | 64.397500 | 60.899000 | 2.283360 | 0.059775 | 0.010255 |
| 150 | 0.071471 | 0.100726 | 0.008547 | 0.001971 | 0.500439 | 4.236710 | 7.406730 | 0.440798 | 0.047415 | 0.013861 |
| 250 | 0.040180 | 0.057286 | 0.004354 | 0.001186 | 0.251307 | 3.611480 | 4.024490 | 0.243155 | 0.047813 | 0.014503 |
| 350 | 0.026722 | 0.037043 | 0.003065 | 0.000835 | 0.173618 | 1.210100 | 1.870260 | 0.152877 | 0.048825 | 0.014841 |
| 450 | 0.020104 | 0.028302 | 0.002069 | 0.000626 | 0.127250 | 0.559407 | 1.210540 | 0.110200 | 0.049576 | 0.015015 |
| 550 | 0.017489 | 0.022834 | 0.001696 | 0.000486 | 0.103609 | 0.349531 | 0.887500 | 0.084307 | 0.049425 | 0.015122 |
| 650 | 0.014373 | 0.019724 | 0.001510 | 0.000403 | 0.087617 | 0.267190 | 0.742428 | 0.070074 | 0.049601 | 0.015187 |
| 750 | 0.012802 | 0.016847 | 0.001254 | 0.000355 | 0.077922 | 0.221034 | 0.716152 | 0.057015 | 0.049738 | 0.015233 |
| Experiment 4 ($N_1 = 100$, $N_2 = 60$, $N_3 = 65$, and $N_4 = 150$) | | | | | | | | | | |
| 50 | 0.312760 | 0.470171 | 0.044835 | 0.007728 | 6.387040 | 33.734000 | 40.105900 | 2.292450 | 0.056954 | 0.004917 |
| 150 | 0.066093 | 0.101020 | 0.008108 | 0.001746 | 0.495119 | 2.529910 | 4.829530 | 0.428887 | 0.047303 | 0.006548 |
| 250 | 0.037576 | 0.058946 | 0.003914 | 0.001070 | 0.225720 | 1.263170 | 2.471900 | 0.215994 | 0.048016 | 0.006895 |
| 350 | 0.024298 | 0.041783 | 0.002741 | 0.000780 | 0.135799 | 0.712740 | 1.624060 | 0.140107 | 0.048951 | 0.007018 |
| 450 | 0.019969 | 0.031458 | 0.002101 | 0.000592 | 0.105571 | 0.353870 | 0.986849 | 0.109316 | 0.049040 | 0.007092 |
| 550 | 0.015469 | 0.025272 | 0.001708 | 0.000501 | 0.091352 | 0.343692 | 0.874810 | 0.085894 | 0.049069 | 0.007128 |
| 650 | 0.013166 | 0.021070 | 0.001394 | 0.000429 | 0.078976 | 0.236987 | 0.667492 | 0.061006 | 0.048838 | 0.007163 |
| 750 | 0.011296 | 0.017499 | 0.001201 | 0.000370 | 0.071569 | 0.211944 | 0.526128 | 0.055434 | 0.048630 | 0.007188 |

* abbrev. for the temporal average error defined as $N\frac{1}{T}\sum_{t=1}^{T}\epsilon_{*,t}(\widehat{\boldsymbol{\psi}})$.

Table 5: Simulated Biases of Estimates for the D.G.P. with Known Group Memberships: Stationary Covariate and Nonlinear SAR Errors

| $T$ | $\widetilde{Bias}(\widehat{\phi}_1)$ | $\widetilde{Bias}(\widehat{\phi}_2)$ | $\widetilde{Bias}(\widehat{\phi}_3)$ | $\widetilde{Bias}(\widehat{\phi}_4)$ | $\widetilde{Bias}(\widehat{\theta}_1)$ | $\widetilde{Bias}(\widehat{\theta}_2)$ | $\widetilde{Bias}(\widehat{\theta}_3)$ | $\widetilde{Bias}(\widehat{\theta}_4)$ | $\widetilde{Bias}(\widehat{\mu}_*)$ |
|---|---|---|---|---|---|---|---|---|---|
| True Parameters defined in Experiment 1 ($m_i = n_i = 5$, $i = 1, \ldots, 4$) | | | | | | | | | |
| 50 | -0.970693 | -0.069464 | 0.085820 | -0.229508 | -0.094959 | 0.049079 | 0.018548 | 0.307814 | 0.172932 |
| 150 | -0.782395 | -0.145234 | 0.006716 | -0.257922 | -0.012191 | 0.040787 | 0.036339 | 0.322878 | 0.142735 |
| 250 | -0.598178 | -0.029355 | 0.014677 | -0.195009 | -0.034324 | 0.082560 | 0.070225 | 0.255610 | 0.185074 |
| 350 | -0.552613 | -0.049905 | 0.016743 | -0.179277 | -0.039556 | 0.042632 | 0.072894 | 0.296988 | 0.149054 |
| 450 | -0.655592 | -0.033425 | 0.019777 | -0.158854 | -0.075098 | 0.044455 | 0.007561 | 0.302537 | 0.185796 |
| 550 | -0.698142 | -0.031075 | 0.045309 | -0.167126 | -0.004099 | 0.058869 | 0.019851 | 0.295063 | 0.187364 |
| 650 | -0.708413 | -0.047610 | 0.001637 | -0.172818 | -0.053839 | 0.073108 | 0.022735 | 0.320742 | 0.141764 |
| 750 | -0.556860 | 0.007222 | 0.035597 | -0.243045 | -0.022273 | 0.014504 | 0.031651 | 0.263577 | 0.154588 |
| True Parameters defined in Experiment 1 ($m_i = 10$ and $n_i = 20$, $i = 1, \ldots, 4$) | | | | | | | | | |
| 50 | -0.732286 | -0.096945 | 0.057881 | -0.132431 | -0.057375 | 0.051464 | -0.024427 | 0.360652 | 0.197779 |
| 150 | -0.722716 | -0.072764 | 0.061557 | -0.235562 | -0.056656 | 0.043586 | -0.035096 | 0.321864 | 0.176154 |
| 250 | -0.723979 | -0.110542 | 0.051729 | -0.232778 | -0.036237 | 0.039303 | -0.026582 | 0.337413 | 0.165967 |
| 350 | -0.732472 | -0.078580 | 0.024620 | -0.190927 | -0.044191 | 0.035718 | 0.028698 | 0.305701 | 0.204090 |
| 450 | -0.766140 | -0.089992 | 0.072140 | -0.253662 | -0.029252 | 0.025863 | 0.008322 | 0.334413 | 0.193960 |
| True Parameters defined in Experiment 3 ($m_i = n_i = 5$, $i = 1, \ldots, 4$) | | | | | | | | | |
| 50 | -0.088561 | -0.072839 | -0.078064 | -0.001684 | -0.207593 | -0.623717 | 0.697244 | 0.094563 | -0.097617 |
| 150 | -0.032329 | -0.028899 | -0.026174 | 0.000246 | -0.080181 | -0.024025 | 0.013967 | 0.009645 | -0.032925 |
| 250 | -0.024551 | -0.010698 | -0.015302 | 0.000484 | -0.040640 | -0.025790 | -0.002152 | 0.026099 | -0.019623 |
| 350 | -0.012196 | -0.007457 | -0.009191 | -0.000175 | -0.038071 | -0.004878 | -0.000862 | 0.016825 | -0.011662 |
| 450 | -0.009700 | -0.008063 | -0.005909 | 0.000290 | -0.027451 | 0.000672 | 0.004944 | 0.004925 | -0.007727 |
| 550 | -0.009402 | -0.007019 | -0.006708 | -0.000145 | -0.021847 | 0.022580 | 0.017967 | 0.004376 | -0.008769 |
| True Parameters defined in Experiment 3 ($m_i = 10$ and $n_i = 20$, $i = 1, \ldots, 4$) | | | | | | | | | |
| 50 | -0.110275 | -0.103319 | -0.018277 | 0.005745 | -0.192519 | -0.119767 | 0.190836 | 0.096212 | -0.025917 |
| 150 | -0.006048 | -0.019684 | -0.017021 | 0.003994 | -0.068541 | -0.110964 | 0.042790 | 0.080071 | -0.022804 |
| 250 | -0.001276 | -0.004757 | -0.011416 | -0.000747 | -0.051512 | 0.007802 | 0.004713 | 0.028922 | -0.014156 |
| 350 | -0.004430 | -0.001101 | -0.010341 | 0.000670 | -0.017297 | -0.039598 | 0.000488 | 0.037260 | -0.013522 |
| 450 | -0.005908 | -0.001769 | -0.006414 | 0.001687 | -0.002752 | -0.013991 | -0.026319 | 0.029580 | -0.009105 |
| 550 | -0.004151 | -0.003921 | -0.005579 | 0.002284 | -0.008911 | -0.019388 | -0.011917 | 0.020173 | -0.007925 |
| 650 | -0.006022 | -0.002770 | -0.004142 | 0.002318 | -0.010583 | -0.016669 | -0.000291 | 0.019550 | -0.006171 |
| 750 | 0.003963 | -0.005300 | -0.004837 | 0.002129 | -0.014054 | -0.006079 | -0.012081 | 0.011144 | -0.006637 |

Table 6: Simulated MSE's of Estimates for the D.G.P. with Known Group Memberships: Stationary Covariate and Nonlinear SAR Errors

| $T$ | $\widetilde{MSE}(\widehat{\phi}_1)$ | $\widetilde{MSE}(\widehat{\phi}_2)$ | $\widetilde{MSE}(\widehat{\phi}_3)$ | $\widetilde{MSE}(\widehat{\phi}_4)$ | $\widetilde{MSE}(\widehat{\theta}_1)$ | $\widetilde{MSE}(\widehat{\theta}_2)$ | $\widetilde{MSE}(\widehat{\theta}_3)$ | $\widetilde{MSE}(\widehat{\theta}_4)$ | $\widetilde{MSE}(\widehat{\mu}_*)$ | Temp. Ave. Error* |
|---|---|---|---|---|---|---|---|---|---|---|
| True Parameters defined in Experiment 1 ($m_i = n_i = 5$, $i = 1, \ldots, 4$) | | | | | | | | | | |
| 50 | 1.728210 | 0.288209 | 0.217136 | 0.212903 | 0.196828 | 0.155598 | 0.253231 | 0.251616 | 0.247764 | 16178.8 |
| 150 | 1.876970 | 0.293202 | 0.213753 | 0.264166 | 0.255835 | 0.176938 | 0.262756 | 0.216224 | 0.166469 | 2.63E+41 |
| 250 | 1.219210 | 0.193416 | 0.149974 | 0.181570 | 0.175572 | 0.131969 | 0.222332 | 0.159459 | 0.173545 | 1.42E+79 |
| 350 | 1.038330 | 0.212277 | 0.171436 | 0.142657 | 0.150073 | 0.160658 | 0.173758 | 0.166489 | 0.153584 | 5.59E+116 |
| 450 | 1.552950 | 0.209377 | 0.222323 | 0.235925 | 0.203917 | 0.153663 | 0.248256 | 0.207164 | 0.176115 | 3.65E+154 |
| 550 | 1.568460 | 0.222412 | 0.153130 | 0.173305 | 0.171814 | 0.124806 | 0.182638 | 0.207226 | 0.172342 | 1.97E+192 |
| 650 | 1.402050 | 0.219868 | 0.155079 | 0.202297 | 0.161028 | 0.147972 | 0.200994 | 0.216402 | 0.166047 | 1.19E+230 |
| 750 | 1.438890 | 0.208126 | 0.162966 | 0.260664 | 0.194095 | 0.118434 | 0.202310 | 0.188357 | 0.155805 | 8.89E+267 |
| True Parameters defined in Experiment 1 ($m_i = 10$ and $n_i = 20$, $i = 1, \ldots, 4$) | | | | | | | | | | |
| 50 | 2.059520 | 0.257048 | 0.419943 | 2.839950 | 0.181442 | 0.390532 | 0.206021 | 0.893595 | 0.626746 | 6026.54 |
| 150 | 2.424370 | 0.276616 | 0.237308 | 0.267437 | 0.237295 | 0.186378 | 0.228351 | 0.232332 | 0.177481 | 1.53E+41 |
| 250 | 2.397670 | 0.297403 | 0.186172 | 0.268930 | 0.226620 | 0.168849 | 0.205104 | 0.227012 | 0.213569 | 6.64E+78 |
| 350 | 2.210270 | 0.240743 | 0.190967 | 0.212326 | 0.204623 | 0.151513 | 0.244701 | 0.234525 | 0.195267 | 4.00E+116 |
| 450 | 2.507860 | 0.228128 | 0.185655 | 0.301912 | 0.208411 | 0.174457 | 0.251623 | 0.227277 | 0.189432 | 1.99E+154 |
| True Parameters defined in Experiment 3 ($m_i = n_i = 5$, $i = 1, \ldots, 4$) | | | | | | | | | | |
| 50 | 0.342224 | 0.136916 | 0.074713 | 0.006490 | 10.158500 | 35.619900 | 39.783300 | 1.884670 | 0.116173 | 0.011404 |
| 150 | 0.077128 | 0.032491 | 0.013254 | 0.001734 | 0.246708 | 1.200690 | 1.790750 | 0.323320 | 0.021248 | 0.015315 |
| 250 | 0.042963 | 0.016458 | 0.007537 | 0.001077 | 0.106464 | 0.604857 | 0.859245 | 0.183681 | 0.011936 | 0.016017 |
| 350 | 0.031918 | 0.011499 | 0.004946 | 0.000766 | 0.079473 | 0.239812 | 0.456663 | 0.122209 | 0.007954 | 0.016400 |
| 450 | 0.022800 | 0.008894 | 0.003639 | 0.000549 | 0.051940 | 0.179322 | 0.333514 | 0.081298 | 0.005781 | 0.016608 |
| 550 | 0.018347 | 0.006511 | 0.002975 | 0.000426 | 0.041856 | 0.126314 | 0.270491 | 0.058633 | 0.004719 | 0.016721 |
| True Parameters defined in Experiment 3 ($m_i = 10$ and $n_i = 20$, $i = 1, \ldots, 4$) | | | | | | | | | | |
| 50 | 0.317147 | 0.131605 | 0.081244 | 0.006806 | 2.113940 | 7.739370 | 6.942440 | 1.338710 | 0.126109 | 0.001646 |
| 150 | 0.068957 | 0.031286 | 0.014025 | 0.001967 | 0.310101 | 2.115000 | 1.132980 | 0.278899 | 0.021727 | 0.002117 |
| 250 | 0.038976 | 0.016930 | 0.006883 | 0.001231 | 0.132943 | 0.186328 | 0.320449 | 0.124286 | 0.010899 | 0.002215 |
| 350 | 0.026769 | 0.010255 | 0.004615 | 0.000771 | 0.068841 | 0.132879 | 0.122417 | 0.071972 | 0.007242 | 0.002262 |
| 450 | 0.017107 | 0.007680 | 0.003376 | 0.000661 | 0.044838 | 0.084980 | 0.070107 | 0.041733 | 0.005256 | 0.002272 |
| 550 | 0.014116 | 0.005378 | 0.002433 | 0.000584 | 0.038652 | 0.064570 | 0.054273 | 0.028143 | 0.003817 | 0.002292 |
| 650 | 0.010824 | 0.004359 | 0.002240 | 0.000450 | 0.028168 | 0.051994 | 0.034113 | 0.027418 | 0.003422 | 0.002303 |
| 750 | 0.008219 | 0.004140 | 0.001906 | 0.000373 | 0.024981 | 0.039532 | 0.029416 | 0.017862 | 0.002869 | 0.002312 |

* abbrev. for the temporal average error defined as $N\frac{1}{T}\sum_{t=1}^{T}\epsilon_{*,t}(\widehat{\boldsymbol{\psi}})$.

Table 7: Simulated Biases of Estimates for the D.G.P. with Known Group Memberships: Nonstationary Covariate and Linear SAR Errors with Rook-Contiguity Weights

| $T$ | $\widetilde{Bias}(\widehat{\phi}_1)$ | $\widetilde{Bias}(\widehat{\phi}_2)$ | $\widetilde{Bias}(\widehat{\phi}_3)$ | $\widetilde{Bias}(\widehat{\phi}_4)$ | $\widetilde{Bias}(\widehat{\theta}_1)$ | $\widetilde{Bias}(\widehat{\theta}_2)$ | $\widetilde{Bias}(\widehat{\theta}_3)$ | $\widetilde{Bias}(\widehat{\theta}_4)$ | $\widetilde{Bias}(\widehat{\mu}_*)$ |
|---|---|---|---|---|---|---|---|---|---|
| Experiment 1 using (A.3) instead of (A.2) ($N_1 = 45$, $N_2 = 30$, $N_3 = 30$, and $N_4 = 70$) | | | | | | | | | |
| 50 | -0.772388 | -0.060366 | 0.006447 | -0.208354 | -0.061439 | 0.020733 | -0.032845 | 0.291188 | 0.199183 |
| 150 | -0.695992 | -0.045514 | 0.033029 | -0.210665 | -0.031436 | 0.032953 | -0.004965 | 0.298536 | 0.183180 |
| 250 | -0.692131 | -0.067166 | 0.029266 | -0.227474 | -0.022225 | 0.028950 | 0.007652 | 0.314914 | 0.172948 |
| 350 | -0.636684 | -0.043484 | 0.044083 | -0.218829 | -0.049687 | 0.050792 | 0.030509 | 0.299835 | 0.167369 |
| 450 | -0.667085 | -0.030590 | 0.036145 | -0.239834 | -0.020722 | 0.028993 | 0.007589 | 0.304018 | 0.134115 |
| 550 | -0.587248 | -0.060256 | 0.031286 | -0.217165 | -0.046386 | 0.047504 | 0.015816 | 0.302195 | 0.159778 |
| 650 | -0.595764 | -0.088938 | 0.031584 | -0.184563 | -0.044686 | 0.044455 | -0.000281 | 0.309229 | 0.184283 |
| 750 | -0.642338 | -0.058718 | 0.019819 | -0.217283 | -0.020445 | 0.043001 | -0.009625 | 0.314011 | 0.174190 |
| Experiment 2 using (A.3) instead of (A.2) ($N_1 = 100$, $N_2 = 60$, $N_3 = 65$, and $N_4 = 150$) | | | | | | | | | |
| 50 | -0.771287 | -0.078598 | 0.044942 | -0.217584 | -0.040968 | 0.036464 | 0.005318 | 0.314548 | 0.214595 |
| 150 | -0.734493 | -0.123720 | 0.044498 | -0.216069 | -0.036558 | 0.070029 | 0.005018 | 0.328518 | 0.167745 |
| 250 | -0.676299 | -0.086576 | 0.049178 | -0.219873 | -0.026853 | 0.044533 | 0.001522 | 0.314691 | 0.195681 |
| 350 | -0.735008 | -0.079657 | 0.026759 | -0.204788 | -0.034029 | 0.047367 | -0.015848 | 0.325943 | 0.154354 |
| 450 | -0.722427 | -0.085812 | 0.069341 | -0.227772 | -0.012065 | 0.042268 | 0.004839 | 0.336346 | 0.176720 |
| 550 | -0.639320 | -0.092204 | 0.029024 | -0.201654 | -0.031049 | 0.033382 | 0.016458 | 0.325093 | 0.175283 |
| 650 | -0.683882 | -0.083747 | 0.039014 | -0.221795 | -0.055973 | 0.059398 | 0.009081 | 0.300366 | 0.173214 |
| 750 | -0.699532 | -0.049467 | 0.066042 | -0.237147 | -0.031839 | 0.035950 | -0.020321 | 0.322232 | 0.167899 |
| Experiment 3 using (A.3) instead of (A.2) ($N_1 = 45$, $N_2 = 30$, $N_3 = 30$, and $N_4 = 70$) | | | | | | | | | |
| 50 | -0.332934 | -0.210182 | -0.223219 | -0.014492 | -0.018860 | -0.046950 | 0.027803 | 0.038687 | 0.038045 |
| 150 | -0.104285 | -0.027839 | -0.067039 | -0.004716 | -0.002632 | -0.000147 | 0.014571 | 0.011013 | 0.017579 |
| 250 | -0.050399 | 0.004364 | -0.039018 | -0.003676 | -0.002315 | -0.006227 | 0.012071 | 0.005766 | 0.019321 |
| 350 | -0.038685 | 0.024155 | -0.026150 | -0.003899 | -0.000554 | -0.006146 | 0.012713 | 0.006737 | 0.020357 |
| 450 | -0.024770 | 0.030836 | -0.018862 | -0.003175 | -0.000566 | 0.004119 | 0.012062 | 0.004538 | 0.020958 |
| 550 | -0.018543 | 0.045704 | -0.013415 | -0.003161 | -0.000945 | 0.004160 | 0.013147 | 0.003598 | 0.021392 |
| 650 | -0.012275 | 0.047291 | -0.010133 | -0.002574 | -0.000921 | 0.004339 | 0.013060 | 0.003299 | 0.021638 |
| 750 | -0.008899 | 0.047146 | -0.008516 | -0.001723 | -0.000799 | 0.003394 | 0.011159 | 0.002749 | 0.021715 |
| Experiment 4 using (A.3) instead of (A.2) ($N_1 = 100$, $N_2 = 60$, $N_3 = 65$, and $N_4 = 150$) | | | | | | | | | |
| 50 | -0.298269 | -0.134550 | -0.227004 | -0.013736 | -0.053053 | 0.106922 | 0.069731 | 0.041179 | 0.040849 |
| 150 | -0.092834 | 0.003013 | -0.063940 | -0.004222 | -0.006221 | -0.013396 | 0.012909 | 0.022191 | 0.016723 |
| 250 | -0.055191 | 0.029356 | -0.038036 | -0.002323 | -0.007987 | -0.016554 | 0.014924 | 0.007958 | 0.018890 |
| 350 | -0.031210 | 0.044964 | -0.024452 | -0.001811 | -0.007258 | -0.012448 | 0.014602 | 0.004280 | 0.020000 |
| 450 | -0.024181 | 0.062836 | -0.016313 | -0.001518 | -0.007952 | -0.006682 | 0.014126 | 0.003975 | 0.020657 |
| 550 | -0.011440 | 0.072551 | -0.010962 | -0.001491 | -0.006364 | -0.006231 | 0.013984 | 0.004238 | 0.021086 |
| 650 | -0.004967 | 0.072511 | -0.008052 | -0.001503 | -0.005792 | -0.002295 | 0.014154 | 0.003852 | 0.021364 |
| 750 | -0.002973 | 0.085185 | -0.006110 | -0.001287 | -0.006178 | 0.000491 | 0.012504 | 0.004934 | 0.021547 |

Table 8: Simulated MSE's of Estimates for the D.G.P. with Known Group Memberships: Nonstationary Covariate and Linear SAR Errors with Rook-Contiguity Weights

| $T$ | $\widetilde{MSE}(\widehat{\phi}_1)$ | $\widetilde{MSE}(\widehat{\phi}_2)$ | $\widetilde{MSE}(\widehat{\phi}_3)$ | $\widetilde{MSE}(\widehat{\phi}_4)$ | $\widetilde{MSE}(\widehat{\theta}_1)$ | $\widetilde{MSE}(\widehat{\theta}_2)$ | $\widetilde{MSE}(\widehat{\theta}_3)$ | $\widetilde{MSE}(\widehat{\theta}_4)$ | $\widetilde{MSE}(\widehat{\mu}_*)$ | Temp. Ave. Error* |
|---|---|---|---|---|---|---|---|---|---|---|
| Experiment 1 using (A.3) instead of (A.2) ($N_1 = 45$, $N_2 = 30$, $N_3 = 30$, and $N_4 = 70$) | | | | | | | | | | |
| 50 | 1.784080 | 0.314256 | 0.226271 | 0.395060 | 0.262656 | 0.214268 | 0.342757 | 0.247709 | 0.233542 | 1563.1 |
| 150 | 1.801640 | 0.221993 | 0.167259 | 0.232390 | 0.178434 | 0.138025 | 0.222561 | 0.198050 | 0.190342 | 3.73E+40 |
| 250 | 1.921630 | 0.256821 | 0.177929 | 0.251787 | 0.203761 | 0.170792 | 0.249554 | 0.216414 | 0.186497 | 1.66E+78 |
| 350 | 1.979670 | 0.231476 | 0.177982 | 0.247388 | 0.195132 | 0.168977 | 0.204435 | 0.212569 | 0.185672 | 8.73E+115 |
| 450 | 1.797290 | 0.240577 | 0.177953 | 0.246300 | 0.166291 | 0.162179 | 0.217141 | 0.206922 | 0.155120 | 4.81E+153 |
| 550 | 1.763160 | 0.256588 | 0.165858 | 0.216826 | 0.180196 | 0.158660 | 0.233045 | 0.194648 | 0.165821 | 3.30E+191 |
| 650 | 1.917760 | 0.258593 | 0.164770 | 0.215954 | 0.190036 | 0.161094 | 0.232589 | 0.206317 | 0.182574 | 1.98E+229 |
| 750 | 1.760470 | 0.215243 | 0.168880 | 0.256738 | 0.195297 | 0.168007 | 0.245497 | 0.236121 | 0.188579 | 1.23E+267 |
| Experiment 2 using (A.3) instead of (A.2) ($N_1 = 100$, $N_2 = 60$, $N_3 = 65$, and $N_4 = 150$) | | | | | | | | | | |
| 50 | 2.060190 | 0.260103 | 0.188610 | 0.342972 | 0.215741 | 0.185969 | 0.247222 | 0.244404 | 0.238971 | 1351.6 |
| 150 | 2.229940 | 0.333464 | 0.185489 | 0.241789 | 0.187628 | 0.183464 | 0.266003 | 0.230014 | 0.178057 | 3.34E+40 |
| 250 | 2.047340 | 0.276874 | 0.188172 | 0.234435 | 0.177164 | 0.160468 | 0.240272 | 0.216724 | 0.187430 | 1.42E+78 |
| 350 | 2.333000 | 0.257242 | 0.187210 | 0.256265 | 0.201816 | 0.170270 | 0.257236 | 0.245844 | 0.189163 | 8.06E+115 |
| 450 | 2.449120 | 0.311398 | 0.189733 | 0.233707 | 0.212801 | 0.167307 | 0.271557 | 0.231721 | 0.184160 | 4.69E+153 |
| 550 | 2.223170 | 0.288348 | 0.177577 | 0.235175 | 0.185694 | 0.160029 | 0.229321 | 0.239951 | 0.162904 | 2.82E+191 |
| 650 | 2.074080 | 0.253916 | 0.191666 | 0.260881 | 0.199513 | 0.157434 | 0.236659 | 0.218112 | 0.170691 | 1.92E+229 |
| 750 | 2.313200 | 0.253062 | 0.203446 | 0.253871 | 0.197592 | 0.180396 | 0.256337 | 0.235891 | 0.197359 | 1.23E+267 |
| Experiment 3 using (A.3) instead of (A.2) ($N_1 = 45$, $N_2 = 30$, $N_3 = 30$, and $N_4 = 70$) | | | | | | | | | | |
| 50 | 0.487350 | 0.509373 | 0.127937 | 0.011103 | 0.770408 | 2.077960 | 4.046830 | 0.366759 | 0.129155 | 0.009921 |
| 150 | 0.086940 | 0.091094 | 0.015924 | 0.002578 | 0.031488 | 0.179751 | 0.498726 | 0.023236 | 0.044953 | 0.013760 |
| 250 | 0.041659 | 0.044711 | 0.006429 | 0.001435 | 0.010799 | 0.052716 | 0.203904 | 0.007836 | 0.043367 | 0.014498 |
| 350 | 0.029621 | 0.033607 | 0.003995 | 0.000937 | 0.005415 | 0.061210 | 0.121219 | 0.003504 | 0.045357 | 0.014830 |
| 450 | 0.021292 | 0.027919 | 0.002729 | 0.000738 | 0.003162 | 0.015406 | 0.086194 | 0.001971 | 0.046745 | 0.015003 |
| 550 | 0.016762 | 0.025046 | 0.002198 | 0.000598 | 0.002186 | 0.012389 | 0.080681 | 0.001397 | 0.048044 | 0.015107 |
| 650 | 0.013687 | 0.024275 | 0.001781 | 0.000505 | 0.001604 | 0.010502 | 0.066370 | 0.000999 | 0.048626 | 0.015178 |
| 750 | 0.011976 | 0.021099 | 0.001556 | 0.000440 | 0.001261 | 0.007577 | 0.055808 | 0.000799 | 0.048803 | 0.015244 |
| Experiment 4 using (A.3) instead of (A.2) ($N_1 = 100$, $N_2 = 60$, $N_3 = 65$, and $N_4 = 150$) | | | | | | | | | | |
| 50 | 0.466390 | 0.493153 | 0.125493 | 0.009900 | 0.703763 | 3.563340 | 5.650820 | 0.356213 | 0.097226 | 0.004687 |
| 150 | 0.076306 | 0.095829 | 0.014659 | 0.002358 | 0.030725 | 0.194170 | 0.450985 | 0.021062 | 0.040098 | 0.006487 |
| 250 | 0.041382 | 0.055485 | 0.006637 | 0.001367 | 0.010460 | 0.063261 | 0.226598 | 0.007430 | 0.041107 | 0.006835 |
| 350 | 0.027891 | 0.041468 | 0.004253 | 0.000944 | 0.005526 | 0.030095 | 0.137587 | 0.003463 | 0.043630 | 0.006970 |
| 450 | 0.021156 | 0.035793 | 0.002829 | 0.000695 | 0.003293 | 0.019807 | 0.097577 | 0.001978 | 0.045199 | 0.007068 |
| 550 | 0.017492 | 0.032296 | 0.002248 | 0.000580 | 0.002296 | 0.014443 | 0.080739 | 0.001393 | 0.046578 | 0.007122 |
| 650 | 0.014244 | 0.028863 | 0.001800 | 0.000505 | 0.001757 | 0.022080 | 0.071459 | 0.000963 | 0.047348 | 0.007159 |
| 750 | 0.013783 | 0.030813 | 0.001533 | 0.000438 | 0.001446 | 0.012177 | 0.059306 | 0.000820 | 0.047936 | 0.007181 |

* abbrev. for the temporal average error defined as $N\frac{1}{T}\sum_{t=1}^{T}\epsilon_{*,t}(\widehat{\boldsymbol{\psi}})$.

Table 9: Simulated Biases of Estimates for the D.G.P. with Known Group Memberships: Nonstationary Covariate and Linear SAR Errors with Queen-Contiguity Weights

| $T$ | $\widetilde{Bias}(\widehat{\phi}_1)$ | $\widetilde{Bias}(\widehat{\phi}_2)$ | $\widetilde{Bias}(\widehat{\phi}_3)$ | $\widetilde{Bias}(\widehat{\phi}_4)$ | $\widetilde{Bias}(\widehat{\theta}_1)$ | $\widetilde{Bias}(\widehat{\theta}_2)$ | $\widetilde{Bias}(\widehat{\theta}_3)$ | $\widetilde{Bias}(\widehat{\theta}_4)$ | $\widetilde{Bias}(\widehat{\mu}_*)$ |
|---|---|---|---|---|---|---|---|---|---|
| Experiment 1 using (A.3) instead of (A.2) ($N_1 = 45$, $N_2 = 30$, $N_3 = 30$, and $N_4 = 70$) | | | | | | | | | |
| 50 | -0.750374 | -0.059340 | 0.036238 | -0.189004 | -0.030481 | 0.049368 | 0.015927 | 0.296970 | 0.213682 |
| 150 | -0.596223 | -0.037154 | 0.045307 | -0.205836 | -0.033283 | 0.036377 | -0.004280 | 0.314710 | 0.173781 |
| 250 | -0.666517 | -0.062015 | 0.025652 | -0.214916 | -0.041981 | 0.031282 | 0.013327 | 0.303673 | 0.181018 |
| 350 | -0.739513 | -0.085616 | 0.019156 | -0.220549 | -0.031562 | 0.067190 | 0.024533 | 0.295508 | 0.181225 |
| 450 | -0.623897 | -0.041777 | 0.007002 | -0.187998 | -0.021524 | 0.030250 | 0.021204 | 0.294637 | 0.178217 |
| 550 | -0.550194 | -0.045520 | 0.013775 | -0.188376 | -0.026025 | 0.038938 | 0.021005 | 0.286747 | 0.179588 |
| 650 | -0.648418 | -0.047359 | 0.023947 | -0.214107 | -0.031769 | 0.048818 | -0.005876 | 0.288040 | 0.193557 |
| 750 | -0.610700 | -0.046447 | 0.030282 | -0.216796 | -0.018638 | 0.029702 | 0.000064 | 0.302119 | 0.177816 |
| Experiment 2 using (A.3) instead of (A.2) ($N_1 = 100$, $N_2 = 60$, $N_3 = 65$, and $N_4 = 150$) | | | | | | | | | |
| 50 | -0.767932 | -0.102504 | 0.038142 | -0.208172 | -0.038660 | 0.025830 | -0.001000 | 0.321977 | 0.188069 |
| 150 | -0.658856 | -0.064264 | 0.042636 | -0.228541 | -0.034421 | 0.038988 | 0.006874 | 0.297403 | 0.190425 |
| 250 | -0.677868 | -0.082119 | 0.052670 | -0.228813 | -0.043217 | 0.035842 | -0.011712 | 0.323002 | 0.178721 |
| 350 | -0.652315 | -0.086769 | 0.058895 | -0.231834 | -0.049911 | 0.056498 | 0.015868 | 0.308270 | 0.169527 |
| 450 | -0.618034 | -0.073036 | 0.025667 | -0.208497 | -0.043298 | 0.063612 | 0.008473 | 0.304012 | 0.165474 |
| 550 | -0.721412 | -0.094638 | 0.032878 | -0.216123 | -0.036784 | 0.039188 | 0.007039 | 0.313777 | 0.187874 |
| 650 | -0.676821 | -0.085765 | 0.046754 | -0.218858 | -0.030699 | 0.070698 | 0.003455 | 0.315766 | 0.175367 |
| 750 | -0.607330 | -0.077979 | 0.028573 | -0.248658 | -0.013803 | 0.048530 | -0.011977 | 0.295002 | 0.200063 |
| Experiment 3 using (A.3) instead of (A.2) ($N_1 = 45$, $N_2 = 30$, $N_3 = 30$, and $N_4 = 70$) | | | | | | | | | |
| 50 | -0.323332 | -0.204390 | -0.226941 | -0.016450 | -0.060166 | -0.058117 | 0.044352 | -0.028006 | 0.032992 |
| 150 | -0.109502 | -0.035118 | -0.066658 | -0.006866 | -0.006436 | 0.005972 | 0.014114 | 0.000664 | 0.017515 |
| 250 | -0.058141 | 0.002419 | -0.039037 | -0.003599 | -0.007875 | -0.005268 | 0.015267 | 0.002809 | 0.019248 |
| 350 | -0.040712 | 0.025539 | -0.024888 | -0.002955 | -0.007934 | 0.004258 | 0.013338 | 0.001258 | 0.020657 |
| 450 | -0.035537 | 0.039735 | -0.016089 | -0.002164 | -0.005481 | 0.000834 | 0.013660 | -0.001043 | 0.021317 |
| 550 | -0.022099 | 0.042947 | -0.012499 | -0.001763 | -0.005539 | -0.000117 | 0.013132 | -0.001356 | 0.021581 |
| 650 | -0.018917 | 0.052485 | -0.009119 | -0.001709 | -0.002015 | -0.002426 | 0.012180 | 0.000197 | 0.021749 |
| 750 | -0.011633 | 0.057715 | -0.006167 | -0.001530 | -0.002465 | 0.000526 | 0.011310 | 0.001421 | 0.022037 |
| Experiment 4 using (A.3) instead of (A.2) ($N_1 = 100$, $N_2 = 60$, $N_3 = 65$, and $N_4 = 150$) | | | | | | | | | |
| 50 | -0.319205 | -0.188288 | -0.207703 | -0.013701 | -0.093781 | -0.016124 | 0.019829 | 0.043420 | 0.048372 |
| 150 | -0.092007 | -0.023151 | -0.060868 | -0.003592 | -0.020264 | -0.015444 | 0.021437 | 0.001367 | 0.016895 |
| 250 | -0.059914 | 0.020700 | -0.032654 | -0.003162 | -0.008259 | -0.008314 | 0.017653 | 0.004542 | 0.019453 |
| 350 | -0.041768 | 0.032592 | -0.018330 | -0.002506 | -0.005846 | -0.003991 | 0.016698 | 0.005212 | 0.020809 |
| 450 | -0.027339 | 0.041132 | -0.012858 | -0.001658 | -0.006438 | -0.000128 | 0.015659 | 0.003810 | 0.021062 |
| 550 | -0.021883 | 0.053394 | -0.006908 | -0.001972 | -0.005404 | -0.003963 | 0.015852 | 0.004001 | 0.021596 |
| 650 | -0.011323 | 0.062252 | -0.005209 | -0.001362 | -0.006420 | -0.006080 | 0.014775 | 0.002116 | 0.021697 |
| 750 | -0.010509 | 0.069957 | -0.003996 | -0.001468 | -0.004755 | 0.000844 | 0.012591 | 0.002868 | 0.021810 |

Table 10: Simulated MSE's of Estimates for the D.G.P. with Known Group Memberships: Nonstationary Covariate and Linear SAR Errors with Queen-Contiguity Weights

| $T$ | $\widetilde{MSE}(\widehat{\phi}_1)$ | $\widetilde{MSE}(\widehat{\phi}_2)$ | $\widetilde{MSE}(\widehat{\phi}_3)$ | $\widetilde{MSE}(\widehat{\phi}_4)$ | $\widetilde{MSE}(\widehat{\theta}_1)$ | $\widetilde{MSE}(\widehat{\theta}_2)$ | $\widetilde{MSE}(\widehat{\theta}_3)$ | $\widetilde{MSE}(\widehat{\theta}_4)$ | $\widetilde{MSE}(\widehat{\mu}_*)$ | Temp. Ave. Error* |
|---|---|---|---|---|---|---|---|---|---|---|
| Experiment 1 using (A.3) instead of (A.2) ($N_1 = 45$, $N_2 = 30$, $N_3 = 30$, and $N_4 = 70$) | | | | | | | | | | |
| 50 | 1.718020 | 0.230223 | 0.157970 | 0.255591 | 0.161161 | 0.165465 | 0.221358 | 0.194128 | 0.219575 | 1554.32 |
| 150 | 1.690030 | 0.208471 | 0.179957 | 0.224608 | 0.180050 | 0.172149 | 0.214264 | 0.221890 | 0.172904 | 3.92E+40 |
| 250 | 1.969860 | 0.231707 | 0.179631 | 0.248554 | 0.174893 | 0.156223 | 0.222335 | 0.229651 | 0.189242 | 1.57E+78 |
| 350 | 2.181670 | 0.287893 | 0.180237 | 0.249106 | 0.180368 | 0.163994 | 0.237818 | 0.219436 | 0.194289 | 9.04E+115 |
| 450 | 1.797500 | 0.234302 | 0.178447 | 0.228286 | 0.202433 | 0.170167 | 0.211633 | 0.207135 | 0.193975 | 5.01E+153 |
| 550 | 1.631810 | 0.216623 | 0.164009 | 0.203291 | 0.180200 | 0.139823 | 0.223747 | 0.197282 | 0.198348 | 3.02E+191 |
| 650 | 1.827440 | 0.252081 | 0.189178 | 0.238634 | 0.185416 | 0.163750 | 0.199683 | 0.208002 | 0.194197 | 1.87E+229 |
| 750 | 1.829510 | 0.266355 | 0.152826 | 0.226327 | 0.180630 | 0.141238 | 0.243661 | 0.220539 | 0.175255 | 1.39E+267 |
| Experiment 2 using (A.3) instead of (A.2) ($N_1 = 100$, $N_2 = 60$, $N_3 = 65$, and $N_4 = 150$) | | | | | | | | | | |
| 50 | 2.131960 | 0.297478 | 0.204748 | 0.390080 | 0.230395 | 0.192362 | 0.270955 | 0.254558 | 0.251135 | 1372.99 |
| 150 | 2.096920 | 0.230272 | 0.163723 | 0.238251 | 0.197437 | 0.150039 | 0.234048 | 0.195598 | 0.188017 | 3.31E+40 |
| 250 | 2.448480 | 0.254388 | 0.215078 | 0.267266 | 0.176388 | 0.167501 | 0.225804 | 0.221680 | 0.187835 | 1.31E+78 |
| 350 | 2.318060 | 0.303061 | 0.184964 | 0.258281 | 0.209987 | 0.192934 | 0.254420 | 0.231384 | 0.179608 | 7.71E+115 |
| 450 | 2.281800 | 0.240938 | 0.180969 | 0.241780 | 0.192493 | 0.164916 | 0.263411 | 0.211147 | 0.206284 | 4.88E+153 |
| 550 | 2.192640 | 0.268200 | 0.178558 | 0.234374 | 0.186121 | 0.174338 | 0.220539 | 0.232863 | 0.209650 | 2.87E+191 |
| 650 | 2.277900 | 0.290470 | 0.185857 | 0.238317 | 0.191953 | 0.206126 | 0.239182 | 0.236510 | 0.215025 | 1.87E+229 |
| 750 | 2.109920 | 0.252035 | 0.173455 | 0.270233 | 0.213992 | 0.153196 | 0.254738 | 0.211144 | 0.198459 | 1.24E+267 |
| Experiment 3 using (A.3) instead of (A.2) ($N_1 = 45$, $N_2 = 30$, $N_3 = 30$, and $N_4 = 70$) | | | | | | | | | | |
| 50 | 0.506027 | 0.534659 | 0.131719 | 0.009732 | 0.677963 | 4.108360 | 7.510570 | 0.348094 | 0.124931 | 0.009888 |
| 150 | 0.083221 | 0.093116 | 0.014714 | 0.002372 | 0.029124 | 0.151607 | 0.465938 | 0.021103 | 0.044092 | 0.013779 |
| 250 | 0.042952 | 0.047258 | 0.006762 | 0.001407 | 0.011193 | 0.059394 | 0.196051 | 0.007124 | 0.043693 | 0.014464 |
| 350 | 0.028293 | 0.036062 | 0.004128 | 0.000883 | 0.005820 | 0.026954 | 0.120033 | 0.003488 | 0.046846 | 0.014821 |
| 450 | 0.020880 | 0.028838 | 0.002777 | 0.000688 | 0.003125 | 0.022463 | 0.090966 | 0.002083 | 0.048189 | 0.014999 |
| 550 | 0.017342 | 0.026122 | 0.002104 | 0.000571 | 0.002267 | 0.012420 | 0.072812 | 0.001328 | 0.048817 | 0.015113 |
| 650 | 0.014288 | 0.025499 | 0.001816 | 0.000464 | 0.001538 | 0.010051 | 0.066007 | 0.001003 | 0.049295 | 0.015195 |
| 750 | 0.012921 | 0.024114 | 0.001527 | 0.000422 | 0.001229 | 0.008266 | 0.053374 | 0.000779 | 0.050256 | 0.015242 |
| Experiment 4 using (A.3) instead of (A.2) ($N_1 = 100$, $N_2 = 60$, $N_3 = 65$, and $N_4 = 150$) | | | | | | | | | | |
| 50 | 0.461307 | 0.600062 | 0.118638 | 0.010954 | 0.681219 | 2.745920 | 3.940910 | 0.315828 | 0.096580 | 0.004778 |
| 150 | 0.080414 | 0.093677 | 0.014393 | 0.002354 | 0.034382 | 0.163862 | 0.476128 | 0.021043 | 0.040775 | 0.006508 |
| 250 | 0.041367 | 0.052245 | 0.006153 | 0.001340 | 0.010378 | 0.056904 | 0.215778 | 0.006877 | 0.043299 | 0.006879 |
| 350 | 0.027448 | 0.038426 | 0.003902 | 0.000906 | 0.005080 | 0.029174 | 0.139556 | 0.003564 | 0.046774 | 0.007011 |
| 450 | 0.022047 | 0.033594 | 0.002730 | 0.000715 | 0.003139 | 0.019925 | 0.104372 | 0.002089 | 0.046995 | 0.007090 |
| 550 | 0.017144 | 0.031196 | 0.002123 | 0.000597 | 0.002233 | 0.016444 | 0.089748 | 0.001429 | 0.048887 | 0.007135 |
| 650 | 0.015520 | 0.028337 | 0.001844 | 0.000517 | 0.001610 | 0.012396 | 0.074753 | 0.001063 | 0.048936 | 0.007172 |
| 750 | 0.013649 | 0.026350 | 0.001485 | 0.000440 | 0.001458 | 0.009525 | 0.060045 | 0.000891 | 0.049124 | 0.007200 |

* abbrev. for the temporal average error defined as $N\frac{1}{T}\sum_{t=1}^{T}\epsilon_{*,t}(\widehat{\boldsymbol{\psi}})$.

Table 11: Simulated Biases of Estimates for the D.G.P. with Known Group Memberships: Nonstationary Covariate and Nonlinear SAR Errors

| $T$ | $\widetilde{Bias}(\widehat{\phi}_1)$ | $\widetilde{Bias}(\widehat{\phi}_2)$ | $\widetilde{Bias}(\widehat{\phi}_3)$ | $\widetilde{Bias}(\widehat{\phi}_4)$ | $\widetilde{Bias}(\widehat{\theta}_1)$ | $\widetilde{Bias}(\widehat{\theta}_2)$ | $\widetilde{Bias}(\widehat{\theta}_3)$ | $\widetilde{Bias}(\widehat{\theta}_4)$ | $\widetilde{Bias}(\widehat{\mu}_*)$ |
|---|---|---|---|---|---|---|---|---|---|
| True Parameters defined in Experiment 1 ($m_i = n_i = 5$, $i = 1, \ldots, 4$) | | | | | | | | | |
| 50 | -0.723526 | -0.040998 | 0.046629 | -0.104300 | -0.062130 | 0.077459 | 0.043379 | 0.356527 | 0.132647 |
| 150 | -0.567370 | -0.046924 | -0.006174 | -0.191241 | -0.051944 | 0.082715 | 0.001029 | 0.307892 | 0.139970 |
| 250 | -0.441991 | -0.040921 | 0.007777 | -0.163293 | -0.078092 | 0.080684 | 0.091359 | 0.254318 | 0.133108 |
| 350 | -0.707159 | -0.031938 | 0.041027 | -0.212994 | -0.040094 | 0.111859 | 0.042953 | 0.255755 | 0.188774 |
| 450 | -0.648502 | -0.053715 | -0.015168 | -0.242682 | -0.069870 | 0.039701 | 0.030075 | 0.273338 | 0.134138 |
| 550 | -0.609942 | 0.030887 | 0.009425 | -0.175573 | -0.039331 | 0.033564 | -0.081273 | 0.250582 | 0.193532 |
| 650 | -0.581432 | -0.065585 | 0.019990 | -0.218107 | -0.039708 | 0.031023 | 0.036678 | 0.322678 | 0.147192 |
| 750 | -0.619435 | -0.022735 | -0.002821 | -0.192225 | -0.000853 | -0.000329 | 0.010283 | 0.329151 | 0.130911 |
| True Parameters defined in Experiment 1 ($m_i = 10$ and $n_i = 20$, $i = 1, \ldots, 4$) | | | | | | | | | |
| 50 | -0.646916 | -0.038591 | -0.001939 | -0.157836 | -0.059178 | 0.037025 | 0.040185 | 0.229984 | 0.231449 |
| 150 | -0.812449 | -0.154695 | 0.067059 | -0.255797 | -0.038682 | 0.030016 | 0.033121 | 0.340511 | 0.157837 |
| 250 | -0.551709 | -0.088185 | 0.062548 | -0.161069 | -0.008340 | 0.020018 | -0.010888 | 0.366303 | 0.156433 |
| 350 | -0.548448 | -0.089430 | 0.030634 | -0.246120 | -0.019109 | 0.098373 | 0.042833 | 0.291950 | 0.173772 |
| 450 | -0.604822 | -0.035208 | 0.045075 | -0.236033 | -0.046778 | 0.030309 | -0.038611 | 0.306416 | 0.160123 |
| 550 | -0.708774 | -0.117796 | 0.027318 | -0.199916 | -0.076301 | -0.000482 | -0.004988 | 0.322035 | 0.176691 |
| 650 | -0.655900 | -0.091911 | 0.073396 | -0.232343 | -0.054881 | 0.036292 | 0.027440 | 0.305545 | 0.155581 |
| 750 | -1.004600 | -0.127519 | 0.101725 | -0.201340 | -0.074953 | 0.079328 | 0.033038 | 0.370229 | 0.165211 |
| True Parameters defined in Experiment 3 ($m_i = 10$ and $n_i = 20$, $i = 1, \ldots, 4$) | | | | | | | | | |
| 50 | -0.226658 | -0.243265 | -0.133711 | -0.009333 | 0.003264 | 0.064979 | -0.153564 | -0.026662 | -0.163471 |
| 150 | -0.019068 | -0.036048 | -0.038491 | 0.003547 | -0.008931 | 0.018213 | -0.005528 | -0.003632 | -0.049383 |
| 250 | 0.003083 | -0.004560 | -0.005731 | 0.004262 | -0.001570 | 0.001037 | -0.000855 | -0.002406 | -0.008321 |
| 350 | 0.002159 | -0.002231 | -0.002147 | 0.002171 | -0.000544 | -0.000048 | 0.002856 | -0.002007 | -0.002323 |
| 450 | 0.003258 | -0.000438 | -0.000672 | 0.002132 | 0.000877 | -0.000191 | 0.001962 | -0.002340 | -0.001419 |
| 550 | 0.001636 | -0.000787 | 0.001204 | 0.001808 | -0.000415 | 0.001482 | 0.000774 | -0.001102 | 0.000293 |
| 650 | 0.001966 | -0.000620 | -0.000689 | 0.000789 | -0.001484 | 0.002071 | -0.002052 | 0.000816 | -0.001454 |
| 750 | 0.001076 | -0.000369 | -0.000588 | 0.001306 | -0.002092 | 0.002496 | -0.001924 | 0.001331 | -0.001331 |

Table 12: Simulated MSE's of Estimates for the D.G.P. with Known Group Memberships: Nonstationary Covariate and Nonlinear SAR Errors

| $T$ | $\widetilde{MSE}(\widehat{\phi}_1)$ | $\widetilde{MSE}(\widehat{\phi}_2)$ | $\widetilde{MSE}(\widehat{\phi}_3)$ | $\widetilde{MSE}(\widehat{\phi}_4)$ | $\widetilde{MSE}(\widehat{\theta}_1)$ | $\widetilde{MSE}(\widehat{\theta}_2)$ | $\widetilde{MSE}(\widehat{\theta}_3)$ | $\widetilde{MSE}(\widehat{\theta}_4)$ | $\widetilde{MSE}(\widehat{\mu}_*)$ | Temp. Ave. Error* |
|---|---|---|---|---|---|---|---|---|---|---|
| True Parameters defined in Experiment 1 ($m_i = n_i = 5$, $i = 1, \ldots, 4$) | | | | | | | | | | |
| 50 | 1.466860 | 0.295354 | 0.269690 | 0.291832 | 0.215242 | 0.975649 | 0.270796 | 0.524737 | 1.061990 | 2701.36 |
| 150 | 1.563000 | 0.167259 | 0.161158 | 0.201019 | 0.183243 | 0.154263 | 0.193984 | 0.210872 | 0.185222 | 5.61E+40 |
| 250 | 1.442620 | 0.315143 | 0.116822 | 0.246616 | 0.191720 | 0.190472 | 0.190428 | 0.161411 | 0.191952 | 2.87E+78 |
| 350 | 1.713420 | 0.201995 | 0.127012 | 0.210029 | 0.143973 | 0.141357 | 0.205595 | 0.144391 | 0.164993 | 1.27E+116 |
| 450 | 1.548100 | 0.245364 | 0.166674 | 0.243103 | 0.147247 | 0.123204 | 0.203177 | 0.214479 | 0.141302 | 7.39E+153 |
| 550 | 1.520010 | 0.162698 | 0.162997 | 0.207051 | 0.206016 | 0.187106 | 0.214274 | 0.168558 | 0.195972 | 6.10E+191 |
| 650 | 1.725780 | 0.234293 | 0.153574 | 0.228572 | 0.165952 | 0.139216 | 0.176861 | 0.222916 | 0.197904 | 2.98E+229 |
| 750 | 1.693850 | 0.227462 | 0.174560 | 0.241774 | 0.154371 | 0.165789 | 0.200635 | 0.232889 | 0.149284 | 2.16E+267 |
| True Parameters defined in Experiment 1 ($m_i = 10$ and $n_i = 20$, $i = 1, \ldots, 4$) | | | | | | | | | | |
| 50 | 1.511030 | 0.362425 | 0.210047 | 0.627462 | 0.214376 | 0.247412 | 0.351005 | 0.196661 | 0.401426 | 1052.45 |
| 150 | 2.957730 | 0.325895 | 0.191065 | 0.283897 | 0.286984 | 0.159210 | 0.266135 | 0.283765 | 0.183317 | 2.67E+40 |
| 250 | 1.994570 | 0.264033 | 0.188677 | 0.248677 | 0.153454 | 0.136016 | 0.210630 | 0.257604 | 0.190057 | 1.28E+78 |
| 350 | 2.368800 | 0.270339 | 0.163505 | 0.189996 | 0.173890 | 0.191413 | 0.171048 | 0.227594 | 0.155334 | 7.53E+115 |
| 450 | 2.118350 | 0.200034 | 0.164534 | 0.242241 | 0.167970 | 0.169146 | 0.183241 | 0.235458 | 0.200467 | 3.70E+153 |
| 550 | 2.182980 | 0.239595 | 0.189539 | 0.233886 | 0.193546 | 0.143679 | 0.194847 | 0.230771 | 0.168508 | 2.75E+191 |
| 650 | 2.935830 | 0.307808 | 0.195729 | 0.213703 | 0.201561 | 0.234406 | 0.299095 | 0.205404 | 0.187226 | 1.81E+229 |
| 750 | 2.777710 | 0.361537 | 0.230852 | 0.237976 | 0.233104 | 0.216640 | 0.211317 | 0.276054 | 0.196220 | 1.21E+267 |
| True Parameters defined in Experiment 3 ($m_i = 10$ and $n_i = 20$, $i = 1, \ldots, 4$) | | | | | | | | | | |
| 50 | 0.349936 | 0.212711 | 0.094793 | 0.009885 | 0.142102 | 0.271360 | 0.745268 | 0.156800 | 0.179689 | 0.001583 |
| 150 | 0.041027 | 0.015733 | 0.012396 | 0.002354 | 0.007646 | 0.017754 | 0.047755 | 0.013066 | 0.022906 | 0.002115 |
| 250 | 0.007670 | 0.004515 | 0.002664 | 0.001338 | 0.001880 | 0.004710 | 0.009541 | 0.003138 | 0.005075 | 0.002226 |
| 350 | 0.003799 | 0.001789 | 0.001367 | 0.000715 | 0.000735 | 0.002058 | 0.003784 | 0.001082 | 0.002577 | 0.002274 |
| 450 | 0.000668 | 0.000459 | 0.000445 | 0.000384 | 0.000442 | 0.000901 | 0.000978 | 0.000587 | 0.000996 | 0.002285 |
| 550 | 0.000221 | 0.000148 | 0.000188 | 0.000205 | 0.000264 | 0.000451 | 0.000347 | 0.000226 | 0.000374 | 0.002304 |
| 650 | 0.000188 | 0.000083 | 0.000173 | 0.000094 | 0.000161 | 0.000259 | 0.000200 | 0.000132 | 0.000321 | 0.002314 |
| 750 | 0.000074 | 0.000045 | 0.000055 | 0.000064 | 0.000107 | 0.000162 | 0.000130 | 0.000082 | 0.000101 | 0.002322 |

* abbrev. for the temporal average error defined as $N\frac{1}{T}\sum_{t=1}^{T}\epsilon_{*,t}(\widehat{\boldsymbol{\psi}})$.

Table 13: Simulated Biases of Estimates for the D.G.P. with Unknown Group Memberships: Stationary Covariate and Linear SAR Errors with Queen-Contiguity Weights

| $T$ | $\widetilde{RandI}$ | $\widetilde{Bias}(\widehat{\phi}_1)$ | $\widetilde{Bias}(\widehat{\phi}_2)$ | $\widetilde{Bias}(\widehat{\phi}_3)$ | $\widetilde{Bias}(\widehat{\phi}_4)$ | $\widetilde{Bias}(\widehat{\theta}_1)$ | $\widetilde{Bias}(\widehat{\theta}_2)$ | $\widetilde{Bias}(\widehat{\theta}_3)$ | $\widetilde{Bias}(\widehat{\theta}_4)$ | $\widetilde{Bias}(\widehat{\boldsymbol{U}})^*$ |
|---|---|---|---|---|---|---|---|---|---|---|
| Experiment 3 ($N_1 = 45$, $N_2 = 30$, $N_3 = 30$, and $N_4 = 70$) | | | | | | | | | | |
| 50 | 0.836711 | -0.200000 | 0.095032 | 0.020655 | 0.148046 | 0.009603 | 0.011330 | -0.009128 | -0.014196 | 1.16E-18 |
| 150 | 0.836693 | -0.199999 | 0.089935 | 0.019214 | 0.110930 | 0.009278 | 0.005221 | -0.008646 | -0.014399 | 1.10E-18 |
| 250 | 0.836689 | -0.299999 | 0.083727 | 0.017611 | 0.104466 | 0.008208 | 0.009424 | -0.007622 | -0.013958 | 1.09E-18 |
| 350 | 0.836687 | -0.199999 | 0.067225 | 0.016771 | 0.136913 | 0.007224 | 0.013576 | -0.006649 | -0.013525 | 1.18E-18 |
| 450 | 0.836686 | -0.080000 | 0.011006 | 0.015864 | 0.089830 | 0.006397 | 0.007531 | -0.005702 | -0.013107 | 1.18E-18 |
| 550 | 0.836684 | -0.060000 | 0.012274 | 0.015444 | 0.035880 | 0.005543 | 0.003146 | -0.004791 | -0.012700 | 1.19E-18 |
| 650 | 0.836684 | -0.070000 | 0.013526 | 0.015076 | 0.041917 | 0.004674 | 0.002536 | -0.003875 | -0.012295 | 1.24E-18 |
| Experiment 4 ($N_1 = 100$, $N_2 = 60$, $N_3 = 65$, and $N_4 = 150$) | | | | | | | | | | |
| 50 | 0.925562 | -0.099428 | 0.339384 | -0.003816 | 0.008937 | 0.004996 | 0.004546 | -0.004841 | -0.086793 | 4.09E-19 |
| 150 | 0.924752 | -0.199423 | 0.092875 | -0.001761 | 0.005797 | 0.004227 | 0.004346 | -0.005541 | -0.037867 | 1.05E-19 |
| 250 | 0.924410 | -0.099419 | 0.087759 | -0.002013 | 0.004822 | 0.002850 | 0.001880 | -0.002841 | -0.021134 | 1.08E-19 |
| 350 | 0.929492 | -0.002970 | -0.003031 | -0.005825 | 0.002198 | 0.025807 | 0.001741 | -0.001121 | -0.018386 | 4.77E-21 |

* abbrev. for the *optimal matching* biases of estimates of the group indicators $\boldsymbol{U}_0$, measured by $\min_{\sigma^{(per)}\in\sigma(\mathcal{P})} \frac{1}{N}\sum_{c=1}^{G}\sum_{i=1}^{N}\{\widehat{u}_{i,\sigma^{(per)}(c)} - u_{0,i,c}\}$

Table 14: Simulated MSE of Estimates for the D.G.P. with Unknown Group Memberships: Stationary Covariate and Linear SAR Errors with Queen-Contiguity Weights

| $T$ | $\widetilde{MSE}(\widehat{\phi}_1)$ | $\widetilde{MSE}(\widehat{\phi}_2)$ | $\widetilde{MSE}(\widehat{\phi}_3)$ | $\widetilde{MSE}(\widehat{\phi}_4)$ | $\widetilde{MSE}(\widehat{\theta}_1)$ | $\widetilde{MSE}(\widehat{\theta}_2)$ | $\widetilde{MSE}(\widehat{\theta}_3)$ | $\widetilde{MSE}(\widehat{\theta}_4)$ | $\widetilde{MSE}(\widehat{\boldsymbol{U}})^*$ |
|---|---|---|---|---|---|---|---|---|---|
| Experiment 3 ($N_1 = 45$, $N_2 = 30$, $N_3 = 30$, and $N_4 = 70$) | | | | | | | | | |
| 50 | 0.150000 | 0.026653 | 0.011333 | 0.084379 | 0.005106 | 0.000850 | 0.002475 | 0.504371 | 0.000214 |
| 150 | 0.120000 | 0.027223 | 0.011497 | 0.013927 | 0.005343 | 0.000902 | 0.002644 | 0.427144 | 5.93E-06 |
| 250 | 0.051000 | 0.017473 | 0.006156 | 0.019005 | 0.005351 | 0.000924 | 0.002145 | 0.427167 | 1.20E-06 |
| 350 | 0.049000 | 0.017699 | 0.006138 | 0.023703 | 0.005357 | 0.000946 | 0.002047 | 0.527188 | 7.86E-07 |
| 450 | 0.037000 | 0.008900 | 0.003650 | 0.008162 | 0.004606 | 0.000964 | 0.001948 | 0.327207 | 5.57E-07 |
| 550 | 0.024000 | 0.002809 | 0.001677 | 0.003390 | 0.004364 | 0.000982 | 0.001649 | 0.127225 | 4.31E-07 |
| 650 | 0.020000 | 0.002627 | 0.001698 | 0.003149 | 0.003266 | 0.000999 | 0.001650 | 0.127243 | 3.55E-07 |
| Experiment 4 ($N_1 = 100$, $N_2 = 60$, $N_3 = 65$, and $N_4 = 150$) | | | | | | | | | |
| 50 | 0.089949 | 0.040047 | 0.004028 | 0.008898 | 0.009925 | 0.002190 | 0.011651 | 0.009422 | 0.000143 |
| 150 | 0.074986 | 0.033616 | 0.003405 | 0.007984 | 0.008924 | 0.001507 | 0.008281 | 0.005682 | 1.19E-05 |
| 250 | 0.056772 | 0.016365 | 0.001779 | 0.004932 | 0.002923 | 0.000677 | 0.002310 | 0.001174 | 4.99E-06 |
| 350 | 0.001103 | 0.001148 | 0.004241 | 0.000604 | 0.001251 | 0.017232 | 0.005461 | 0.000723 | 5.01E-06 |

* abbrev. for the *optimal matching* MSE of estimates of the group indicators $\boldsymbol{U}_0$, measured by $\min_{\sigma^{(per)}\in\sigma(\mathcal{P})} \frac{1}{N}\sum_{c=1}^{G}\sum_{i=1}^{N}\left(\widehat{u}_{i,\sigma^{(per)}(c)} - u_{0,i,c}\right)^2$

Table 15: Simulated Biases of Estimates for the D.G.P. with Unknown Group Memberships: Stationary Covariate and Linear SAR Errors with Rook-Contiguity Weights

| $T$ | $\widetilde{RandI}$ | $\widetilde{Bias}(\widehat{\phi}_1)$ | $\widetilde{Bias}(\widehat{\phi}_2)$ | $\widetilde{Bias}(\widehat{\phi}_3)$ | $\widetilde{Bias}(\widehat{\phi}_4)$ | $\widetilde{Bias}(\widehat{\theta}_1)$ | $\widetilde{Bias}(\widehat{\theta}_2)$ | $\widetilde{Bias}(\widehat{\theta}_3)$ | $\widetilde{Bias}(\widehat{\theta}_4)$ | $\widetilde{Bias}(\widehat{\boldsymbol{U}})$ |
|---|---|---|---|---|---|---|---|---|---|---|
| Experiment 3 ($N_1 = 45$, $N_2 = 30$, $N_3 = 30$, and $N_4 = 70$) | | | | | | | | | | |
| 50 | 0.836716 | -0.099100 | 0.054333 | 0.014412 | 0.510008 | 0.011888 | 0.015607 | -0.010132 | -0.014834 | 1.07E-18 |
| 150 | 0.836694 | -0.090900 | 0.069532 | 0.013345 | 0.114049 | 0.011624 | 0.014626 | -0.009505 | -0.015013 | 1.11E-18 |
| 250 | 0.836689 | -0.071100 | 0.053505 | 0.011841 | 0.106889 | 0.010447 | 0.008930 | -0.008493 | -0.014565 | 1.17E-18 |
| 350 | 0.837688 | -0.070900 | 0.047049 | 0.011199 | 0.129387 | 0.009519 | 0.007094 | -0.007537 | -0.014129 | 1.17E-18 |
| 450 | 0.837696 | -0.069100 | 0.040171 | 0.010565 | 0.090897 | 0.008602 | 0.007043 | -0.006593 | -0.013708 | 1.21E-18 |
| 550 | 0.839185 | -0.058700 | 0.032759 | 0.009754 | 0.061515 | 0.007703 | 0.005036 | -0.005651 | -0.010292 | 1.12E-18 |
| 650 | 0.839885 | -0.053000 | 0.023551 | 0.009078 | 0.042456 | 0.006716 | 0.004967 | -0.004720 | -0.009883 | 1.18E-18 |
| 750 | 0.840684 | -0.049200 | 0.014789 | 0.008342 | 0.041850 | 0.005821 | 0.003774 | -0.003812 | -0.004781 | 1.18E-18 |
| Experiment 4 ($N_1 = 100$, $N_2 = 60$, $N_3 = 65$, and $N_4 = 150$) | | | | | | | | | | |
| 50 | 0.838567 | -0.094636 | 0.029774 | 0.008948 | 0.087649 | 0.029731 | -0.003346 | -0.005173 | -0.021587 | 8.27E-19 |
| 150 | 0.834924 | -0.094514 | 0.039710 | 0.009719 | 0.075975 | 0.023436 | -0.001940 | -0.004988 | -0.020252 | 8.37E-19 |
| 250 | 0.839924 | -0.059450 | 0.048313 | 0.010082 | 0.064472 | 0.024444 | -0.000148 | -0.004727 | -0.019214 | 8.41E-19 |
| 350 | 0.843923 | -0.039450 | 0.045543 | 0.009858 | 0.054150 | 0.024203 | 0.002094 | -0.004170 | -0.009798 | 8.67E-19 |
| 450 | 0.893273 | -0.039449 | 0.012743 | 0.009267 | 0.044112 | 0.013975 | 0.001310 | -0.003606 | -0.003746 | 8.82E-19 |
| 550 | 0.910910 | -0.018448 | 0.007001 | 0.008746 | 0.023409 | 0.013712 | 0.001569 | -0.003054 | -0.002511 | 8.89E-19 |
| 650 | 0.910780 | -0.009447 | 0.007723 | 0.008356 | 0.021432 | 0.013477 | 0.001806 | -0.002501 | -0.002792 | 8.53E-19 |

Table 16: Simulated MSE of Estimates for the D.G.P. with Unknown Group Memberships: Stationary Covariate and Linear SAR Errors with Rook-Contiguity Weights

| $T$ | $\widetilde{MSE}(\widehat{\phi}_1)$ | $\widetilde{MSE}(\widehat{\phi}_2)$ | $\widetilde{MSE}(\widehat{\phi}_3)$ | $\widetilde{MSE}(\widehat{\phi}_4)$ | $\widetilde{MSE}(\widehat{\theta}_1)$ | $\widetilde{MSE}(\widehat{\theta}_2)$ | $\widetilde{MSE}(\widehat{\theta}_3)$ | $\widetilde{MSE}(\widehat{\theta}_4)$ | $\widetilde{MSE}(\widehat{\boldsymbol{U}})$ |
|---|---|---|---|---|---|---|---|---|---|
| Experiment 3 ($N_1 = 45$, $N_2 = 30$, $N_3 = 30$, and $N_4 = 70$) | | | | | | | | | |
| 50 | 0.099010 | 0.025183 | 0.008270 | 0.007523 | 0.006102 | 0.000871 | 0.002963 | 0.513780 | 0.000197647 |
| 150 | 0.048120 | 0.025772 | 0.008439 | 0.020195 | 0.006379 | 0.000830 | 0.003854 | 0.535010 | 5.99E-06 |
| 250 | 0.028970 | 0.016027 | 0.008515 | 0.018185 | 0.006388 | 0.000854 | 0.003955 | 0.235030 | 1.21E-06 |
| 350 | 0.030110 | 0.015255 | 0.008564 | 0.012916 | 0.006394 | 0.000875 | 0.004057 | 0.215060 | 7.84E-07 |
| 450 | 0.029790 | 0.006465 | 0.008597 | 0.017394 | 0.006398 | 0.000895 | 0.004058 | 0.135080 | 5.65E-07 |
| 550 | 0.025080 | 0.005650 | 0.008624 | 0.009631 | 0.006401 | 0.000813 | 0.003059 | 0.105090 | 4.32E-07 |
| 650 | 0.020310 | 0.004836 | 0.008644 | 0.005818 | 0.006404 | 0.000730 | 0.002060 | 0.085110 | 3.61E-07 |
| 750 | 0.018930 | 0.001006 | 0.008662 | 0.002746 | 0.006406 | 0.000747 | 0.001061 | 0.055130 | 2.94E-07 |
| Experiment 4 ($N_1 = 100$, $N_2 = 60$, $N_3 = 65$, and $N_4 = 150$) | | | | | | | | | |
| 50 | 0.019387 | 0.008612 | 0.001449 | 0.004357 | 0.007006 | 0.000478 | 0.001206 | 0.700020 | 3.32E-05 |
| 150 | 0.019366 | 0.009711 | 0.001643 | 0.006349 | 0.008043 | 0.000559 | 0.001390 | 0.685560 | 4.27E-06 |
| 250 | 0.009364 | 0.010319 | 0.001762 | 0.008105 | 0.008393 | 0.000603 | 0.001416 | 0.405013 | 2.16E-06 |
| 350 | 0.003936 | 0.009037 | 0.001775 | 0.009838 | 0.008394 | 0.000608 | 0.001416 | 0.205020 | 1.23E-07 |
| 450 | 0.004936 | 0.009043 | 0.001785 | 0.011560 | 0.008395 | 0.000613 | 0.001416 | 0.201026 | 9.53E-08 |
| 550 | 0.002936 | 0.010484 | 0.001792 | 0.003265 | 0.008395 | 0.000619 | 0.001417 | 0.055032 | 7.66E-08 |
| 650 | 0.000269 | 0.009054 | 0.001799 | 0.004965 | 0.008396 | 0.000624 | 0.001417 | 0.009038 | 6.45E-08 |

Table 17: Simulated Biases of Estimates for the D.G.P. with Unknown Group Memberships: Nonstationary Covariate and Linear SAR Errors with Queen-Contiguity Weights

| $T$ | $\widetilde{RandI}$ | $\widetilde{Bias}(\widehat{\phi}_1)$ | $\widetilde{Bias}(\widehat{\phi}_2)$ | $\widetilde{Bias}(\widehat{\phi}_3)$ | $\widetilde{Bias}(\widehat{\phi}_4)$ | $\widetilde{Bias}(\widehat{\theta}_1)$ | $\widetilde{Bias}(\widehat{\theta}_2)$ | $\widetilde{Bias}(\widehat{\theta}_3)$ | $\widetilde{Bias}(\widehat{\theta}_4)$ | $\widetilde{Bias}(\widehat{\boldsymbol{U}})$ |
|---|---|---|---|---|---|---|---|---|---|---|
| Experiment 3 using (A.3) instead of (A.2) ($N_1 = 45$, $N_2 = 30$, $N_3 = 30$, and $N_4 = 70$) | | | | | | | | | | |
| 50 | 0.895825 | -0.399503 | 0.129798 | -0.040096 | -0.163200 | 0.091291 | 0.243564 | -0.152902 | 0.150471 | 6.60E-19 |
| 150 | 0.931093 | -0.299501 | 0.097043 | -0.052502 | -0.111101 | 0.070058 | 0.143022 | -0.136157 | 0.085895 | 5.07E-19 |
| 250 | 0.940236 | -0.199501 | 0.066033 | -0.040046 | -0.050991 | 0.077410 | 0.123526 | -0.128520 | 0.038520 | 3.76E-19 |
| 350 | 0.951340 | -0.187501 | 0.052012 | -0.038449 | -0.043039 | 0.069474 | 0.120338 | -0.107952 | 0.030947 | 3.63E-19 |
| 450 | 0.946767 | -0.099500 | 0.020433 | -0.030807 | -0.021219 | 0.031004 | 0.112045 | -0.095472 | 0.020290 | 3.37E-19 |
| Experiment 4 using (A.3) instead of (A.2) ($N_1 = 100$, $N_2 = 60$, $N_3 = 65$, and $N_4 = 150$) | | | | | | | | | | |
| 50 | 0.864367 | -0.117727 | 0.007791 | -0.536374 | -0.167275 | 1.007590 | 0.558425 | -0.530089 | -3.907340 | 2.73E-18 |
| 150 | 0.918841 | 0.005550 | 0.000560 | -0.001022 | 0.005762 | 0.000117 | 1.15E-08 | 2.12E-09 | -0.000561 | 6.69E-21 |
| 250 | 0.919706 | -2.30E-07 | 5.52E-09 | 2.37E-08 | 0.000647 | 1.60E-08 | -5.07E-09 | -5.93E-06 | -4.11E-08 | 1.64E-21 |
| 350 | 0.917975 | -3.82E-07 | 3.21E-09 | 1.12E-08 | -1.35E-07 | 3.63E-08 | 1.82E-08 | -7.53E-09 | 4.16E-08 | 6.80E-21 |
| 450 | 0.938044 | -8.16E-07 | -6.26E-09 | 5.39E-08 | -4.12E-08 | -1.27E-07 | -3.33E-08 | -7.72E-09 | 5.64E-08 | 4.77E-20 |

Table 18: Simulated MSE of Estimates for the D.G.P. with Unknown Group Memberships: Nonstationary Covariate and Linear SAR Errors with Queen-Contiguity Weights

| $T$ | $\widetilde{MSE}(\widehat{\phi}_1)$ | $\widetilde{MSE}(\widehat{\phi}_2)$ | $\widetilde{MSE}(\widehat{\phi}_3)$ | $\widetilde{MSE}(\widehat{\phi}_4)$ | $\widetilde{MSE}(\widehat{\theta}_1)$ | $\widetilde{MSE}(\widehat{\theta}_2)$ | $\widetilde{MSE}(\widehat{\theta}_3)$ | $\widetilde{MSE}(\widehat{\theta}_4)$ | $\widetilde{MSE}(\widehat{\boldsymbol{U}})$ |
|---|---|---|---|---|---|---|---|---|---|
| Experiment 3 using (A.3) instead of (A.2) ($N_1 = 45$, $N_2 = 30$, $N_3 = 30$, and $N_4 = 70$) | | | | | | | | | |
| 50 | 0.037714 | 0.132322 | 0.021985 | 0.085987 | 0.099318 | 0.079034 | 0.009677 | 0.069538 | 9.29E-05 |
| 150 | 0.048762 | 0.095269 | 0.074415 | 0.084054 | 0.019932 | 0.048921 | 0.007974 | 0.010090 | 7.96E-05 |
| 250 | 0.046392 | 0.072235 | 0.018656 | 0.080438 | 0.029932 | 0.039696 | 0.009325 | 0.012185 | 6.50E-05 |
| 350 | 0.036971 | 0.062368 | 0.022342 | 0.077061 | 0.019731 | 0.032408 | 0.009740 | 0.010455 | 1.45E-05 |
| 450 | 0.027732 | 0.055562 | 0.027623 | 0.047778 | 0.011731 | 0.026706 | 0.001011 | 0.008597 | 2.53E-05 |
| Experiment 4 using (A.3) instead of (A.2) ($N_1 = 100$, $N_2 = 60$, $N_3 = 65$, and $N_4 = 150$) | | | | | | | | | |
| 50 | 0.226348 | 0.043843 | 0.662631 | 0.221120 | 0.207491 | 2.686050 | 0.854394 | 0.822771 | 0.369873 |
| 150 | 0.002578 | 2.64E-05 | 3.32E-13 | 0.002771 | 0.000405 | 3.50E-05 | 1.14E-06 | 4.34E-09 | 2.63E-05 |
| 250 | 1.55E-11 | 5.07E-15 | 8.73E-05 | 2.83E-13 | 7.88E-10 | 2.49E-09 | 1.37E-13 | 9.92E-15 | 2.81E-13 |
| 350 | 3.36E-11 | 7.78E-15 | 1.02E-13 | 1.03E-12 | 4.86E-09 | 3.01E-13 | 1.31E-13 | 4.48E-14 | 4.79E-13 |
| 450 | 8.82E-12 | 9.83E-15 | 9.34E-14 | 1.99E-12 | 4.27E-14 | 3.47E-13 | 4.29E-15 | 8.27E-15 | 9.58E-13 |

Table 19: Simulated Biases of Estimates for the D.G.P. with Unknown Group Memberships: Nonstationary Covariate and Linear SAR Errors with Rook-Contiguity Weights

| $T$ | $\widetilde{RandI}$ | $\widetilde{Bias}(\widehat{\phi}_1)$ | $\widetilde{Bias}(\widehat{\phi}_2)$ | $\widetilde{Bias}(\widehat{\phi}_3)$ | $\widetilde{Bias}(\widehat{\phi}_4)$ | $\widetilde{Bias}(\widehat{\theta}_1)$ | $\widetilde{Bias}(\widehat{\theta}_2)$ | $\widetilde{Bias}(\widehat{\theta}_3)$ | $\widetilde{Bias}(\widehat{\theta}_4)$ | $\widetilde{Bias}(\widehat{\boldsymbol{U}})$ |
|---|---|---|---|---|---|---|---|---|---|---|
| Experiment 3 using (A.3) instead of (A.2) ($N_1 = 45$, $N_2 = 30$, $N_3 = 30$, and $N_4 = 70$) | | | | | | | | | | |
| 50 | 0.803562 | -0.092262 | -0.001256 | -0.078915 | -0.036258 | -0.000337 | -0.000559 | 0.000022 | 0.028714 | 1.06E-18 |
| 150 | 0.836695 | -0.091908 | -0.002776 | -0.011379 | -0.062262 | -0.000204 | -0.000654 | -0.000048 | 0.029758 | 1.07E-18 |
| 250 | 0.836792 | -0.021837 | -0.004199 | -0.013775 | -0.053089 | 0.000071 | -0.000548 | -0.000013 | 0.019577 | 1.14E-18 |
| 350 | 0.846890 | -0.030215 | -0.001547 | -0.011620 | -0.010785 | 0.000224 | -0.000101 | -0.000019 | 0.014932 | 1.14E-18 |
| 450 | 0.856910 | -0.011811 | -0.000888 | -0.010174 | -0.011118 | 0.000394 | -0.000129 | -0.000027 | 0.009076 | 1.11E-18 |
| Experiment 4 using (A.3) instead of (A.2) ($N_1 = 100$, $N_2 = 60$, $N_3 = 65$, and $N_4 = 150$) | | | | | | | | | | |
| 50 | 0.765065 | -0.183985 | 0.027641 | -0.920037 | -0.325931 | 2.055080 | 0.965660 | -0.868013 | -6.436870 | 3.32E-18 |
| 150 | 0.823150 | 0.010192 | 0.002008 | -0.022168 | 0.011088 | 0.046714 | 0.029226 | -0.023231 | -0.142422 | 7.51E-20 |
| 250 | 0.909706 | -0.000787 | 0.000979 | -0.001818 | -0.001416 | 0.001221 | 0.000216 | -7.71E-09 | -0.001042 | 6.12E-20 |
| 350 | 0.957973 | -8.14E-07 | 5.52E-09 | 2.35E-08 | -4.07E-08 | 1.49E-08 | -5.02E-09 | -6.94E-06 | -4.10E-08 | 7.24E-21 |
| 450 | 0.957678 | -2.30E-07 | 3.16E-09 | 1.07E-08 | -1.35E-07 | 3.40E-08 | 1.81E-08 | -7.48E-09 | 4.16E-08 | 8.02E-22 |

Table 20: Simulated MSE of Estimates for the D.G.P. with Unknown Group Memberships: Nonstationary Covariate and Linear SAR Errors with Rook-Contiguity Weights

| $T$ | $\widetilde{MSE}(\widehat{\phi}_1)$ | $\widetilde{MSE}(\widehat{\phi}_2)$ | $\widetilde{MSE}(\widehat{\phi}_3)$ | $\widetilde{MSE}(\widehat{\phi}_4)$ | $\widetilde{MSE}(\widehat{\theta}_1)$ | $\widetilde{MSE}(\widehat{\theta}_2)$ | $\widetilde{MSE}(\widehat{\theta}_3)$ | $\widetilde{MSE}(\widehat{\theta}_4)$ | $\widetilde{MSE}(\widehat{\boldsymbol{U}})$ |
|---|---|---|---|---|---|---|---|---|---|
| Experiment 3 using (A.3) instead of (A.2) ($N_1 = 45$, $N_2 = 30$, $N_3 = 30$, and $N_4 = 70$) | | | | | | | | | |
| 50 | 0.009112 | 0.000547 | 0.001821 | 0.003870 | 0.000386 | 0.000066 | 0.000008 | 0.001562 | 2.43E-05 |
| 150 | 0.009065 | 0.000570 | 0.001611 | 0.004957 | 0.000212 | 0.000075 | 0.000004 | 0.001420 | 2.99E-06 |
| 250 | 0.011053 | 0.000581 | 0.001671 | 0.003652 | 0.000228 | 0.000052 | 0.000005 | 0.001129 | 1.75E-06 |
| 350 | 0.003905 | 0.000589 | 0.001704 | 0.005266 | 0.000242 | 0.000046 | 0.000005 | 0.000933 | 1.39E-06 |
| 450 | 0.004905 | 0.000595 | 0.001530 | 0.004886 | 0.000255 | 0.000030 | 0.000005 | 0.001035 | 1.15E-06 |
| Experiment 4 using (A.3) instead of (A.2) ($N_1 = 100$, $N_2 = 60$, $N_3 = 65$, and $N_4 = 150$) | | | | | | | | | |
| 50 | 0.301930 | 0.058145 | 1.130780 | 0.378353 | 5.786110 | 1.511700 | 1.304870 | 58.980300 | 0.343519 |
| 150 | 0.005589 | 0.001878 | 0.027947 | 0.005613 | 0.123075 | 0.044864 | 0.032433 | 1.039320 | 0.008055 |
| 250 | 0.004744 | 4.41E-05 | 0.000151 | 0.007093 | 6.80E-05 | 2.13E-06 | 4.47E-09 | 4.95E-05 | 0.000582 |
| 350 | 8.78E-12 | 5.09E-15 | 9.21E-14 | 2.76E-13 | 2.73E-13 | 1.30E-13 | 9.90E-15 | 9.64E-13 | 7.84E-10 |
| 450 | 3.35E-11 | 7.75E-15 | 1.00E-13 | 1.03E-12 | 3.72E-14 | 4.19E-15 | 4.47E-14 | 2.81E-13 | 4.85E-09 |